\documentclass{report}
\usepackage{amsmath}
\usepackage[a4paper,inner=30mm,outer=30mm,top=20mm,bottom=40mm]{geometry}
\usepackage{graphicx}
\usepackage{textcomp}

\usepackage{customisations}

\title{Minimal models of invasion and clonal selection in cancer}
\author{Chay Paterson}

\begin{document}

\maketitlepage

\pagenumbering{gobble}

\chapter*{Declaration}

I declare that:

\begin{enumerate}

\item This thesis has been composed by myself and that the work has not been
submitted for any other degree or professional qualification. I confirm that the work 
submitted is my own, except where work which has formed part of jointly-authored 
publications has been included. My contribution and those of the other authors to this 
work have been explicitly indicated below. I confirm that appropriate credit
has been given within this thesis where reference has been made to the work of others.

\item The work presented in chapter \ref{ch:minimal} was previously
published in Nature - Scientific Reports as ``An exactly solvable, spatial
model of mutation accumulation in cancer'' by myself (Chay Paterson, primary
author), Prof. Martin A. Nowak and Dr. Bartek Wac\l{}aw (corresponding author).
The model therein was conceived by all the authors. I derived the analytical
results, and Bartek Wac\l{}aw and I produced the figures. All authors wrote
the manuscript, which I edited and restructured to make appropriate for
inclusion as a chapter in this thesis.

\item The data presented and analysed in chapter \ref{ch:clinical} was obtained in a clinical
study carried out by Dr. Luis Diaz and his research group at the Johns Hopkins
Kimmel Cancer Center in Baltimore, USA. The data analysis and interpretation
detailed in chapter \ref{ch:clinical} and the text of chapter \ref{ch:clinical} are 
entirely my own work. Other contributions from colleagues are explicitly referenced 
in the text.

\end{enumerate}

\par\vspace{1in}
\begin{flushright}
    (\emph{Chay Paterson, October 16, 2017})
\end{flushright}

\chapter*{Acknowledgements}

In addition to my supervisor Dr. Bartek Wac\l{}aw and my co-supervisor Prof.
Martin Evans for their direction and generosity with their time, I would also
like to thank Prof. Martin Nowak for his interest and support, and 
Prof. M. E. Cates for several illuminating conversations. 

I would like to thank both my parents for nurturing my interest in science from an
early age, and my sisters for encouraging me to continue in difficult times, and
for their unwavering belief in me. I would also like to thank my late
grandmother for her enthusiastic support of my decision to pursue this
particular area of study.

So many of my friends here in Edinburgh have contributed with stimulating
questions or their help in learning new skills that it would be impossible to list
them all individually without omitting someone: I owe my thanks to them all.

But in particular, I would like to dedicate the following to my wife Philippa,
without whom this would not have been possible.

\begin{abstract}

One of the defining features of cancer is cell migration: the tendency of
malignant cells to become motile and move significant distances
through intervening tissue. This is a necessary precondition for metastasis,
the ability of cancers to spread, which once underway permits more rapid growth
and complicates effective treatment. In addition, the
emergence and development of cancer is currently believed to be an
evolutionary process, in which the emergence of cancerous cell lines and the
subsequent appearance of resistant clones is driven by selection.

In this thesis we develop minimal models of the relationship between
motility, growth, and evolution of cancer cells. These should be simple enough to be
easily understood and analysed, but remain realistic in
their biologically relevant assumptions. We utilise simple simulations of
a population of individual cells in space to examine
how changes in mechanical properties of invasive cells and their surroundings 
can affect the speed of
cell migration. We similarly examine how differences in the  
speed of migration can affect the growth of tumours. From this we conclude that 
cells with a higher elastic stiffness experience
stronger resistance to their movement through tissue, but this resistance is
limited by the elasticity of the surrounding tissue. We also find that the growth rate
of large lesions depends weakly on the migration speed of escaping cells, 
and has stronger and more complex dependencies on the rates of
other stochastic processes in the model, namely the rate at which cells
transition to being motile and the reverse rate at which cells cease to be
motile.

To examine how the rates of growth and evolution of an ensemble of 
cancerous lesions depends on their geometry and underlying fitness landscape,
we develop an analytical framework in which the spatial structure is coarse
grained and the cancer treated as a continuously growing system with
stochastic migration events. Both the fully stochastic realisations of the
system and deterministic population transport approaches are studied. 
Both approaches conclude that the whole
ensemble can undergo migration-driven exponential growth regardless of the
dependence of size on time of individual lesions, and that the
relationship between growth rate and rate of migration is determined by the
geometrical constraints of individual lesions. We also find that linear fitness landscapes
result in faster-than-exponential growth of the ensemble, and we can
determine the expected number of driver mutations present in several
important cases of the model.

Finally, we study data from a clinical study of the effectiveness of a new
low-dose combined chemotherapy. This enables us to test some
important hypotheses about the growth rate of pancreatic cancers and the speed
with which evolution occurs in reality. We test a moderately successful simple
model of the observed growth curves, and use it to infer how frequently drug
resistant mutants appear in this clinical trial. We conclude that the main shortcomings
of the model are the difficulty of avoiding over-interpretation in the face of
noise and small datasets. Despite this, we find 
that the frequency of resistant mutants is far too
high to be explained without resorting to novel
mechanisms of cross-resistance to multiple drugs. We outline some speculative
explanations and attempt to provide possible experimental tests.
\end{abstract}

\pagenumbering{arabic}

\tableofcontents


\chapter{Introduction}
\label{ch:intro}



The impact that cancer has had on human society is difficult to exaggerate. The
earliest surviving record of the disease dates from ancient Egypt \cite{hajdu2011note},
and with ageing populations across the developed world, most
families have some familiarity with the suffering that accompanies some 
types. All cancers are difficult to treat in their advanced stages, and some
types are difficult to diagnose due to their non-specific symptoms in early stages.

It is widely understood that cancers are overgrowths of cells, and that there are many
different types. Different types have different symptoms and treatments depending on where they
appear, and different degrees of survivability. But less commonly
appreciated are the underlying similarities between these different types and
the mechanisms at work in all cancers. Common features of all true cancers, 
regardless of their type or tissue of origin, are abnormalities in the 
growth and movement of cells \cite{Hallmarks,hanahan2011hallmarks}.
They also have a shared nature as evolutionary 
processes, which involve changes in some of the most fundamental molecular machinery 
of complex animal life \cite{MolBioCell}.

In this thesis, we ask how quickly cancers can evolve and grow, how this may be affected
by the migration of cancer cells, and how cancers' evolution might affect the
effectiveness of therapy.

\section{Background}

\begin{figure}[ht]
\begin{center}
    \includegraphics[width=0.66\textwidth]{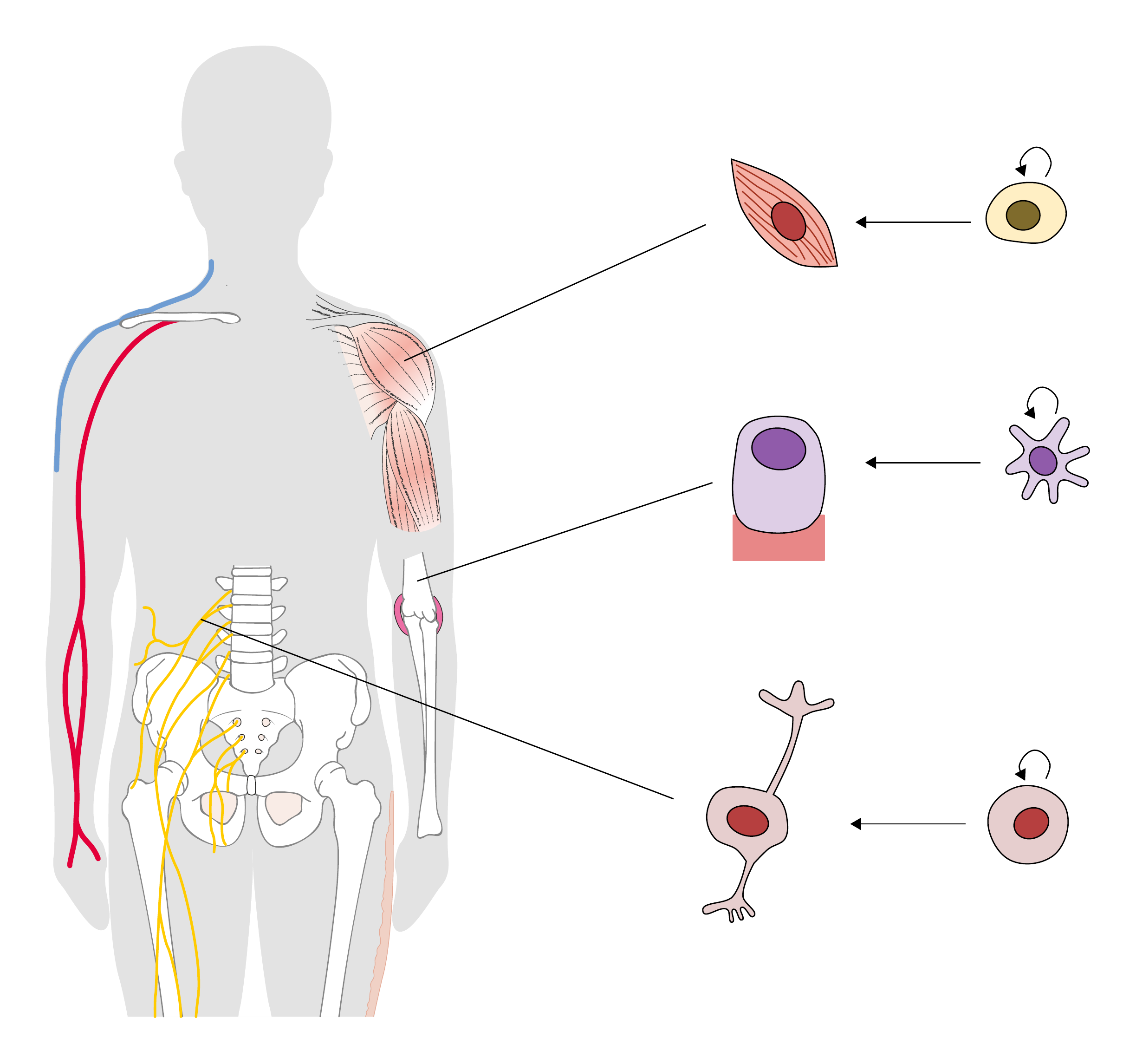}
    \caption{\label{fig:celltypes} Examples of three different normal cell types in 
    tissues of the human body, from top to bottom: a muscle cell (myocyte), bone-producing cell (osteoblast)
    producing bone (reddish square), and a nerve cell (neuron). Each of these
    have an associated type of precursor cell that maintains their normal
    populations in the appropriate balance (right). Modified from \cite{CRUKDiagrams}.}
\end{center}
\end{figure}

Cancer cells don't just divide more often.
Rather, they survive for longer periods of time without undergoing 
programmed cell death. A cell that divides more often but also dies more often
won't result in an overgrowth as such. In fact, having a population of rapidly 
dividing cells, in balance with the rate at which other cells die is an essential 
part of the normal function of tissues. Animal cells have different degrees of
specialisation, such as muscle, or bone, or nerve cells (see figure \ref{fig:celltypes}),
and as a rule of thumb the more specialised a cell type's functions are, the less often it will
divide. Less specialised cells divide more often, and produce more
specialised cells in a hierarchical manner \cite{GBPierce,MolBioCell}. The balance between the different
division and death rates, and the rate at which incrementally more specialised 
(or \emph{differentiated}) cells are produced, are what
determine the long term stability of cell populations in all tissues in the body.
To understand precisely how this balance is upset requires a closer look at
how it is controlled in normal cells.

To get a sense for what exactly goes wrong as a cancer emerges, let us take as 
a simple example a population of precursor cells (less specialised cells that
divide and die often), which produce another population of specialised gland
cells (that secrete some substance, say, and which don't divide). If these
populations are both balanced over time, this means that a typical precursor
cell has to produce, on average, one new precursor cell and a few more 
specialised cells over the course of its lifetime \cite{morrison2006asymmetric,ASCDormancy2,ASCDormancy,frede2016single}.
There are three basic processes which determine
what happens: the rate at which precursor cells divide, the rate at which they
differentiate and turn into gland cells and the rate at cells of either type are
removed.

There are therefore three possible points of failure here: cells can divide too
quickly, they can be removed too slowly, or they can differentiate too slowly. The
control of cell division, programmed cell death, and cell specialisation are
critical in the normal maintenance of tissue
(\emph{homoeostasis}), and the loss of control of these is similarly critical in
the development of cancer \cite{MolBioCell,morrison2006asymmetric,alcolea2014differentiation}.

The type and form of a given cancer is determined by the type of cell from
which it derives: that is, which sub-population of specialised cells
it descended from. There are as a result approximately as many different 
possible cancers as there possible types of cell (including both specialised
cells and their precursor cells). Despite
this, all cancers have a common set of properties or ``hallmarks'' that 
distinguish cancers from similar diseases involving differentiation and
growth. These six hallmarks are: the resistance to normal \emph{controlled} cell death (``apoptosis''), sustained
proliferation, replicative immortality, a muted response to growth-suppressing
signals, triggering the formation of capillaries, and invasion 
(``metastasis'') \cite{Hallmarks,hanahan2011hallmarks}. 

\subsection{Cancer as an evolutionary process}
\label{sec:evolution}

Cancers are fundamentally evolutionary processes, in the very precise sense 
that some population of cells gets a replicative
advantage over their neighbours \cite{nowell1976clonal}.
The changes in cancer cells that enable them to escape the usual control of
cell division and type are due to genetic mutations \cite{nowell1976clonal,
Vogelstein2004, BVogelstein2013}. Genetic mutations are heritable, being
passed down to daughter cells when a cell carrying the mutation divides \cite{MolBioCell}.
What is crucial in the process of uncontrolled cell division, and what allows
it to form an overgrowth (i.e. a tumour) rather than simply increasing the
speed with which cell populations turn over, is not that cells in the tumour
divide \emph{faster} as such, or even that they die slower, but that they 
produce more offspring on average than normal cells. This concept of
the effective number of offspring, especially if it is tied to heritable
mutations, is mathematically speaking just the same as fitness in evolutionary
biology, except that instead of talking about individuals and species we are
talking about populations of cells in a single human (or animal) body
\cite{nowell1976clonal,durrett_evolutionary_2010,Mutation3}.

The early development of cancer can therefore be seen as an
evolutionary process, as successive genetic mutations enable related cells to
produce a slightly higher number of daughter cells than before, and ``fitter''
lines gradually overtake the others. This process includes the earliest
stages of tumour initiation from a single mutant cell, and continues 
throughout the course of the disease \cite{nowell1976clonal}. This picture 
has since become widely accepted, and forms a significant landmark in the
current understanding of how cancers emerge \cite{fearon1990genetic,Hallmarks}.

\begin{figure}[ht]
\begin{center}
    \includegraphics[width=0.8\textwidth]{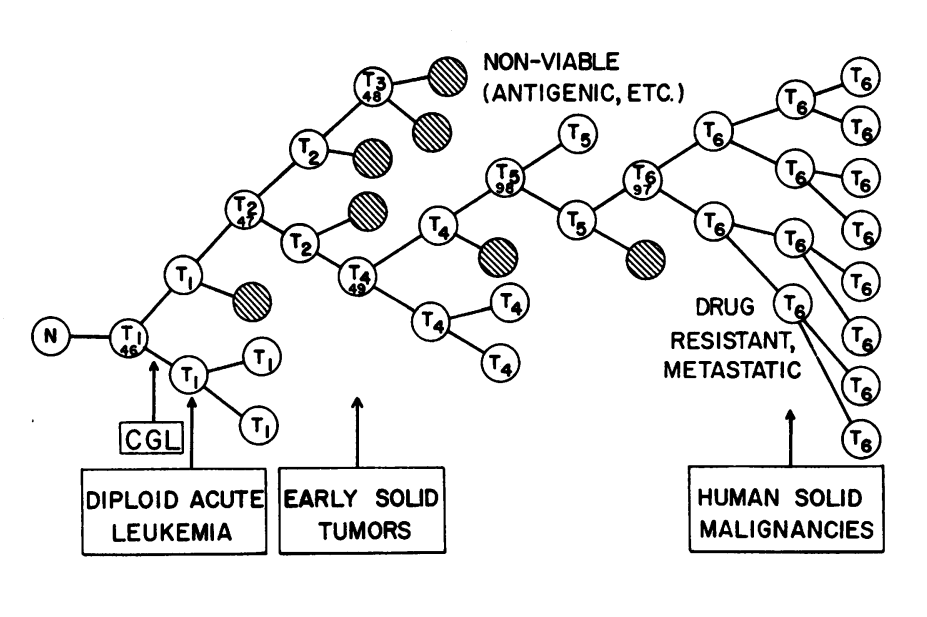}
    \caption{\label{fig:clonalselection} A hypothetical family tree of
    different mutant lineages that emerge in the earliest stages of a
    typical cancer, illustrating which alterations are associated with which advantages. 
    The single clonal mutation that initiates the entire cancer is labelled
    $N$. From P. Nowell 1976 \cite{nowell1976clonal}.}
\end{center}
\end{figure}

This evolutionary picture of cancer can be summarised as follows: first, one
common progenitor cell acquires a mutation that causes it to produce slightly
more offspring than other surrounding normal cells. The descendant cell line
then goes on to produce more descendants with additional mutations which are
slightly fitter in turn (see Figure \ref{fig:clonalselection}). The whole population of cells gradually acquires
uncontrolled growth and other hallmarks of full-fledged cancer in an
evolutionary process, which may have been underway for many years before the
cancer is finally discovered.

Most of the mutations that occur in the course of a tumour's ``life story''
don't carry any additional advantages. In between the very rare mutations that
actually bring about an advantage over surrounding cells, many
neutral mutations will gradually accumulate in populations. Neutral mutations
also accumulate faster when growth is sped up by advantageous mutations, and 
can be thought of as ``passengers'' riding the occasional
waves of growth brought about by the rare advantageous mutations. The
precise number of known advantageous (or ``driver'') mutations across a wide
variety of known types of cancer has
been estimated at 140: of these, two to eight will be present in any given
tumour \cite{BVogelstein2013}. Compare this with the mean number of new neutral point
mutations that appear during a given cell division event, which has been
estimated at about $0.02$, or about one every fifty divisions. Clearly, a cancer that consists of a billion cells
will have around twenty million neutral mutations, far in
excess of the typical two to eight driver mutations \cite{bozic2010accumulation}.

As well as explaining the earliest stages of initiation, evolution by clonal
selection is also
one possible hypothesis for drug resistance in late stage metastatic cancers.
Although it should be noted that the mechanism that underlies drug
resistance is not as well-understood as the earlier mutations that actually get 
the tumour started, and establishing that evolution is responsible for
resistance to chemotherapy can only be done by ruling out alternative
explanations, at the time of writing this is a plausible explanation with the additional point in its
favour that evolutionary processes are already known to occur \cite{diaz_jr_molecular_2012}.

Although our main areas of interest will be the relationship 
between invasion and evolutionary change, it is worth summarising some
important discoveries of the broader underlying mechanisms that have been
discovered in recent years.


\subsection{Cell signalling and kinase cascades}

\begin{figure}[ht]
\begin{center}
    \includegraphics[width=0.7\textwidth]{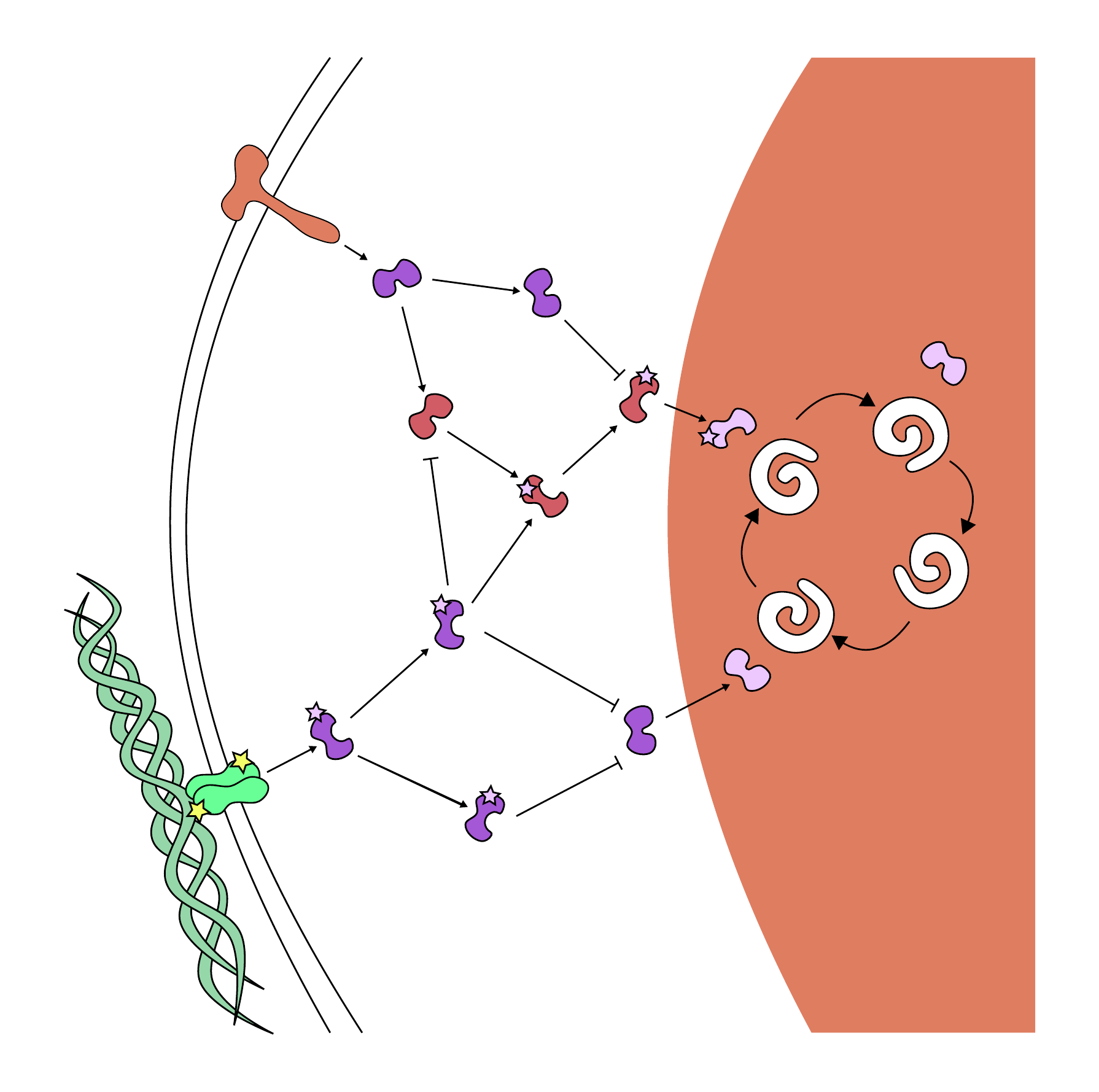}
    \caption{\label{fig:signalingcartoon}A cartoon of a fictitious signalling
    pathway, showing transmembrane proteins (green and red structures on membrane) adhering to extracellular matrix
    (green fibres), and triggering a kinase cascade (kinases being B-shaped red and
    purple blobs) that propagates into the nucleus (red oval, right). This interferes 
    with the cell cycle through the binding of cyclin-dependent kinases (pink kinases) 
    to cyclins (white spirals). In this example, the cell adhesion molecules
    function as receptors, although many other specialised receptors
    exist (see \cite{MolBioCell} for a comprehensive review).}
\end{center}    
\end{figure}

The type of a cell, that is, its form and function in terms of what it
secretes and how it interacts with the cells around it, is determined by what
proteins it expresses. What proteins it expresses is determined by which genes
are switched on and actively ``transcribed'', read by the cellular machinery
that builds proteins from amino acids according to the sequence contained in
the genome. The genetic information needed to code for any protein in the body
is contained in the DNA of every single cell in the body, but is only
expressed and actively used when the relevant gene is switched on. Differences
in cell type are thus due to differences in expression of genes found in all
cells \cite{MolBioCell}.

The process by which genes are activated or deactivated by stimuli is complex,
but involves a network of chemical reactions between proteins inside the cell
that result in the relevant gene becoming ``exposed'' and
accessible to the transcriptional machinery. The proteins involved may be
classified into transcription factors, signalling proteins and receptors. 
Transcription factors directly interact with the gene by sticking to it, either exposing it to
transcriptional apparatus or hiding it. Signalling proteins relay information about stimuli (or
signals) from receptors elsewhere in the cell to the transcription factors in the 
nucleus \cite{MolBioCell}.

Many of the signalling proteins are kinases, enzymes which catalyse the
attachment of a phosphate group to a specific target molecule. This 
brings about a structural change in the target molecule.
The products of many of these reactions are also protein kinases, which
catalyse a similar reaction for another protein kinase. This means that
phosphorylation reactions between these kinases can form a long branching chain
of similar reactions. If one of the kinases near the ``top'' of the chain is
activated by a receptor, this triggers a cascade of phosphorylation reactions
that cascades down the chain of kinases like a burning fuse, with the ultimate result that
the transcription factors in the nucleus are activated, triggering the 
production of other proteins in response (see figure \ref{fig:signalingcartoon}).
Protein kinases are extremely important in development and cancer initiation, as well
as a host of other functions of the cell: more than 500 different protein
kinases are known, and the number of known interactions between them increases
from week to week  \cite{manning2002protein,lahiry2010kinase}.

Another important group of regulatory proteins are cyclins, and
the protein kinases that interact with them, the cyclin-dependent kinases. 
In cells which have not yet fully differentiated and are still precursors 
or stem cells, the cell cyclically progresses
through different stages of cell division. This cyclical behaviour is driven
by an oscillating biochemical reaction between cyclins, and may be paused
depending on the presence or absence of these other cyclin-dependent kinases, which
interact with the cell's larger signal transduction network in turn. In this way, 
what happens to a receptor on, for example, the surface of the cell, can trigger 
a cascade of events that pauses or un-pauses the process of cell division  \cite{nigg1995cyclin}.

The true picture is rather more complex than this, not least because of the
fact that some signal molecules inhibit rather than simply activate
others, and the resulting dynamics can have multiple steady states, possibly
oscillatory. Nonetheless, this picture of information about stimuli being
carried through the cell to trigger responses in regulation and expression 
by a kind of domino effect has gradually become widely accepted, and mutated
proteins in this network are implicated in many known
cancers \cite{Hallmarks,lahiry2010kinase}.

\subsection{Adhesion}

At the cell membrane, the presence of intermolecular forces between the
molecules present on the membrane and those in the external environment
determine the strength of the cell's adhesion to that environment.
As in other materials, the strength of adhesion between two surfaces $i$ and $j$ is measured by the
surface free energy $\gamma_{ij}$ associated with the interface: how much work is
needed to create a given area $A_{ij}$ of that interface. 
This is essentially how sticky $i$ and $j$ are. It is closely related to
surface tension, and corresponds to the change in free energy $\Delta F$ 
when peeling the surfaces apart over a given area \cite{foty1994liquid},

\begin{equation}
    \gamma_{ij} = \frac{\partial \Delta F}{\partial A_{ij}}\;.
\end{equation}

Despite the variety of molecules usually present, the interfacial free energy
between two cells of different types is currently believed to be determined by
the gross quantity of cell adhesion molecules present rather than their
specific types. Cells tend to sort into regions of different homogeneous types
with similar surface energies under the action of adhesion, which provides a
purely mechanical contributing mechanism to pattern formation during
morphogenesis \cite{DAHtime,DAH}.

\subsection{Cytoskeletal stiffness and migration}

\begin{figure}[ht]
\begin{center}
    \includegraphics[width=0.8\textwidth]{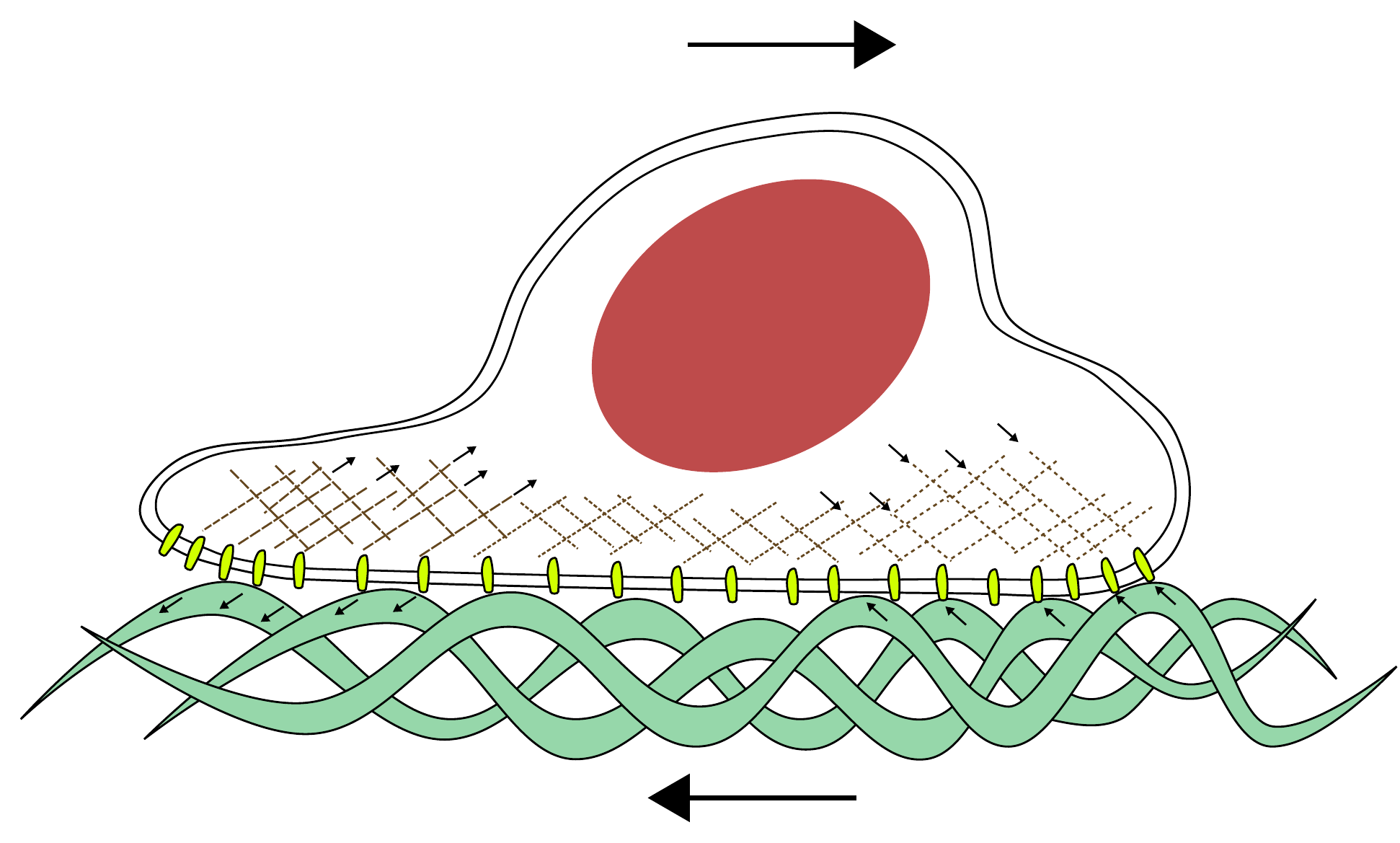}
    \caption{\label{fig:crawling}A cartoon to show the generation of forces on
    a motile cell by dynamical polymerization of actin and interactions with
    cell adhesion proteins, showing forces due to polymerization of actin
    filaments and the resultant tractive force on the cell and equal and
    opposite force acting on the surrounding matrix.}
\end{center}
\end{figure}

Cells' resistance to mechanical deformation consists of both the resistance to
applied pressure common to all materials, the bulk modulus, and from the
tensile and compressive resistance of the cytoskeleton. The cytoskeleton is
complex and active in structure, and is composed of a network of actin
filaments. Actin is an active polymer, and in addition to contributing to the
mechanical stiffness of the cell plays important roles in the dynamics of
signal transduction and motility \cite{MolBioCell}.

The properties of most immediate interest to us are the contribution to
mechanical stiffness and the generation of forces during motility. Forces
between the cell and the surrounding matrix are generated by 
chemical reactions in the actin network, and its connection to transmembrane cell
adhesion proteins. The depolymerisation and shrinkage of actin filaments near
the leading edge of the cell ``pulls'', while the polymerisation and expansion at the trailing
edge ``pushes'' \cite{insall2009actin}. See Figure \ref{fig:crawling} for a
sketch of the process.

\section{Why mathematical models?}

Many mathematical and computational models in this field are scrupulous in
their inclusion of all imaginable complications that can be found in real
biological systems. Many different systems are coupled together in ways
observed to occur experimentally, at many different scales: such multi-scale
models, although they can be successfully fitted and in many cases can be used
to predict the behaviour of individual systems, sometimes requiring more
than a dozen adjustable or empirically set free parameters. 

I deliberately
want to avoid developing models that are ``too realistic'' in the course of my research
here for a number of reasons, some practical and some philosophical.

Firstly, models with a large number of complex systems which are themselves complex,
having dozens of free parameters, are opaque. Their investigation and
empirical fitting can usually only be carried out computationally, often
involving elaborate numerical estimation procedures in addition to the complex
numerical integration of the underlying differential equations or master
equations. The freedom that comes from having so many adjustable parameters, a
situation that is almost inevitable in any field with a large number of
experimental unknowns, means that often a model will have a good fit, and
can even turn out to have a decent amount of predictive power, but the
complexity makes the models essentially impossible to interpret in simple
terms. This approach encourages a ``black box'' approach to science, in which
verifiable results are arrived at, along with means for predicting them, with
little or no increase in human understanding of the things being studied.

Secondly, computational models with an abundance of degrees of freedom are
very difficult to falsify. Enormous differences in qualitative behaviour can
be achieved with relatively small parameter changes in many such models: if we
take the view that the scientific method has to do with ruling out one of a
number of different alternative hypotheses through experiment, then models
which display many different qualitative behaviours, and can be fitted to a
variety of different empirical curves in each regime, are very difficult to
rule out. 

On the other hand, deliberately simplistic toy models have the opposite set of
problems. Simple and tractable models which share only a qualitative
resemblance to the system they are ostensibly a theory of can produce the
feeling of understanding in the scientist who studies them, and can be coaxed
into providing predictions about behaviour which are eminently falsifiable.

There seems to be a gap in the literature of theoretical frameworks that fall
in between: imperfect but decent in predictive power, and decent but
not exhaustive in explanatory power. It is this particular void that we want
our research to fill: a simple theory which explains and predicts the effect of a few
specific phenomena well. Here, we attempt to develop such a framework and set of
models for the problem of determining the dynamics of cancer evolution in
terms of the biophysical properties of their constituent cells, models which are 
simple enough to be understood fully, but which
still attempt to maintain a testable correspondence to experiment: as simple 
as possible but no simpler.

With this goal in mind, we will set out to develop and elucidate a theoretical framework
within which we can begin to understand and predict the interaction between
evolutionary dynamics and invasion in late-stage cancers, and which can serve
as a starting point for future work in the field.
I will now take a moment to discuss the most commonly
encountered mathematical models of tumour growth in oncology, before offering
a critique of their specific relevance to full-blown invasive cancer and
beginning my development of analytical and computational models which attempt
to bridge the gap and provide a framework for the modelling of evolution during
invasive cancers.

\subsection{Basis of commonly used growth models}



There are a variety of mathematical models currently in use to describe the growth
curves of individual tumours, or the total tumour burden in advanced cases of
metastatic cancers. An exhaustive survey of current models, and their
applicability to different cases and diseases, can be found elsewhere \cite{miklavvcivc1995mathematical, 
stein2011dynamic, benzekry2014models}. Here, I
will provide a summary of the different classes of models, their empirical
basis, and a brief discussion of their limitations.


The purpose of a mathematical model of growth is to provide some
description of observable growth curves in terms of a few underlying
parameters. Ideally, the parameters and terms
of the model interpreted in terms of physiological mechanisms at play during
the process of biological growth. Models with more parameters will obviously
tend to provide better fits to data, but this does not always correspond to
stronger predictive power, in that future extrapolations of the curve do not
agree with experimental data for very long after the parameters have been set.
Many models with a large number of parameters are prone to numerical
stiffness, having many local minima in fitting procedures, as well as
tending not to have a clear interpretation in terms of physiological
mechanisms \cite{benzekry2014models,rodriguez-brenes_tumor_2013}.

Models can be aggregated, very crudely, into groups whose growth curves
display long-term slowdown to stable maximum sizes (plateaus) and those that
instead display unbounded growth. The simple reason to expect
long-term slowdown and plateaus in the growth rate is that a given individual
solid tumour will be limited in terms of the supply of nutrients (namely, oxygen and
glucose) that it has access to, and in terms of how much space is available
for it to occupy. Growth can can only continue as long as both nutrients and volume
are available for the tumour to
expand \cite{benzekry2014models,bonate2011modeling}.

It is not obvious that a model for the growth of non-solid, diffuse tumours or
more highly invasive cancers should have a slow-down in terms of growth rate ---
in fact, it may not be true at all. I will return to this point after
discussing the most commonly encountered growth models in the literature.

\begin{figure}
\begin{center}
    \begin{tabular}{ll}
    A & B \\
    \includegraphics[width=0.4\textwidth]{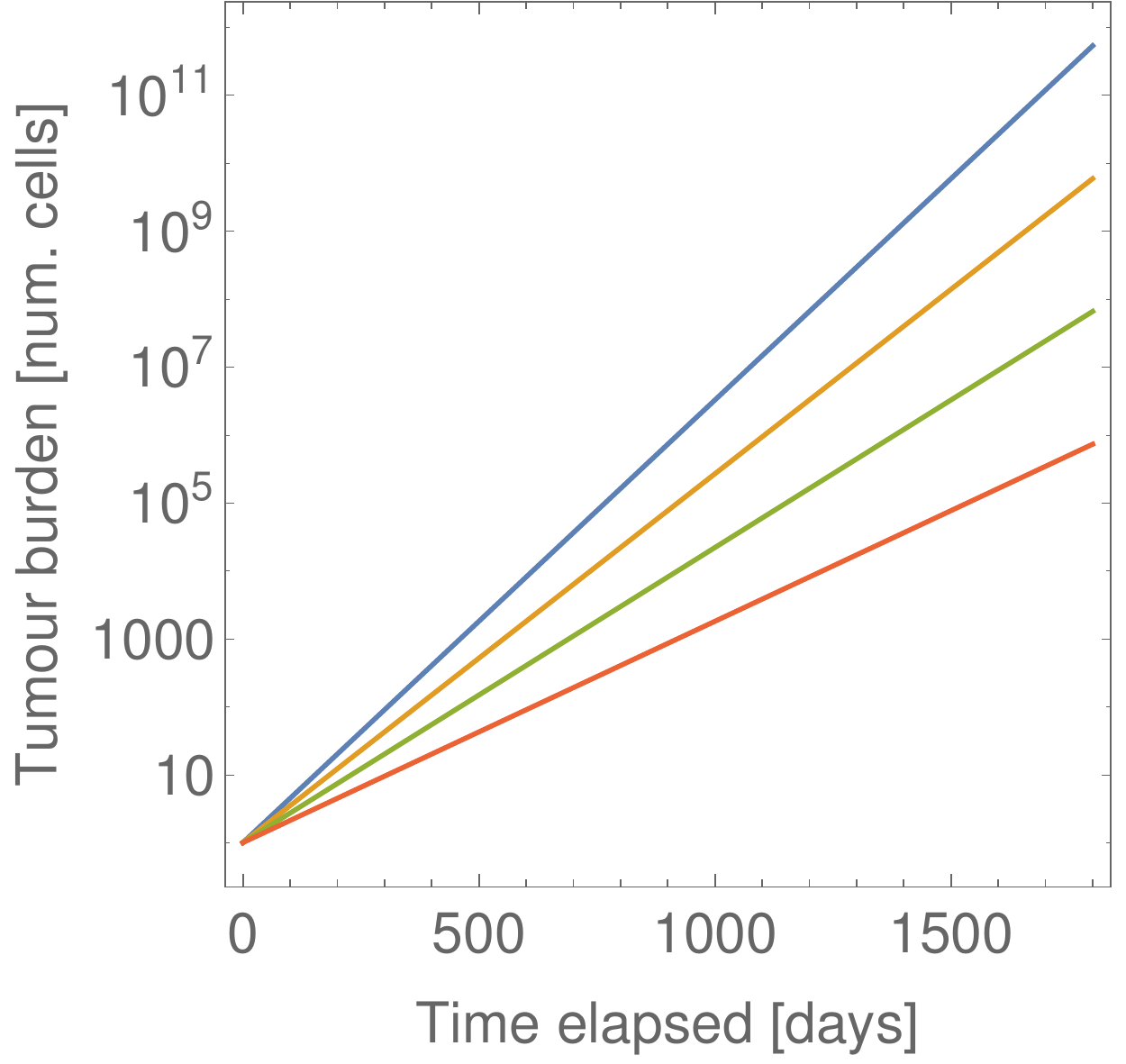} &
    \includegraphics[width=0.4\textwidth]{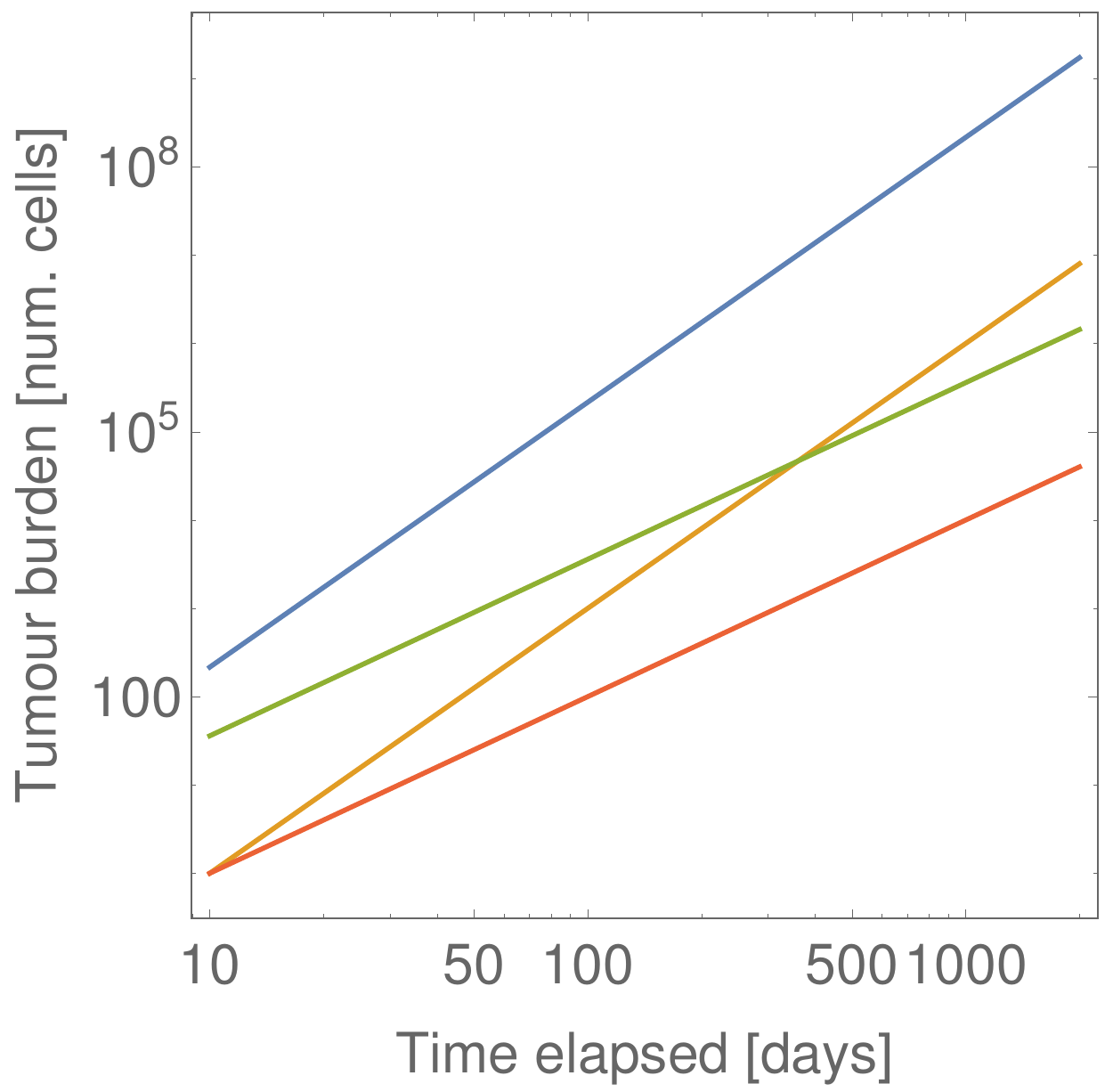} \\
    C & D \\
    \includegraphics[width=0.4\textwidth]{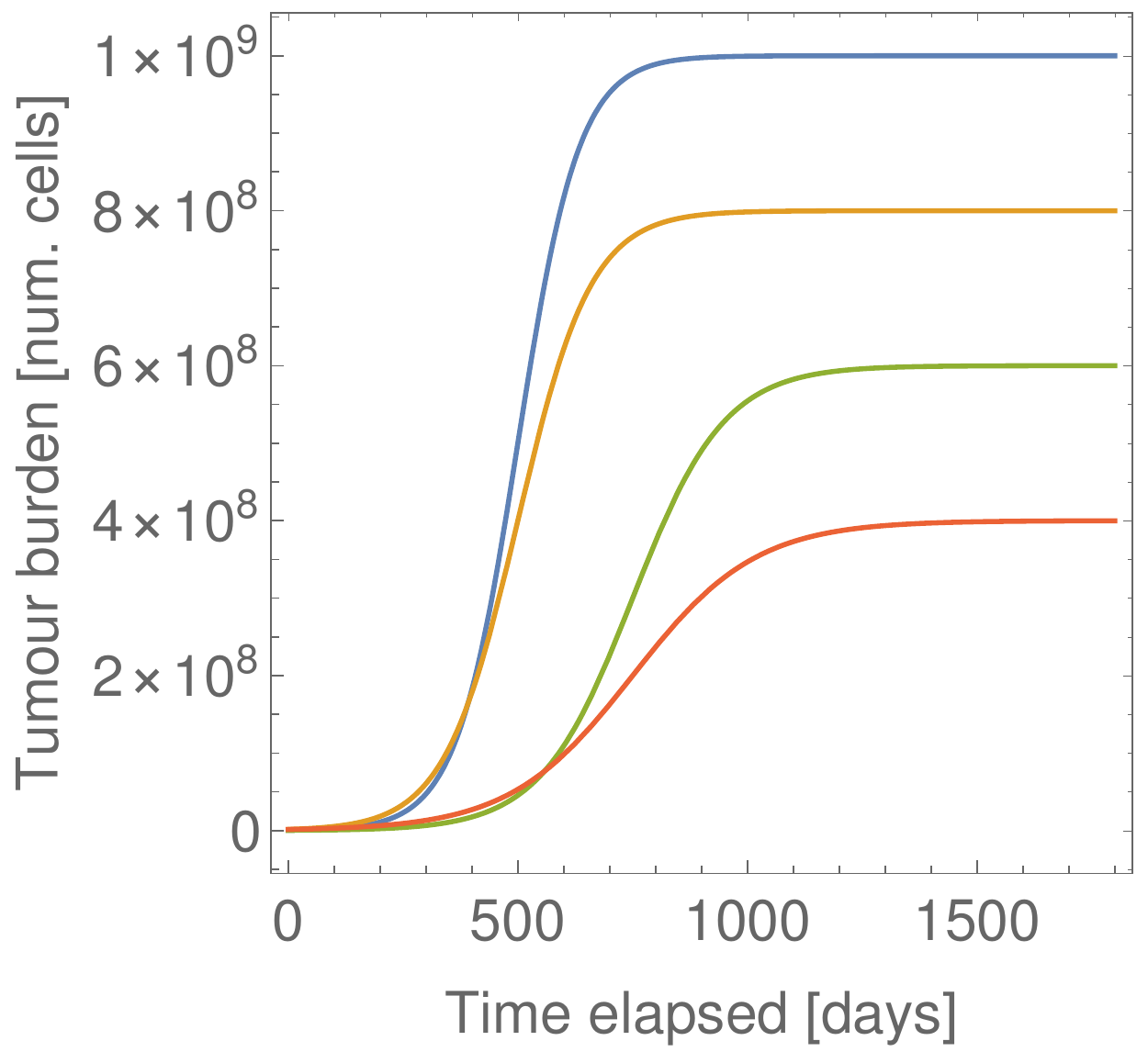} &
    \includegraphics[width=0.4\textwidth]{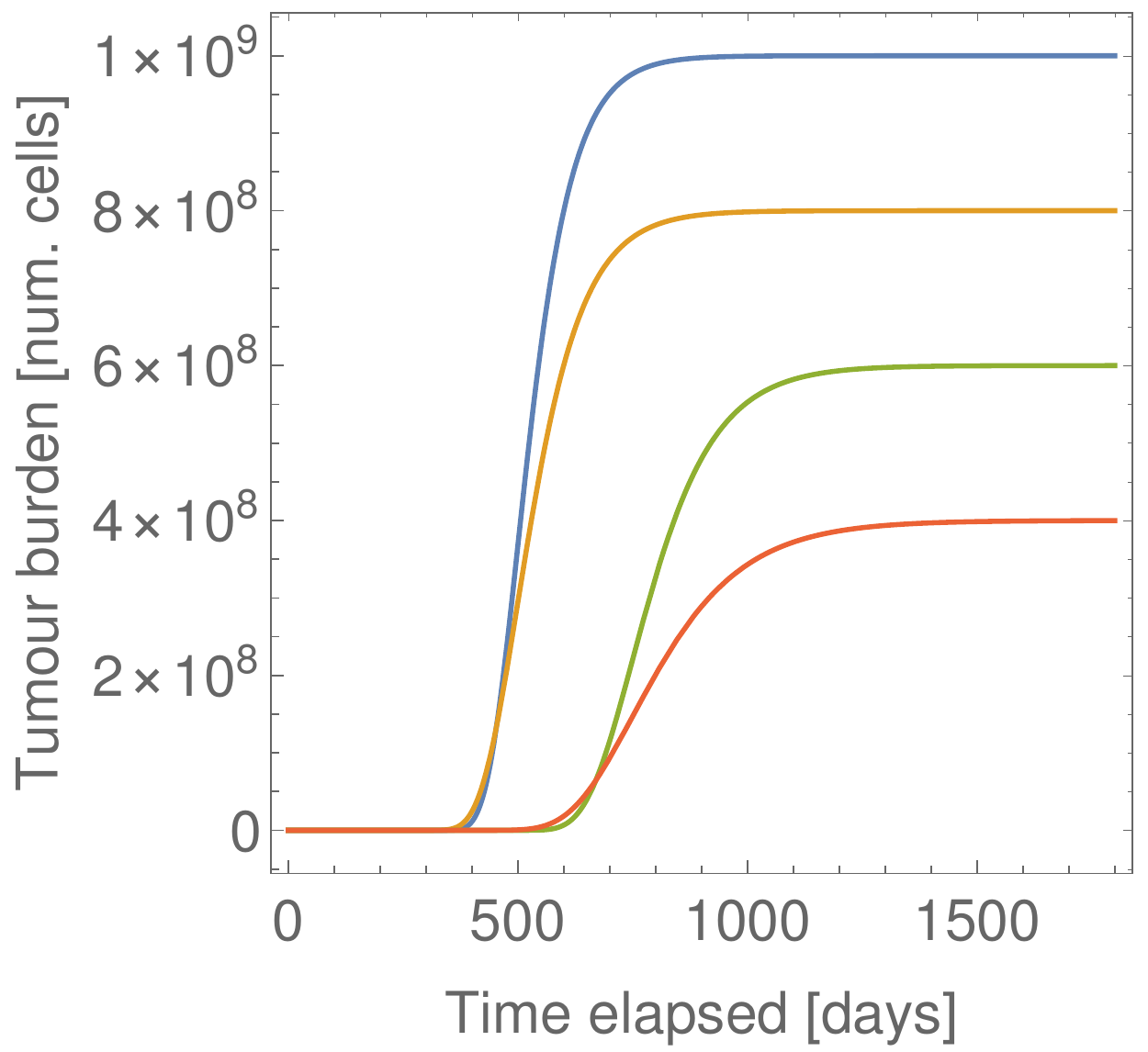} \\
    \end{tabular}
\end{center}
    \caption{\label{fig:curves} A comparison of example growth curves for the
    two- and three-parameter models discussed in this section. A: four
    solutions of the exponential growth model with identical initial
    populations $n(0)=1$ and growth rates $G=\{0.015,0.0125,0.01,0.0075\}$.
    Note the logarithmic scale. B: four representative solutions of the power-law
    growth model with parameter values of $C=0.216$,$p=2/3$ (blue curve);
    $C=10^-3$,$p=2/3$ (yellow curve); $C=0.216$,$p=1/2$ (green
    curve); and $C=10^-3$, $p=1/2$ (red curve). Note the log-log scale.
    C: four examples of logistic growth, with $n(0)=5.53\times 10^5$, $G=0.015$, and $K=10^9$ (blue curve);
    $n(0)=1.54\times 10^6$, $G=0.0125$, and $K=8\times 10^8$ (yellow curve);
    $n(0)=3.31\times 10^5$, $G=0.01$, and $K=6\times 10^8$ (green curve);
    $n(0)=1.43\times 10^6$, $G=0.0075$, and $K=4 \times 10^8$ (red curve).
    D: four representative solutions of the Gompertz differential equation,
    with $b=1808$, $\alpha = 0.015$, and $K=10^9$ (blue curve);
    $b=518$, $\alpha = 0.0125$, and $K=8 \times 10^8$ (yellow curve);
    $b=1808$, $\alpha=0.01$, and $K=6 \times 10^8$ (green curve);
    and $b=277$, $\alpha=0.0075$, and $K = 4\times 10^8$ (red curve).
    }
\end{figure}

\subsubsection{Two parameter models}

The simplest two models which display unlimited growth are exponential and
power-law growth. Writing the initial size as $n(0)$, and the size at time $t$
as $n(t)$, exponential growth corresponds to the differential equation

\begin{equation}
    \frac{dn}{dt} =  G n(t)
\end{equation}

which may be interpreted as every cell in the population dividing at a
frequency equal to $\ln(2) G$, with the resulting growth curve given by 
the solution

\begin{equation}
    n(t) = n(0) \exp(G t)
\end{equation}

with $G$ the exponential growth rate of the model. See figure \ref{fig:curves}
for representative solutions. The initial size and growth
rate $G$ are the defining two parameters of the model \cite{benzekry2014models}.

The power-law model corresponds to the differential equation

\begin{equation}
    \frac{dn}{dt} = a n^p
\end{equation}

with $p$ some dimensionless number between $0$ and $1$, and $a$ some
constant. Solutions of this equation,

\begin{equation}
    n(t) = C (t-t_0)^{\frac{1}{1-p}}
\end{equation}

with $C=\left((1-p) a\right)^{\frac{1}{1-p}}$ and $t_0$ determined by initial conditions,
display unlimited growth, and are
exhaustively characterised by two parameters (see figure \ref{fig:curves} 
for representative solutions), but also have the property that
the size of the tumour is always $0$ at some point in time. The power-law model is therefore
not suitable for very small tumours. The parameter $p$ often has an
interesting interpretation in terms of the geometry or perfusion (how 
efficiently blood vessels can penetrate the tumour) of individual
tumours. If all of the nutrients that flow into the tumour get used for
additional growth, and perfusion is perfect (by which I mean that every cell
in the tumour has good access to nearby blood supply, and the capillaries
reach throughout the tumour perfectly efficiently) then the coefficient
$p=1$. If, on the other hand, the capillaries can only reach cells on the
surface of the tumour, and the tumour is spheroidal, then $p$ will be
substantially lower, at $p=2/3$. If perfusion falls somewhere in between
these values, then $p$ will fall somewhere in between $2/3$ and
$1$ \cite{benzekry2014models}.

Despite the simplicity of the exponential growth model, it should be
emphasised that it is, in practice, the most popular model for tumour growth
by a substantial margin, both among clinicians in need of a quick and simple
estimate for the purposes of extrapolating growth or characterising it in
exploratory studies in which dynamical data may be sparse, in studies of drug
effectiveness, and also among mathematical biologists in the context of a basis for more complex
models \cite{benzekry2014models,rodriguez-brenes_tumor_2013}. For example, one can readily generalise the underlying equations of
the model to a more complex system of linear differential equations,
representing different sub-populations of cancer cells, and analyse the
resulting dynamics very easily, often even giving some meaningful
interpretation to the underlying parameters.
I will return to this paradoxical popularity shortly.

However, even though this is the most commonly encountered model in practice,
individual tumours often show a gradual slow-down in their growth, tending
towards some sort of plateau. Although it should be pointed out that plateaus
are not always observed clinically \cite{benzekry2014models}. The basic inability of simple exponential growth
to capture this motivates the development of more complex models.

\subsubsection{Three parameter models}

The simplest models displaying such a plateau in growth are the logistic model

\begin{equation}
    \frac{dn}{dt} = G n (1-n/K)
\end{equation}

with a carrying capacity $K$, and the Gompertz model

\begin{equation}
    \frac{dn}{dt} = \alpha \log(K/n) n
\end{equation}

with some ``growth rate'' $\alpha$, respectively \cite{benzekry2014models,rodriguez-brenes_tumor_2013}.
Solutions to these models can be expressed analytically,

\begin{equation}
    n(t) = \frac{K}{1+\mathrm{e}^{-G(t-t_0)}}
\end{equation}

for the logistic model, with $t_0$ a constant set by initial conditions; and

\begin{equation}
    n(t) = K \mathrm{e}^{-b \mathrm{e}^{-\alpha t}}
\end{equation}

for the Gompertz model, where $b$ is a constant that is set by initial
conditions, and can be related to the initial population size $n(0)$ by

\begin{equation}
    b = \log \left(\frac{K}{n(0)}\right)
\end{equation}

where $\log$ is the natural logarithm. Examples of solutions are displayed in
figure \ref{fig:curves}.

Both of these models' growth curves show an initial period of
growth, which then decelerates as the tumour grows larger, as is easily seen in
all examples given in figure \ref{fig:curves}. Both models' growth
curves are characterised by three parameters, displaying an initial period of
rapid growth followed by a slow-down and plateau. The main visible difference
between the two is that the logistic growth curves are symmetric, 
whereas the Gompertz model's growth curves are asymmetric, showing 
fast initial growth and a very gradual slow-down. Of the two, Gompertzian growth
tends to perform somewhat better than logistic empirically, and was historically more
popular than many competing models, but has a number of serious problems,
among which are the behaviour of the model at small sizes. When $n\ll K$, the replication
rate of cells blows up to infinity, indicating discontinuities in growth.
Furthermore,
the difficulty of interpreting $\alpha$ or justifying the strange dependence
on population size on physiological grounds. For these reasons, and more
importantly for its empirical failures, Gompertzian growth is decreasingly
popular \cite{benzekry2014models}.

\subsubsection{Four and more}

More complex models, such as Gomp-ex, can be constructed from these simple
models in a piecewise manner, typically introducing new adjustable parameters
with each ``switch'' between different models. One may use Gompertzian growth
for tumours larger than $n_c$, and exponential for tumours smaller than this
size, treating $n_c$ as an adjustable parameter \cite{benzekry2014models}.

One type of model which is not constructed in this way, and which has the
additional advantage that it has a straightforward physiological
interpretation, is the von Bertalanffy growth model characterised by the
differential equation

\begin{equation}
    \frac{dn}{dt} = a n^p - b n
\end{equation}

with $a$ and $p$ constants characterising uptake and use of nutrients and
oxygen, and $b$ a rate related to turnover and the breakdown of cells and biomass 
(catabolism) which also controls plateau size, as this is the size at which all nutrients 
taken in are used simply to maintain the tumour at its current size.

These may be interpreted physiologically as follows, in terms of simple 
conservation of mass and the energy budget of an organism (or tumour, in our case). 
The rate at which mass of nutrients can be taken in and used in metabolic processes
should be bounded from below by the surface
area (if perfusion is very poor, and capillaries don't penetrate into the tumour 
very effectively at all) and from above by the total
volume (if perfusion is very thorough). A simple model therefore has metabolic 
rate $= a n^p$, with $p$ between $2/3$ and $1$. $p$ can
therefore be interpreted as how well vascularised a tumour is.

This biomass can be used either for
additional growth or for the maintenance of the system at a given size. The
mass used to compensate for catabolic processes should be proportional to the amount
of living material present (i.e. tumour size), with the remainder resulting in
growth. Conservation of mass then implies

\begin{equation}
    \underbrace{a n^p}_\text{metabolic rate}
    = \underbrace{b n}_\text{catabolic rate} +
    \underbrace{\frac{dn}{dt}}_\text{new growth}
\end{equation}

and the von Bertalanffy equation comes from a simple rearrangement \cite{benzekry2014models}.

For $p<1$, interpretable as less-than-perfect perfusion, this model
displays an initial burst of power-law growth, tending towards a plateau
determined by $a$, $b$ and $p$.

One remarkable feature of this model is that it contains both simple
exponential growth and power-law as limits. The first, in the limit that
$p=1$, and the second in which $b=0$.

It is also worth noting that in many cases the Bertalanffy growth model performs much
better than other four-parameter models such as Gomp-ex, and has the
additional advantage that the parameters have a straightforward
interpretation. Despite these points in its favour, it is still somewhat
obscure at the time of writing, and sees a rather limited use. It is rarely used in 
studies of drug effectiveness despite its many advantages \cite{benzekry2014models}.

\subsection{How applicable are these to invasive cancers?}

Comparisons of these models (and a preponderance of others) are typically
carried out on observations of single tumours in mouse models.
A serious limitation here is that it is not obvious that models which are
well-suited to single tumours will carry over to collections of multiple,
diffuse tumours. It isn't very hard to see that a collection of metastatic or
locally invasive lesions could have a very different growth curve than an
individual lesion. A single lesion might only grow for a limited time up to
some plateau size, but  continue to send out invasive cells to other
regions in which growth can continue. The whole may grow differently from the
individual parts.

Imagine a population of rabbits. An individual rabbit can only grow for an
initial short period of time, for about six months, after which they have
basically reached their adult size and will not grow any larger. A field of
rabbits, on the other hand, can support a continuously growing number of
rabbits for a much longer length of time, and it's possible to see exponential
growth of the population. This is despite the fact that individual rabbits can
only grow for a certain limited period of time. 
The important idea here is that a \emph{population} can experience
exponential growth despite quite different growth curves for its
\emph{individual} constituents. 

Exponential growth can still describe the growth of
the whole population, even when individuals in it only grow a limited amount.
Similarly, in the case of metastatic cancers, the growth curve of the entire
tumour burden may be very different from the growth curve of an isolated
tumour. The feedback due to migration and metastasis may have a complex
effect. We analyse and extend models of this feedback mechanism to account for
evolutionary dynamics in chapter \ref{ch:minimal}.

As previously noted, despite widespread knowledge of its inability to display
plateaus commonly observed in single-tumour growth curves and many other
complex behaviour, the exponential growth model is still the most commonly
encountered in practice, both among clinicians and applied mathematicians.
That such a simple and ``stupid'' model should see such widespread use,
despite not being taken seriously by the majority of those who do use it is
surprising, and bears some explanation.


Two basic reasons why this is the case are the fact that the model is easy
to fit and apply, having only two parameters, and the fact that it is often
impossible to get long-term dynamical data about tumour burden. It is also easy and
tractable to use, and due to a scarcity of data for ethical reasons is usually
difficult to rule out.

This results in the remarkable situation that two scientific communities
widely use and apply a
model which, despite its popularity and debatable success, is not taken ``too
seriously'' by any member of these communities. Considering the fact that it is
clearly useful in many cases owing to its tractability and simplicity, it is
my belief that the paradoxical popularity of exponential growth warrants
a serious investigation. How might exponential growth arise in the
context of more detailed and realistic models? I address this in
chapters \ref{ch:Lattice} and \ref{ch:minimal}. 

\subsection{Mathematical modelling of cancer evolution}

The picture of cancer as an evolutionary process has become well established \cite{BVogelstein2013,weinberg_book}. Many genes have been found which code for
proteins which, in their mutated form, are important in the start and
progression of cancer. Identifying their ordinary functions is an active and
rewarding program in current research, and fits well into the picture of an
accumulation of mutations which give abnormal cells an ``advantage'' over
normal cells.
In addition to the initiation of tumours, the emergence of resistance to
therapy is also suspected by many to involve a Darwinian process of
selection \cite{bozic2010accumulation,diaz_jr_molecular_2012}.

However, detailed measurements of the dynamical course of cancer are extremely
difficult to obtain. For obvious ethical reasons, collecting an exhaustive
life history of the untreated disease is usually impossible due to the suffering that
this entails. As a result, although the idea that the underlying process
can be described as evolutionary is quite widely accepted, the timing and
detailed dynamics underlying these processes have not been definitively
established.

There are several current proposals for the nature of the underlying dynamical
process. These proposals can usually be described as variations on one of a
few hypotheses, which differ first in the relative importance of selective 
pressure, and second in the timing of selective sweeps (if they occur). A
selective sweep is an evolutionary event in which a new, fitter (``driver'') mutant rapidly
overtakes other lines. This event pulls neutral (or passenger) mutations up in
frequency of occurrence along with the driver mutation 
\cite{bozic2010accumulation}.

To summarise the current hypotheses:

\begin{itemize}
    \item \emph{Late evolution models} posit that additional driver mutations
    occur, and each carry some very small advantage. These successive drivers
    increase in frequency in a series of sweeps, which occur at an
    accelerating rate. As a result, most evolutionary change happens late in
    the life history of the cancer\cite{yachida2010distant}.
    \item \emph{Early evolution models}, like late evolution models, posit that
    additional driver mutations occur, but carry diminishing returns. Only the
    first few drivers increase in frequency at an accelerating rate, after
    which the pace of evolutionary change slows dramatically. Most evolution
    occurs early in the course of the cancer in these
    models\cite{ling2015early,durrett_evolutionary_2010}.
    \item \emph{Neutral models} attempt to explain high
    levels of intra-tumour heterogeneity by positing that only the very earliest,
    initiating mutations carry a significant advantage, and subsequent
    mutations are essentially neutral, with fixation due to the effects of
    rapid early expansion. The so-called ``Big Bang'' model falls into this
    class \cite{sottoriva_2015}.
\end{itemize}

In a little more detail, these models can also be related to different assumptions
about the underlying fitness landscape available to mutant
clones in a tumour, and what other factors affect the probability of
fixation. However, how exactly underlying fitness landscapes should affect
growth and the dynamics of selection if spatial constraints and geometry are
important is an open question: one which we hope to address at least in part
in chapter \ref{ch:minimal}.

In the case of neutral models, selection pressure is hypothesised to be
swamped by rapid growth. Selection due to small
differences in cell survival needs to be stronger than random fluctuations,
which in practice means that populations have to be sufficiently large that
these ``founder effects'' can be
avoided \cite{murray2001mathematical,mayr1942systematics}. But if only cells near
the growing surface are able to divide, the active population is much
smaller than the true population size, and stochastic effects like genetic
drift are much more important. Neutral and even somewhat deleterious mutations
can reach fixation on the surface of expanding tumours in theory
\cite{Timetocancer,hallatschek2007genetic}.

Recent proposals such as the ``Big Bang'' model for intratumour heterogeneity
postulate that the majority of genetic variation in tumours is due to early
fluctuations. These enable fixation of almost neutral alterations. While notable for its
novel prediction of characteristic backgrounds of heterogeneity that are
independent of selective pressure, phenomena such as the appearance of 
resistance to chemotherapy in late stage cancers are easier to explain
in terms of acquired selective advantages. Spatial constraints are not
necessarily as relevant for metastatic cancers which are able to evade local
factors that would otherwise inhibit their growth \cite{sottoriva_2015}.

Quantitative experiments on bacterial colonies and numerical simulations of
single tumours support the picture that spatially constrained growth can
drastically increase the probability of fixation for non-advantageous mutant
lines, even for somewhat deleterious mutations. A plausible example of this
occurring in cancer would be a solid tumour with poor vascularisation, in 
which growth can only occur in a narrow rim at the
surface \cite{hallatschek2007genetic,waclaw_spatial_2015}.

We will therefore treat neutral models as essentially null models. They are an important 
class of alternatives to selective pressure for the purposes of empirical tests, 
which we will occasionally revisit during the development of our theories of 
selection during metastasis. We should emphasise that the fixation due to
early expansion and ``freezing in'' of primordial neutral mutations
is a prediction which is hard to avoid on the basis of simulations and other
experiments on bacteria, and which is an important component of our later,
developed theories in chapter \ref{ch:minimal}.

The first important class of models for selection in cancer are models which
assume fitness increases to be linear, which corresponds to an accelerating
pace of selection. Crucially, these models tend to assume well-mixed growth,
analogous to the simple exponential growth model described in the previous
section. The relevance of this class of models to spatially constrained growth
with invasion has not been well studied.

One important recent mathematical model which attempts to predict driver
accumulation, and furthermore relate this to measurable proxies of dynamical
information akin to the molecular clock, is the work of Bozi\v{c} \emph{et al.}'s
\cite{bozic2010accumulation}. This is a well-mixed model, which assumes that
individual lineages grew exponentially with characteristic cell division time
$T$, differences in fitness $s$ determined purely by differences in survival 
and turnover probability (so that $s=1-b/d$, $b$ division rate and $d$ death rate), and
driver mutation occurrence rate $u$. The core prediction of this framework is an
exponential increase in the frequency of driver mutations over time. Although 
many models of tumour growth are
dynamical, data on the course of growth over time is difficult to
obtain for ethical reasons. To address this, this model also includes the
accumulation of passenger mutations. Passenger mutations by definition do not
alter the rate of growth of tumours, and as such the number $n_p$ of passengers present
grows at a constant rate over time. Namely

\begin{equation}
    n_p = \nu t /T
\end{equation}

where $\nu$ is the product of the point mutation rate per base pair and the
number of base pairs at which mutations are neutral, and $T$ is the time in
between divisions. This assumption of a molecular clock enabled the authors to
attempt to test their model empirically, which is a significant achievement.
But the assumption of a constant driver accumulation rate is not obviously
consistent with the genetic hitch-hiking of passenger mutations,
in which they increase in frequency due to their coincidental association 
with drivers. The assumption of a constant rate can be justified on the basis
that in this model, the cell cycle time is constant. As a result, only changes in the
probability $d$ that a cell survives to divide more than once can affect fitness.
The frequency of actual divisions, and hence the rate at which new passenger
mutations occur, must therefore also be constant. There is evidence that 
passengers accumulate broadly linearly prior to cancer initiation, though,
since tumours in 90-year old patients have roughly twice as many mutations as in 
45-year old patients at the same stage of tumour progression \cite{BVogelstein2013}.

This model is a good example of a steady, incremental increase in fitness
that is reminiscent of classic Darwinian selection, and predicts an
exponential accumulation of drivers \cite{bozic2010accumulation}. This exponential increase in the average
number of drivers corresponds to a so-called ``late evolution'' model, in line
with what we might expect. However, this model's tractability hinges on its
assumption of a well-mixed model, and the derivation of the time constant of
the accumulation depends on the assumption that sweeps are ``hard'' and widely
separated in time. This requires only one clone to be prominent in the population 
at any one time, which is a rather strong requirement. In chapter \ref{ch:minimal}, 
we propose a theoretical framework in which
similar predictions about the accumulation of drivers can be made with weaker 
assumptions about the dynamics.

Like many others, this model is ``well-mixed'' --- it does not consider the
effects of shape on growth. One
attempt to include geometrical constraints on the emergence of new drivers is
the model by Antal \emph{et al.} of a single spheroidal tumour \cite{antal_spatial_2015}. Two important results from
this work are the time for a new driver mutation to reach fixation on the
surface of a tumour, and the probability that a cell at a chosen radius
belongs to the wild-type or the mutant line. Equivalently, the shape and
size of the spatially contiguous mutant sector. This geometrical model treats
the nucleation of new diver mutations as stochastic and the growth of existing
lines as deterministic, which heavily influences the analytical model
developed in chapter \ref{ch:minimal}. Growth in this model only occurs on the surface
of the (initially spherical) tumour: the rate at which new driver mutations
occur is therefore also proportional to the area of this growing surface. This
geometrical constraint strongly restricts the effective population size, and
growth and mutation occur more slowly than in well-mixed
models \cite{antal_spatial_2015}. 


Despite being a significant step away from the widespread assumption of
well-mixed models, this is strongly constrained in two regards. The first is
the idealised picture of perfectly smooth spheroids. Real tumours are unlikely
to be perfectly smooth. It has been found in the case of bacterial
colonies that roughness can strongly affect the probability of survival 
of mutant clones on the surface \cite{hallatschek2007genetic}. This may 
substantially revise one of this model's core
predictions. The second limitation for our purposes is the model's lack of
attention to migration. We revisit and attempt to address these issues in
chapter \ref{ch:minimal}.

Finally, the possibility of more complex fitness landscapes and evolutionary
processes cannot be conclusively ruled out \cite{greulich2012mutational}. In this case, the pace of selection
could intermittently accelerate and decelerate. This results in a picture of
selective sweeps occurring in a few evolutionary bursts between long
periods of slow evolutionary change, similar to what are known in
evolutionary biology as punctuated equilibria \cite{gould1972punctuated}. 
One plausible and simple scenario is one in which only a few
initial drivers after initiation carry a substantial advantage, followed by
diminishing returns. This would result in an early, rapid succession of
selective sweeps, followed by a period of relative stasis. This picture is
intermediate between neutral models and the gradualistic late-evolution models.
However, we should emphasise that this connection between the fitness
landscape and evolutionary dynamics is based largely on expectations from
well-mixed models. Studying the combined effects of space and invasion will be
a major theme of this thesis.

\subsection{Our research in context}

The interpretation of any experiment can only be as clear as the theory 
that it sets out to test. There is a clear need for
a theoretical scheme which enables the evolutionary dynamics of metastatic
cancers to be connected to the statistical behaviour of individual cells. 
Statistical properties of the large population should be related to
mechanisms at play in the smallest members of the population. This search for
a connection between aggregate behaviour and individual dynamics is a
long-standing theme in both statistical physics and cell biology.

The work presented in this thesis was carried out in the spirit of connecting
the world of the microscopic with that of the macroscopic. 
This should contribute to making possible the eventual 
prediction of the dynamics of cancer evolution from known properties of the 
cells that comprise it. Another outcome is that observations of the course of cancers
could be used to draw inferences about the mechanisms at play on a smaller
scale. While the achievement of both of these will in all likelihood take many
more years of intensive research, we hope that our work here represents an
incremental advance towards making mathematical models which are simple enough
to be straightforwardly interpreted, whilst also being brought closer to
realistic situations.

To summarise, two important open problems in the dynamics of cancers are the
growth curve of the tumour burden over time, and the timing of
evolutionary events such as those suspected to produce resistance (see section
\ref{sec:evolution}). Whilst there are 
too many unknowns in our current knowledge of cancer to model these 
phenomena in full, we will 
attempt to advance our understanding of how growth and the speed of 
evolution are affected by tumour geometry and local invasion.

In chapter \ref{ch:Lattice}, we use a set of numerical simulations to
investigate how the growth curve of an ensemble of tumours may be affected by
changes in the properties of individual motile cells. Namely, how changes
in elastic stiffness affect the speed of cells' migration through tissue; and
how changes in cell type and speed affect the form of growth curves of
tumours. We also study whether it is possible, and if so under what
conditions, for growth curves to be truly exponential even when individual
lesions in the ensemble grow much slower than exponentially.

We elaborate substantially on the last question in chapter \ref{ch:minimal},
in which we present a stochastic theory for the growth of cancers with several
constituent genotypes. The goal is to relate the fitness landscape of the
different mutant types to the relative frequencies and average number of mutations 
present after a given time. We find a sophisticated way of doing so, and show
that the mean-field limit of our theory is analytically solvable with complex
analysis. The analytical approach enables some fundamental results about the
growth of populations to be generalised, and permits exact solutions in a
number of cases.

Finally, in chapter \ref{ch:clinical} we outline our contributions to an
experimental study of a novel form of combined chemotherapy. The new therapy is based on
a hypothesis that resistance to chemotherapy is acquired through several
independent mutations. We
discover that a simple model of tumour growth based on two independent
populations provides a decent description
in several cases. The results seem to refute the original
hypothesis, and we provide some possible explanations as to why this may be.

\chapter{Individual-based models of local invasion}
\label{ch:Lattice}

\section{Introduction}



Cancer involves the escape and migration of cells from primary tumours to
form new tumours elsewhere in the body. This
process is termed \emph{metastasis} and is one of the most important features of
advanced cancer \cite{Hallmarks}. Most experiments studying the migration of
cancer cells around the body involve the injection of mice with human cancer
cells, and measuring the locations and sizes of the resulting tumours 
\cite{benzekry2014models,miklavvcivc1995mathematical}. Since the results show 
new tumours forming in similar ways to untreated metastatic
cancers, the current theory of how metastasis occurs is that a small
number of cells become invasive and enter the bloodstream, before escaping again
to start a new tumour elsewhere \cite{chiang2008molecular,miklavvcivc1995mathematical}.

However, the presence of cells in the margin surrounding tumours is another
important indicator of the success of surgical therapy and forms part of
the standard process of tumour biopsy. In addition to distant migration, escaping cells can also be found
wandering around the margins of existing tumours, not yet having escaped to
remote locations \cite{stefaniejefrey1995importance,porpiglia2005assessment}. It is routine practice
to check that the margins of a tumour are clear of invasive cells, as this is
an indicator of how likely the tumour is to have produced metastases
either locally (in the same organ or tissue) elsewhere in the body \cite{maxwell2015early}.

\begin{figure}[ht]
\begin{center}
    \includegraphics[width=0.9\textwidth]{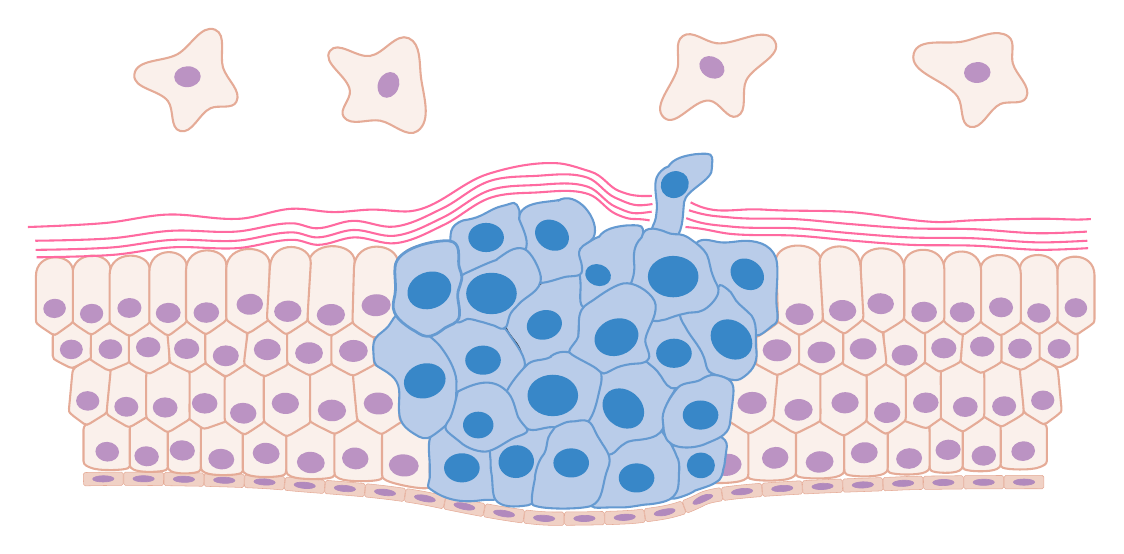}
    \caption{\label{fig:membrane} Diagram of the structure of a cancer of
    epithelial tissue (dense, orderly layer), showing the tumour itself (in blue), surrounding
    connective tissue (above, scattered cells) and the basement membrane
    separating them (red lines). A tumour cell is shown escaping through a
    hole in the basement membrane. Modified, from \cite{CRUKDiagrams}. }
\end{center}
\end{figure}

We will explicitly \emph{not} attempt to model the process of escape
and metastasis in full detail, and will deal primarily with abstracted models
that attempt to answer specifically how local migration affects growth.
It is nonetheless worth giving an overview of
what is currently known about the process of \emph{distant} metastasis. This is to put our work into
context, and to show what questions we can and cannot answer with our
emphasis on local migration.

For distant metastasis to occur, invasive cells have to escape the lesion and
surrounding tissue that they come from. The most common types of cancers are
carcinomas, which are cancers of specialised epithelial tissue: this is tissue
which forms layers of cells, like in glands or skin or in the lining of 
the digestive tract. Supporting any epithelial tissue is a larger bulk of
connective tissue (see figure \ref{fig:membrane}, upper structures). 
The two are separated by a thin layer called the basement
membrane, which is mostly made of collagen fibres \cite{liotta1980metastatic}.
After escaping into the connective tissue, the invasive cells have to squeeze back through another basement
membrane, and through layers of cells lining blood vessel walls, to escape
into the bloodstream. From here, the cells undergo the reverse process,
escaping first the bloodstream, travelling through connective tissue, and
escaping another basement membrane, to found a new tumour
\cite{Hallmarks,chiang2008molecular}. 

By ``lesion'' or ``micro-lesion'', we mean a single, small mass of cancer
cells which is basically uniform in composition. By ``tumour'', we mean
instead larger and more complex structures, which possibly include new vasculature
(capillaries), and some normal and necrotic tissue interspersed with the cancerous 
cells. Tumours can then be seen as aggregates of smaller microscopic lesions.
Neither of these terms has a strict, hard-and-fast definition in the
literature, and this choice of meanings is purely for technical convenience.

We are not interested in modelling every step of this process explicitly, as
the number of unknowns alone would force us to include a large number of adjustable
parameters. We shall instead focus on local migration, consisting of lateral
movement through the same layer of tissue before invasive cells escape the
basement membrane. This is intended to capture the process by which cells
arrive in the surgical margin, the region surrounding malignant tumours. In particular, we will
investigate how changes in individual cell properties affect the speed of the
cell's movement through layers of tissue, and how this affects the growth
of tumours over time. 

Changes in these properties could imaginably be achieved
with novel drugs that make migratory cells stiffer \cite{waclaw_spatial_2015,schenk2015salinomycin}. 
However, what effect this has
on the rate of migration and escape is not obvious, let alone the knock-on
effects on growth rates, and extremely difficult to
measure in living test organisms such as mice. This is the main motivation for
our set of virtual experiments, which attempt to measure the effect on tumour
growth of changes in various factors that affect cell invasion.

\section{Mechanistic simulation}
\label{sec:mech}

Whether or not cells are invasive or stay in place is connected to their
\emph{type}, how specialised the cell is. Cell specialisation consists of which functional
proteins it produces, and what these do. For example, a cell that formed the
lining of a gland may produce protein complexes that speed up the synthesis of
some hormone, which the gland secretes into the body; muscle cells produce
larger amounts of actin and myosin, which form fibbers which enable the cell to
contract. Other types of cells produce other functional proteins
\cite{MolBioCell}. This specialisation also manifests as differences in
physical and mechanical properties of cells. How stiff cells are mechanically,
and how many adhesion molecules are present on their membranes, are two
examples particularly relevant to us.

The question that we are trying to answer here is how quickly invasive cells
are able to move through layers of epithelial tissue (or monolayers of cancer
cells), and how this is affected by mechanical stiffness. Put more
mathematically, these simulations are to measure how the speed of cell
migration depends on the elastic modulus of invasive cells. While it is
possible to study individual cancer cells' migration in isolation on a growth
medium, it is much harder to experimentally investigate how their migration is
affected by mechanical interactions with other cells around them in dense
tissues.

To this end, we write and perform several set of centre-dynamical simulations which
resolve the forces acting on each cell in a virtual monolayer of cells,
treating each cell as a deformable sphere with a radius $R_i$. We adopt a fairly
standard approach to other centre-dynamical models in the field, but make a
few choices regarding different types of cell and how neighbouring cells
interact that distinguish our simulation from others. For the unfamiliar with
centre-dynamical models, we will briefly summarise the approach and the most
unusual of our choices about what to model before going
into details about the equations of motion and other technical choices.

The drag force acting on each cell is simply proportional to their
velocity and in the opposite direction, and we use the classical expression
for Stokes' drag on a sphere. This is a standard assumption in
centre-dynamical models \cite{lighthill1975mathematical}. The surrounding
interstitial fluid is assumed to have a viscosity similar to water. This
means that at these small length-scales viscous drag is considerably larger
than inertia, allowing for substantial simplifications to the equations of
motion.

Cells experience
forces due to interactions with their neighbours in addition to the viscous 
drag force, and an additional traction force if they are of
the motile type. Cells can switch between a static type, which do not migrate
and which divide; and a motile type, which migrate and which do not divide.
These two types of cell correspond roughly to what are called epithelial and
mesenchymal types in the literature, and the transition between them is a
simple model of epithelial-mesenchymal transition (EMT)
\cite{EMT1,EMTsarc,EMTnoise}.

As we re-use our model of EMT in later models essentially unchanged, we will
summarise it before moving on to the mechanical model.

\subsection{Equations of motion and algorithm}

\subsubsection{Epithelial-mesenchymal transition}
\label{sec:EMT}

We model epithelial-mesenchymal transition as a Markov process: there is a
rate $k_{EMT}$ at which an epithelial cell transitions to a mesenchymal one,
and a reverse rate $k_{MET}$ at which mesenchymal cells switch back to being
epithelial again. After an interval of time $\Delta t$, there is thus a 
conditional probability $P_{E\rightarrow E}$ for $E$ cells to remain $E$ cells of

\begin{equation}
    P_{E\rightarrow E}(\Delta t) = e^{-k_{EMT} \Delta t}
\end{equation}

and a conditional probability to have switched to $M$-type $P_{E\rightarrow M}$ of

\begin{equation}
    P_{E\rightarrow M}(\Delta t) = 1-e^{-k_{EMT} \Delta t}\;.
\end{equation}

Similar formulae hold for $M$ types to switch back to $E$ types, substituting
$k_{MET}$ for $k_{EMT}$ in the formulae above. These
transition probabilities imply that the overall probabilities $P_E$ and $P_M$ 
that a given cell will be a found to be type $E$ or $M$ after a period of
time $t$ obey a set of simple ordinary differential equations,

\begin{equation}
    \frac{d P_E}{dt} = -k_{EMT} P_E + k_{MET} P_M
\end{equation}

\begin{equation}
    \frac{d P_M}{dt} = -k_{MET} P_M + k_{EMT} P_E
\end{equation}

and since there are only two possible types of cell, $P_E+P_M = 1$.

In the limit that we have very large numbers of cells, these probabilities
$P_E$ and $P_M$ correspond to the proportions of the population of each type.
These proportions can be seen to relax exponentially towards an equilibrium
point, regardless of the choice of initial conditions. This steady state
corresponds to the solution of the two linear equations

\begin{equation}
   -k_{EMT} P_E + k_{MET} P_M = 0
\end{equation}

and

\begin{equation}
    P_E + P_M = 1\;.
\end{equation}

These have the unique solution

\begin{equation}
    P_E = \frac{k_{MET}}{k_{EMT}+k_{MET}}\;,
\label{eq:fracM}
\end{equation}

\begin{equation}
    P_M = \frac{k_{EMT}}{k_{EMT}+k_{MET}}\;,
\label{eq:fracE}
\end{equation}

and as a result, the relative numbers of $M$ and $E$ cells can (in principle)
be tuned by choosing different values of the forward and reverse rates of EMT.

This choice of model was primarily for simplicity, as differentiation
hierarchies can be very complex and involve several different populations, but
very similar Markov chain models have been found to provide a remarkably good
description of type switching behaviour in breast cancer cells \emph{in vitro} \cite{EMTnoise}.

\subsubsection{Simulation units and parameter values}

Motility is assumed to be due to adhesion of motile cells to a substrate such
as the basement membrane or a culture medium, along which motile cells drag
themselves (see figure \ref{fig:crawling}). We make the deliberate choice not
to model the mechanics of the underlying substrate or actin treadmilling
\cite{insall2009actin}, and
instead simply apply a constant traction force to each motile cell, and treat
the population of cells as a simple two-dimensional layer of interacting
spheres. We then vary the motile cell's elasticity to see how this affects the
motile cell's speed. The simple expectation is that the more strongly motile
cells interact with the non-motile epithelial cells surrounding them, the
slower they will be able to move.

The choice of a plausible traction force can be done by reviewing experimental
results on how quickly motile cells of the same size move, and calibrating the
traction force on motile cells in the simulation in the absence of other cells
until the simulated speed and experimental speeds agree. Ideally these experiments
would use cancer cells, but in many experiments amoebas or immune cells called 
neutrophils are also used. This results in speeds in the range of
$0.03$---$1.0 \mathrm{\hbox{\textmu} m \cdot s^{-1}}$ 
\cite{ferguson2007pi,insall2009actin,li2015cancer,hoang2013measuring}.

We
should point out that many of these experiments involve migration of cells on
clean surfaces rather than dense tissue cultures, and so the speeds 
involved will likely be much higher than in our simulations.

To judge how relevant inertial effects are, we can use the above typical
scales and combine them to determine an upper bound on the Reynolds' number for our system, 
a dimensionless quantity representing the relative importance of viscous
damping (in this case due to the interstitial fluid) and inertia (due to
density).

Given that the interstitial fluid surrounding these cells is composed of a
complex fluid containing water and other much more viscous substances and
proteins, it will be at least as
viscous as water: so, $\mu \sim 10^{-3}\; \mathrm{Pa\cdot s}$ \cite{lighthill1975mathematical}.

Both the interstitial fluid and the cellular substructure
will, as they are composed of both water and proteins which in bulk are less
dense than water, be somewhat less dense than water overall. The density of
either the cells or the interstitial fluid will therefore be 
$\rho \approx 10^3\;\mathrm{kg\cdot m^{-3}}$ \cite{lighthill1975mathematical}. 

Finally, the cells of interest will be between $L=10-20\mathrm{\hbox{\textmu} m}$ in diameter, and
neutrophils move at around $U=1 \mathrm{\hbox{\textmu} m \cdot s^{-1}}$ \cite{ferguson2007pi,insall2009actin,li2015cancer,hoang2013measuring},
which puts the Reynolds number of the flow around the cell at 

\begin{equation}
    \mathrm{Re} = \frac{\rho U L}{\mu} = {10}^{-5}
\end{equation}

which is indeed much smaller than $1$, meaning that inertial effects are $10^{-5}$ 
times the strength of viscous and other effects. This justifies the use of the 
over-damped limit in which cell motility is usually analysed \cite{lighthill1975mathematical,DAH}.

The most ``natural'' choice of units for these simulations is one in which the
most commonly encountered quantities are all close to $1$. We are also not
interested in electromagnetic effects, or other ``exotic'' physics: these
simulations and models concern mechanics and motion of cells. Base units for
our simulations should therefore consist of a mass $\mathrm{M}$, a length 
$\mathrm{L}$, and a time $\mathrm{T}$.

An obvious choice for $\mathrm{L}$ is based on the typical size of eukaryotic
cells, at around $10\mathrm{\hbox{\textmu} m}=10^{-5}\mathrm{m}$ \cite{li2015cancer}, so that average cell sizes
are all around $1.0\;\mathrm{L}$. Cell speeds should also,
conveniently, be on the order of $1.0\;\mathrm{L} \mathrm{T}^{-1}$ , which
given a typical speed of $0.1\mathrm{\hbox{\textmu} m \cdot s^{-1}}$ from above 
suggests a time unit of $1\mathrm{T}=100 \mathrm{s}$.

The mass of a typical
cell is probably not a good choice for $\mathrm{M}$, as inertia is much less
important than drag here: we shall therefore choose units so that the 
viscosity of the interstitial fluid (assumed to be similar to that of water) 
$\mu=1.0\;\mathrm{M}\mathrm{L}^{-1}\mathrm{T}^{-1}$, 
which determines drag on these microscopic scales, is
close to $1$. It follows from straightforward dimensional analysis that
the corresponding mass scale is $\mathrm{M}=\mu \mathrm{L T}$.

Given a viscosity of $\mu \approx 10^{-3}\mathrm{Pa\cdot s}$ and cell size of
$R\approx5\mathrm{\hbox{\textmu} m}$, a velocity of $1 \mathrm{\hbox{\textmu} m \cdot {s}^{-1}} = 10
\mathrm{L\cdot T^{-1}}$ on the upper end of the range of possible speeds
requires a traction force of about $9.4\times 10^{-14}\mathrm{N}$, or $94
\mathrm{M L \cdot T^{-2}}$.

The various different units and adjustable parameters in these simulations can
then be expressed as:

\begin{center}
    \begin{tabular}{| l | c | r |}
    \hline
    Quantity & Magnitude & Typical values \& reference \\ \hline
    Mass unit $\mathrm{M}$ & $=\mu \mathrm{L} \mathrm{T} = 10^{-6} \mathrm{kg}$ &
    Calculated, see above. \cite{lighthill1975mathematical}\\ \hline
    Length unit $\mathrm{L}$ & $1\mathrm{L}=10^{-5}\mathrm{m}$ & 
    Cell diameter, \cite{li2015cancer}\\ \hline

    Time unit $\mathrm{T}$ & $100 \mathrm{s}$ & Calculated, see text. \\ \hline

    Traction force $\|\vec{F}_{T}\|$ & $94 \mathrm{M L \cdot T^{-2}}$ &
    Calculated. \cite{ferguson2007pi,insall2009actin,li2015cancer,hoang2013measuring} \\ \hline

    Interstitial fluid viscosity $\mathrm{\hbox{\textmu}}$ &
    $1.0\;\mathrm{M}\mathrm{L}^{-1}\mathrm{T}^{-1} = 10^{-3} \mathrm{Pa\cdot
    s}$ & $10^{-3}\;\mathrm{Pa\cdot s}$ \cite{lighthill1975mathematical}\\ \hline

    Cell elastic modulus $E^\star$ & $0$---$10\;\mathrm{kPa}$ for $M$ cells,
    $1\;\mathrm{kPa}$ for $E$ cells & 
    $1$---$13$ $\mathrm{kPa}$ \cite{lekka1999elasticity} \\ \hline

    Work of adhesion $\gamma$ & $10^{-2}\;\mathrm{N\cdot m^{-2}}$ &
    $1$---$9\times 10^{-3}$ $\mathrm{N\cdot m^{-2}}$ \cite{DAH,foty1994liquid}  \\ \hline

    Cell bulk modulus $B$ & $10\;\mathrm{kPa}$ & Same as $E$-type elastic
    modulus\\ \hline

    Critical pressure $p_C$ & $1\;\mathrm{M\cdot L^{-1}\cdot T^{-2}} = 10
    \;\mathrm{\hbox{\textmu} Pa}$ & Estimated, see text. \\ \hline

    Simulation area & $40\;\mathrm{L}\times40\;\mathrm{L}$ & N\/A \\ \hline
    \end{tabular}
\end{center}

\subsubsection{Equations of motion}

As in other centre-dynamics models of tissue layers, inertia is neglected
\cite{DAH,Drasdo2000,schaller2005multicellular}.
Otherwise, our equation of motion for each cell $i$ would be

\begin{equation}
    \sum_j \vec{F}_{ij} - b_i \vec{v}_i = m_i \vec{a}_i\;,
\end{equation}

where $\vec{v}_i$ is the velocity of cell $i$, $\vec{a}_i$ the acceleration of
cell $i$, $b_i = 6\pi \mu R_i$, and $\sum_j \vec{F}_{ij}$ the resultant of all forces \emph{other
than drag} acting on cell $i$. This is considerably simplified by taking
$\mathrm{Re} \approx 0$, resulting in the quasi-static approximation \cite{lighthill1975mathematical}

\begin{equation}
    \sum_j \vec{F}_{ij} - b_i \vec{v}_i = 0\;,
\end{equation}

which can be more conveniently written in terms of cell mobility $\zeta_i =
b_i^{-1}$,

\begin{equation}
    \vec{v}_i = \zeta_i \sum_j \vec{F}_{ij}
    \;.
    \label{eq:mecheom}
\end{equation}

The idea is that at such low Reynolds' numbers, the balance of forces is in
instantaneous equilibrium with viscous drag, to an approximation as good as
$\mathrm{Re}$ is small.

To integrate these equations, we use Euler integration, the simplest possible
scheme. With a time step size of $\Delta t$, the position of each cell $\vec{r}_i$ evolves
according to 

\begin{equation}
    \vec{r}_i(t+\Delta t) = \vec{r}_i(t) + 
    \Delta t \zeta_i \sum_j \vec{F}_{ij}
    \label{eq:eomimplicit}
\end{equation}

The forces $\vec{F}_{ij}$ represent the interactions between cells and their
neighbours, and the traction force experienced by motile cells. The
interactions between cells and their neighbours consist of attractive and
repulsive forces which act on lines between their centres: since we have
abstracted away all of the details of a cell's shape except for its ``size'',
given by its radius $R_i$, the physics of our layer of virtual tissue seems
vaguely reminiscent of the physics of a system of particles suspended in a
viscous solution. The crucial differences consist of the growth of cells, to
which we will return shortly, and the active ``driving'' of motile cells,
represented with this constant traction force $\vec{F}_T$.

With the exception of this traction force, the interaction forces can be 
calculated from a potential $U_{ij}$ which encodes all information about the
physical interaction between cell $i$ and its neighbour $j$,

\begin{equation}
    \vec{F}_{ij} = -\nabla U_{ij}\;,
\end{equation}

which we represent as a sum of attractive and repulsive components,

\begin{equation}
    U_{ij} = U^\mathrm{(adh.)}_{ij} + U^\mathrm{(rep.)}_{ij} + U^\mathrm{(press.)}_{ij}\;, 
\end{equation}

arising from adhesive and elastic interactions respectively. As in other
centre-dynamical models, each of these is interpreted as due to a distinct
physical interaction between the cells, and the aggregate potential $U_{ij}$
should have a minimum at some distance between the two cells close to the
average radius $\bar{R}$. This is so that
the tissue has a stable mechanical equilibrium.

\subsubsection{Adhesion}

We assume that the only attractive forces which act between cells are those
due to cell adhesion. Most adhesion between cells is caused by the formation of 
chemical bonds between molecules on the cell membranes, and is primarily
determined by the concentration of cell adhesion molecules called cadherins on
the cell membrane \cite{DAH,DAHtime,foty1994liquid,foty2005differential}.

This contribution $U^\mathrm{(adh.)}_{ij}$ to the interaction potential $U_{ij}$ is 
simply the work done by adhesion, or the interfacial free
energy due to bonds between these cell adhesion molecules.
This interfacial free energy between two cells $i$ and $j$ depends on their
respective types, but if the surface free energy (work of adhesion per unit
area, analogous to surface tension) of their interaction is $\gamma_{ij}$, and
the contact area between them $A_{ij}$, then the work done by adhesion is

\begin{equation}
    U^\mathrm{(adh.)}_{ij} = -\gamma_{ij} A_{ij}\;,
\end{equation}

and what remains is to specify $\gamma_{ij}$ for the different possible cell
types and calculate $A_{ij}$.

Different types of cadherins do not bind to each other very specifically. One
type of cell with a high concentration of one type of cadherin and another
type of cell with a similar concentration of a different type of cadherin will 
stick together with similar $\gamma_{ij}$ than they would to each other. The
interfacial free energy $\gamma_{ij}$ is therefore believed to depend only on
the concentration of adhesion molecules present on each type of cell, and not
on the type of cadherin present \cite{DAH,DAHtime,foty1994liquid,foty2005differential,foty2004cadherin}.

However, direct measurements of the different possible $\gamma_{ij}$ are
extremely difficult, and despite many studies of adhesion between cells of the
same type, we are not aware of any simple model of adhesion that is both
founded on quantitative measurements of the work of 
adhesion between cells of different types and generally agreed upon
\cite{foty2004cadherin,jones2012modeling}.

In our simulations, we choose simply set all $\gamma_{ij}$ to be the same value,
so that
$\gamma_{ME}=\gamma_{MM}=\gamma_{EE}=10\times10^{-3} \mathrm{N\cdot m^{-2}}$.
This is on the same order of magnitude as many measurements of the strength of
cell adhesion \cite{DAH,DAHtime,foty1994liquid,foty2005differential,foty2004cadherin}.

\begin{figure}[ht]
\begin{center}
    \includegraphics[scale=0.6]{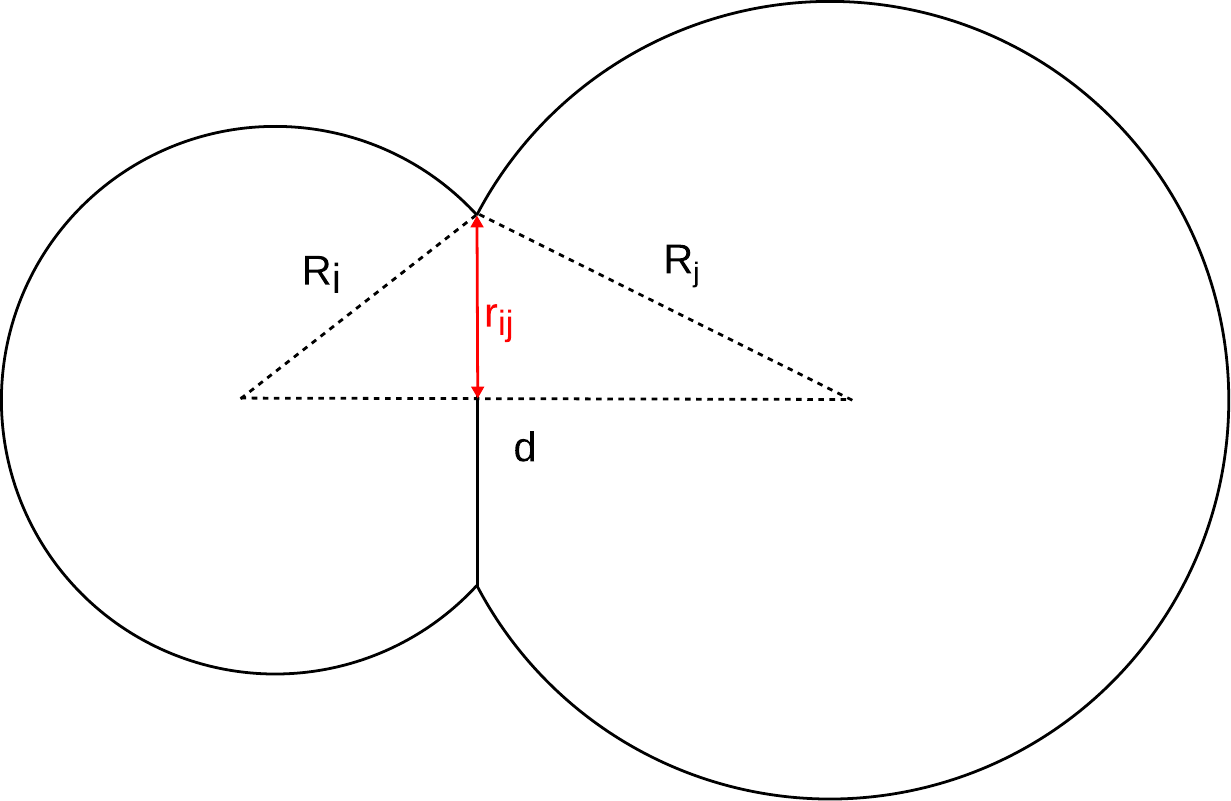}
\end{center}
\caption{\label{fig:cellgeom}Cross-sectional view of the assumed geometry for the contact area
(radius marked by red arrow) of two neighbouring cells $i$ and $j$ of radii $R_i$ and $R_j$.}
\end{figure}

Now, to calculate $A_{ij}$. Suppose two spherical cells $i$ and $j$ with radii $R_i$
and $R_j$ come into contact each other: we take the contact area to be
circular and flat, with radius $r_{ij}$. (See figure \ref{fig:cellgeom}.) We
also assume for convenience that the shape and area $A_{ij}$ of contact are
unaffected by the cell's contacts with other neighbours, and that there are no
complex effects like hysteresis. This is purely for convenience and
simplicity: other authors do take into account hysteresis, or use Voronoi
contact areas \cite{galle2005modeling,drasdo2005single}. However, we are not
interested in the effect of hysteresis \emph{et cetera} on cell motility, so
this will suffice.

Call the distance between the two cell centres is $d$, and define

\begin{equation}
\label{eq:epsilon}
    \epsilon = R_i + R_j - d
\end{equation}

and

\begin{equation}
    R = \left( R_i^{-1} + R_j^{-1} \right)^{-1}
    \label{eq:harmr}
\end{equation}

to be the overlap $\epsilon$ and harmonic mean radius $R$. The contact area $A_{ij}$ is then given by

\begin{equation}
    A_{ij}(\epsilon) = \pi r_{ij}^2(\epsilon) =
    2\pi R\epsilon -\frac{\pi}{4}\epsilon^2
\end{equation}

and so the adhesion interaction potential is 

\begin{equation}
    U^\mathrm{(adh.)}_{ij}(\epsilon) =-\gamma_{ij} A_{ij}
    = -\gamma_{ij}\left( 2\pi R \epsilon -\frac{1}{4}\epsilon^2\right)
\end{equation}

for an interaction force of

\begin{equation}
    \vec{F}^\mathrm{(adh.)}_{ij} = - \nabla_\epsilon U^{\mathrm{(adh.)}_{ij}}
    = \gamma_{ij} (2\pi R  -\frac{1}{2}\pi \epsilon) \vec{n}
    \label{eq:adhforce}
\end{equation}

where $\vec{n}$ is the unit vector pointing from the centre of $i$ to that of
$j$.

\subsubsection{Hertzian repulsion}
\label{sec:hertz}

In addition to adhesive forces, there must also be repulsive forces due to the
cells' elastic resistance to applied stresses \cite{landau1986theory}. When any two elastic bodies are
brought into contact, they must be deformed at least slightly, and this
requires mechanical work. The simplest such model of the work required to
bring two spheres closer than possible in their undeformed state is Hertzian
repulsion:

\begin{equation}
    U^\mathrm{(rep.)}_{ij} = \frac{2 E^\star_{ij} R^{1/2}}{5} \epsilon^{5/2} 
\end{equation}

where $E^\star_{ij} = (E_i^{-1} + E_j^{-1})^{-1}$ is an ``effective'' elastic
modulus accounting for the fact that the more elastic of the two bodies will
do more of the deformation. For example, a rubber ball squashed into a much
more rigid concrete surface will be strained much more dramatically than the
concrete. Because most of the work will go into straining the rubber ball, 
the force required to squash the two together will be determined by
the elastic modulus of the ball, rather than the concrete. 

$R$ is once again the
harmonic mean of the radii of the two cells, given by equation \eqref{eq:harmr}.

With $\epsilon$ defined as above in equation \eqref{eq:epsilon}, 
the repulsive force between $i$ and $j$ due to Hertzian repulsion is 

\begin{equation}
    \vec{F}_{ij}^\mathrm{(rep.)} = - E^\star_{ij} R^{1/2} \epsilon^{3/2} \vec{n}
    \label{eq:hertz}
\end{equation}

where $\vec{n}$ is again the unit vector pointing from the centre of $i$ to
the centre of $j$. One can easily verify that the corresponding reaction force
on $j$ due to $i$ is equal and opposite \cite{landau1986theory}.

\subsubsection{Internal pressure}

In addition to Hertzian repulsion due to elastic deformation, we also include
repulsion due to bulk compression. Even balloons full of any given 
liquid will avoid being forced into one another, despite the lack of
resistance to shear stress. All media, whether elastic or fluid, have a bulk
modulus $B$ that describes their resistance to compression.

The internal pressure $p$ of the cell is

\begin{equation}
    p = B \frac{\Delta V(\epsilon)}{V}
\end{equation}

in terms of the change in volume $\Delta V$ as a function of the overlap
$\epsilon$ from equation \eqref{eq:epsilon},

\begin{equation}
    \Delta V(\epsilon) = \frac{\pi}{4} R \epsilon^2 -
    \frac{\pi}{24}\epsilon^3\;,
\end{equation}

with $R$ as before from equation \eqref{eq:harmr}, and

\begin{equation}
    V = \frac{4\pi}{3} R_i^3 \;.
\end{equation}

The true value of the bulk modulus,
along with Poisson's ratio, is typically unknown for most cell types
\cite{lekka1999elasticity}. 
We take the bulk modulus $B$ to be the same as the elastic modulus of $E$-type
cells,

\begin{equation}
    B = E^\star_{E} = 10 \;\mathrm{kPa}
\end{equation}

corresponding to a Poisson's ratio of $\nu\approx1/3$ \cite{landau1986theory}.

The force corresponding to this internal pressure is, to leading order,

\begin{equation}
    \vec{F}^\mathrm{(press.)}_{ij} = - p A_{ij} \vec{n}
    = -\frac{3\pi}{8} B \frac{R^2 \epsilon^3}{R_i^3} \vec{n}\;,
\end{equation}

which is cubic in overlap $\epsilon$. This means that this term will be
dominated by Hertzian repulsion at $\epsilon\ll R$, only becoming relevant at
extremely dense packings of cells. However, although the internal pressure $p$
is not relevant to mechanical interactions, it does provide a useful way to
track which cells are overcrowded and should not divide.

We will only allow cells with internal pressure $p<p_C =
10^{-5}\;\mathrm{\hbox{\textmu} Pa}$ to undergo cell division. This is to ensure that the
population of cells in the tissue stay at a constant, predictable density:
cell division will be naturally suppressed in regions where the internal
pressure $p$ is too high, but will continue if a cell dies, filling the void
it leaves behind. We will elaborate on this shortly.

\subsubsection{Traction}

In addition to their distinct mechanical and adhesive properties, $M$-type cells
do not divide, and also experience a tractive force $\vec{F}_T$. This force is constant in
time and oriented in the $x$-direction. This is partially for convenience of
calculation, but can be readily interpreted as the motile cells' being
attracted up a concentration gradient, perhaps of oxygen, glucose or another
chemical signal. Such sensing of chemical gradients and migration in more
favourable directions is very well-established
\cite{hoang2013measuring,kowalczyk2005preventing,insall2009actin}.

We assume that the traction force that $M$ cells experience is constant in
magnitude and direction, and equal to

\begin{equation}
    \vec{F}_T = + 94 \;\mathrm{M L\cdot T^{-2}}\; \vec{e}_x,
    \label{eq:traction}
\end{equation}

where $\vec{e}_x$ is a unit vector pointing along the $x$ axis.

As a cartoonish simplification of this process, one can picture the
mechanical effect of the active stresses in the cytoskeleton, both contractile
and extensile forces, as a taut rope connecting the cell to its external
environment under tension $F_{T}$. The actively generated tension in the ``rope'' manifests as a
traction force on the motile cell, $\vec{F}_T$, and a reaction force
$-\vec{F}_T$ acting on the cells microenvironment (see figure \ref{fig:crawling}).

In our simulations, the reaction force is assumed to act on the non-cellular,
rigid substrate on which the cellular monolayer grows and subsists: it does
not explicitly enter our simulations. Some of the limitations of this treatment
of tractive forces are discussed later, in section \ref{sec:discussmech}.

\subsubsection{Cell division and algorithm overview}

\begin{center}
\noindent\fbox{
    \parbox{\textwidth}{
        \begin{tabbing}
            xxxx\=xxxx\=xxxx\=xxxx\=xxxx\=xxxx\=\kill
            tt = 0 \\
            ttmax = 150 days \\
            $\Delta t$ = 1 second \\
            while (tt $<$ ttmax): \\
            \> for each cell: \\
            \> \> if ($p<p_C$) then \\
            \> \> \> grow cell: \\
            \> \> \> cell$\rightarrow R_i = R_i + \Delta t \times
            \mathrm{growth\_speed} $ \\
            \> \> \> if (cell$\rightarrow R_i$ $>$ $R_i^{\mathrm{final}})$ then \\
            \> \> \> \> divide: \\
            \> \> \> \> $R_{new}=R_{i}/2$\\ 
            \> \> \> \> cell$\rightarrow R_i = R_{new}$ \\
            \> \> \> \> newcell$\rightarrow$position =
            cell$\rightarrow$position $+ R_{new}\times$ random-direction \\
            \> \> \> \> newcell$\rightarrow$type $=$ cell$\rightarrow$ type \\
            \> \> if (rand $<$ switching\_probability) then \\
            \> \> \> switch type \\
            \> \> calculate forces: \\
            \> \> force = sum(forces from neighbours) \\
            \> \> move cell: \\
            \> \> mobility = $1/(6\pi R_i)$ \\
            \> \> cell$\rightarrow$position = cell$\rightarrow$position + mobility $\times$ force$\times \Delta t$ \\
            \> tt += $\Delta t$ \\
            \> save velocities and populations to file \\
        \end{tabbing}
    }
}
\end{center}

Each time step, we loop over all cells present, and for each cell implement 
growth, type switching, cell division, resolve the forces on each cell, and move each cell.
Although the choice of the precise order for these is arbitrary, cell type
information and cell radii are needed to specify the forces acting on each
cell, and are most conveniently calculated together.

Cells gradually expand over time until they reach a critical size, at which
point they divide into two cells with half the volume of the parent cell each
if the internal pressure $p$ is not too high. If $p$ is larger than a critical
pressure $p_C$, which we set to $1.0\;\mathrm{M \cdot L^{-1} \cdot{T}^{-2}} =
1 \mathrm{\hbox{\textmu} Pa}$. This ensures that cells only divide when there is plenty of
room for additional cells, avoiding unrealistic ``clumping'' of the kind 
detailed in \cite{jones2012modeling}. Again, we should emphasise that we are
not interested in modelling cell adhesion or division very realistically: we
just want to know how fast $M$-cells can move as their stiffness changes.

We include a small random variability in the threshold for division when the
cell is initialised,
although the expansion of the cell during this interphase
period \cite{MolBioCell} and the sizes and types of the new cells are
deterministic. In this way, the average size of cells falls in a set range, and
the small random variability in sizes at division causes a gradual
desynchronization of cell cycles between cells over time.

This is implemented by gradually increasing each cell's radius $R_i$ by
a constant amount $\Delta R_i$ each time step. When $R_i$ reaches a threshold value
$R_i^{\mathrm{final}}$ set when the cell was initialised, the cell divides.
Both new cells have radii equal to $2^{-1/3} R_i^{(\mathrm{old})}$ to conserve
volume, and the new cell has a threshold

\begin{equation}
    R_i^{\mathrm{final}}=\bar{R}(0.9+0.2\times\mathrm{rand})\;,
\end{equation}

where $\mathrm{rand}$ is a random number in between $0$ and $1$ and $\bar{R}$ is the
specified mean radius. This insures that all cells have a threshold division
size within ten percent of the specified average, and that their sizes all
fall within the same reasonable range. In practice, $\bar{R}$ is set to
$5\mathrm{\hbox{\textmu} m}$ to match typical cell sizes.

After changing the cell's size and performing division if it is large enough
and not overcrowded, we calculate the net force acting
on the cell,

\begin{equation}
    \label{eq:eomfull}
    \vec{v}_i = \zeta_i \left( \sum_j \left( 
    \vec{F}^{\mathrm{(adh.)}}_{ij}+\vec{F}^{\mathrm{(rep.))}}_{ij}
    +\vec{F}^\mathrm{(press.)}_{ij} \right)
    + \delta_{\mathrm{T}_i,M} \vec{F}_T \right)
\end{equation}

where the sum is taken over all neighbours in contact with the cell $i$ and
$\delta_{\mathrm{T}_i,M}=1$ if the type $\mathrm{T}_i$ of cell $i$ is $M$-type 
and $0$ otherwise. The position $\vec{r}_i$ is then updated to

\begin{equation}
    \vec{r}_i(t+\Delta t) = \vec{r}_i(t) + \Delta t \vec{v}_i
\end{equation}

in accordance with equation \eqref{eq:eomimplicit}, with $\vec{v}_i$ from
equation \eqref{eq:eomfull}.

The determination of which cells should be considered neighbouring cells is
not obvious, given that cells can move continuously around.
One non-trivial optimization to accelerate the time taken to calculate all
possible contact forces acting on a cell is the inclusion of a data structure
that records which cells currently lie in which of a grid of larger boxes
overlaid on top of the simulated area: when calculating the contact forces in
equation \eqref{eq:eomfull}, only cells in the same box as $i$ and neighbouring boxes
are checked. If a pair of cells are found not to be in contact ($\epsilon<0$,
from equation \eqref{eq:epsilon}), then the contact force between the two is zero.

This is a simple example of a bounding volume method: the bounding volume in
all cases consists of the grid square containing the cell and its neighbouring
squares \cite{koziara2005bounding}.

\subsection{Results}
\label{sec:discussmech}


\begin{figure}
\begin{center}
    \includegraphics[width=0.7\textwidth]{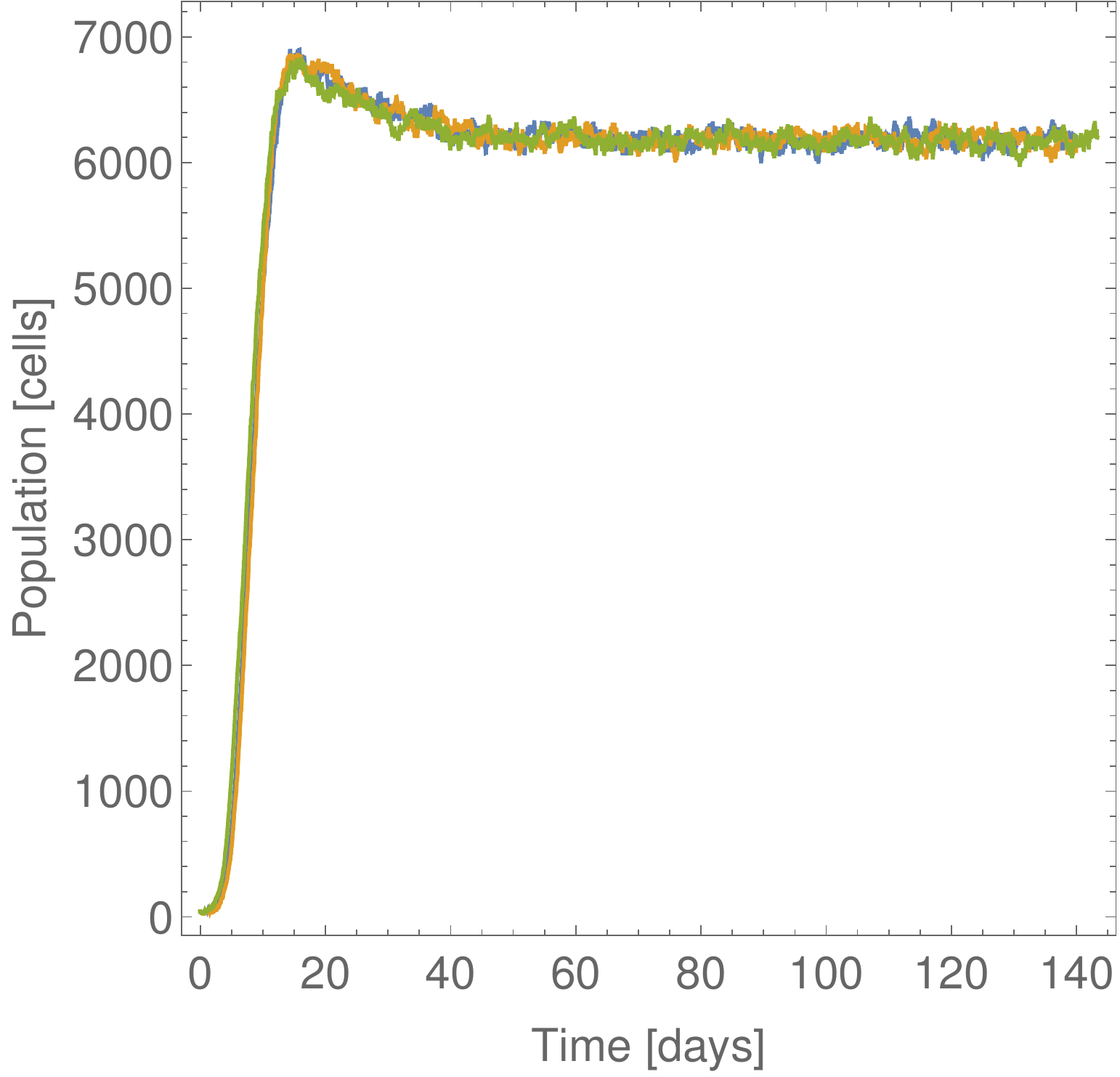}
    \caption{\label{fig:confluence}A plot of the change in population of the
    virtual monolayer, showing the approach to confluence and fluctuations
    around equilibrium population for three realisations with different values of effective
    elasticity of motile cells (blue: $E^{\star}_{MM}=0.0$, yellow:
    $E^{\star}_{MM}=1.0$, green: $E^{\star}_{MM}=10.0$). Note the initial
    overshoot before relaxation to the equilibrium population density, and the
    fact that all three profiles look qualitatively similar. The overshoot and
    approach to confluence is not influenced by the properties of the motile
    cells.}
\end{center}
\end{figure}

The total population of cells tends towards a stable equilibrium after an
initial overshoot at about $20$ days. As is visible in figure
\ref{fig:confluence}, the profile of the population's growth is not affected
by the elasticity of the motile $M$ cells.

\begin{figure}[ht]
    \begin{tabular}{cc}
    A & B \\
    \includegraphics[width=0.5\textwidth]{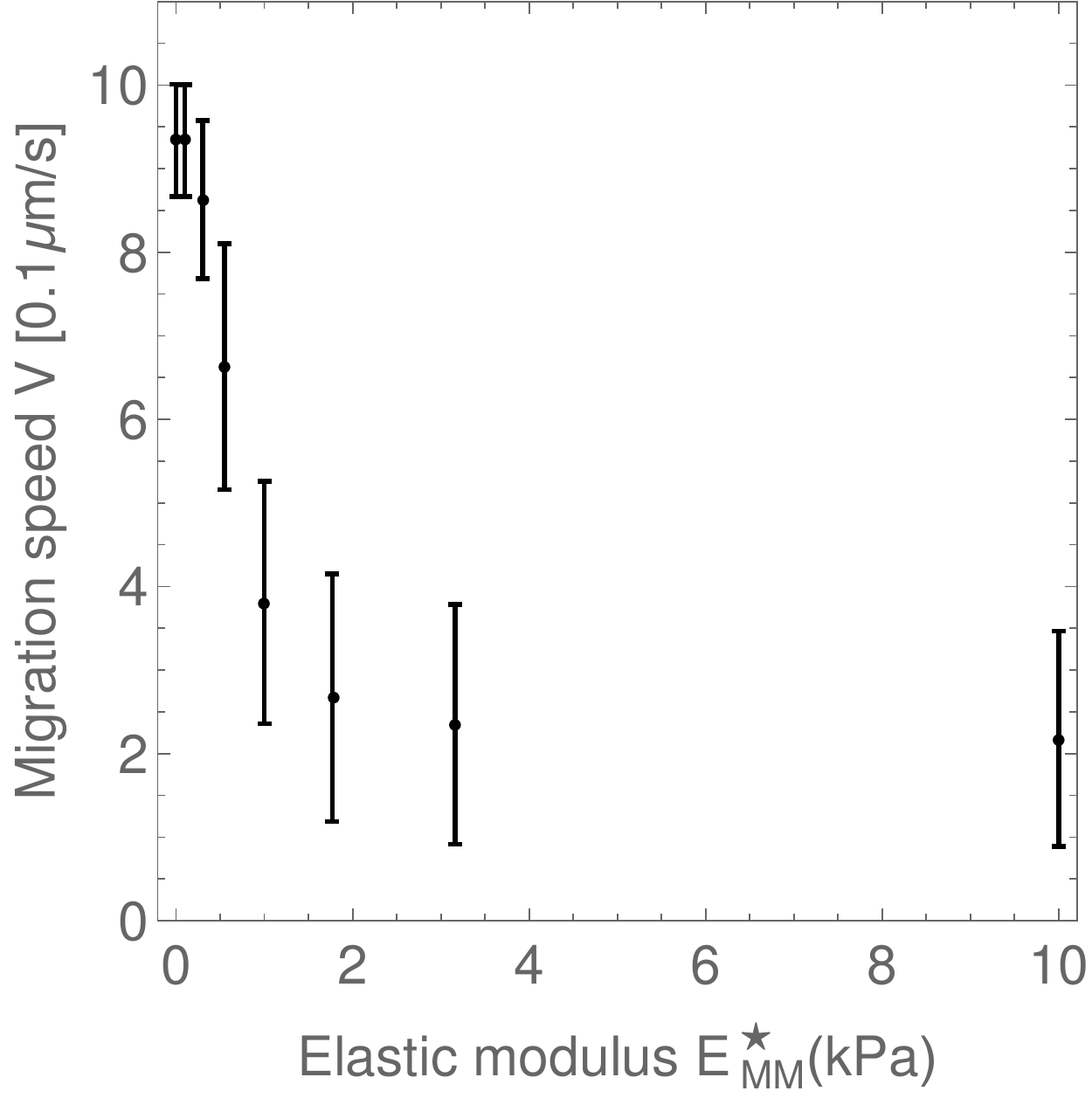} &
    \includegraphics[width=0.5\textwidth]{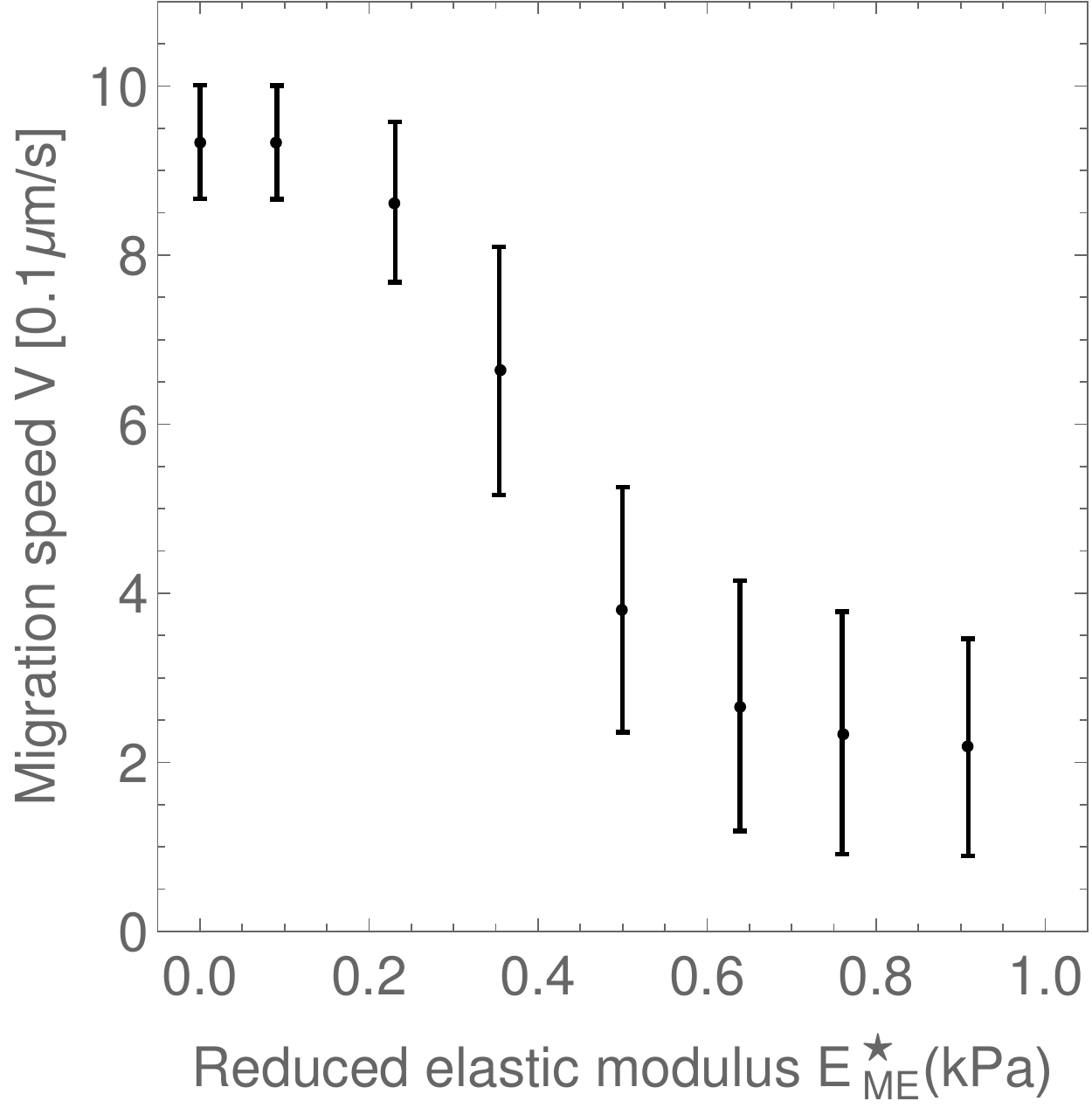} \\
    \end{tabular}
    \caption{\label{fig:vstiff}Plots showing the average speed of motile
    cells as (A) a function of the effective elastic stiffness of motile cells
    $E^{\star}_{MM}$, and (B) a function of the reduced elastic stiffness of
    motile cells $E^{\star}_{ME}$, corresponding to the interaction strength
    between epithelial and motile cells. Note the initial decline in $|\vec{v}|$
    followed by a levelling off at high values of $E^{\star}_{MM}$, and that
    the interaction strength saturates at the stiffness of surrounding tissue, 
    $1.0$ in these units.}
\end{figure}

As can be seen from figure \ref{fig:vstiff}(A), stiffer migrating cells with higher
$E^\star_{MM}$ experience stronger resistance 
to their motion than cells with weaker interactions, and move slower under a constant 
traction force $\vec{F}_T$. 
As the stiffness $E^\star_{MM}$ of the motile cells increases, however, the speed at which
the $M$-cells migrate levels off asymptotically. The underlying reason for this is the fact
that the strength of Hertzian repulsion is determined by the 
more deformable of the interacting materials, as the more elastic material will
deform more and consequently absorb more of the elastic strain energy.
This is basically in line with the ``rubber ball'' argument described in subsection \ref{sec:hertz},
and the dependence of speed on $E^\star_{ME}$ can be seen in see figure
\ref{fig:vstiff}(B). 

In theory, even a perfectly rigid $M$ cell would only see
a limited finite resistance to its motion, as its elastic interactions with
the surrounding tissue would still be finite, and controlled by the magnitude
of $E^\star_{EE}$.

All this strongly suggests that the transition from stiff, slow cells to
flexible and fast cells is controlled by the relative strengths of elastic
interactions between $M$ and $E$ cells via the effective elastic modulus
$E^\star_{ME}$. Although this was not intended to model long-range migration
through the extracellular medium, we can speculate that similar elastic
interactions between the elasticity of cells and the elasticity of the fibrous
extracellular matrix affect the speed of cell motility over much longer
distances. Whether or not this is the case of course depends on experimental
falsification or verification.

One important shortcoming of this model is that cells are assumed to be
internally homogeneous and isotropic continua, and so their mechanical
interactions can be described entirely in terms of the strength of adhesion
and their elastic moduli. However, cells are known to not be internally
homogeneous, and the cell nucleus can have a very different elastic response from
the cytoplasm and cytoskeleton that surround it \cite{thiam2016perinuclear}.
If it is the case that the degrees of freedom of the nucleus are important,
centre-dynamical approaches that take cells to be homogeneous spheres may be
inadequate, and could be extended to include the position and shape of the nucleus as
well. Many of the ``standard'' approaches in the field, even more
sophisticated vertex-dynamical models or those which treat individual cells 
as continuously deformable bodies also ignore nuclear degrees of freedom \cite{jones2012modeling,neilson2011modeling}: 
although there has been some progress in including the mechanics of the
nucleus in lattice models \cite{scianna2013modeling}, we are not aware of any
successful attempts to do so in off-lattice, continuous models.. 

Another shortcoming is that this approach was also very computationally intensive, requiring six months of
runs to produce useful quantities of data (seen in figures
\ref{fig:confluence} and \ref{fig:vstiff}), and being limited to populations
of a few thousand cells in two dimensions rather than the many millions in 
complex three-dimensional configurations that real tumours are
often composed of. Some rather aggressive coarse-graining is necessary to
generalise our results to large and realistic tumour sizes.

Despite these shortcomings, this simple model shows that increases in elastic
stiffness of motile cells can enhance the resistance to motile cell's motion
through epithelial tissue. It also shows that the degree by which motile cells
can be slowed down by elastic interactions is limited by the elasticity of the
surrounding tissue rather than the stiffness of the migrating cell. To answer
what --- if any --- effect these changes in motile speed actually have on the
growth rate of tumours composed of larger ensembles of lesions requires a look
at a much larger length-scale.

\section{Spatial lattice model of local migration}
\label{sec:local}


The off-lattice mechanistic model studied in section \ref{sec:mech} is too
computationally demanding to scale easily to tumours that contain millions of
cells, and in which cells are free to migrate in three spatial dimensions. But
it is on these scales that most clinically relevant tumours exist and grow, so
to understand how changes in the speed of cell motility affect growth, we need
to develop a simpler model that can nonetheless preserve basic insights and
phenomena from the more complex model.

To analyse how migration affects the growth curve of large tumours, we
developed a simpler, lattice model that incorporated cell motility with an 
adjustable migration speed $|\vec{v}|$, kept the same model of changes in cell type as
in \ref{sec:EMT}, and which included some degree of randomness in the control
of cell motility. We explicitly abstract away details such as the mechanical
properties of cells and cell adhesion, and treat cell division and motility as
stochastic processes.

\begin{figure}[ht]
\begin{center}
    \includegraphics[width=0.4\textwidth]{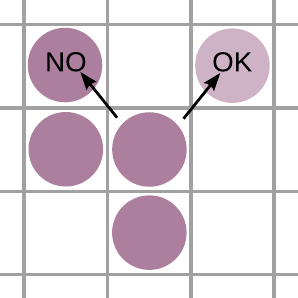}
    \caption{\label{fig:edenrule}A figure showing the basic growth rule in the
    Eden model of growth. With a probability $b\Delta t$, the cell in the
    centre of the diagram attempts to divide, choosing a neighbouring lattice
    site at random. The upper-left lattice site (marked NO) is occupied, so an attempt to
    divide here will fail. The upper-right lattice site (marked OK) is unoccupied, 
    so an attempt to divide here will succeed, and a new cell (light colour)
    will be placed here.}
\end{center}
\end{figure}

In this lattice model, cells are particles which occupy single sites on a
three-dimensional cubic lattice. Cells can be of type $E$ or type $M$, as in
section \ref{sec:EMT}, and transition between the two types stochastically
with forward and reverse rates $k_{EMT}$ and $k_{MET}$ respectively.
The dynamics of growth and migration are treated with a combination of Eden-model
growth for $E$-type cells and a persistent random walk to migration for $M$-type
cells \cite{eden1961two,angelani2015run}. In the Eden model of growth, a given cell
will attempt to divide at a stochastic rate $b$, by attempting to place a new
cell at a neighbouring lattice site chosen uniformly at random. The division
is only successful if the chosen lattice site is free: otherwise, the cell
does not divide this turn (see figure \ref{fig:edenrule} for an illustration). 
Only type $E$ cells are able to divide.

An $E$-type cell that is completely surrounded will effectively not divide at
all, whereas one which is not entirely surrounded will divide at a
progressively slower effective rate as its neighbours become full
\cite{eden1961two}. In this model of growth, constraints from overcrowding of
cells naturally emerge from a very simple set of rules. The growth of large
tumours is naturally slower than exponential, and tumours are roughly
spheroidal (see figure \ref{fig:visualisation} in the results section).

Cells of type $M$ do not divide, but are motile. Each motile cell has the same
characteristic speed $|\vec{v}|$ and an individual \emph{polarity} $\vec{P}$, corresponding to
its instantaneous direction of motion. The polarity of a cell corresponds to 
the direction of one of its neighbouring sites on the lattice. Each time step $\Delta t$, the $M$-cell
attempts to move $V\Delta t$ times by hopping to the next lattice site in its
direction of motion: if the lattice site it is attempting to
move to is occupied, it stays still. These cells also ``tumble'' with a rate
$\alpha$, meaning that there is an expected probability to tumble of $\alpha\Delta t$ 
per time step $\Delta t$. With a probability of $\alpha/V$ per hop, cells
tumble, meaning that they choose a new polarity from the set of possible
neighbouring lattice sites.

This probability to tumble per hop $\alpha/V$ can also be expressed as an
average length $L=V/\alpha$ between tumbling events.

This model of the motion of an active particle belongs to the
``run-and-tumble'' class of models \cite{angelani2015run}. Although more commonly used as models of
bacterial locomotion \cite{angelani2015run,tailleur2008statistical}, 
here we are using it to describe invasive mesenchymal cancer cells. Many of
the details of the precise dynamics of how cancer cells move through tissues
in reality is unknown: our use of this type of model is intended to capture the idea that
cells move with some characteristic speed $|\vec{v}|$, and the environment
surrounding them is heterogeneous, which causes their motion to gradually
decorrelate. One plausible justification for why cells might ``tumble'' in
this way would be that they are following shallow chemical gradients (\emph{chemotaxis} \cite{neilson2011modeling}) 
which fluctuate over time, causing the cell's direction of motion to wander.

\begin{figure}[ht]
\begin{center}
    \includegraphics[width=0.6\textwidth]{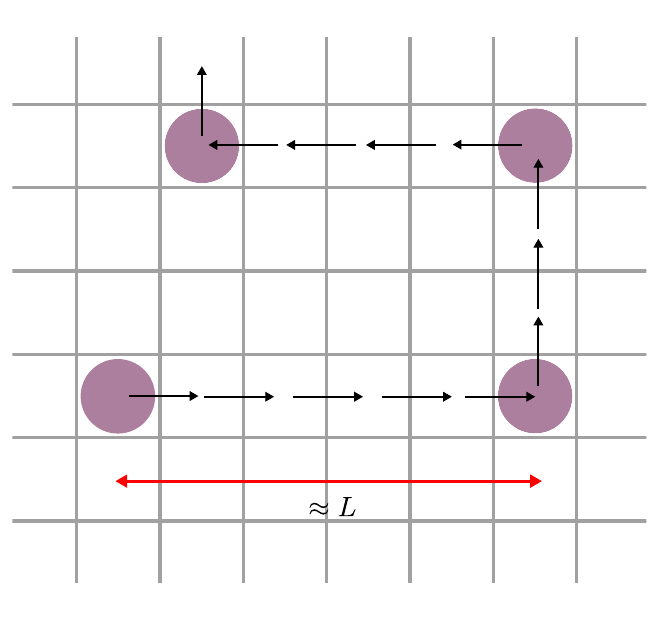}
    \caption{\label{fig:tumble} Cartoon of cell motility in the ``run-and-tumble''
    lattice model. A motile $M$-type cell has a ``polarity'' or instantaneous
    direction of motion, marked with an arrow, and proceeds in this direction
    at a speed $|\vec{v}|$. This is modelled by having the cell move $V\Delta t$ hops
    per time step $\Delta t$. Each hop, there is a probability $\alpha/V=1/L$ (see main text) that the
    cell will tumble, re-orienting so that its polarity points in in a
    different, randomly chosen direction.}
\end{center}
\end{figure}

We measured the total population of cells of both types over time, running
multiple replicates for different values of the type transition rates $k_{EMT}$ 
and $k_{MET}$, tumbling rate $\alpha$, and speed $|\vec{v}|$. Our expectation at the
outset was that the rate of tumbling $\alpha$ and the speed of migration $|\vec{v}|$
would both have a relatively weak effect on the observed growth curves of the
tumour, on the basis that neither of these should affect the cell division
rate $b$ or the population of motile cells, which from \ref{sec:EMT} we
expected to be controlled by $k_{EMT}$ and $k_{MET}$.

\subsection{Algorithm}

\begin{center}
\noindent\fbox{
    \parbox{\textwidth}{
        \begin{tabbing}
        xxxx\=xxxx\=xxxx\=xxxx\=xxxx\=xxxx\=xxxx\=xxxx\=xxxx\=\kill
        tt = 0 \\
        ttmax = 150 days \\
        $\Delta t$ = 1 second \\
        while (tt $<$ ttmax): \\
        \> for each cell: \\
        \> \> if (type \=\= $M$) then \\
        \> \> \> for (i=0, i$<$speed$\times \Delta t$, i++) \\
        \> \> \> \> \> newposition = position + polarity \\
        \> \> \> \> \> if newposition is empty then \\
        \> \> \> \> \> \> position = newposition \\
        \> \> \> \> if (rand $<$ tumbling\_probability) then \\
        \> \> \> \> \> polarity = random\_neighbour \\
        \> \> \> if (rand $<k_{MET}\Delta t$) then \\
        \> \> \> \> change type to $E$ \\
        \> \> \> \> polarity = $(0,0,0)$ \\
        \> \> else if (type \=\= $E$) then \\
        \> \> \> if (rand $< b\Delta t$) then \\
        \> \> \> \> try to divide to random neighbour: \\
        \> \> \> \> x = position + random\_neighbour \\
        \> \> \> \> if (x is empty) then \\
        \> \> \> \> \> add new cell at x \\
        \> \> \> if (rand $<k_{EMT}\Delta t$) then \\
        \> \> \> \> polarity = random\_neighbour \\
        \> tt $+= \Delta t$ \\
        \> if (tt mod 10 $==$ 0) then \\
        \> \> save populations to file \\
        \end{tabbing}
     }
}
\end{center}

A lattice sites' neighbours consist of the sites in an octahedral
neighbourhood surrounding the site, a choice known in the cellular automata
research community as a three dimensional von Neumann neighbourhood
\cite{breukelaar2005using}. Polarities of $M$-type cells are represented a
position vector $\vec{P}$ which points to one of these neighbours, and
corresponds to which neighbour the cell will attempt to hop to next. For
example, a cell with $\vec{P}=(0,0,1)$ will attempt to move in the
$z$-direction.

Each time step, the set of all cells, which consisted of their positions,
types, and polarities, is iterated over. For simplicity, cell type is not
stored separately from polarity. Instead, non-motile $E$-type cells are
considered to have a polarity of $\vec{P}=(0,0,0)$, with any other polarity
indicating that the cell is of type $M$.

If the cell is type $M$, it attempts to hop to a new position $V\Delta t$
times per time step, succeeding only if the new lattice site is unoccupied.
With probability $\alpha/V$ per hop, the cell tumbles, adopting a new polarity
from the set of von Neumann neighbours with uniform probability each.
Finally, with a probability $k_{MET}\Delta t$ it switches to an $E$-type 
cell, with $\vec{P}=(0,0,0)$.

If the cell is instead type $E$, then with probability $b\Delta t$, the cell
attempts to divide. A lattice site at one of this cell's neighbours is chosen,
and if free, becomes occupied by a new $E$-type cell. The cell division
otherwise fails.

Every $10$ time steps, the elapsed time, total size $n$, numbers of non-motile and motile
cells, and instantaneous growth rate (additional cells per time step, $\Delta n/\Delta t$) 
are calculated. 

The mode of growth can be determined by plotting the instantaneous growth 
rate against the total population: that is, what does $\Delta n/\Delta t$ 
look like as a function of total size $n$? We did not assume that growth 
would take a particular form, and are interested in how the resulting 
growth curve emerges from the underlying rule set. If the growth is
exponential, we would expect $\Delta n/\Delta t$ to be proportional to $n$.
Otherwise, in the case that growth can only occur on the surface and the
tumour is spheroidal, we might expect

\begin{equation}
    \Delta n/\Delta t \propto n^{2/3} \;,
\end{equation}

instead.

\subsubsection{Choice of data structures}

During initial tests of the simulations, it was found that one of the slowest
steps of the algorithm was the procedure in which a cell's neighbouring sites
were checked for the presence of additional cells, which was performed during
both division and cell movement. This test is possibly related to the
collision detection problems encountered in the off-lattice model, but a
different way around the problem was found.

Although it would be the fastest in terms of algorithmic complexity, it would
have been far too memory intensive to store every possible lattice site,
including empty ones. This meant that in early versions of the simulations, a
cell's neighbouring lattice sites were not explicitly stored. Instead, all information on cells was
stored in a vector containing the position, type and polarity of a cell. New
cells would always be added to the end of this vector, which necessitated that
tests to find a cell at any given lattice point had to iterate over the
whole vector of cells, which took $\mathcal{O}(n)$ to achieve for a population
of $n$ cells. This
significantly slowed down the simulations, as similar checks would have to be
repeated for a constant fraction of all $E$-type cells present.

Attempted solutions included various ways to store and access every cell's
set of neighbouring lattice sites directly, but these ran into the problem
that memory requirements were greatly increased: those solutions which managed
to reduce the time complexity to $\mathcal{O}(1)$ tended to have a much larger
memory requirement of at least $\mathcal{O}(n)$, strongly limiting the maximum
size of tumours that could be simulated effectively. While related approaches
may be able to solve the problem, this was eventually abandoned in favour of a
compromise solution which reached a somewhat better time complexity without
severely increasing memory requirements.

The solution which I eventually settled upon was to store information about
all present cells in an associative array
that took a cell's position as a key and
returned the cell's type/polarity --- but \emph{only} if a cell was present 
at that location, as the position would otherwise be absent from the array. As the
structure underlying maps is a binary tree, operations involving search,
lookup, addition and removal are all of order $\mathcal{O}(\mathrm{log}(n))$ in
time. This method did not necessitate the storage of every cell's neighbour's,
which would have exacerbated memory requirements, but nonetheless allowed an
appreciable speed-up \cite{mehlhorn2007algorithms}.

Whether or not a cell was present at any given position could be determined by 
a $count$ operation, which counted the number of entries in the map with a
given key (in this case, a given position). If no cell was present at a given 
position, the $count$ operation would return $0$, and movement or division
could proceed to move or add a cell at the tested site.

Since all of the $count$, lookup, addition and removal operations have time
complexity $\mathcal{O}(\mathrm{log}(n))$, the division and motility
sub-procedures in the algorithm have the same time complexity, and iterating
through all cells in one time step in the algorithm results in a total time
complexity of $\mathcal{O}(n \mathrm{log}(n))$.

\subsection{Results}

\begin{figure}[ht]
\begin{center}
    \includegraphics[width=0.8\textwidth]{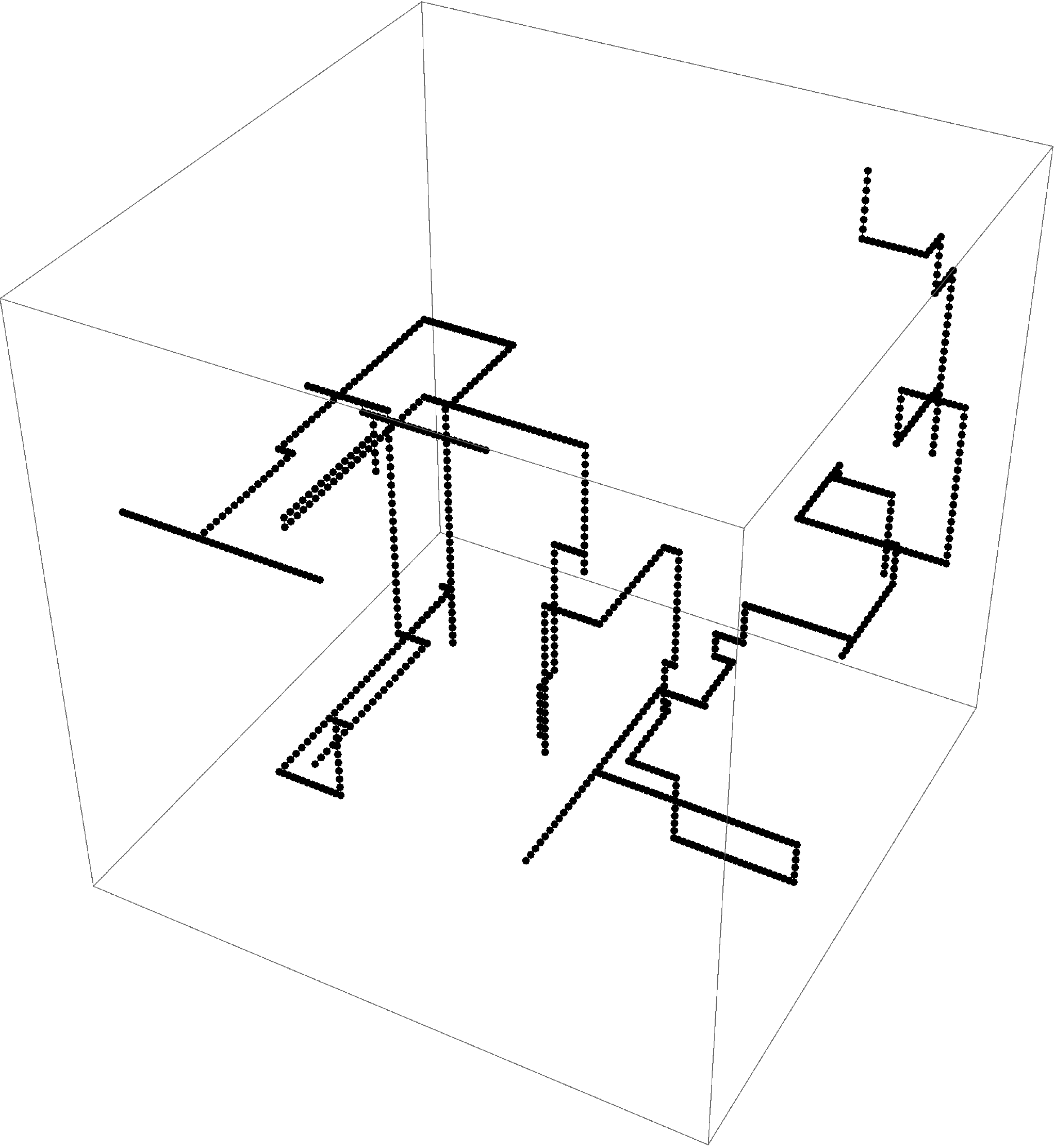}
\end{center}
    \caption{\label{fig:wander}A trajectory of a single motile cell with tumbling rate 
    $\alpha=0.1\;\mathrm{day}^{-1}$ and speed $V=1\;\mathrm{sites}\cdot\mathrm{day}^{-1}$, 
    corresponding to mean run length $L$ of $10$ sites. There was no switching to
    the non-motile cell type or growth dynamics.}
\end{figure}

\subsubsection{Single motile cell}


Run-and-tumble models are closely related to the Telegrapher's equation
\cite{goldstein1951diffusion,tailleur2008statistical,angelani2015run}. The
probability distribution $\rho(x)$ for the position $x$ of a
particle undergoing a persistent random walk in one spatial dimension
with speed $|\vec{v}|$ and tumbling rate $\alpha$ obeys the partial differential equation

\begin{equation}
    \frac{\partial^2 \rho}{\partial t^2} 
    + \alpha \frac{\partial \rho}{\partial t}
    = V^2 \frac{\partial^2 \rho}{\partial x^2}\;,
    \label{eq:tele}
\end{equation}

as detailed in \cite{goldstein1951diffusion}. While it would seem that a
one-dimensional random walk won't provide a good description of a particle
moving through a three-dimensional space, our motile cells are not allowed to
take on arbitrary orientations. The choice of von Neumann neighbourhood means
that they are forced to move along the axes of the lattice (see figure
\ref{fig:wander}) and the random walk is not isotropic.

Without solving it explicitly, as long as the boundary conditions

\begin{equation}
    \lim_{\|x\|\rightarrow \infty}\rho = 0
\end{equation}

and

\begin{equation}
     \lim_{\|x\|\rightarrow \infty}\frac{\partial \rho}{\partial x} = 0
\end{equation}

hold, and $\rho$ is normalized so that

\begin{equation}
    \int \rho dx = 1\;,
\end{equation}

then we can define the mean-squared displacement to be

\begin{equation}
    \langle x^2 \rangle = \int x^2 \rho d x
\end{equation}

and it follows from the application of integration by parts to equation \eqref{eq:tele}
that $\langle x^2 \rangle$ obeys

\begin{equation}
    \frac{\partial^2 \langle x^2 \rangle}{\partial t^2}
    + \alpha \frac{\partial \langle x^2 \rangle}{\partial t}
    = 2 V^2\;.
\label{eq:msqdisp}
\end{equation}

With the initial condition that, if the cell's initial position is already
known, $\langle x^2 \rangle = 0$ when $t=0$, it follows that

\begin{equation}
    \langle x^2 \rangle(t) = \frac{2
    V^2}{\alpha^2}\left(1-e^{-\alpha t}\right) + \frac{2 V^2}{\alpha} t
    \label{eq:msqdisp2}
\end{equation}

solves equation \eqref{eq:msqdisp}. This provides a theoretical mean-squared
displacement to which we can compare the observed mean-squared displacement of
an individual motile cell \cite{goldstein1951diffusion}.

Two phenomena are clear from equation \eqref{eq:msqdisp2}: on time-scales $t$ much
shorter than $\alpha^{-1}$, the mean-squared
displacement is quadratic, indicating that the motile cell is
initially moving ballistically in a straight line at constant speed. Due to
tumbling, on time-scales much longer than $\alpha^{-1}$, $\langle x^2 \rangle$
slows down to growing linearly in time $t$. This indicates that after tumbling
has set in, the cell begins to move diffusively through the available space,
approximating a random walk.

\begin{figure}[ht]
    \begin{center}
        \includegraphics[width=0.6\textwidth]{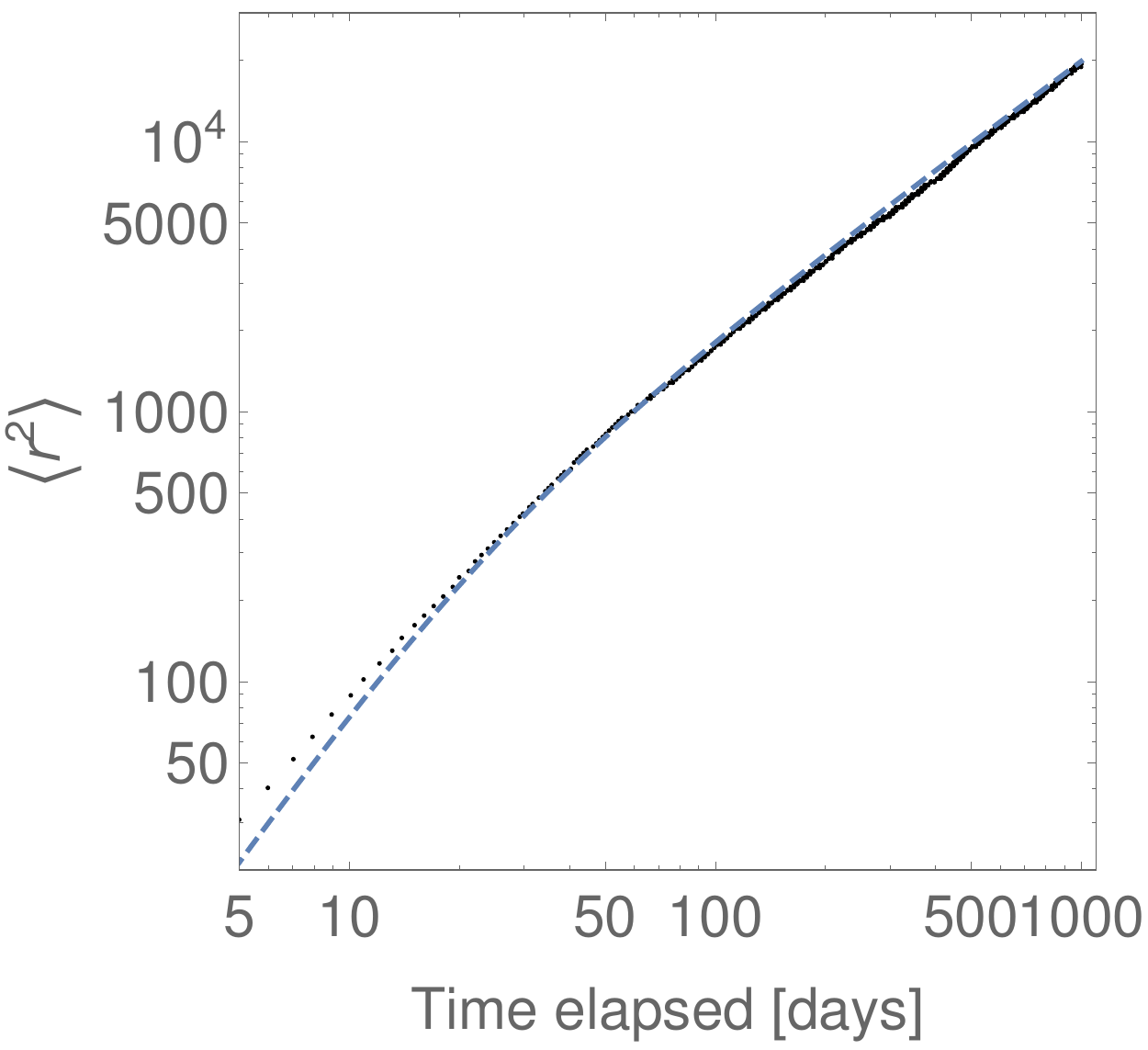}
    \end{center}
    \caption{\label{fig:msqd}Plot of the mean square displacement averaged over 1000
    replicate simulations of the motion of a single
    motile cell with the same parameters as above (black dots) compared to the theoretical mean
    square displacement from equation \eqref{eq:msqdisp2} (blue dashed line). Parameter values for both 
    the simulations as theoretical curve were $V=10\; \mathrm{sites/day}$ and $\alpha=0.1$ $\mathrm{day}^{-1}$. 
    Note the initial quadratic increase followed by a later diffusive tendency.}
\end{figure}

As can be seen from figure \ref{fig:msqd} showing the mean square displacement
averaged over a thousand replicate simulations of a single $M$ cell, this
classical theory shows a very good agreement with numerical results.

\subsubsection{Aggregate tumour}



\begin{figure}[ht]
\begin{center}
    \includegraphics[width=0.7\textwidth]{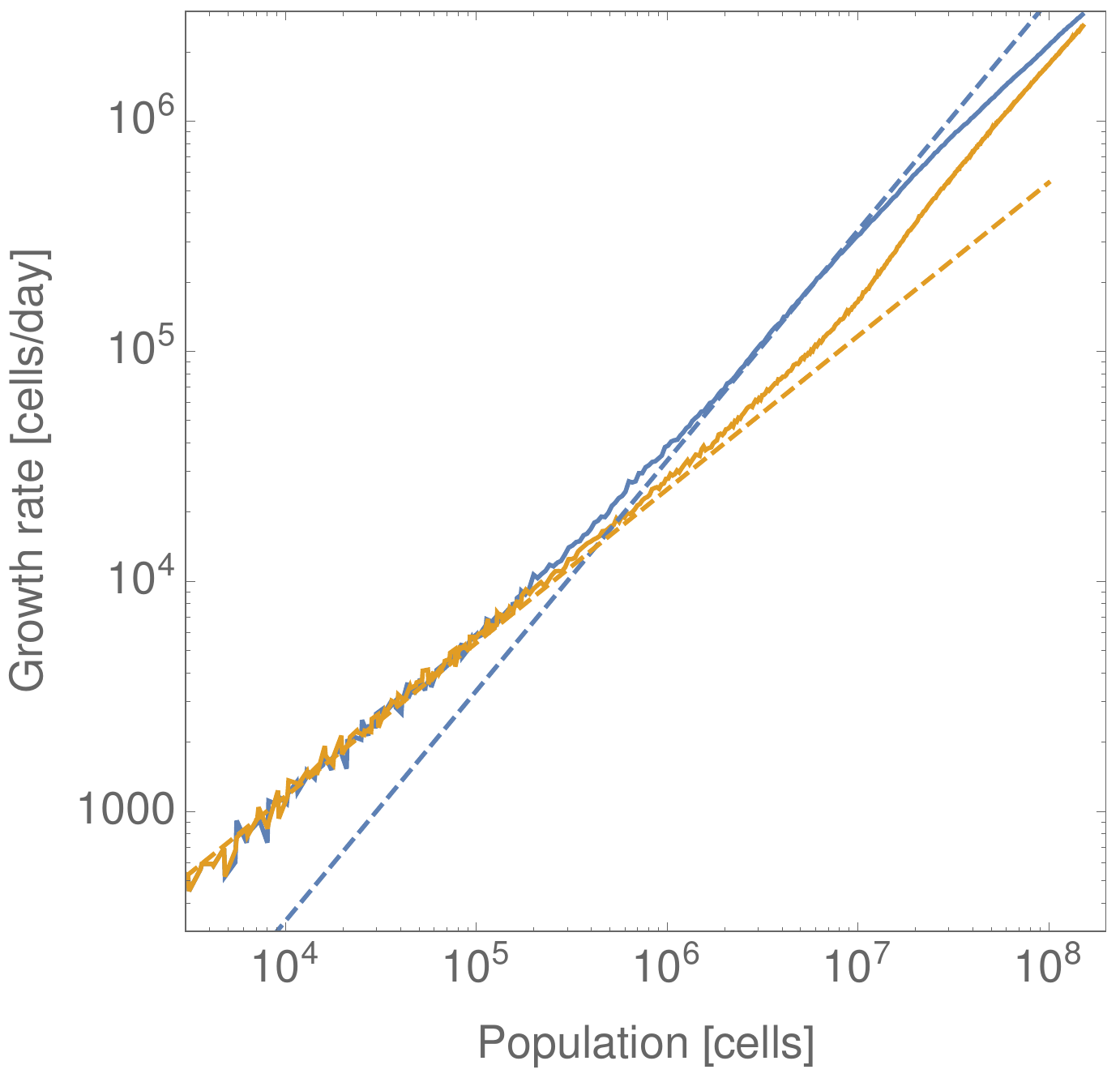}
    \caption{\label{fig:RTMgrowth} A plot of the instantaneous growth rate
    (rate of cell division in cells per day) against the tumour size (in
    cells) for two different values of tumbling rate in the run-and-tumble
    model, $\alpha=0$ day${}^{-1}$ in blue and $\alpha=0.5$ day${}^{-1}$ in
    yellow. Theoretical relationships for exponential growth (growth rate
    $\propto$ size, steep dashed curve) and surface growth (growth rate
    $\propto$ size${}^{2/3}$, shallow dashed curve) are plotted for
    comparison, showing that in both cases surface growth initially dominates,
    before the tumour grows faster than for a single spheroidal lesion for a
    short period of time, before slowing down again. It is not clear that
    the tumour ever grows ``truly'' exponentially in this model, although
    growth is faster than for a single spheroid without migration. Other
    parameter values used were $k_{EMT}=10^{-2}\;\mathrm{day}^{-1}$,
    $k_{MET}=10^{-4}\;\mathrm{day}^{-1}$ and $b=1.0\;\mathrm{day}^{-1}$.}
\end{center}
\end{figure}

Plots of $\Delta n/\Delta t$ against $n$ for different values of tumbling rate
$\alpha$, $|\vec{v}|$, $k_{EMT}$ and $k_{MET}$ all show that initially, tumours grow
slower than exponentially, with $\Delta n/\Delta t \propto n^{2/3}$. This can
be easily interpreted as growth being dominated by a single lesion which grows
only on its surface, in accordance with other results for the Eden model
\cite{eden1961two}. After some length of time controlled by the tumbling rate
$\alpha$, growth accelerates for a while before settling down again to a
similar dependency  $\Delta n/\Delta t \propto n^{2/3}$, albeit with a larger
prefactor (see figure \ref{fig:RTMgrowth}).

\begin{figure}
\begin{center}
    \includegraphics[width=0.6\textwidth]{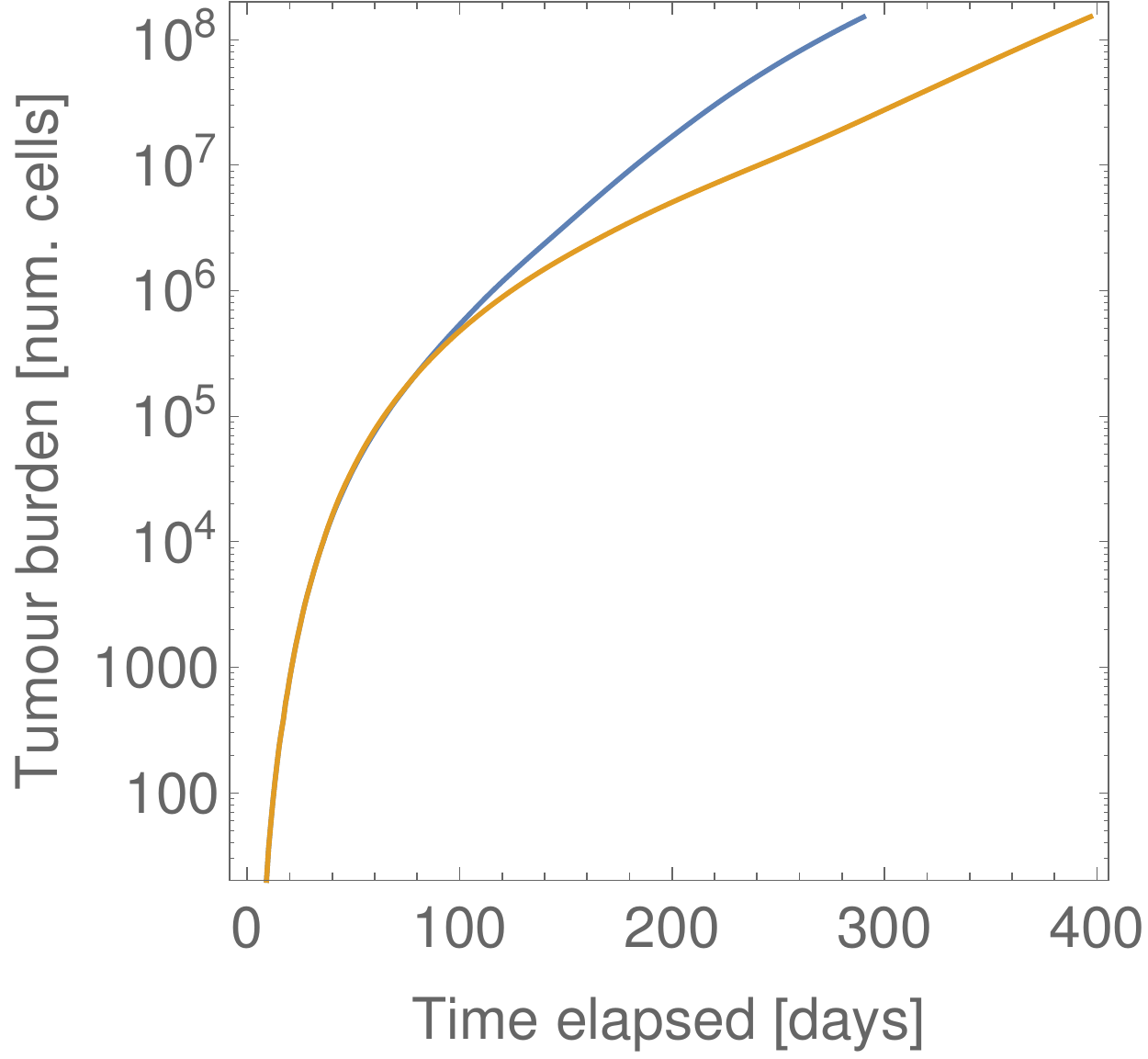}
\end{center}
    \caption{\label{fig:threephase}A figure showing the effect of changes in the
    mesenchymal-epithelial transition rate $k_{MET}$ on the growth of the volume of the ensemble
    over time, in both cases 
    with $k_{MET}=0.1\;\mathrm{day}^{-1}$ 
    (blue curve) and $k_{MET}=10^{-3}\;\mathrm{day}^{-1}$ (lower, yellow curve): initial 
    sub-exponential growth characterised by one primary, followed by a cascade 
    of micro-metastases driving exponential growth, and in the case with 
    $k_{MET}=0.1\;\mathrm{day}^{-1}$, later slow-down  and return to
    sub-exponential growth caused by coalescence of 
    distinct lesions. Other parameters were set to 
    tumbling rate $\alpha=0.01\;\mathrm{day}^{-1}$, rate of EMT
    $k_{EMT}=10^{-4}\;\mathrm{day}^{-1}$, speed
    $V=30\;\mathrm{sites}\cdot\mathrm{day}^{-1}$ and birth rate 
    $b=1.0\;\mathrm{day}^{-1}$.
    }
\end{figure}

\begin{figure}[ht]
    \begin{center}
        \includegraphics[width=0.8\textwidth]{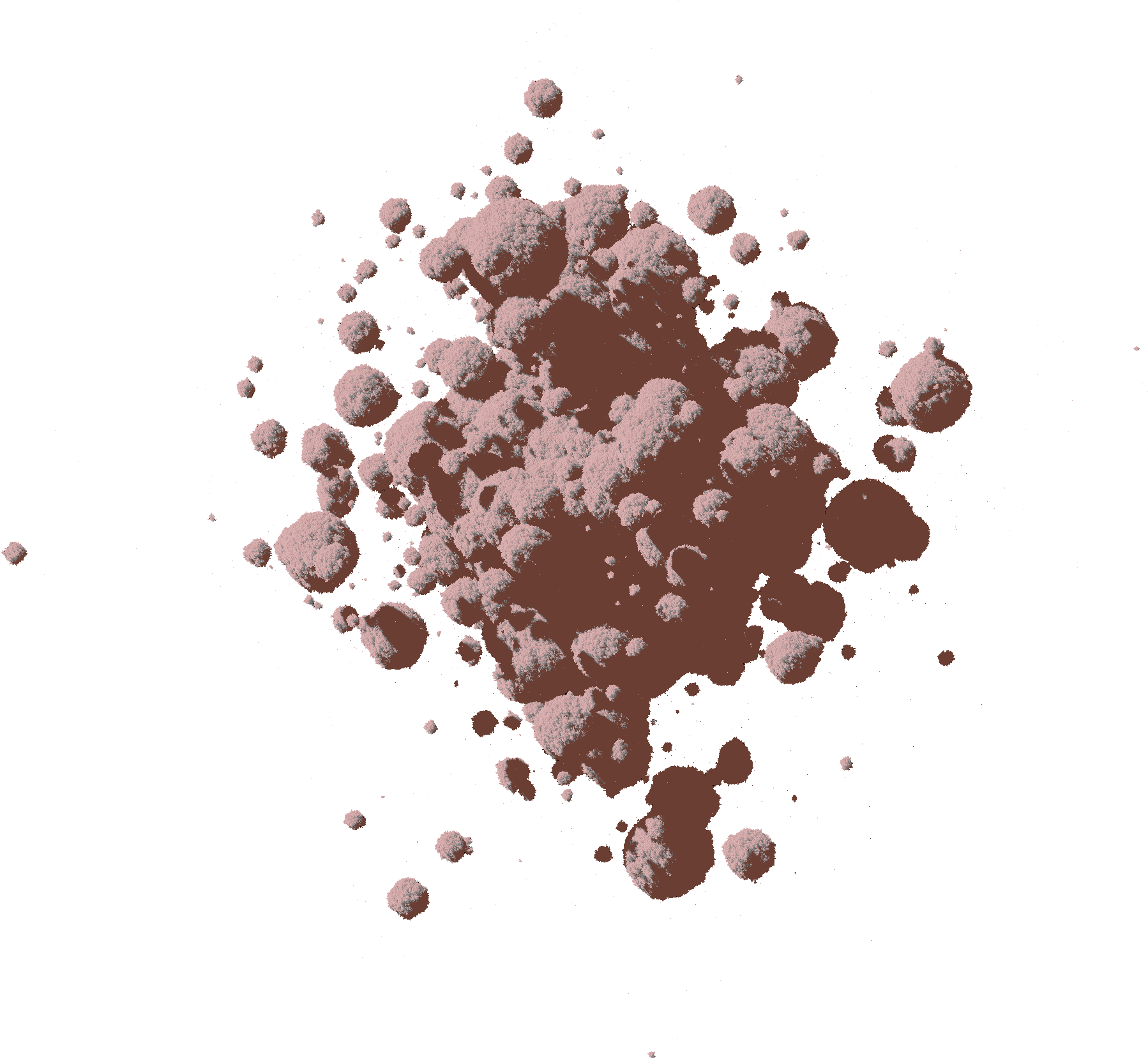}
    \end{center}
    \caption{\label{fig:visualisation}A visualisation of the positions of cells in the ensemble of
    lesions, showing the emergent cascade of migration and new growths that the
    model entails. Escaping motile cells can be seen circulating around the larger
    primary tumour and some of its secondaries. Parameter values used were
    $b=1\;\mathrm{day}^{-1}$, $k_{EMT}=10^{-4}\;\mathrm{day}^{-1}$,
    $k_{MET}=10^{-2}\;\mathrm{day}^{-1}$, $V=10\;\mathrm{sites/day}$ and
    $\alpha=0.1\;\mathrm{day}^{-1}$.}
\end{figure}

It is very tempting to try to interpret this brief acceleration in growth as a
period of exponential growth caused by the appearance of secondary lesions as
motile cells switch back to $E$ types and begin dividing again: it certainly 
seems from figure \ref{fig:visualisation} that something like this is
occurring. But it is difficult to pick out a clear and definite ``exponential growth''
regime either on any of the plots
exemplified by \ref{fig:RTMgrowth} in which a simple
proportionality between $\Delta n/\Delta t$ and $n$, or a clear line on a
logarithmic plot of $n$ against elapsed time $t$ as in \ref{fig:threephase}. 
Although growth definitely
seems to be accelerated by migration, as can be seen by increasing $k_{MET}$
(see figure \ref{fig:threephase}), it's fairly clear from elementary calculus that 
one can draw a line tangent to any smooth enough curve and measure the gradient,
so we should be very cautious before interpreting the rapid growth as
``genuinely'' exponential.

Nevertheless, for the purposes of comparing how much growth is accelerated by
changes in the different microscopic processes, it's useful to talk about an
average growth rate. We find this by fitting a linear function 

\begin{equation}
    \frac{\Delta n}{\Delta t} = G  n + c
\end{equation}

to data of the instantaneous growth rate $\Delta n/\Delta t$ against size $n$
by simple linear regression.
The coefficient $G$ \emph{would} correspond to the exponential growth rate
\emph{if} growth was truly exponential. The use of simple linear regression
means that $G$ will be equal to the mean slope of $\Delta n/\Delta t$ against
$n$, so will in some sense be an ``average'' growth rate regardless of the
true form of the growth curve. 

Measuring the average growth rate $G$ for various different values of $k_{MET}$,
$k_{EMT}$, and $|\vec{v}|$ for many different replicates revealed several interesting
dependencies of $G$ on each of these variables. 

\begin{figure}
    \begin{tabular}{cc}
        A & B \\
        \includegraphics[width=0.5\textwidth]{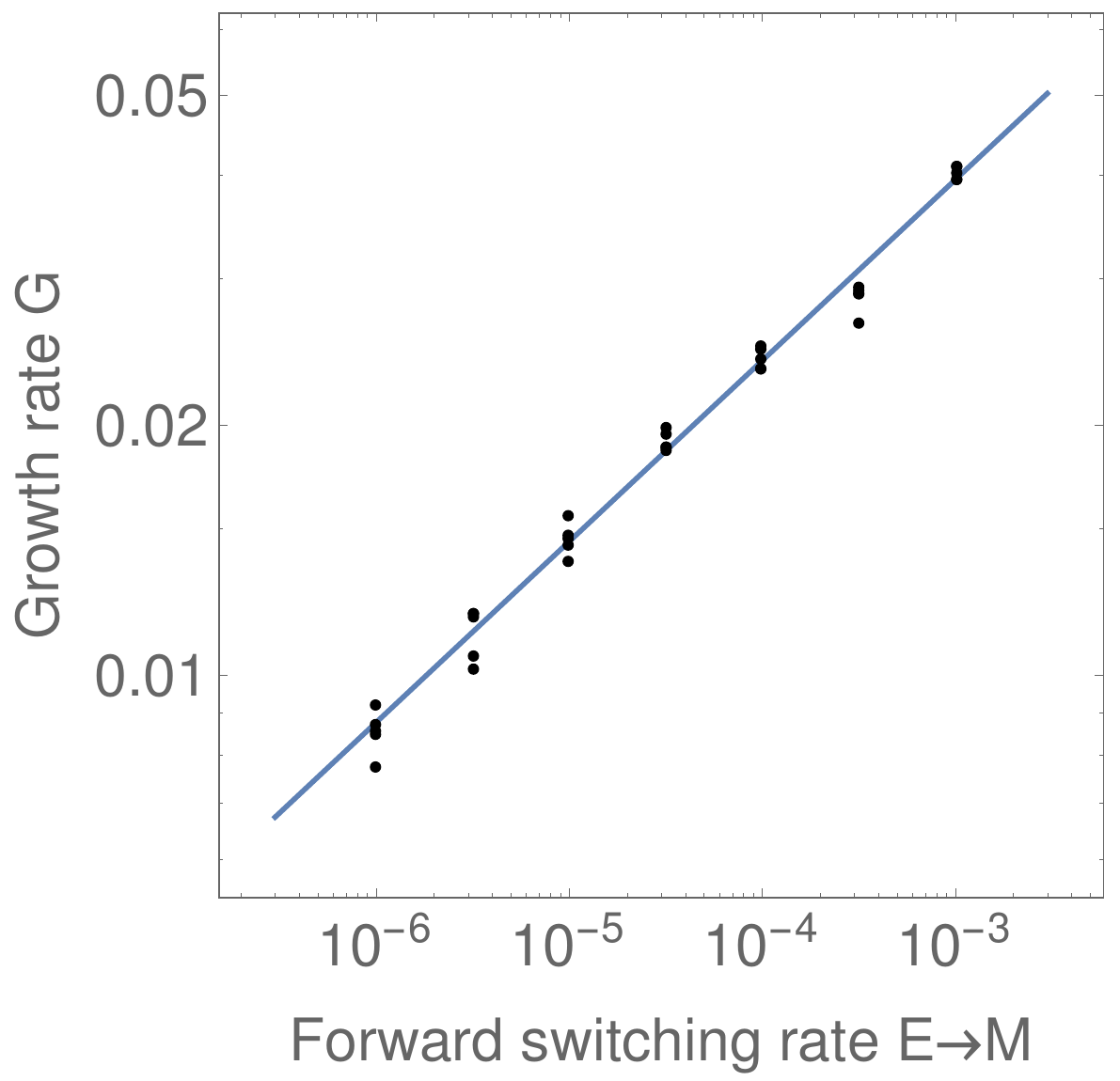} &
        \includegraphics[width=0.5\textwidth]{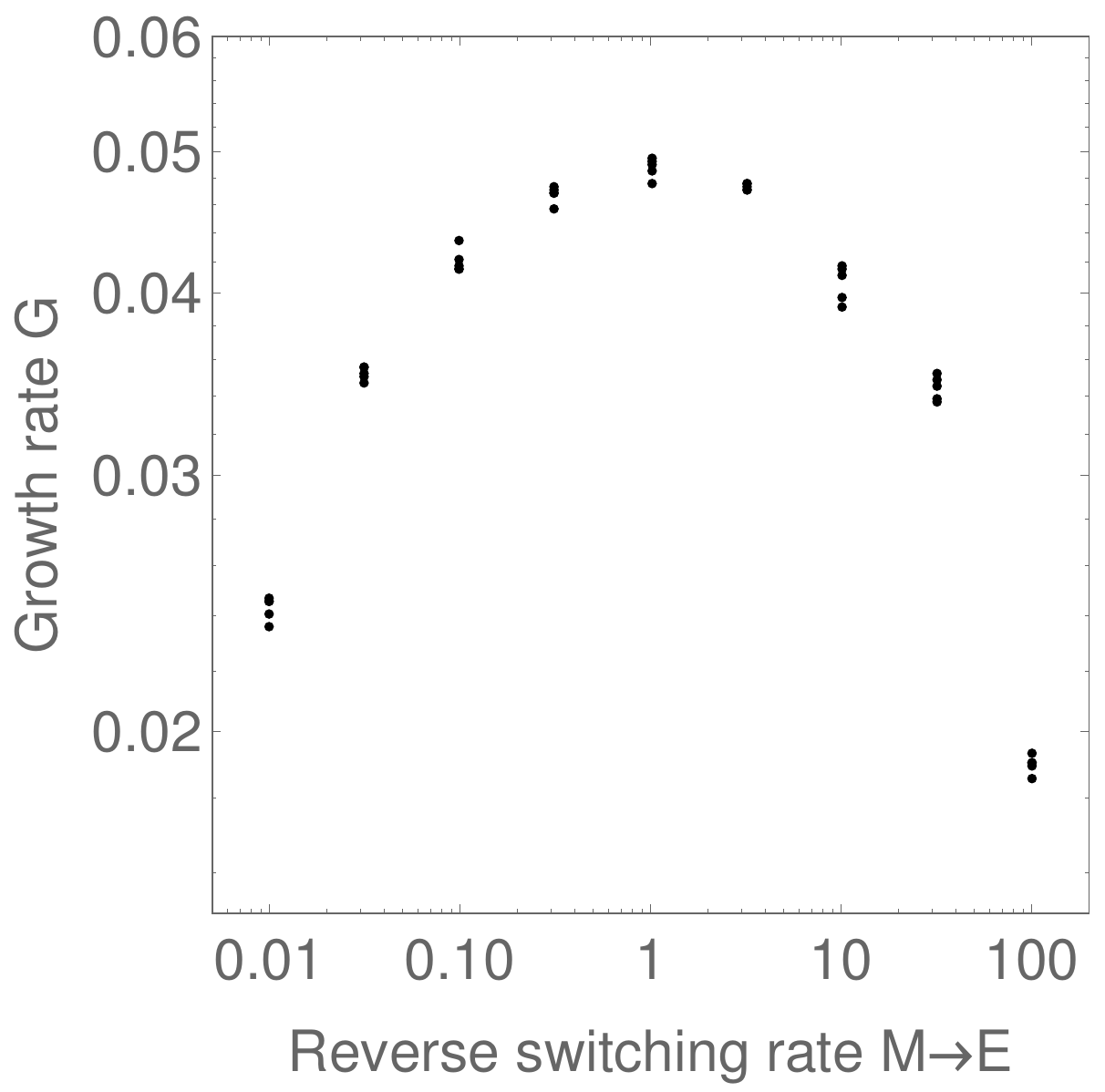} \\
        C &   \\
        \includegraphics[width=0.5\textwidth]{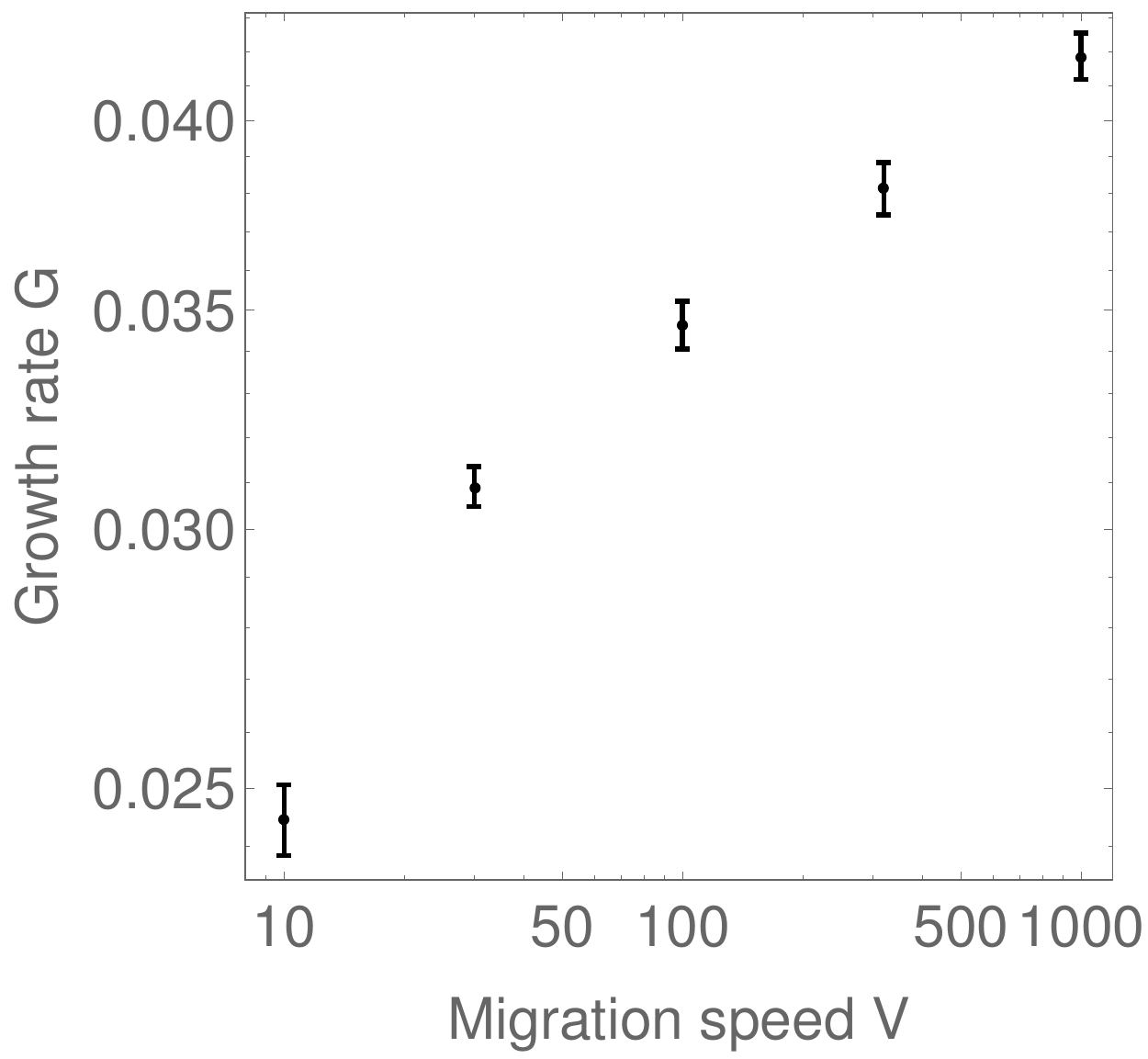} & \\
    \end{tabular}
    \caption{\label{fig:Veffect}
    Plots showing the dependence of the average growth rate on (A) the
    rate of the EMT transition $k_{EMT}$, with an approximate power law
    dependence on $k_{EMT}$ showing the measured average growth rate with
    black points and the power-law of best fit in blue, $G = c k_{EMT}^{0.22}$ %
    (other parameter values were $k_{MET}=0.01$ day${}^{-1}$,
    $V=10$ sites/day, and $\alpha=0.01$ day${}^{-1}$); 
    (B) the rate of the reverse MET transition 
    $k_{MET}$, holding $k_{EMT}=10^{-4}$ day${}^{-1}$ (other parameter values
    were $V=10$ sites/day and $\alpha=0.01$ day${}^{-1}$); 
    and (C) the speed of
    migrating cells $|\vec{v}|$, holding $k_{EMT}=10^{-4}$ day${}^{-1}$ and
    $k_{MET}=0.01$ day${}^{-1}$.
    }
\end{figure}

As can be seen from figure 
\ref{fig:Veffect}(A), increases in $k_{EMT}$ correspond to increases in the
average growth rate $G$. This is not very difficult to understand: from
equation \eqref{eq:fracE}, the proportion of $M$-type cells increases with
$k_{EMT}$. So, as we increase $k_{EMT}$, the number of motile $M$-cells
present also increases, and the rate at which secondary lesions form (see
figure \ref{fig:visualisation}) increases as well.

Somewhat more curious is the
particular power-law dependence of $G$ on $k_{EMT}$: approximately of the form
$G \propto k_{EMT}^p$, with $p=0.22$. Dimensionally, $G$ and
$k_{EMT}$ are both rates with dimensions of $[\mathrm{T}^{-1}]$.
The parameter $k_{EMT}$ is theoretically related to the fraction $P_M$ of motile
cells $M$ (see equation \eqref{eq:fracE}, also chapter \ref{ch:minimal}),

\begin{equation}
    P_M = \frac{k_{EMT}}{k_{EMT}+k_{MET}}\;.
\end{equation}

which is almost simply proportional to $k_{EMT}$ when $k_{EMT}\ll k_{MET}$.
 We can hypothesise that
the growth rate $G$ in these cases is therefore empirically related to $M$ as

\begin{equation}
    G \sim b\;M^p\;.
\end{equation}

To explain why this should be the case requires a more sophisticated analytical approach,
which is developed in chapter \ref{ch:minimal}. In a few special cases of
related models of migration and growth, explicit analytical formulae for the
exponent $p$ can be derived.

Figure \ref{fig:Veffect}(B) shows that despite an initial increase
when $k_{MET}$ is small, $G$ decreases with increasing $k_{MET}$ once 
$k_{MET}$ is higher than about $1\mathrm{day}^{-1}$, which seems to correspond
to the cell division rate $b$ in the model. The reverse switching
rate therefore has two competing effects on $G$: increasing the growth rate
when small, and decreasing it again when it takes on sufficiently large
values. This may be related to the effects that $k_{MET}$ has on the length of
time motile cells remain in the motile phase, and the proportion of all cells
that are motile.

Let's remind ourselves that $k_{MET}$ represents the rate at which motile
cells switch back to being non-motile tumour cells. Slower rates $k_{MET}$
will mean that cells spend longer as type $M$ before transitioning back to
type $E$: slower rates therefore induce a delay between becoming motile and
switching back. This means that when $k_{MET}$ is very low, there will be a
long delay before cells which escape form new lesions elsewhere: this delay is
shorter when $k_{MET}$ is faster. By decreasing the delay between cells
escaping and their founding new lesions, increases in $k_{MET}$ can accelerate growth.

A countervailing effect is that the proportion $M$ of cells which will be motile
will correspond to $k_{EMT}/(k_{EMT}+k_{MET})$, so when $k_{MET}$ is high, $M$ will 
be small (all else being the same). When the reverse switching
rate $k_{MET}$ is much faster than the forward switching rate $k_{EMT}$,
relatively few cells will be motile, and capable of founding new lesions. By
decreasing the relative number of invasive cells, increases in $k_{MET}$ can
therefore also decelerate growth: there are two countervailing effects from
$k_{MET}$, as visible in \ref{fig:Veffect}(B).

The average growth rate seems to be more sensitive to
$k_{EMT}$ than to $k_{MET}$: that is, the rate at which cells become motile
and escape is more important than the rate at which motile cells revert to
tumour cells and form new lesions. In order for cells to be predominantly non-motile, and
thus for a very small proportion of cells to be motile, the reverse rate
$k_{MET}$ must be much larger than the forward rate $k_{EMT}$. This means
cells are removed from the ``pool'' of circulating motile cells very quickly,
after a mean time of around $k_{MET}^{-1}$, almost as fast as they can be
added to the pool. As a result, the rate at which cells are added to the pool,
determined by $k_{EMT}$, will be almost the same as the rate at which they are
removed from the pool. 

The rate at which cells are released from lesions will be proportional to the
number of cells on the free surface $n_{surf}(t)$, which is in turn
determined by the growth of the underlying lesion,

\begin{equation}
    J_{\mathrm{release}}  = \alpha k_{EMT} n_{surf}(t)
\end{equation}

with $\alpha$ some factor representing geometrical constraints, as only some fraction
of cells will be able to face the right direction to escape. The rate at which
motile cells revert back into epithelial-type cells, and in so doing form new
tumours, will be proportional to the number of circulating motile cells, $x$,
and $k_{MET}$

\begin{equation}
    J_{\mathrm{revert}} = k_{MET} x\;.
\end{equation}

The rate at which new micro-lesions are formed, $\phi$, is simply equal to the
rate at which motile cells revert,

\begin{equation}
    \phi(t) = J_{\mathrm{revert}} = k_{MET} x 
\end{equation}

and the dynamics of the population of motile cells will be determined by

\begin{equation}
\label{eq:circulatefast}
    \dot{x} =J_{\mathrm{release}} - J_{\mathrm{revert}}
    = \alpha k_{EMT} n_{surf}(t) - k_{MET} x \;.
\end{equation}

In terms of chemical kinetics, equation \eqref{eq:circulatefast} can be taken to be
instantaneously ``in equilibrium'' if $k_{MET} x \gg \alpha k_{EMT} n_{surf}$, and
consequently $\phi \approx \alpha k_{EMT} n_{surf}$. More rigorously, however,
one can show that

\begin{equation}
    x(t) = \int_0^t n_{surf}(t-\epsilon) \alpha k_{EMT} e^{-k_{MET}\epsilon}
    d\epsilon
\end{equation}

solves equation \eqref{eq:circulatefast}. Writing $n_{surf}$ as a Taylor series, one
can then find

\begin{equation}
    \phi(t) = k_{MET} x(t) = \alpha k_{EMT} \sum_{j=0}^\infty
    \frac{(-1)^j}{k_{MET}^j} n_{surf}^{(j)}(t) 
    = \alpha k_{EMT} n_{surf}(t) - \alpha k_{EMT} \frac{n'_{surf}(t)}{k_{MET}}
    +\mathcal{O}(k_{MET}^{-1})
\end{equation}

which shows that the rate at which new lesions can form is proportional to
$k_{EMT}$, but asymptotically insensitive to $k_{MET}$. The effect of changing
$k_{MET}$ is to effectively introduce a delay in $\phi$ of $k_{MET}^{-1}$,
which will be as small as $k_{MET}$ is large: this is consistent with the
behaviour observed in the second run of simulations, in which $k_{MET}$ only
had a strong affect on $G$ when $k_{MET}$ was relatively small. 

Intuitively,
this can be understood as the fact that as $k_{MET}$ controls the delay
between invasive cells' release and their founding of new micro-lesions,
slower $k_{MET}$ rates correspond to longer delays, and exponential growth
will take longer to start.
This explains how $\phi$, the rate at which new secondary tumours are founded,
is generally more sensitive to $k_{EMT}$ than $k_{MET}$. As the period of
rapid growth is driven by a cascade of secondary micro-metastases, $\phi$ in turn
indirectly determines the growth rate: the exponential growth rate is thus
more sensitive to changes in $k_{EMT}$ in a related way.

A more detailed treatment of how precisely 
the rate at which new lesions are seeded by older primary tumours $\phi$
determines the exponential growth rate is developed in chapter
\ref{ch:minimal}, but in essence the relationship is determined by the
Euler-Lotka equation

\begin{equation}
\label{eq:eulerlotka}
    \int_0^\infty \phi(t) e^{-G t} dt = 1
\end{equation}

at the growth rate $G$: $G$ can thus be written as a functional of $\phi$,
although not usually in terms of elementary functions.
One effect that this argument misses out is that at very high rates $k_{MET}$,
motile cells will only be present for a vanishingly short time, not long
enough to actually escape, and therefore above a certain value increases in
$k_{MET}$ can serve to slow growth down rather than accelerate it. 

Finally, one can see from figure \ref{fig:Veffect} that the average growth
rate $G$ also depends on the speed of migrating cells $|\vec{v}|$. We did not expect
to find any interesting dependence between the two, and I did not find either an
approximate functional form that could convincingly describe the dependency or
a convincing simple argument as to why this should be the case. It might be
important that since $\alpha$ was held constant at 
$0.01$ day${}^{-1}$ in these runs, increasing $|\vec{v}|$ had the effect of increasing $L$: 
as to \emph{why} this should be important, I am unfortunately at a loss.

An outstanding issue raised by these simulations is if the period of rapid
growth enabled by migration that is observed here does indeed correspond to a
period of \emph{exponential} growth as such. While the tumours observed here
seem to decelerate again after a brief period of faster growth, the strict
answer seems to be no: but whether or not some phenomenon of exponential
growth of an ensemble of small tumours could somehow \emph{underlie} the acceleration
observed here is quite a different question.

For growth to be truly exponential, the additional secondary lesions would
have to grow totally independently of one another: in this model, they all
grow on the same lattice, finite distances from each other, and may inhibit
each others' growth when they coalesce. To answer what would happen if they
were truly independent requires another, even more coarse-grained model.

\section{Lattice model with instantaneous migration}


Although the lattice model developed in section \ref{sec:local} allowed the
effect of migration on the growth of tumours consisting of up to $10^8$ cells
to be studied, the results raised a number of questions that simulations of
the model were not able to answer. Namely, whether truly exponential growth
can be enabled by cell migration. Growth that is confined to a limited region
of space, such as the region that can be explored by motile cells in the
previous lattice model, or the volume of an isolated tumour or cluster in the
Eden model \cite{eden1961two}, can never be exponential for an arbitrarily
long period of time if that region expands slower than exponentially.

For example, $M$ cells in the previous model can travel at a speed $|\vec{v}|$. The
largest possible volume that could be explored by a time $t$ is therefore a
sphere (really an octahedron due to lattice anisotropy) of radius $V t$. If
this volume is fully occupied by cells, the number of cells $n$ after a time
$t$ can be at most on the order of $(V t)^3$: really, as the volume will in
practice be much emptier than this, this is a large upper bound. No
exponential curve can stay under $\mathcal{O}(t^3)$ for an indefinite period
of time, so growth which is confined to a finite region which grows slower
than exponentially must also be sub-exponential in the long run.

But what if we don't impose the constraint that the cells have to migrate to
some finite distance away? Biologically, of course, bodies and tissues are 
only so big, but distinct lesions could plausibly be essentially
non-interacting. Perhaps they are separated by some glandular structure,
separated from each other by tissue in a duct \cite{suzuoki2002impact}.

This motivates a further abstraction to the previous model: to address how the
growth of tumours was affected by migration to more distant regions (but still
under similar environmental conditions), we developed a lattice model that
treated local invasion as an instantaneous event, that founded a new lesion
that did not interact with its founder.

\subsection{Algorithm}

The cell division dynamics were identical to to the previous model. 
Cells attempt to divide at some 
rate $b$: each time step $\Delta t$, with probability $b\Delta t$, cells 
attempt to place a new daughter cell at a randomly
chosen neighbouring lattice site. If the neighbouring lattice site is
unoccupied then the attempt succeeds, otherwise it fails. Once again, cells
were considered neighbours if they were von Neumann neighbours \cite{breukelaar2005using}.

In contrast to the previous model, cells do not have an explicit phenotype,
and migration does not involve the detailed, explicit movement of individual 
cells through space. New cells on the surface instead
have some aggregated overall probability $M$ to escape immediately after being
produced in a replication event,
to found a new lesion far removed from the original. These new lesions are
indefinitely far away from the lesions that originated them: they do not
interact with each other at all, indefinitely prolonging the intermediate
phase of exponential growth in the model above.

Only new cells on the surface were capable of escaping and founding new
lesions in this way. This is to reflect the fact that in the model developed
in section \ref{sec:local}, $M$-cells can only migrate to free lattice sites,
so $M$-cells which happen to find themselves embedded deep inside a tumour are
stuck. Only $M$-type cells on the free surface were able to escape.

This model abstracts away the process of phenotypic switching and migration
into a one-step process: cells do not have an explicit type, but new cells may
escape and form a new, distant lesion instantaneously with probability $M$ per
division.
All cells in this model are able to divide and produce new cells, providing
there is enough room in their immediate environments: there are no distinct
types as such, as the migration process is instantaneous, and as such cells
are assumed to spend a vanishingly small time as motile $M$-type cells.
Each given site in the model may therefore be occupied or unoccupied, but if
occupied by a cell does not, as previously, have an explicit type as such.

\subsection{Results}

This model shows some clear qualitative resemblances to the previous lattice
model: an initial stage of sub-exponential growth in which the original
primary tumour dominates the mass of the whole ensemble, and a second later
stage in which the secondary and additional metastases permit sustained,
exponential growth. The re-coalescence exhibited in the earlier lattice model
does not occur simply because the individual micro-metastases in this model are
explicitly assumed to be non-interacting: the exponential nature of this
latter stage of growth is readily apparent, and straightforward to measure. 

The growth rate $G$ of these ensembles depends on the probability
of migration $M$. Furthermore, $G$ seems to have a power-law-like dependence
on $M$, so we can hypothesise that

\begin{equation}
   G \approx c M^p
   \label{eq:powerlaw}
\end{equation}

where $c$ and $p$ are parameters to be determined.

\begin{figure}[ht]
\begin{center}
    \includegraphics[width=0.6\textwidth]{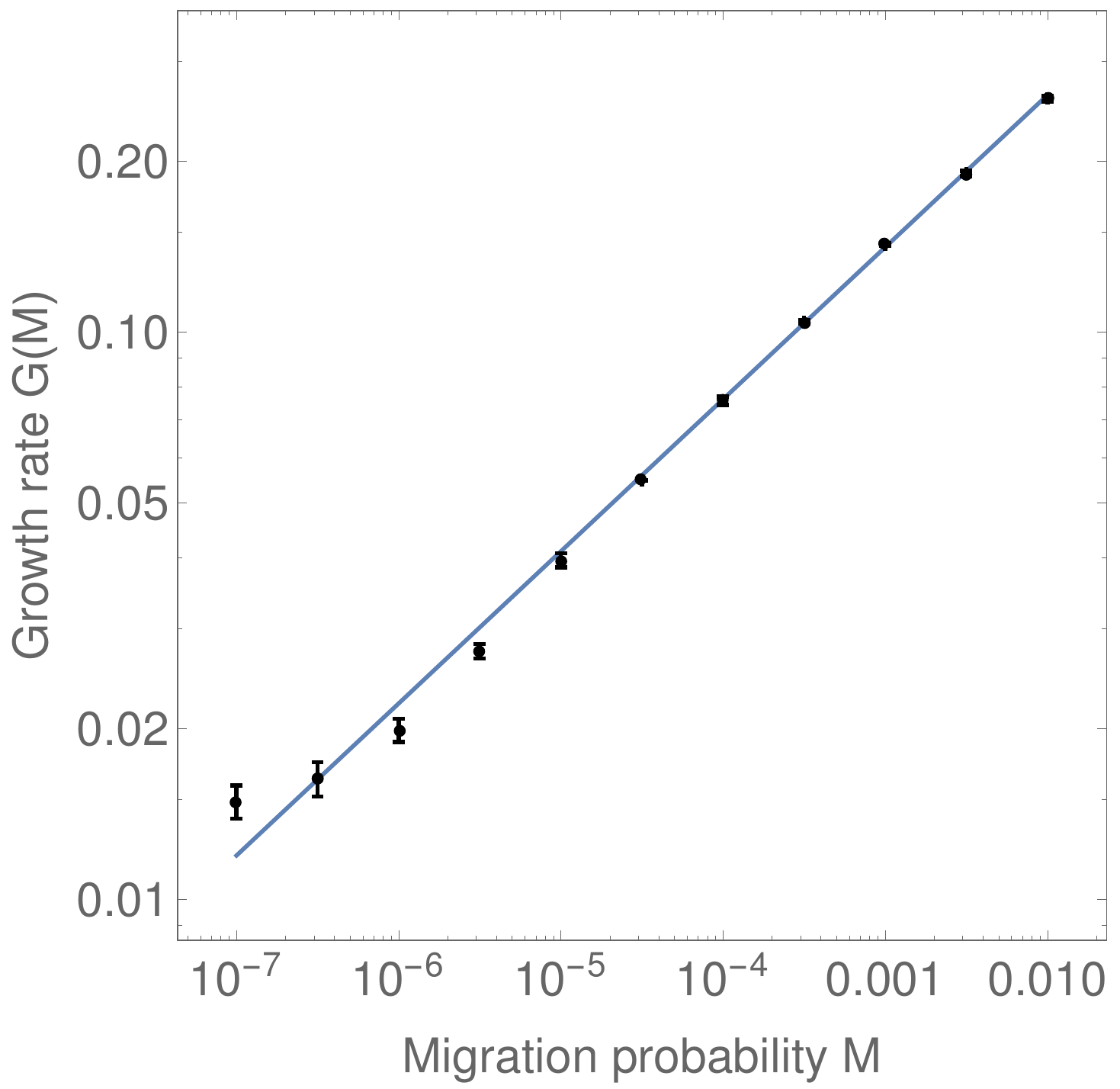} 
    \caption{\label{fig:Eden}A plot to show the power law dependence of the
    exponential growth rate $G$ of an ensemble of metastatic tumours 
    on the migration probability $M$ for the Eden-like lattice model with
    instantaneous migration, showing data from simulations (blue
    points and bars) alongside the curve of best fit, $G = c M^p$ with
    $c=0.898\pm0.007$ and $p=0.268\pm0.001$.
    Parameter values for the simulations were division rate
    $b=1.0\;\mathrm{day}^{-1}$, death rate $d=0$: there were no other free
    parameters in this minimalistic model. Finite size effects become visible
    at the lowest values of $M$.}
\end{center}
\end{figure}

We elaborate on this in more detail in chapter \ref{ch:minimal}, but a
qualitative argument as to why this should be the case is due to Prof. M.E.
Cates is as follows:

If growth is sustained and exponential, it must be determined by the average
frequency with which new micro-metastases occur. If the growth of one lesion
follows $n(t)$, and any new cell has a probability $M$ to escape and found a new lesion,
the expected number of new lesions by time $t$ is $M n(t)$. The frequency $G$
with which new lesions appear will scale as $G \sim T^{-1}$ with the time $T$ 
between these occurring: so $M n(G^{-1}) \approx 1$. For
spheroids which expand cubically in time as $n(t) \propto t^3$, this implies $G \propto
M^{1/3}$. Although this argument (and the more detailed analytical models
which support it) does not precisely fix the value of the exponent $p$, it
does demonstrate how the power-law dependence emerges from geometrically
constrained growth: other growth curves will presumably entail other dependencies,
which will require a more rigorous derivation to characterise (see chapter
\ref{ch:minimal}).

The situation in actual simulations is of course a little more complicated: the
exponent $p$ is significantly lower at around $p=0.26(6)1$. The underlying
reason why this differs from $1/3$ is surface roughness: only new cells
present on the surface are able to escape, so the rate at which
motile cells escape is proportional to the surface area. If the surface is
very rough, the number $n_S$ of cells on the surface will scale approximately as

\begin{equation}
    n_S \sim n^q
\end{equation}

with $q>2/3$. Given $n\propto t^3$, the expected number of lesions present
after a time $t$ will then be $\propto \int M n_S dt \propto M t^{3q+1}$. By
similar arguments to above, the growth rate $G$ should follow

\begin{equation}
    G = c M^\frac{1}{3q+1}
\end{equation}

with $2/3<q<1$: the exponent $p$ should therefore take values in between $1/4$
and $1/3$, consistent with the results of simulations.

An interesting point to note is that the proportion $M$ of cells that escape
to form new micro-lesions is closely related to the frequency of EMT in the
previous model. A simple argument inspired by chemical kinetic arguments 
as to why this should be is that since the reverse rate of switching $k_{MET}
\gg k_{EMT}$, the rate at which motile cells are released and switch back to
non-motile types should be determined by $k_{EMT} n_{surf}$, where $n_{surf}$
is the number of cells on the surface: given that in this set of models, the
rate at which new metastases occur will also be proportional to the number of
cells on the surface, around $M b n_{surf}$.

So, to the extent that $k_{MET} \gg k_{EMT}, $we have

\begin{equation}
    k_{EMT} n_{surf} \approx M b n_{surf}
\end{equation}

so

\begin{equation}
    M \approx k_{EMT}/b\;.
\end{equation}

There are several things that this crude analysis does not take into account:
effects of the finiteness of $k_{MET}$, and the delays in founding of new
lesions, are two of these. However, the basic conclusions that the speed of
cell movement does not strongly affect growth rates, 
and that $k_{EMT}$ and $M$ are the strongest
determinants of $G$ in both models are both supported.

As has already been observed, the growth rate $G$ has a different
dependence on $M$ in the two models: although a power law provides a
decent approximation in both cases, the exponent is not the same. For the
model with instant fixation, the exponent is very close to $1/3$, whereas in
the earlier model with explicit spatial structure and non-instantaneous
migration, it was closer to $0.21(8)^4$, significantly lower. This may be a
finite-size effect: tumour sizes were limited by constraints on computational
resources to around $10^8$. However, it may also be explained by the
occurrence of migration from deep inside solid tumours: the Eden model
simulations also showed a considerably lower exponent when cell death and
migration occurred throughout the volume of the tumour, and not only at
the surface, so it may also be the case that due to roughness and porosity of
tumours, cells are able to escape from much deeper into tumours than is
immediately obvious.

What is especially interesting about the observed importance of the rates of
EMT and MET is that these dynamics of
have not seen as drastic degrees of simplification as
have the mechanics, motility and spatial structure of cells in the course of
progressive abstraction from the off-lattice mechanical model up to this point.
Although, the description of phenotypic plasticity is already much simplified
in the underlying model, and the extent to which this level of simplification
is justified is a subject of active research \cite{EMTnoise}.

\section{Concluding remarks}

We established that the elastic stiffness of migrating cells could affect the
speed with which they can migrate through epithelial monolayers. (see figure
\ref{fig:vstiff}) 
This is potentially relevant to treatments which target the cell's
cytoskeleton, as these may have knock-on effects on the cell's elastic
stiffness \cite{schenk2015salinomycin,MolBioCell}.

We also
established that the rates at which cells switched between invasive,
mesenchymal types and static, epithelial types controlled how quickly tumours
could grow (see figures \ref{fig:RTMgrowth}). In essence, the forward and reverse rates with which cells switched 
between the different types controlled how many invasive cells were present, and how 
long it took them to form new lesions. Furthermore, the growth of tumours was 
also affected by the speed with which invasive cells move. We have not
arrived at a satisfying explanation for this latter observation.

Finally, more coarse-grained simulations supported the idea that exponential
growth could be enabled by migration (see figure \ref{fig:Eden}). But only as long as this migration
resulted in new lesions that were essentially non-interacting, and grew
independently of their primaries and of each other. This raised a number of
interesting questions about how this exponential growth could be related to
how frequently migration occurred which could not be easily answered. To
address these, and bridge the gap between abstract microscopic models and
macroscopic tumours, we will need to develop new analytical tools.
\chapter{Structured population models of clonal selection}
\label{ch:minimal}

\section{Introduction}
\label{sec:stochast}

In the previous and following chapters, we explore how lattice models of evolution and invasion
in cancer could be used to make predictions about the rate of growth and of
the accumulation of advantageous drivers, and also what forms of growth and
resistance are observed in clinical settings. In
this chapter, we will develop a theoretical approach which makes as few
assumptions about the underlying model as necessary, and which enables us to
draw connections between the large-scale behaviour of the computational models
and the observables of the experimental studies.

The main phenomenon we want to capture
here is the feedback between growth and migration that enables metastatic
cascades: growing tumours produce new tumours, which begin new growth, and
produce more new tumours in an accelerating process.

Computational models have recently been proposed  \cite{enderling_migration_2009,
waclaw_spatial_2015,hanin2006stochastic} in which cancerous tumours
are conglomerates which consist of one large primary tumour with many surrounding
microscopic lesions, presumed to form when cancer cells detach from the
primary lesion, migrate or are passively advected through intervening tissue,
and resume growth at an adjacent site. This process is similar to metastasis -
the ability to invade form new micro-lesions elsewhere in the body - which is
arguably the defining characteristic of cancer: however, these models 
 \cite{enderling_migration_2009,waclaw_spatial_2015,hanin2006stochastic}are also
applicable to short-range invasiveness, perhaps as short as a few micrometers
from the edge of the primary lesion. The presence or absence of cancer cells 
in a narrow region surrounding the primary tumour called the surgical margin 
has in fact long been used as a predictor of recurrence and indicator of the 
success or failure of surgical excision of solid tumours, and is a standard
histological practice \cite{stefaniejefrey1995importance,suzuoki2002impact}.

The movement of cells in these models is usually understood to be caused by
cells undergoing epithelial-to-mesenchymal transition and local invasion \cite{mcdonald_genome-scale_2011}, but
may also possibly be due to ``sprouting'' caused by mechanical instability \cite{Drasdo2000,Basan2011}:
nonetheless, the models tend to be neutral as to the precise underlying mechanism.
Having left the primary lesion, cells may enjoy better access to nutrients and
oxygen and can proliferate faster. The tumour has a complex structure which
can enable faster expansion that if it were made of a homogeneous and simple
cluster of cells \cite{iwata2000dynamical,michor_stochastic_2006}.


In this chapter, we study a simplified, mathematical counterpart of the models
in previous chapters and  \cite{enderling_migration_2009,waclaw_spatial_2015,hanin2006stochastic}
which can be solved exactly and
used to make predictions about the accumulation of drivers and the growth
curves of the conglomerate tumours. 


To this end, we will begin by studying a stochastic model which abstracts away
the behaviour of individual cells and treats individual micro-lesions as
continuous masses, whose growth and dynamics over time are determined
by their geometry and size. The underlying growth dynamics is thus
deterministic rather than stochastic as in the case of the lattice models, and
in this model only migration is a stochastic process. The underlying
motivation for this was the hypothesis that migration was relatively rare
compared to division and death, otherwise tumours would be rather diffuse.

The entire tumour is modelled as a collections of micro-lesions made of cancer
cells. Distinct micro-lesions do not interact with one another: this is key to
the analytic solubility of the model. This rather strong assumption could
possibly be justified by interpreting micro-lesions as spatially-separated
metastases, but even if micro-lesions are fragments of the same primary tumour,
numerical simulations indicate that as long as micro-lesions are separated by
normal tissue, their growth rate is minimally affected by these interactions,
 \cite{waclaw_spatial_2015}, and data from the lattice simulations with
detailed cell motility (chapter 2) support the conclusion that interactions
only become significant once the tumour has become much larger than the
average run distance.

One plausible scenario in which lesions will typically be separated by normal
tissue is in the case of ductal carcinomas. In these cases, the primary tumour
forms deep within a duct, and secondary tumours in the same gland will tend to
be separated by some amount of distorted basement membrane and supporting
extracellular matrix (see figure \ref{fig:membrane}).


This stochastic model will be compared to an analytical treatment of its
average or expected behaviour, and the degree of disagreement quantified and
tentatively explained. The expected behaviour will be analysed by means of
interpreting the rates of the stochastic model as parameters in a set of
differential equations governing a frequency distribution in the number of
lesions: the average behaviour thus results in yet another even more abstract
model which is entirely deterministic, and which will be shown to be exactly
solvable in a few important special cases.

This latter, entirely deterministic model belongs to the class of
age-structured population models, which originate from population ecology and
epidemiology  \cite{keyfitz1997,perthame2006transport}, but which have since
also been applied to modelling metastasis in cancer with some degree of empirical
success  \cite{iwata2000dynamical,baratchart2015computational}.

These two contrasting approaches, the simplified stochastic model and the
averaged-out deterministic model, nonetheless both make predictions regarding
the total volume of the entire tumour burden, and of the relative frequencies
of mutant strains with some increasing relative selective advantages. We will
use these two quantities to compare the predictions of the averaged,
deterministic model to both the simplified stochastic model with deterministic
growth, and later to the fully stochastic lattice models. This is done with a
view to gauging the predictive power of the deterministic model, and the
various quantities that can be extracted with a few powerful analytical
methods, but also to attempt to better understand phenomena which are
qualitatively striking in the lattice model: in particular, the possibility of
sustained exponential growth of the entire tumour even when individual
micro-lesions grow much slower than exponentially, and if this phenomenon
depends on the sustained growth of these individual micro-lesions or not.

\section{Stochastic model}

Let us begin by recalling some qualitative observations of the previous
computational models, which can be easily verified visually (see figure
\ref{fig:visualisation}).

In both the Eden-type lattice model with instantaneous, distant migration, and
in the lattice model with local migration, distinct micro-lesions were
approximately spherical. The growth of each of these micro-lesions was slower
than exponential after an initial burst of fast, well-mixed growth: contact
and spatial exclusion of cells meant that only cells near the surface of the
spheres could grow, and the spheres' diameters increased linearly in time.
This can be expressed equally accurately by stating that each spheroid
expanded with some characteristic speed.

In the local migration model and in the off-lattice mechanical model, some 
proportion of the cells in any given lesion were motile, and moved with
some characteristic speed, in the former case faster than the expansion speed 
of the lesion they were in. These cells could stochastically switch back to
a non-motile type which could divide and produce additional offspring.

In the instant migration model, this corresponded to a process in which
any individual cell on the surface could escape and found a new micro-lesion
instantaneously, with some probability $M$ per time step of doing so.

In all models, there was initially only one cancer cell: one cell which had
accumulated just enough genetic damage to decisively out-compete surrounding
tissue, and to propagate in an uncontrolled manner. This is a well-established
hypothesis in the experimental literature.

Finally, in the instant migration model, there was some probability per
division that each daughter cell could gain a new mutation. This mutation
could either be neutral, or give some advantage to the daughter cells: cells
with advantageous mutations could divide faster overall (that is, not
necessarily absolutely faster: this could equally result from dying less
before having the chance to divide) and would typically form an expanding
sector in spheroidal micro-lesions which grew at a faster rate than the
surrounding cells without these mutations.

These qualitative observations about the geometry, and the stochastic
processes which govern rare but important events in the growth of individual
spheroids, can be captured in a simple model.

At any given time, the state of each micro-lesion is fully characterized by two
numbers: its age $a$, and its sub-line $n$ which reflects the genetic make-up
of the lesion and for simplicity is taken to be equal to the number of driver
mutations in the cell which initiated that lesion. This convention doesn't
require all micro-lesions to be genetically identical: in fact, we should
expect a large number of selectively neutral passenger mutations to be present
in all micro-lesions, many of which will be unique to lineages present in
single lesions. But as these passenger mutations are selectively neutral, they
should not effect the growth dynamics at all, and only the driver mutations
should affect growth. We shall study only the case in which the total number of driver mutations
affects fitness: more complex functional dependencies are of course possible,
but are not our object of study at this early stage.

\begin{figure}[ht]
    \begin{tabular}{lll}
    A & B & C \\
        \includegraphics[width=0.3\textwidth,angle=90]{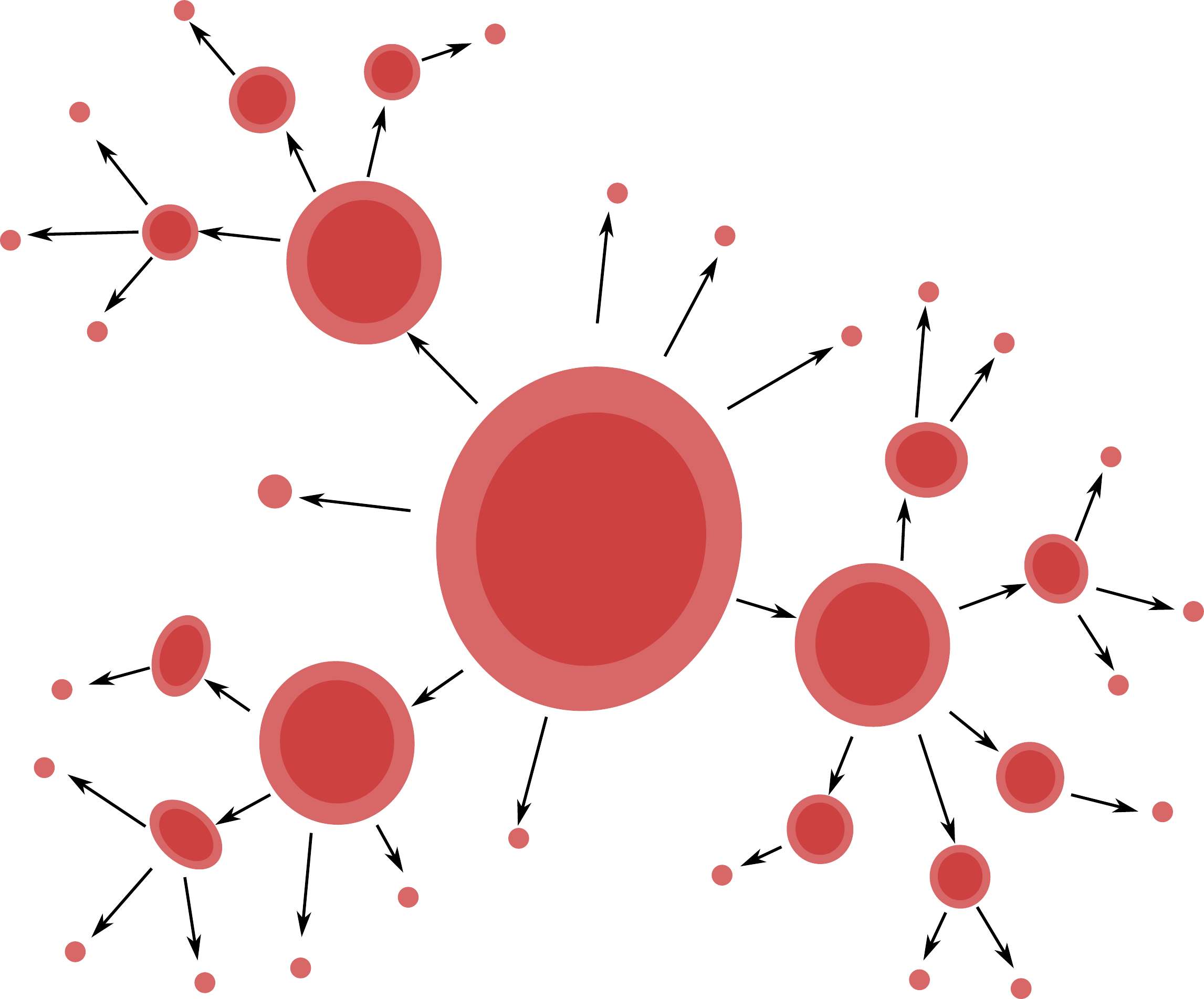} \hspace{1cm} &
        \includegraphics[width=0.2\textwidth]{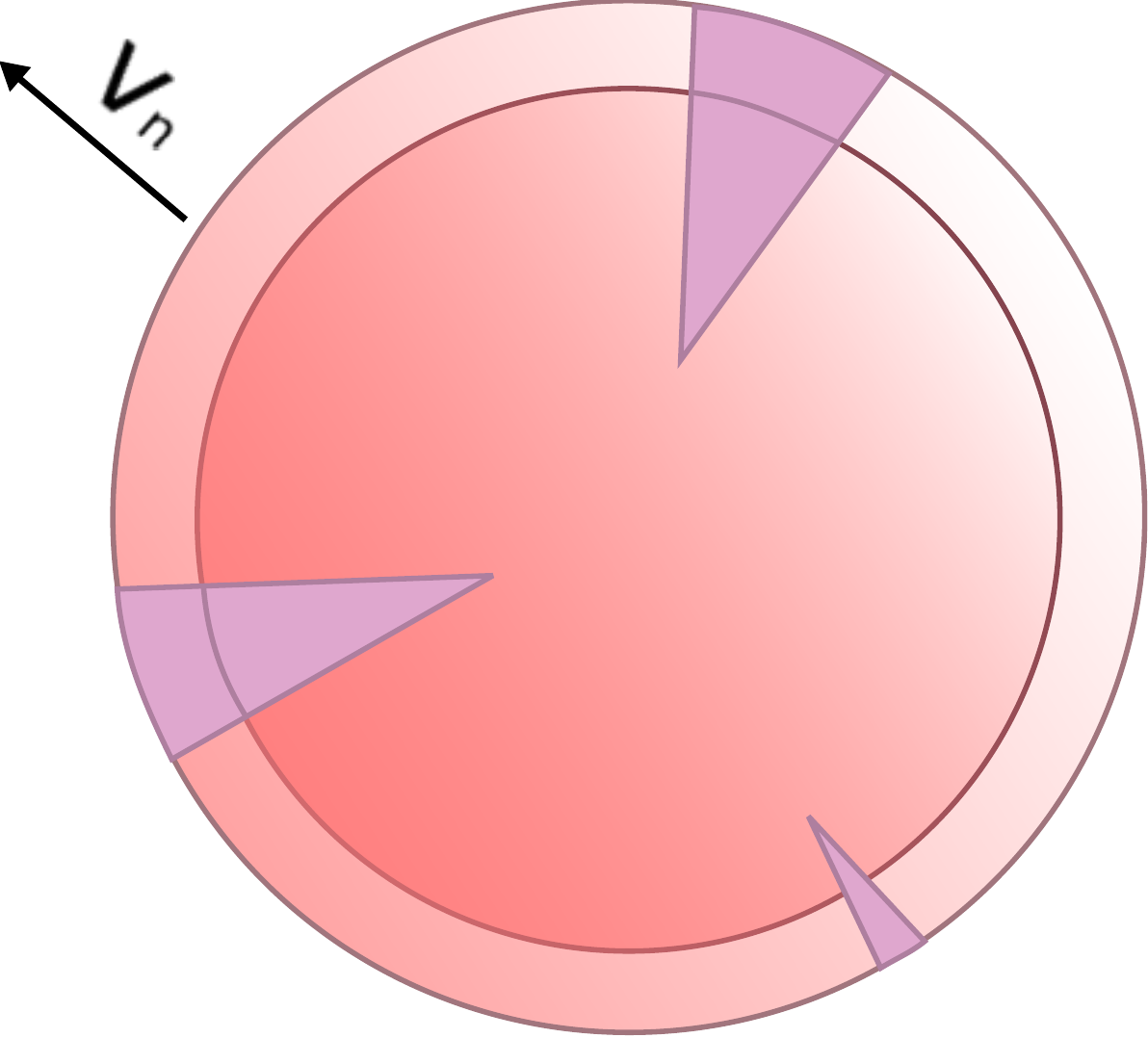} \hspace{1cm} &
        \includegraphics[width=0.2\textwidth]{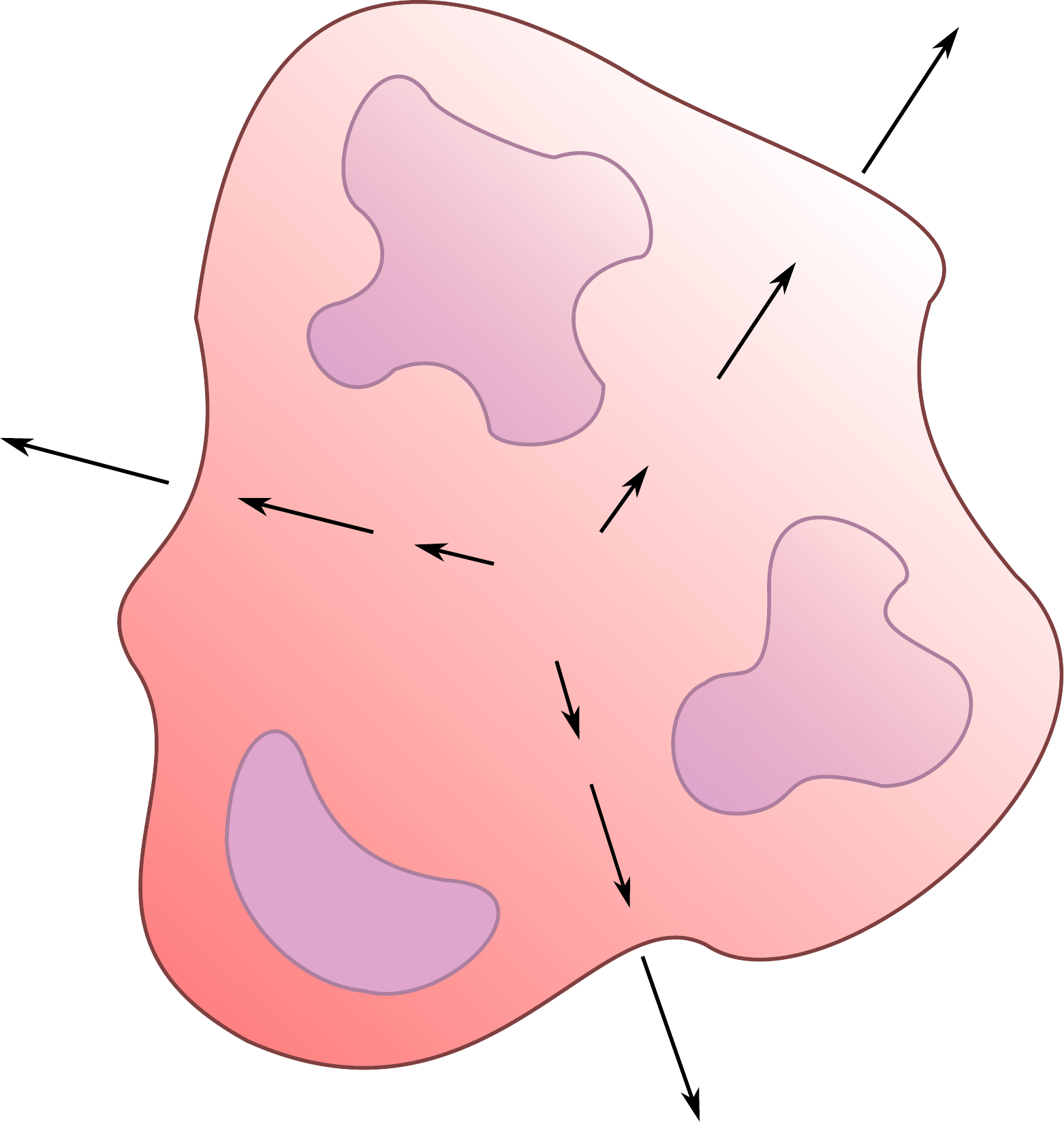} \\
    \end{tabular}
    \caption{\label{fig:model}A: The model assumes that the tumour is made of discrete micro-lesions. Cells migrate from micro-lesions (arrows) and establish new microscopic lesions. All lesions increase in size over time. B,C: two different growth models of individual micro-lesions. In the surface growth model (B), cells replicate only in a narrow layer of constant thickness near the surface, and the radius of the lesion increases with velocity $v_n$. In the volumetric growth model (C), replication occurs everywhere and the micro-lesion grows exponentially which causes the whole lesion to ``inflate''. Arrows show the expansion velocity which is small close to the centre and increases towards the surface. Purple areas correspond to new mutations.}
\end{figure}

In the following, we will also assume that any lesion ${n,a}$ contains at most
two different genotypes: $n$ and $n+1$, with relative abundances among those
cells that can migrate (a distinction whose relevance which shortly become
clear) of $r_n(a)$ and $1-r_n(a)$ respectively. Note that individual cells are not 
modelled explicitly: only the aggregate behaviour of these
microscopic lesions is studied, and these may contain thousands or even
millions of individual cells.

The following processes are accounted for:

\begin{itemize}
    \item Lesions grow in a deterministic way: their volume and area increase
    with their age (the time since the lesion was founded).
    \item Cells rarely gain benign mutations: most cells in any given lesion
    will have the same number of driver mutations as the cell which founded
    the lesion, and some fraction near the surface have one additional driver.
    \item Cells escape from the surface of these lesions stochastically: the
    rate at which new lesions appear is proportional to the area of the lesion
    they were seeded by.
    \item All cells in the tumour descend from one originating clone.
\end{itemize}

In this scheme, the common process of cell division and growth in the
underlying models occurs relatively frequently, and is abstracted to a
deterministic trend. The rare processes of cell mutation and migration occur
much less frequently: although, since mutation has a relatively large effect
when it does produce an advantageous strain, this is summarised as a
deterministic process by which an increasing proportion of cells escaping from
the surface are mutants, and all of the randomness in this model falls into
the rate at which migration happens.


These assumptions may be expressed with some convenient notation:

\begin{itemize}
    \item \emph{Growth.} Lesions' growth is determined by the number of driver mutations in
    the founding cell, $n$, and their age, $a$: their volume is denoted
    $V_n(a)$, their surface area $S_n(a)$, and their radius $\rho_n(a)$ increases
    as $\rho_n(a) = v_n a$. The spatial character of the model manifests itself
    in our choice of these functions: $V_n(a)$ is very different for
    three-dimensional, spatially constrained growth in which replication can
    occur only on the surface of growing spheroids, than it is for
    ``well-mixed'' growth in which growth may occur throughout the entire
    volume and all cells may replicate with identical rates.
    \item \emph{Driver accumulation.} New lesions of type $n$ are produced by other lesions: lesions of 
    type $n$ and age $a$ produce micro-lesions of the same type with probability 
    $r_n(a)$, and lesions of type $n+1$ with probability $1-r_n(a)$.
    \item \emph{Migration.} The production of new lesions via the escape and re-settlement of
    tumour cells occurs in a inhomogeneous Poisson process. 
    The rate with which new lesions of either type are produced by a
    lesion of type $n$ and age $a$ may be denoted $\phi_n(a)$, and is
    proportional to the instantaneous rate of growth of these lesions, $\phi_n \propto \dot{V}_n$.
    \item \emph{Initial conditions.} There is initially one lesion of type $n=1$ and age $a=0$.
\end{itemize}

We will later examine other forms of growth than expanding spheres, such as
plateaus or well-mixed exponential (``volumetric'') growth.

The precise fitness landscape, and what advantage is actually imparted by
each additional driver, is a matter of what the precise form each of these
functions takes: in particular, if the advantage manifests through a more rapid
net growth rate, through the form of $v_n$.

By ``fitness'', we mean 

\begin{equation}
    s = \frac{b_{n+1}}{b_{n}}-1 
\end{equation}

with $s$ the relative advantage imparted by the $n+1^{\mathrm{th}}$ driver
over the $n^{\mathrm{th}}$.

Several such landscapes may be investigated: of particular interest are
landscapes with a continuous increase is fitness, in which one expects to see
straightforward selection for fitter drivers, and landscapes in which only a
few specific drivers carry a substantial advantage.

Two distinct possibilities for the case in which additional drivers
monotonically increase fitness are the case in which $v_n$ increases
exponentially with $n$ (i.e. having a constant relative advantage) and the
case in which $v_n$ increases linearly in $n$ (i.e. having a diminishing
relative advantage) \cite{antal_spatial_2015,bozic2010accumulation}.

There are many imaginable cases in which only the first few drivers after
tumorigenesis increase fitness and growth rate further, as might be expected
if tumours have already accumulated sufficient mutations to out-compete
surrounding normal tissue, but there are still remaining opportunities for
further adaptation, such as to new environmental niches in the body, or to
adverse conditions (such as therapy). For simplicity, we will investigate here
the cases in which only the first two additional drivers increased
$v_n$ substantially, in our cases by increasing it rapidly and linearly at
first, and any additional drivers after these impart only a small advantage.

\subsection{Basic algorithm}
\label{sec:algorithm}

In the following, we will use $n$ to index the number of driver mutations
present instead of the population of cells in the tumour or lesion. The size
of the tumour will be expressed in terms of its volume, $V_{\mathrm{tot}}$.

The rules of the model can be straightforwardly simulated once $v_n$ and
corresponding parameter values are specified.

The basic algorithm is as follows:

\begin{enumerate}
    \item One tumour with $\rho=0$ and $n=1$ is initialised: this is stored in a
    data structure $lesions$ with all other tumours and micro-lesions.
    \item Each time step: every tumour in $lesions$ grows, its radius $\rho$ 
    is incremented by $v_n \;dt$. Its area $S_n$ and fraction of mutants on the
    surface $1-r_n$ are recalculated.
    \item The number of escaping cells both of type $n$ and $n+1$ are
    calculated. These are both Poisson-distributed random variables $dN_n$ and $dN_{n+1}$ with means
    $r_n \phi_n \;dt = M r_n v_n S_n \;dt$ and $(1-r_n)\phi_n = M (1-r_n) v_n S_n \; dt$ 
    respectively. After these Poisson-distributed random
    variables are calculated, $dN_n$ and $dN_{n+1}$ new lesions are
    initialised with $\rho=0$ of types $n$ and $n+1$ respectively.
    \item Time advances by $dt$.
\end{enumerate}

The total volume of all tumours, the volume of tumours of each type, and the
average number of drivers present in the entire ensemble may then be
calculated and recorded.

The following forms were assumed for the geometry, migration rate $\phi_n$, the
expansion speed $v_n$ and the non-mutant surface function $r_n$ as functions
of age $a$:

\begin{itemize}
    \item A tumour is a sphere with radius $\rho=v_n a$. Consequently, the surface 
    area $S_n = 4\pi v_n^2 a^2$, and the volume $V_n = 4 \pi v_n^3 a^3 /3$.
    \item New lesions are established with rate $\phi_n = 4\pi M v_n^3 a^2$.
    Of these, $r_n = \exp(-\mu a)$ are of type $n$, the same as the
    originating lesion, and $1-r_n = 1-\exp(-\mu a)$ are of type $n+1$,
    carrying one additional driver mutation.
    \item Only the first three driver mutations carried a substantial
    selective advantage. As such, $v_n = n v_1$ for $n\leq 3$, and $v_n = 3
    v_n + (n-3)\epsilon v_n$ for $n>3$.
\end{itemize}


\begin{figure}[ht]
\begin{center}
    \begin{tabular}{lll}
    A & B & C \\
    \includegraphics[width=0.33\textwidth]{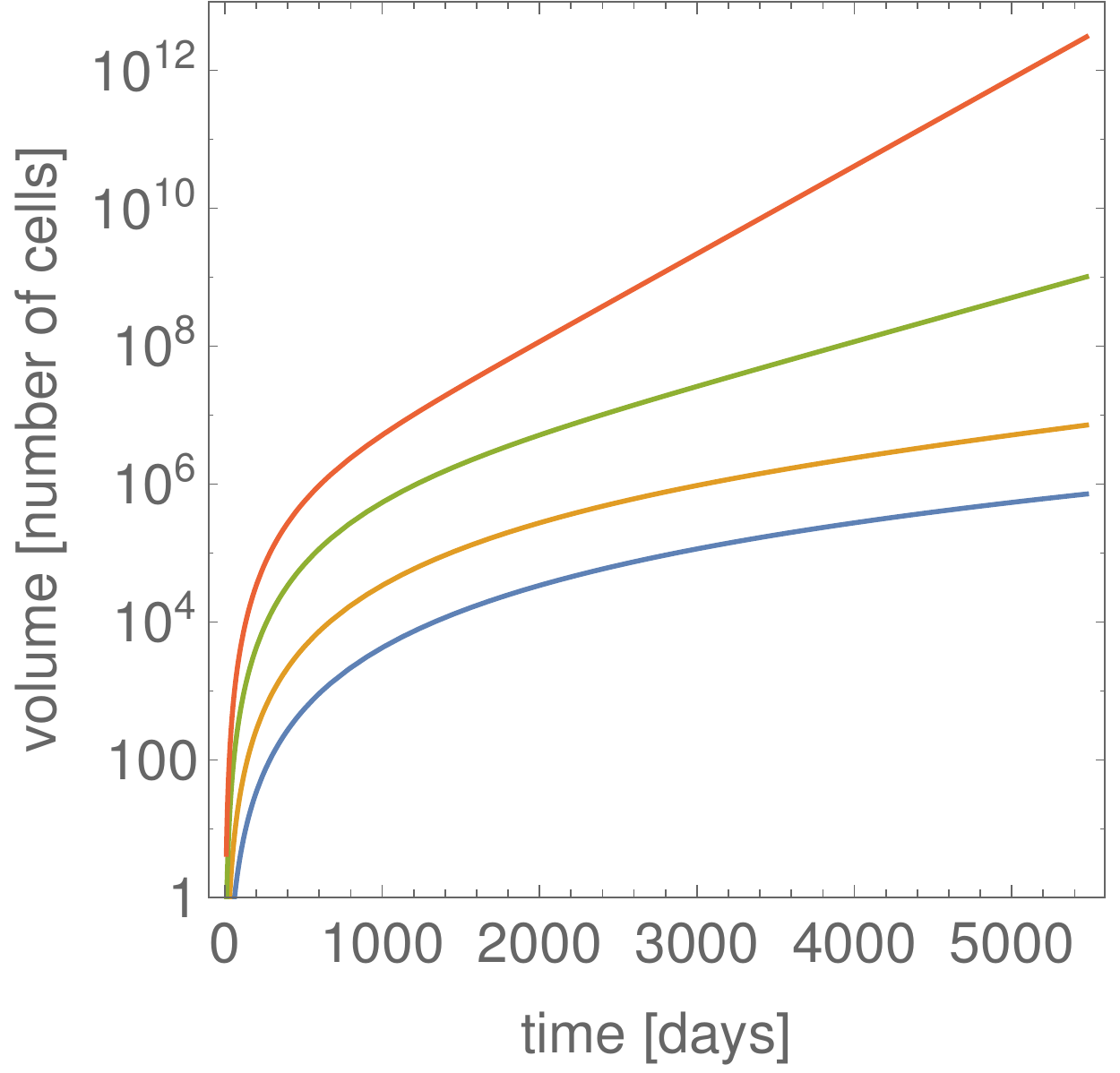} &
    \includegraphics[width=0.33\textwidth]{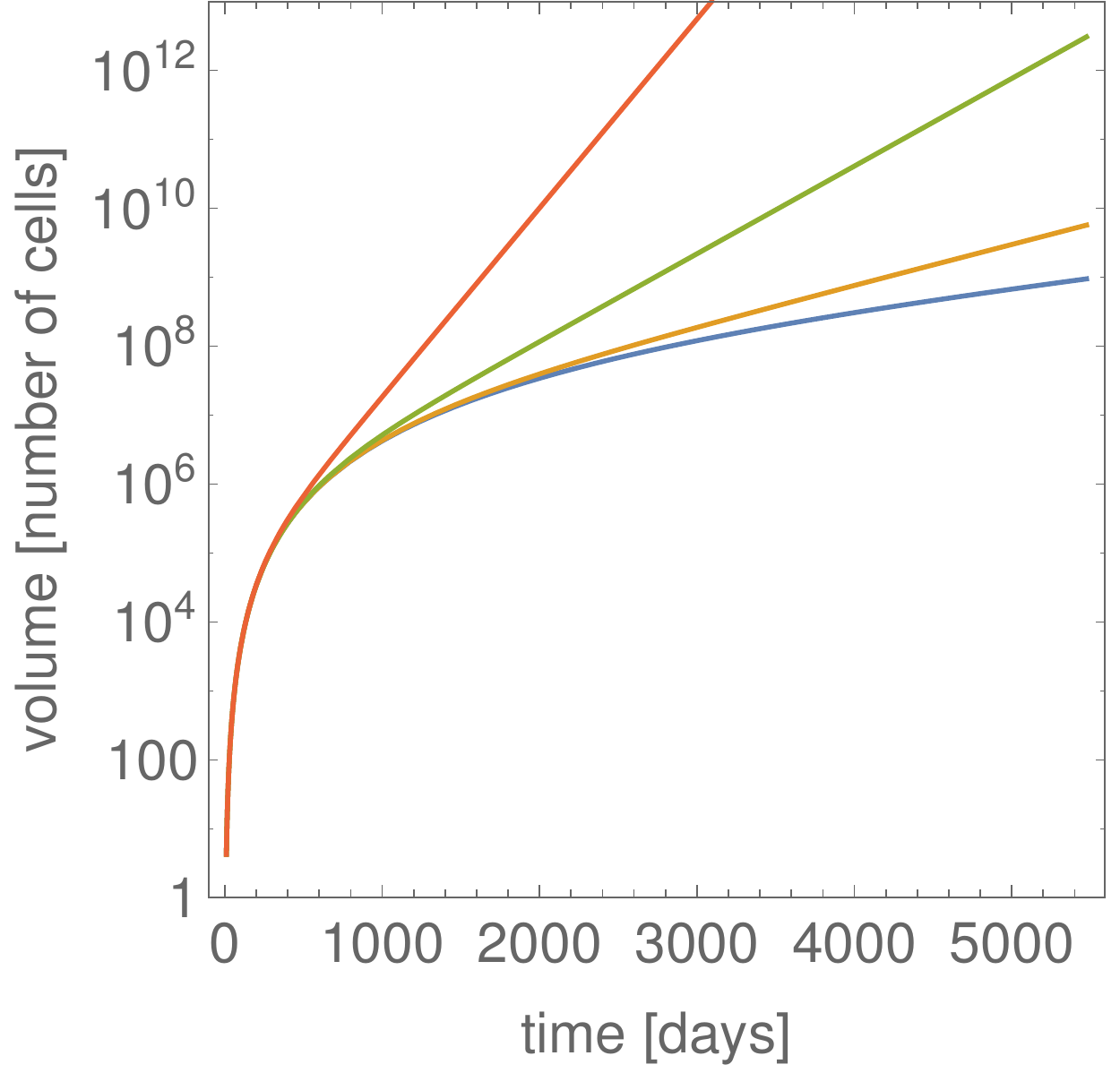} &
    \includegraphics[width=0.33\textwidth]{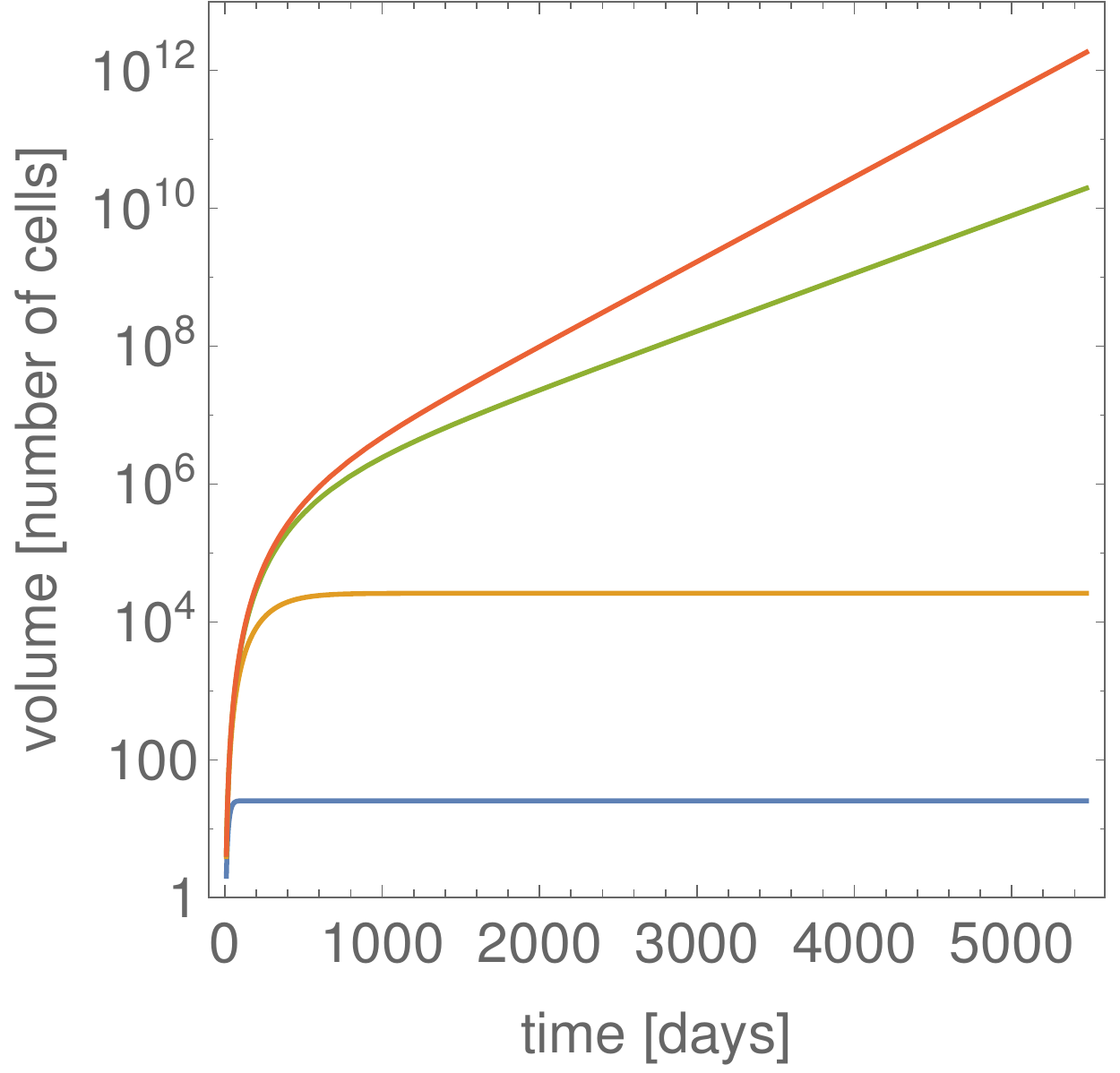}
    \end{tabular}
    \caption{\label{fig:single1}A: total tumour volume $V_{\rm tot}(t)$ versus time for different single-lesion expansion velocities $v=0.01$ (blue), $v=0.02$ (yellow), $v=0.05$ (green) and $v=0.1$ (red), and the same migration probability $M=10^{-6}$. B: $V_{\rm tot}(t)$ for $v=0.1$ and four different $M=10^{-8},10^{-7},10^{-6}, 10^{-5}$ (blue, yellow, green and red, respectively). C: $V_{\rm tot}(t)$ for the slow-down model, $M=10^{-6},v=0.1$ and four different $\lambda=0.1,0.01,10^{-3},10^{-4}$ (blue, yellow, green and red).}
\end{center}
\end{figure}

Some basic phenomena are readily apparent from figure \ref{fig:single1}: in
simulations with no mutation (and only one driver, the first), after an
initial phase of growth in which metastases and new micro-lesions are
infrequent and the primary lesion forms most of the mass of the tumour, the
cloud of micro-metastases increases in frequency and mass exponentially,
despite the sub-exponential growth of individual lesions. With some
mutation rate incorporated via $r_n(a)=\exp(-\mu a)$, this exponential growth
is accelerated to super-exponential, driven by an exponential accumulation of
driver mutations.

An extended discussion of these phenomena, an analytical theory of what
determines them, and a detailed comparison of the two can be found in section
\ref{sec:comparison}.


The questions that our analytical theory has to address are several: how can
the growth of the ensemble be exponential whilst the growth of individual
components is considerably slower? What determines the precise form of
the accumulation of driver mutations? Can these be related to quantitative
determinants of the microscopic behaviour?


The cumulative frequency distribution of the resulting lesions' ages can also
be calculated: by this, we understand the number of lesions of a given type
$n$ which are younger than a certain age $a$. An empirical cumulative age
distribution can be calculated in any given realisation of the model, and the
proportion of cells younger than a given age will
clearly increase with age: this distribution, denoted $\mathrm{CDF}_n(a,t)$, is simply
the number of lesions which are younger than $a$, and will therefore start
from zero

\begin{equation}
    \mathrm{CDF}_n(a,t) = 0 \;\|\; a \leq 0
\end{equation}

since there are no lesions with negative age, and will flatten out at the
total number of type-$n$ lesions present at any given time, $N_n$,

\begin{equation}
    \mathrm{CDF}_n(a,t) = N_n \;\|\; a > t
\end{equation}

since there are no lesions older than the oldest lesion, whose age is always
equal to the total elapsed time.

This is primarily an auxiliary quantity: useful in order to define other
distributions. It is nonetheless relatively easy to
calculate for both this stochastic model and other analytical models we will
later analyse, and as such provides a way to compare the two approaches meaningfully.

\section{Analytical methods}
\label{sec:analytic}

\subsection{Definitions and exact results}

The mean behaviour of the stochastic model can be readily analysed in terms of
the expected distribution of the number of lesions of a given type and age: by
``mean'' and ``expected'' in the following, we understand these means and
expectations to be over realisations of the underlying stochastic process. A
smooth, expected distribution can be given which is governed by a system of
differential equations in age and time.

The stochastic process of escape and self-metastasis is replaced with a
deterministic drift term, with a rate equal to the rate of the underlying
Poisson process. The average number of new metastases to appear stochastically 
in a short interval will consequently agree with the number (or weight) of new
metastases that appear in the model of the lesion size distribution. Since 
this process accounts for all stochasticity in the underlying model, the
evolution of the expected lesion size distribution is purely deterministic.

It is important to note that the underlying model, as well as most simulations
of the lattice models, exhibits exponential growth enabled by the migration of
escaping cells. It is therefore reasonable to expect to observe exponential
growth.

The dynamics of the model can be expressed in terms of the distribution of lesions' ages
and types $n$: in this deterministic model, the cumulative distribution of
lesion ages and types will be expected to have a smooth, analytic appearance
rather than a jagged and noisy appearance, but will
nonetheless follow the same obvious tendency of increasing from zero to the
total number of lesions $N_n$ of type $n$ at a given time.

From this function, one can define an age distribution function as the
density or mass function corresponding to $\mathrm{CDF}_n$:

\begin{equation}
    f_n(a,t) = \partial_a \mathrm{CDF}_n(a,t)
\end{equation}

which, as lesions are assumed to grow deterministically and are not removed,
will evolve in a very simple manner for $a \neq 0$, $t \neq 0$

\begin{equation}
\label{eq:goveqs}
    \partial_t f_n(a,t) + \partial_a f_n(a,t) = 0\;,
\end{equation}

so that if we wait for any time interval $\epsilon$, the distribution $f_n$ will
``age'' to $f_n(a,t+\epsilon) = f_n(a-\epsilon,t)$. This ageing process
corresponds to the underlying growth dynamics of each individual tumour, 
and given the function $V_n(a)$ the total size can be calculated to be

\begin{equation}
    V_n(t) = \int_0^\infty V_n(a) f_n(a,t) da\;.
\end{equation}

The processes of local invasion and self-metastatic seeding corresponding to
migration and driver accumulation are represented as a boundary condition

\begin{equation}
\label{eq:def2}
    f_n(0,t) = \int_0^\infty \left[ r_n(a)\phi_n(a)f_n(a,t) +
    (1-r_{n-1}(a))\phi_{n-1}(a)f_{n-1}(a,t) \right] da
\end{equation}

for $n>1$, which represents the production of new lesions of type $n$ from both other
type $n$ lesions (first term in integral) and from mutations in type-$n-1$
lesions (second term in integral). Since we start numbering lesions at $n=1$
in this model, there are no lesions of type $0$, and the equivalent of
equation \eqref{eq:def2} is just

\begin{equation}
    \label{eq:def3}
    f_1(0,t) = \int_0^\infty r_1(a)\phi_1(a) f_1(a,t) da\;.
\end{equation}

The initial condition that there is one micro-lesion of type-$n$ and age $a=0$ is simply

\begin{equation}
    f_n(a,0) = \delta_{n,1} \delta(a).
\end{equation}

The growth of individual lesions over time, encoded in the functions $V_n(a)$
and $S_n(a)$, do not directly appear in these equations, which are expressed
entirely in terms of time and lesion age, is nonetheless implicitly present.
The rate of metastasis $\phi_n(a)$ is directly proportional to $S_n(a)$, as
before,

\begin{equation}
    \phi_n(a) = M v_n S_n(a),
\end{equation}

and the volume of individual tumours appears both in expressions for the total
volume of tumours of type-$n$,

\begin{equation}
    V_{\mathrm{tot},n} = \int_0^\infty f_n(a,t) V_n(a) da\; ;
\end{equation}

for the total volume of the tumour,

\begin{equation}
    V_{\mathrm{tot}} = \sum_n V_{\mathrm{tot,n}} = \sum_n \int_0^\infty f_n(a,t) V_n(a) da\; ;
\label{eq:totvol}
\end{equation}

and for the average number of drivers per cell,

\begin{equation}
    \left<n(t)\right> = \frac{1}{V_{\rm tot}(t)} \sum_{n=1}^\infty n
    \int_0^\infty f_n(a,t) V_n(a) da.
\label{eq:avern0}
\end{equation}

Despite the generality of the equations (\ref{eq:goveqs}) and (\ref{eq:def2}),
the infinite genome approximation enforces a strict hierarchical structure on
them: lesions of type $n$ can only produce micro-lesions of type $n+1$, never
the reverse \cite{kimura1969number}. In addition to this --- crucially --- the growth of individual
lesions is assumed to be independent. These two properties of the underlying
stochastic model carry over into equations equation \eqref{eq:goveqs}-equation \eqref{eq:def3},
ensuring that they are linear in age and time, and that they form a strict
hierarchy which can be closed once we solve for the type-$1$ lesion size distribution
$f_1(a,t)$. This last case should be possible as no other
distribution $f_n$ should appear in equation \eqref{eq:def3}.

\subsection{Only one driver}

Since only $f_1$ appears in equation \eqref{eq:def3}, it is essential to analyse the case in which there is
only one strain, with no mutations. Without loss of generality, let us take
$r_n = 1$ at all ages $a$. Three special cases will be considered after the
general formal solution is found.

In this case, the governing equations reduce to McKendrick-von Foerster
equation without removal  \cite{perthame2006transport}, 

\begin{equation}
\label{eq:mvf1}
    \partial_t f_1(a,t) + \partial_a f_1(a,t) = 0
\end{equation}

\begin{equation}
\label{eq:mvf2}
    f_1(0,t) = \int_0^\infty \phi_1(a)f_1(a,t) da
\end{equation}

for $t\neq a$, and $f_1(a,0) = \delta(a)$.

This set of equations are better known as the McKendrick-von Foerster, or
age-structured population equations, well-known in demography, epidemiology,
and mathematical ecology: here, the ages do not correspond to individual
organisms, but rather to individual cancerous micro-lesions in a diffuse
cancerous tumour composed of a conglomerate of such micro-lesions.

The first equation, equation \eqref{eq:mvf1}, implies that $f_1(a,t)$ can be written as
a simpler function

\begin{equation}
    f_1(a,t) = F_1(t-a)
\end{equation}

which should consist of some smooth part, corresponding to a diffuse mass of
secondary and tertiary micro-lesions (and so on), and a non-smooth
Dirac delta corresponding to the presence of the initial, primary lesion,

\begin{equation}
    F_1(t-a) = F^{\mathrm{sm.}}_1(t-a)\Theta(t-a) + \delta(t-a)
\end{equation}

with $F^{\mathrm{sm.}}_1$ the smooth part of $F_1$. This means that the
Laplace transforms of $F_1$ and $F^{\mathrm{sm.}}_1$ are related by

\begin{equation}
    \tilde{F}_1 = \tilde{F}^{\mathrm{sm.}}_1 + 1\;.
    \label{eq:convenientlap}
\end{equation}

Since at $t>0$, the non-smooth part of 
$F_1$ will drop out of the left side of equation \eqref{eq:mvf2}, it can then be
written

\begin{equation}
    F^{\mathrm{sm.}}_1(t) = \int_0^t \phi_1(a) (F^{\mathrm{sm.}}_1(t-a)+\delta(t-a))da\;.
\end{equation}

This equation may be Laplace-transformed in time to yield

\begin{equation}
    \tilde{F}^{\mathrm{sm.}}_1(s) = \frac{\tilde{\phi_1}(s)}{1-\tilde{\phi_1}(s)}
\end{equation}

so that \eqref{eq:convenientlap} implies

\begin{equation}
    \tilde{F}_1(s) = \frac{\tilde{\phi_1}(s)}{1-\tilde{\phi_1}(s)} +1 =
    \frac{1}{1-\tilde{\phi_1}(s)}\;. \label{eq:gensol}
\end{equation}

Providing that the shedding rate $\phi_1(a)$ has a sufficiently tractable
Laplace transform (that is, meromorphic in $s$), the lesion age distribution
$F_1(a,t)$ can be given explicitly, not merely as a formal exact solution.
This is the case for several important special cases that will be discussed
shortly.

However, even if the full distribution cannot be obtained this way, Eq.
equation \eqref{eq:gensol} can still provide us with a means to infer qualitative
insight into the large-$t$ behaviour of $f_1(a,t)$. Note that $\tilde{F_1}(s)$
in the general solution equation \eqref{eq:gensol} becomes singular wherever
$\tilde{\phi_1}(s) = 1$: near any of the roots of this equation, which can
equivalently be written

\begin{equation}
\label{eq:euler}
    \int_0^\infty \phi_1(a) e^{-G a} da = 1\;,
\end{equation}

the Laplace transform of the distribution has a pole, and goes as
$\tilde{F_1} \sim 1/(s-G)$. The inverse of this Laplace transform will
therefore be of the form

\begin{equation}
    F_1(t-a) \propto \exp(G(t-a)) + \dots
\end{equation}

and contain as many exponential terms as there are distinct poles of
$\tilde{F}_1(s)$: or equivalently, roots of equation \eqref{eq:euler}.

Not only should exponential growth be a \emph{generic} feature -- one which is
enabled by migration regardless of the specific form of the seeding rate $\phi_1(a)$ or even 
the lesion volume $V_1(a)$ -- equation \eqref{eq:euler} provides a direct link between the rate at which new
lesions are seeded and the growth rate of the ensemble.
This could also go a long way to explaining why so
many of the underlying models showed a sudden, qualitative change to their
growth curves once any model of migration was incorporated.
This is a really remarkable result!

Equation \eqref{eq:euler} is better known as the Euler-Lotka
equation \cite{keyfitz1997,perthame2006transport},
previously discovered by both Euler and Alfredo Lotka in the context of
quite different of ecological models, and has been independently rediscovered a 
number of times \cite{murray2001mathematical}. The importance of this remarkable result is that it
shows how populations and ensembles of individuals can grow exponentially even while
individuals themselves grow much slower than exponentially (or not at all),
and enables the growth rate of the population as a whole to be determined in
terms of the mean rates at which new individuals (in our case, new
micro-lesions) are created. 

This result is so useful that its generalization in section \ref{sec:geneuler}
and related techniques is very much the keystone for all subsequent mathematical results in this chapter.

If one can think of the Laplace transform as
being, roughly speaking, the representation of a function in terms of
frequency rather than time or space, this growth rate can be thought of as the
average \emph{frequency} with which new micro-lesions appear.

Alternatively, as $\phi_n$ represents the rate of a stochastic process,
the Laplace transform of this rate gives the number of expected
events at a given frequency. If the number of expected events at the growth
rate $G$ is $1$, then this equation can also be interpreted as telling
us that the growth rate of a population corresponds to how often one individual in the population
produces one more individual in the population.

This is a process of positive feedback: larger, mature lesions produce new,
small lesions, which in turn grow into larger, mature lesions and go through a
similar process again. The overall mechanism is very similar to the growth of
animal populations, or the spread of infectious diseases, in the context of
which closely related models have been proposed \cite{keyfitz1997}.

The total number of lesions $N_1(t)$ and the total volume of these lesions
$V_{\mathrm{tot}}(t)$ also show very similar behaviour:

\begin{equation}
    N_1(t) = \int_0^\infty f(a,t) da \propto e^{Gt}
\end{equation}

and

\begin{equation}
    V_{\mathrm{tot}}(t) = \int_0^\infty f(a,t)V(a) da \propto e^{Gt}\;.
\end{equation}

We can now use these general results to investigate three special cases of the
growth of individual tumours and the effect of these variations on the growth
of the ensemble of tumours.

\subsubsection{Surface growth}

As before, we assume that micro-lesions are spherically symmetric, and that all
growth can occur only on the surface. The centre of the lesion is static in
this case: it does not contribute to growth at all.

Let the volume of the lesion increase as it grows as

\begin{equation}
\label{eq:growsas}
    V(a) = \frac{4\pi}{3}(v_1 a)^3
\end{equation}

where $v_1$ is the speed with which the radius of this lesion expands in time.
The rate of migration $\phi_1$ should, as before, be proportional to the
surface area,

\begin{equation}
\label{eq:oldphi}
    \phi_1(a) = M v_1 (4\pi v_1^2 a^2) = 4\pi M v_1^3 a^2
\end{equation}

The above method for deriving exact, formal solutions can be applied here,
yielding

\begin{equation}
    \tilde{F}_1(s) = \frac{s^3}{s^3-8\pi M v_1^3}
\end{equation}

which can be inverted using Mellin's inverse formula

\begin{equation}
    F_1(z) = \mathcal{L}^{-1}\{\tilde{F}_1\}
        = \frac{1}{2\pi i} \oint_\Gamma e^{s z} \tilde{F}_1(s) ds\;,
\end{equation}

where the contour $\Gamma$ encircles all poles of $\tilde{F}_1(s)$
anticlockwise. The residue theorem then yields

\begin{equation}
\label{eq:easilyinverted}
    F_1(z)=\delta (z)+\frac{1}{3} G e^{-\frac{G z}{2}} \left[e^{\frac{3 G
    z}{2}}-\sqrt{3} \sin \left(\frac{\sqrt{3}}{2}  G z\right) -\cos
    \left(\frac{\sqrt{3}}{2}  G z\right)\right]\Theta(z)
\end{equation}

where

\begin{equation}
    G = 2 \pi^{1/3} M^{1/3} v
\end{equation}

from equation \eqref{eq:euler}.

This allows expressions for the total number $N(t)$ and volume $V_1(t)$ to be
derived by substituting equation \eqref{eq:easilyinverted} and equation \eqref{eq:growsas} into equation \eqref{eq:totvol} and
equation \eqref{eq:avern0},

\begin{equation}
   N(t) = \int_0^t F(t-a) da = \frac{1}{3} \left[e^{G t}+2 e^{-\frac{G t}{2}}
   \cos \left(\frac{\sqrt{3}}{2}  G t\right)\right]
\end{equation}

and

\begin{equation}
   V_{tot}(t) = \frac{1}{3M} e^{G t}
   + \frac{2}{3 M} e^{\frac{-G t}{2}}\cos\left(\frac{\sqrt{3}}{2}Gt\right)
   -\frac{1}{M}.
   \label{eq:singletotvol}
\end{equation}

both of which increase exponentially in time after an initial period during
which the growth predominantly consists of the originating primary tumour.

For the tumour to grow to a size typical of those that have been detected and
warrant treatment, say 10$\mathrm{cm}^3-100\mathrm{cm}^3$ after a long period
of progression prior to intervention of 15 years, $V_{tot}$ should be
$10^{11}-10^{12}$ cells, and it follows that $M$ and $v$ must be sufficiently
large. However, $M$ and $v$ only have to be $10^{-6}$ and
$0.1\;\mathrm{cells/day}$ respectively for the tumour burden to reach
macroscopic sizes and continue to expand exponentially in this time frame.

However, individual spheroidal lesions can't be expected to expand
indefinitely: there are a number of factors which act to constrain their
growth, such as limited diffusion and availability of oxygen and glucose, or
mechanical pressure from surrounding tissue. As a result, many tumours never
reach clinically detectable sizes, and can remain undetected and asymptomatic
for many years \cite{folkman_what_1990,naumov_tumor_2008}.

\subsubsection{Surface growth with slow-down}

As a simple but somewhat artificial case, consider a scenario in which the
volume undergoes an initial burst of growth before slowing down. One model
which will show this behaviour, but which will also bear a close resemblance
to the case we previously studied, is if the volume follows

\begin{equation}
    \dot{V}(a) = 4\pi v^3 a^2 e^{-\lambda a}
\end{equation}

where $\lambda > 0$ is the time-scale after which the growth rate decreases.
Consequently,

\begin{equation}
    V(a) = \frac{8\pi v^3}{\lambda^3} \left( 1 - \left(1+\lambda a +
    \frac{(\lambda a)^2}{2}\right) e^{-\lambda a}\right)\;.
    \label{eq:consequently}
\end{equation}

At first glance, this does not look like it generalises equation \eqref{eq:growsas}.
But it can be verified by performing a Taylor expansion of
equation \eqref{eq:consequently} that it is in fact cubic in age $a$ to leading order,
as is equation \eqref{eq:growsas}.

This function has a sigmoidal shape, and saturates at the volume $V(\infty) =
8\pi v^3\/\lambda^3$. Finally, if we also assume that the rate of migration 
$\phi$ also declines exponentially after the same characteristic time-scale, so
that

\begin{equation}
    \phi(a) = 4 \pi M v^3 a^2 e^{-\lambda a}
\end{equation}

one can easily see that these formulas reduce to equations equation \eqref{eq:oldphi}
for $\lambda=0$. $\tilde{F}_1(s)$ is easily calculated through the above
method to be

\begin{equation}
    \tilde{F}_1(s) = \frac{(s+\lambda)^3}{(s+\lambda)^3-8\pi M v^3}
\end{equation}

which corresponds to

\begin{equation}
    F_1(z) = \delta(z) +\frac{G+3\lambda}{3} e^{-\frac{G+\lambda}{2} z} \left[
    e^{\frac{3(G+\lambda)}{2} z} - \sqrt{3}
    \sin\left(\frac{\sqrt{3}}{2}(G+\lambda) z\right) 
    - \cos \left(\frac{\sqrt{3}}{2}(G+\lambda) z\right) \right] \Theta(z)
\end{equation}

where $z=t-a$, and

\begin{equation}
    G = 2 \pi^{1/3} \; M^{1/3} v -\lambda
\end{equation}

which corresponds to the long-term exponential growth or decline rate of the tumour.

\subsubsection{Volumetric growth}

Finally, one can consider the almost trivial case in which all cells from the
lesion are able to replicate and migrate, not only those on the surface, and
that all lesions increase in size exponentially. If the rate with which calls
replicate is $b$ and the rate at which they migrate is $m$, then

\begin{equation}
    V(a) = \exp[(b-m)a]
\end{equation}

\begin{equation}
    \phi(a) = m V(a)
\end{equation}

and as might be expected, the ensemble of all micro-lesions comprising the
tumour grows exponentially in time with rate 

\begin{equation}
    G=b\;.
\end{equation}

\subsection{Multiple driver mutations}
\label{sec:geneuler}

These results can be extended in a more-or-less straightforward manner to the
case in which there are arbitrarily many cell lines, denoted by the number of
drivers in their founding clone, $n$. The underlying reason for the ease of
this extension is the conjunction of two assumptions regarding growth and
mutation: that individual lesions do not affect one another's growth, and that
mutation is an essentially irreversible event. The latter assumption is
reasonably uncontroversial, although as previously discussed there are several
important criticisms that can be made of the former, and several justifications of it.

The key additional ingredient in the mix in the case of many driver mutations
is, of course, the mutant surface fraction $1-r_n$. Some obvious constraints 
can be put
on it: there should initially be no mutants on the surface at all, so
$r_n(0)=1$, and since the new mutant strains we are interested in carry
active, driver mutations, they will eventually take over the lesion in which
they appear, so $\lim_{a\rightarrow\infty}(r_n) = 0$. Beyond this, there are
no specific forms which appear especially likely for all models \emph{a
priori}, and no specific forms suggested by experiment at this time, as the
dynamics of the emergence and expansion of driver mutations and their
sub-lines within tumours is as extremely difficult object of study at present
given that there are few ways to identify specific driver mutations and their
frequencies within actual tumours.

Nonetheless, several forms of $r_n$ can be investigated that do show some
reasonable decline, with determining parameters which can be tentatively
connected to experimental observables for individual cells, but which can
nonetheless be used for tractable mathematical calculations due to their
simple form. This will be the case for any
form of $r_n$ which falls off exponentially in lesion age or faster.

For most of the following results and analysis, we will use an exponentially
decreasing surface fraction

\begin{equation}
    r_n(a) = \exp(-\mu a)
\label{eq:exprn}
\end{equation}

because it declines sufficiently quickly and has a simple Laplace transform
that is easily invertible (roughly speaking, meromorphic). There are a few
possible
physiological and geometrical justifications that can be made for this choice,
but it is primarily for convenience.

This form for $r_n(a)$ be rationalized by assuming that cells on the surface
of the lesion with $n+1$ drivers appear with some small rate $\mu$ and that
their selective advantage can be neglected, so that the dynamics is
effectively neutral and dominated by drift and geometrical effects. A mutant
sector that arose at time $a_0$ will then have on the order of $(a/a_0)^2$
cells on the surface, and the fraction of mutant cells $1-r_n(a)$ then obeys

\begin{equation}
    \frac{\int_0^a 4\pi a_0^2 \mu r_n(a_0) (a/a_0)^2 da_0}{4\pi a^2} = 1-r_n(a)
\end{equation}

which given $r_n(0)=1$, implies that $r_n(a)=\exp(-\mu a)$.

Other than affecting $r_n$, a problem which we will revisit later, the effect
of these driver mutations will be encapsulated by specifying $v_n$, the
expansion speed of lesions with $n$ drivers. We will investigate several such
cases: a continuous, steady increase in fitness, consisting of linear and
exponential increases of $v_n$ with $n$, and the case in which only a few
additional mutations $n$ carry substantial advantages, after which their
effect is small.

However, for the assumed $v_n=n v_1$, the selective advantage carried by novel
driver mutations is $(n+1)/n-1=1/n>0$ and the assumption that population
dynamics are mostly neutral is explicitly not true for small $n$. The dynamics
of non-neutral mutant sectors in spatial models is complex
 \cite{antal_spatial_2015,lavrentovich_survival_2015} and leads to formulae for
$\phi_n r_n$ whose Laplace transforms are not analytically tractable. We will
therefore use the simpler and admittedly less realistic formula for now.

The effect of these driver mutations should in general be to accelerate the
expansion of lesions: lesions with additional advantageous mutations should
obviously grow faster than lesions without, at least under conditions in which
these mutations are advantageous. We will not consider the accumulation of
mutations which are ordinarily not advantageous but become so under certain
conditions: that is, mutations which confer some resistance to specific
therapies but which otherwise slow down cell division or obstruct cell
survival before a cell has a chance to divide. Under the conditions of long,
steady growth prior to discovery and clinical intervention, these mutations
can only accumulate by genetic drift, rather than by clonal selection, and
this stochastic process is not what we are interested in studying.

The effect of each new driver mutation is therefore assumed to be an increase
in the expansion speed of the micro-lesion in which it is present. The main
motivation for this is that replication and death should balance exactly in
normal tissue in the absence of wounds and suchlike in order to maintain
homeostasis. A driver mutation increases the rate of cell division, or
decreases the rate of cell death, and hence the growth rate becomes more
positive \cite{weinberg_book}. In the case of lesions which can grow only on their surfaces due to
inefficient perfusion or spatial constraints this is assumed to translate into
a change in the radial expansion rate $v_n$.


Let us recall equation \eqref{eq:goveqs} 

\begin{equation}
    \partial_t f_n(a,t) + \partial_a f_n(a,t) = 0\;,
\end{equation}

and its boundary conditions

\begin{equation}
    f_n(0,t) = \int_0^\infty \left[
    (r_n(a) \phi_n(a) )f_n(a,t) + (1-r_{n-1}(a)) \phi_{n-1}(a) f_{n-1} (a,t)
    \right] da
\end{equation}

for $n>1$, and

\begin{equation}
    f_1(0,t) = \int_0^\infty r_1(a) \phi_1(a) f_1(a,t) da
\end{equation}

for $n=1$, and

\begin{equation}
    f_n(a,0) = \delta_{n,1} \delta(a)\;.
\end{equation}

The method for solving these in full generality follows that for the case in
which there is only one strain quite closely: in particular, the solution for
$f_1$ is formally identical,

\begin{equation}
    \tilde{F}_1(s) = \frac{1}{1- \mathcal{L}[r_1 \phi_1](s)}
\end{equation}

Equation \eqref{eq:goveqs} implies that $f_n$ can be written

\begin{equation}
    f_n(a,t) = F_n(t-a)
\end{equation}

for some $F_n$ to be determined. The Laplace transform of the corresponding
boundary condition then reads

\begin{equation}
    \tilde{F}_n(s) = \mathcal{L}[r_n \phi_n](s) \tilde{F}_n(s) +
    \mathcal{L}[(1-r_{n-1})\phi_{n-1}](s) \tilde{F}_{n-1}(s)
\end{equation}

for $n>1$, or

\begin{equation}
    \tilde{F}_n(s) = \frac{\mathcal{L}[(1-r_{n-1})\phi_{n-1}](s)}
    {1-\mathcal{L}[r_n \phi_n](s)} \tilde{F}_{n-1}
\end{equation}

from which one can determine $\tilde{F}_n$ for arbitrary $n$ recursively,
given that $\tilde{F}_1$ is known, closing the hierarchy.

Explicitly,

\begin{equation}
    \tilde{F}_{n} (s) = \frac{1}{1-\mathcal{L}\left[r_1 \phi_1\right](s)}
    \prod_{m=2}^n \frac{\mathcal{L}\left[(1-r_{m-1})\phi_{m-1}\right](s)}{1-\mathcal{L}\left[r_m \phi_m\right](s)} . 
    \label{eq:fngen}
\end{equation}

This general solution clearly has poles at every $s=G_m$ which satisfies

\begin{equation}
    \int_0^\infty r_m(a)
    \phi_m(a) e^{-G_m a} da = 1
    \label{eq:gmn}
\end{equation}

which is a generalised version of the classical Euler-Lotka equation for
structured population models, holding for each strain $m$. At sufficiently
long times, those for which $G_m t \gg 1$ or once the mass of secondary
and tertiary micro-lesions is substantial, the number of micro-lesions of any
type $n$ will increase exponentially. The rate of this exponential growth
corresponds to the largest real root in the set ${G_j}$ with $1\leq j \leq n$: 
this is distinct from the real root $G_n$ only when $v_n$ is less than or 
equal to some other $v_j$ of a previous strain.

In the more likely and interesting case that each additional driver increases
the net replication rate of cells in the lesion, the number and mass of
micro-lesions of type $n$ will increases exponentially with rate $G_n$.

\subsubsection{Linear increase in fitness}

\begin{figure}
    \begin{tabular}{lll}
        \hspace{2cm}Surface growth & \hspace{1.3cm}Slow-down surface growth & \hspace{2cm}Volumetric growth\\
        & & \\
        A & B & C \\
        \includegraphics[height=0.3\textwidth]{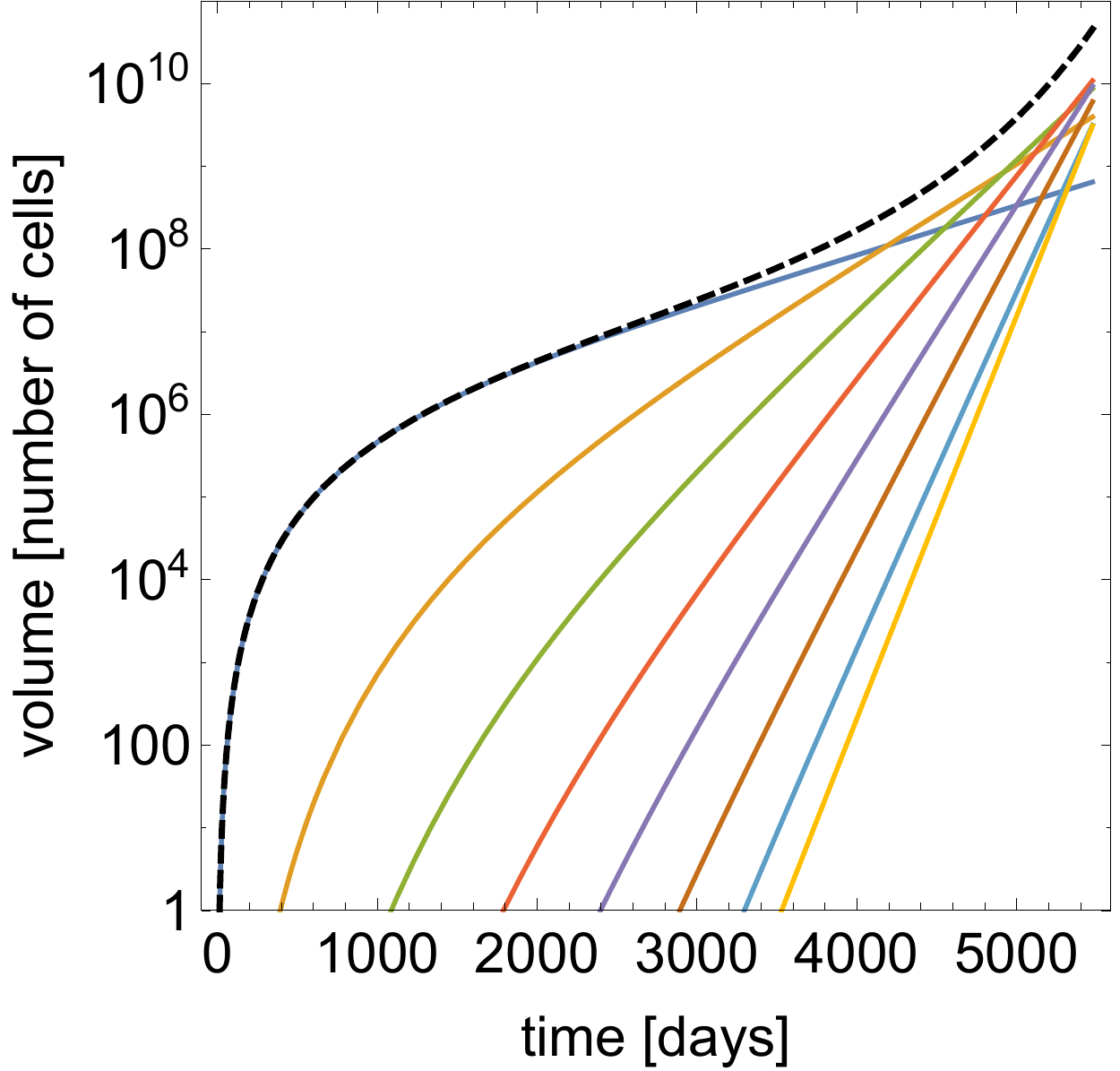} & \includegraphics[height=0.3\textwidth]{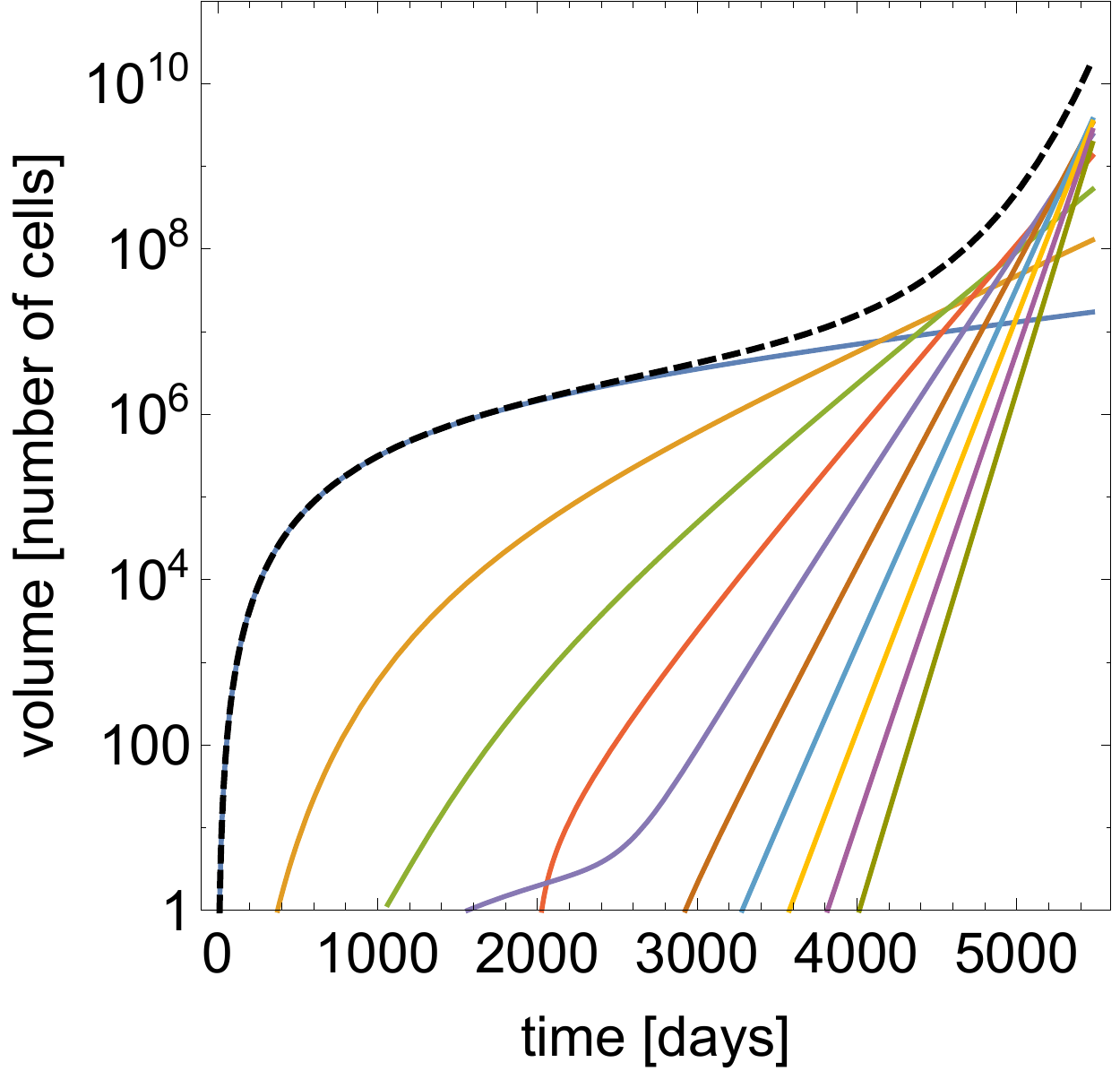} &
 \includegraphics[height=0.3\textwidth]{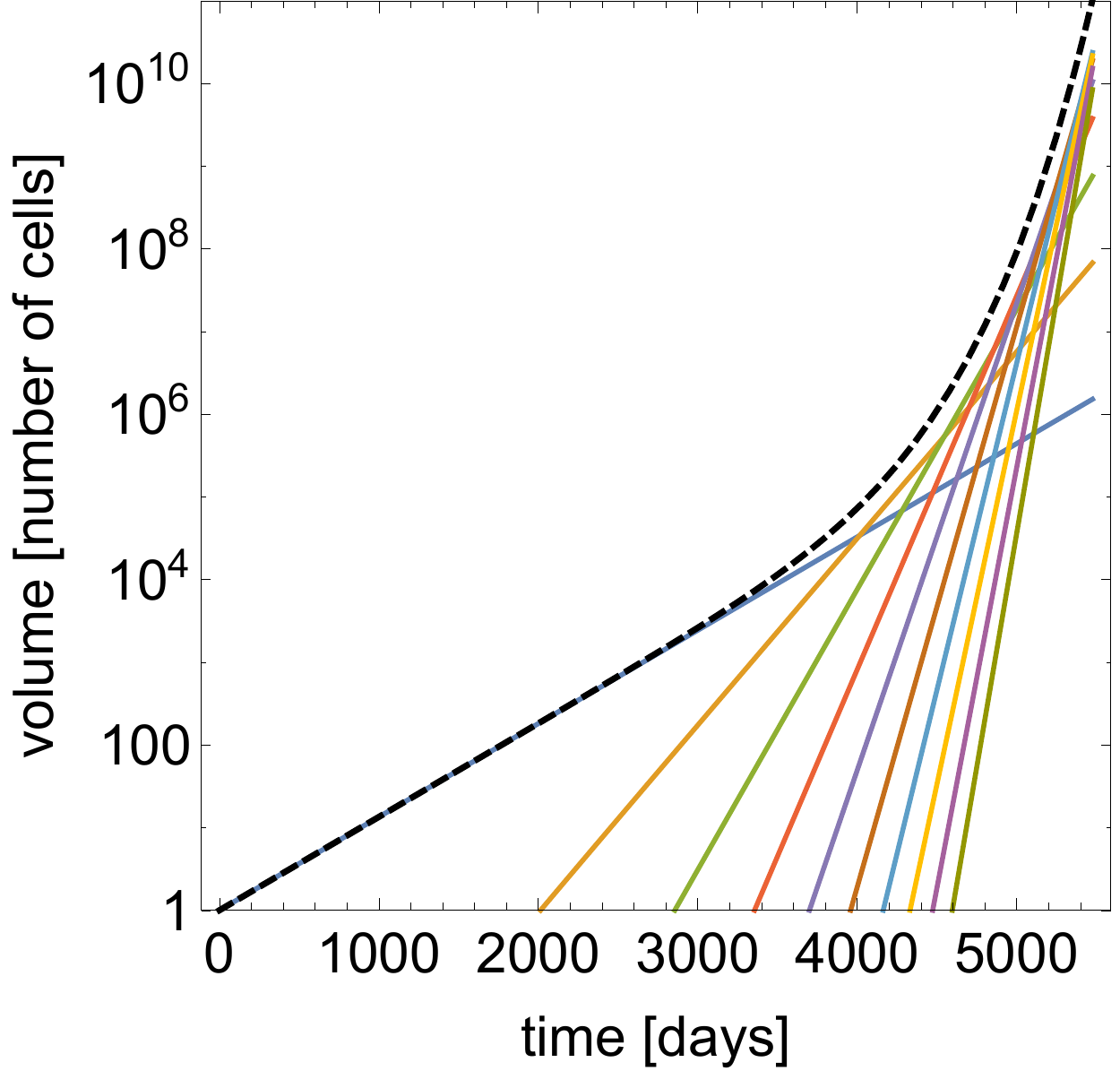} \\
        D & E & F \\
        \includegraphics[height=0.3\textwidth]{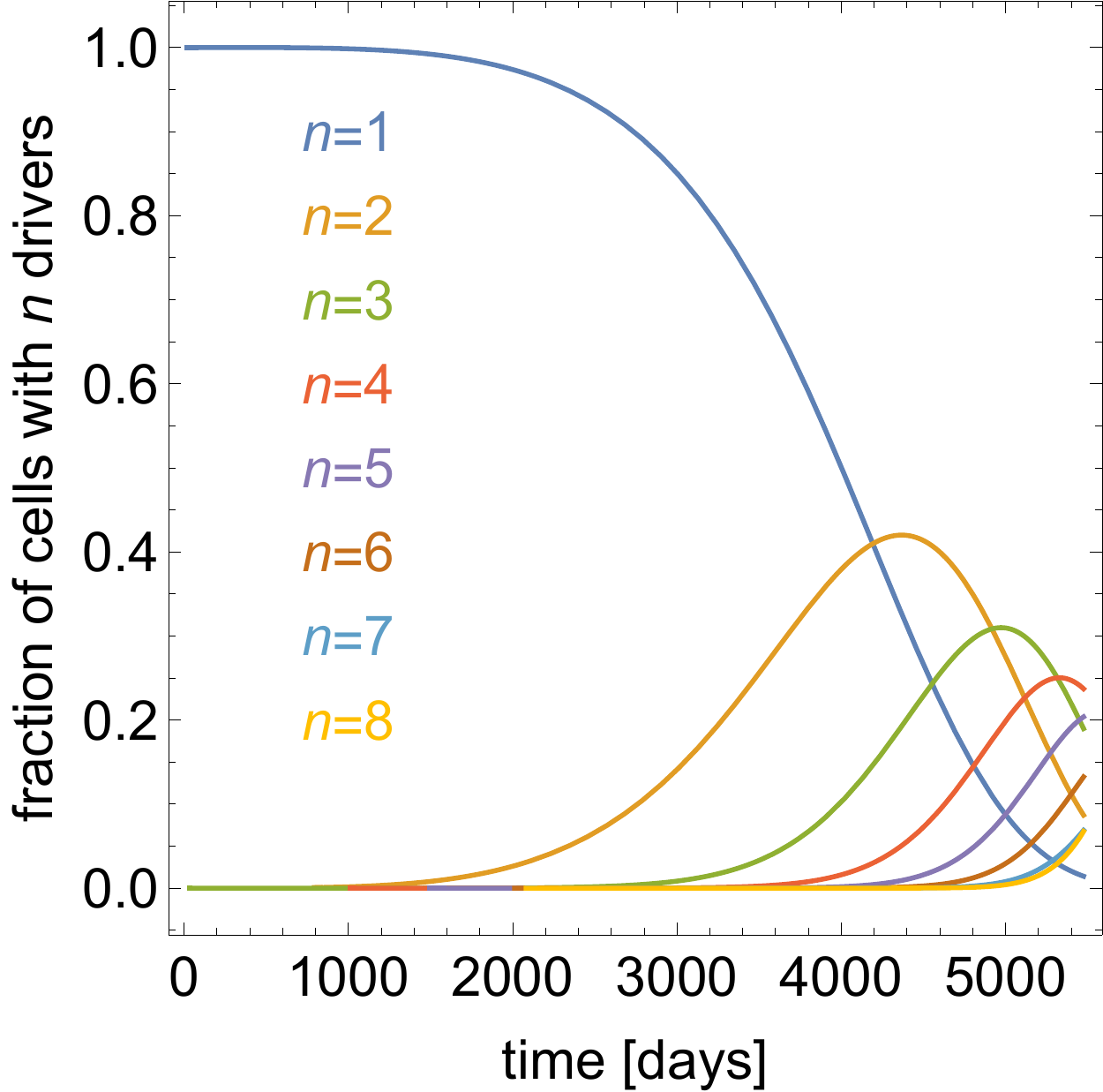} & \includegraphics[height=0.3\textwidth]{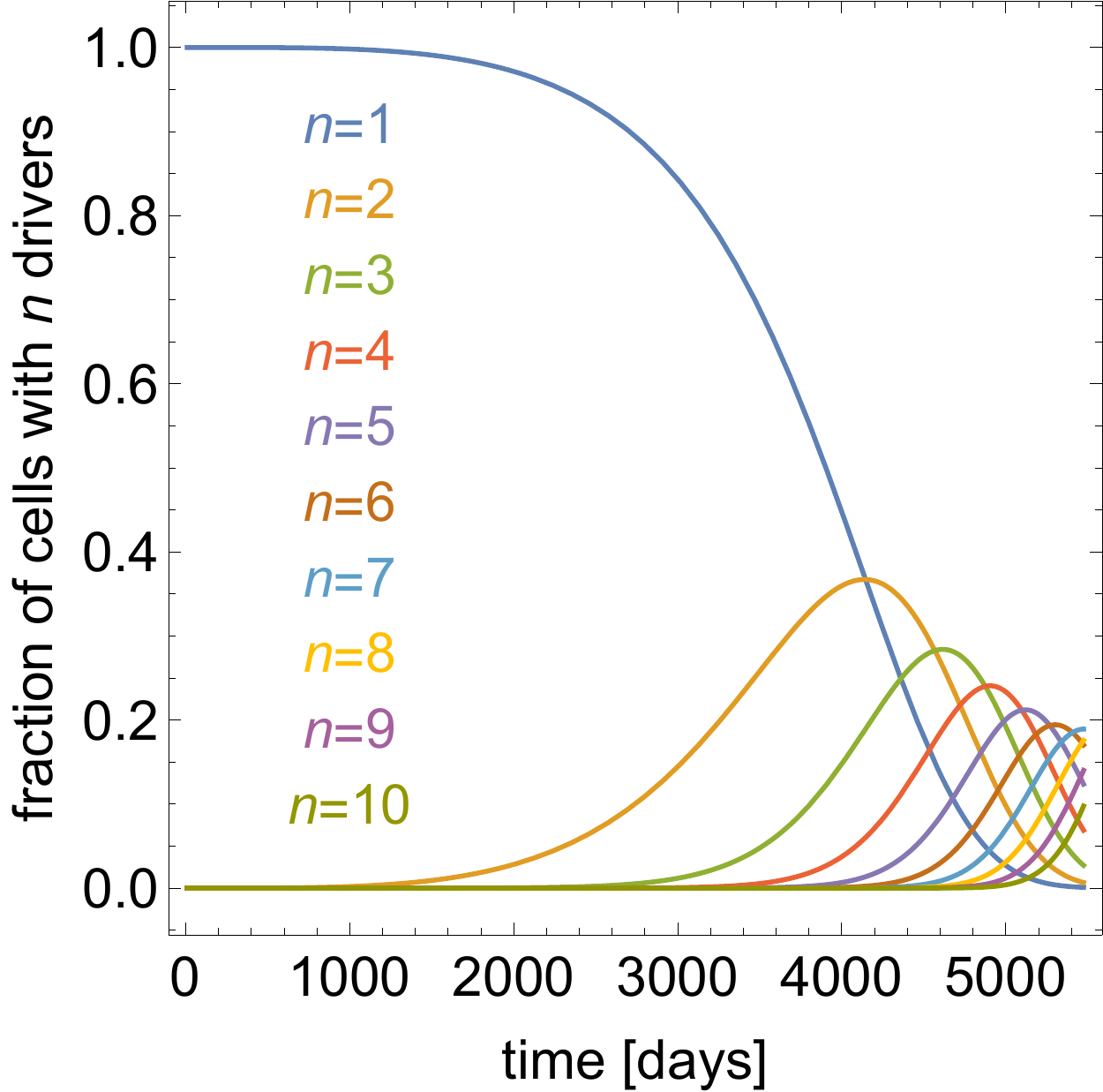} & \includegraphics[height=0.3\textwidth]{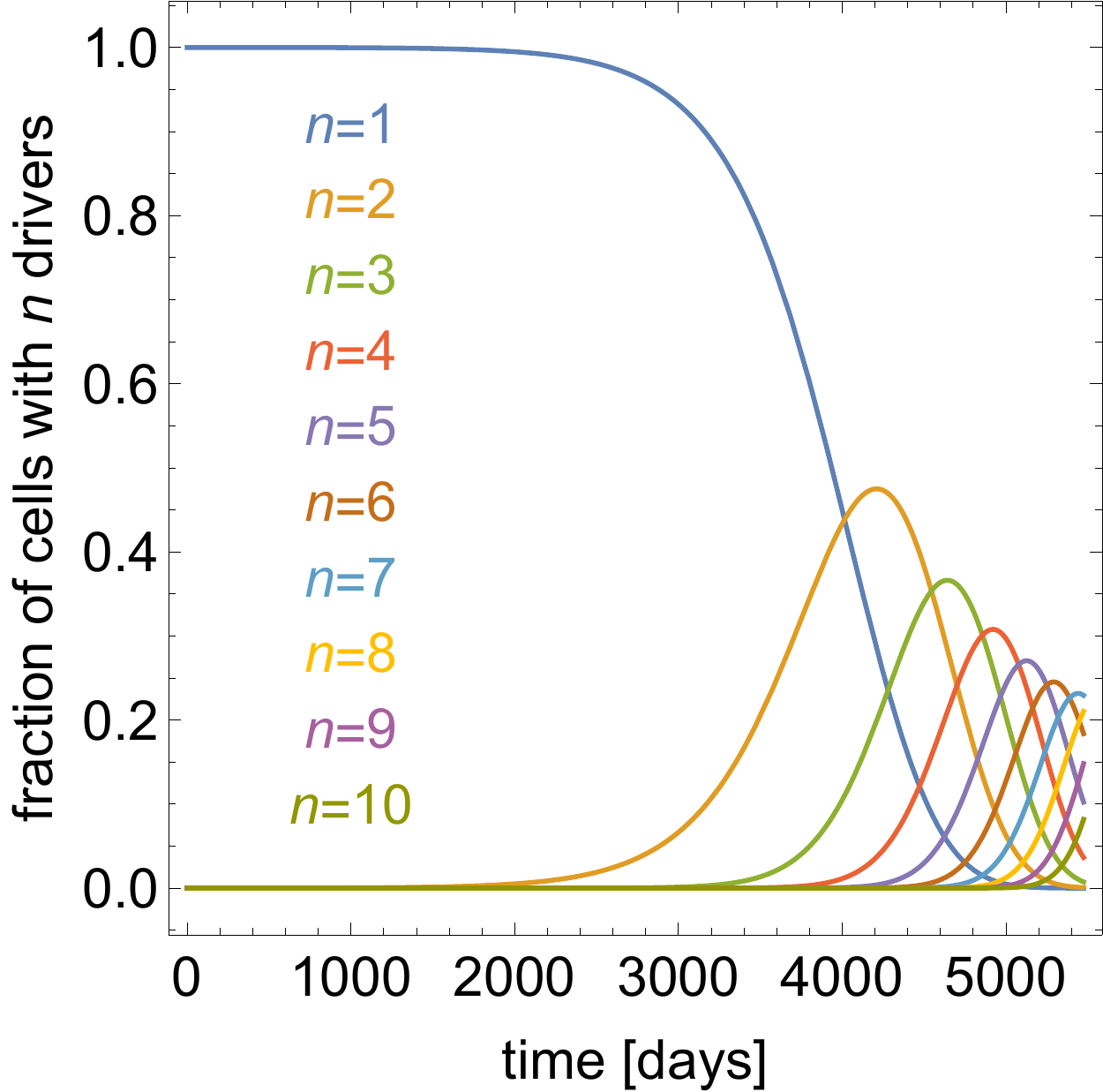} \\
        G & H & I \\
        \includegraphics[height=0.3\textwidth]{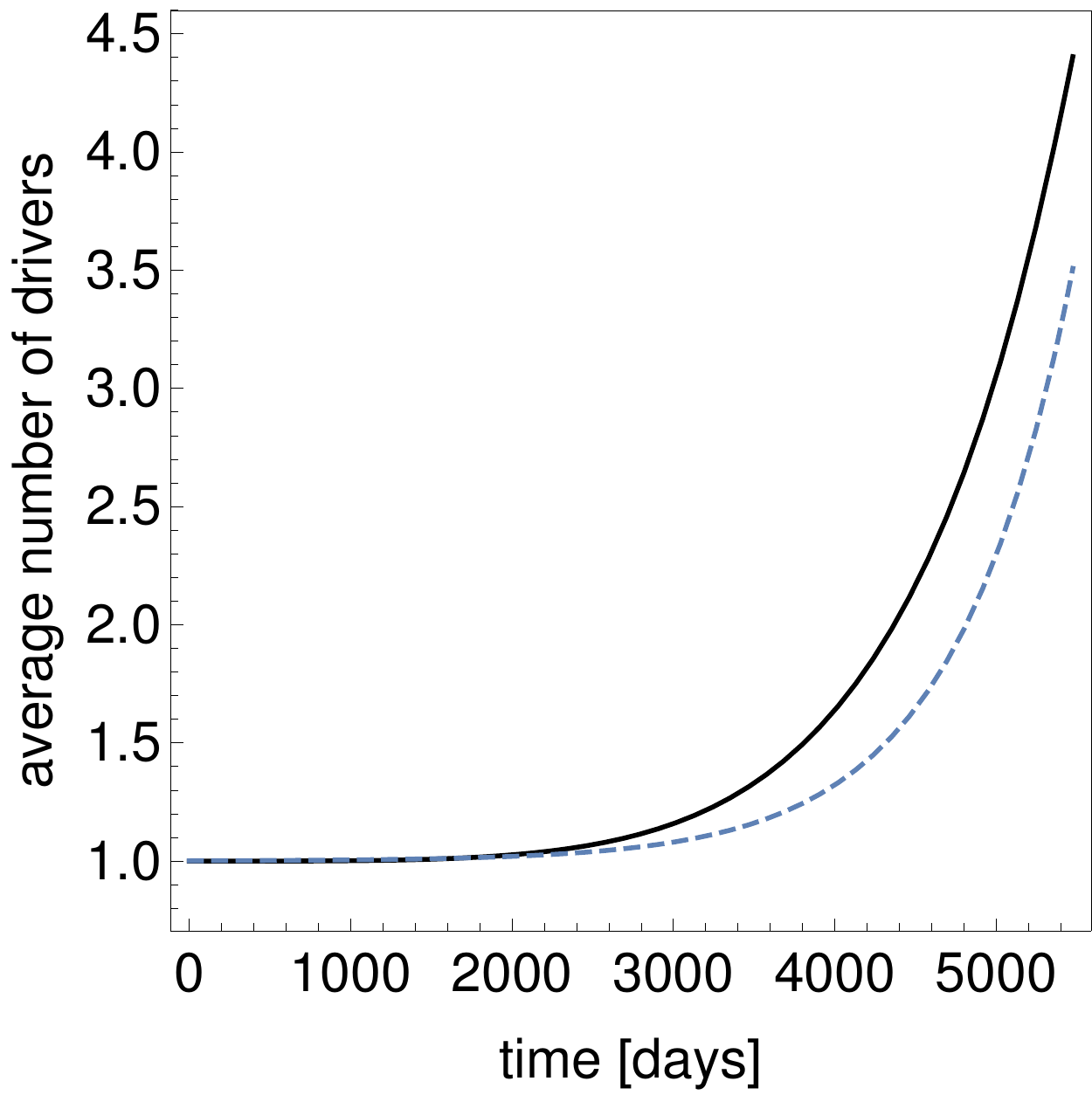} & \includegraphics[height=0.3\textwidth]{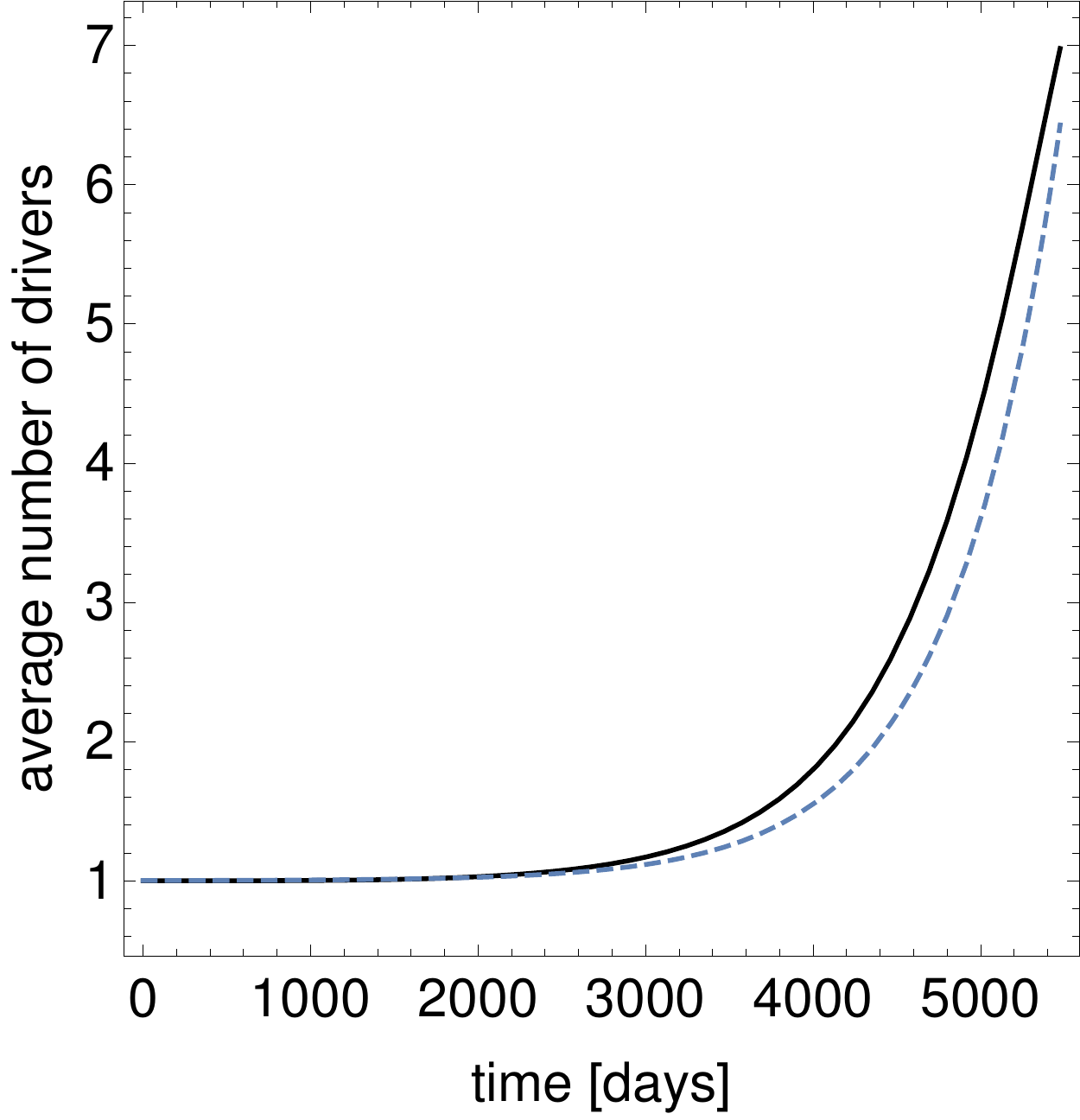} & \includegraphics[height=0.3\textwidth]{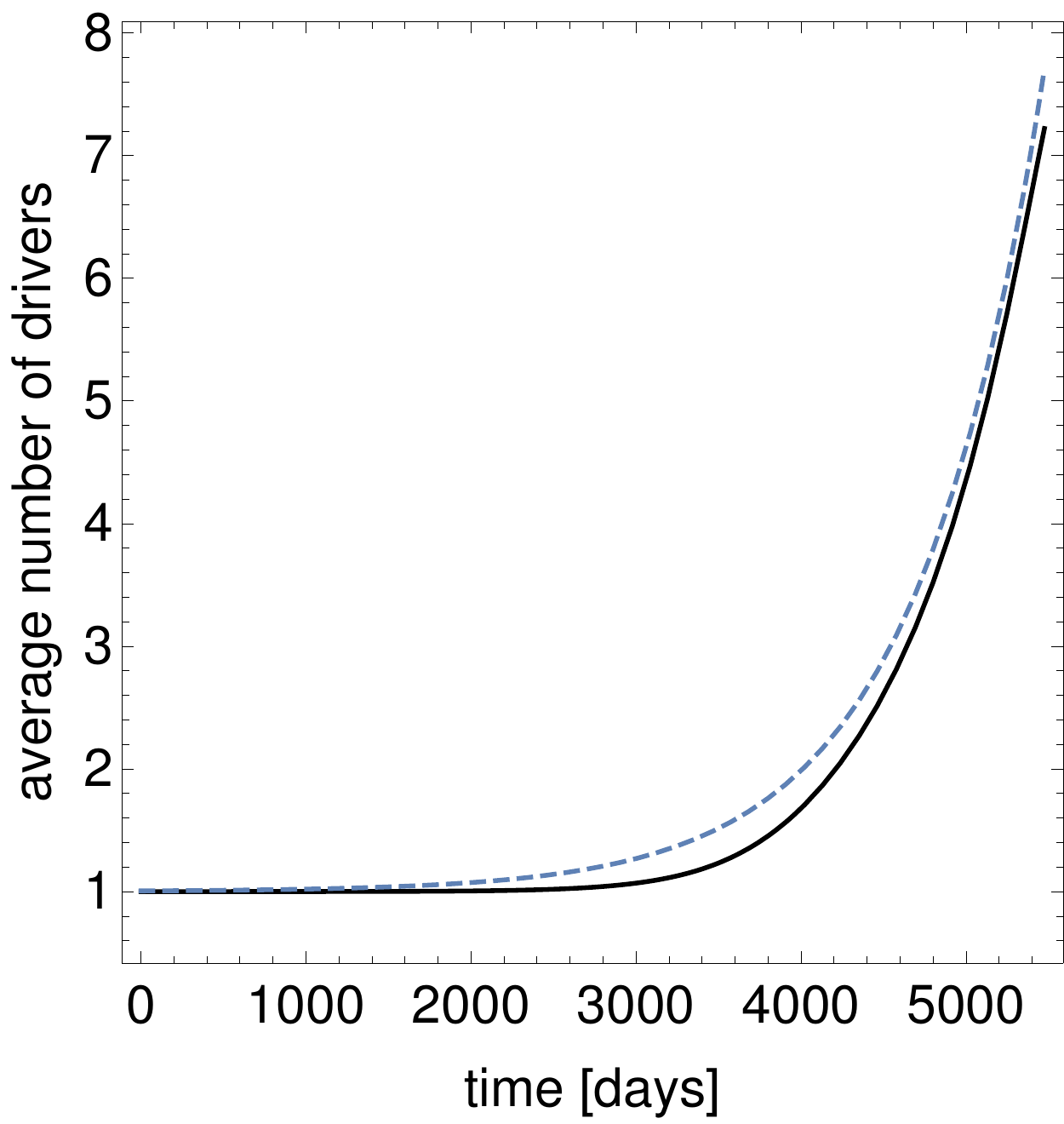}
    \end{tabular}
\caption{\label{fig:multn1}Plots of different quantities (rows of panels) characterizing the tumour as a function of time (days), for different growth scenarios (columns of panels). A,B,C: the total volume (number of cells) of micro-lesions with $n$ driver mutations (coloured curves) and the total volume $V_{\rm tot}$ of the whole tumour (black dashed curve). D,E,F: the fraction of cells with $n=1,2,3,\dots$ driver mutations in the whole tumour (colours as in A,B,C). G,H,I: the average number of drivers $\langle n(t) \rangle$: exact calculation (black) and asymptotic formula (dashed blue). The columns are as follows. 
Panels A,D,G: from the section on the ``Surface growth'' model with parameters $v_1 = 0.0475,\; M = 10^{-6},\; \mu = 2\cdot 10^{-5}$ (all rates in units $[\mathrm{day}]^{-1}$). The parameters $v_1,M$ have been chosen such that the tumour reaches about $10^{11}$ cells in 15 years ($\sim 5500$ days) and accumulates $3-4$ drivers, $\mu$ was assumed to be the same as in Ref.  \cite{bozic2010accumulation}. Black dashed curve in A is the total volume $V_{\rm tot}$ derived from Eqs. (\ref{eq:totvol}) and (\ref{eq:fngen}) with $F_n(z)$ given by (\ref{eq:fnzexact}). Solid black curve in G is the exact analytical calculation from  Eqs. equation \eqref{eq:avern0} and equation \eqref{eq:fnzexact}, blue curve the asymptotic approximation equation \eqref{eq:asymp1}.
Panels B,E,H: corresponding to the section on the ``Surface growth with decreasing replication rate'' model) for $v_1 = 0.053,\; M = 10^{-6},\; \mu = 2\times10^{-5},\; \lambda = 10^{-3}$. Blue curve in panel H is the asymptotic approximation equation \eqref{eq:gnslow}.
Panels C,F,I: volumetric growth for $b = 0.0026,\; M = 10^{-5},\; \mu = 2\cdot 10^{-5}$. Solid black curve is the exact analytical calculation derived from Eqs. equation \eqref{eq:totvol} and equation \eqref{eq:fngen} with $F_n(z)$ given by (\ref{eq:fnvol}). Blue curve is the asymptotic approximation derived from equation \eqref{eq:nvol}.
}
\end{figure}

\paragraph{Surface growth}

First, we study the case in which each new driver contributes the same
advantage to an individual cell, and $v_n = n v_1$: that is, each additional
driver after tumorigenesis increases cell survival and division by the same
amount as the last driver during tumorigenesis. In this case,

\begin{equation}
    V_n(a) = \frac{4\pi}{3} (n v_1 a)^3
\end{equation}

and

\begin{equation}
    \phi_n(a) = 4 \pi M (n v_1 )^3 a^2
\end{equation}

with an exponentially declining $r_n(a)=\exp(-\mu a)$ as discussed previously.

These can be
substituted into equations equation \eqref{eq:goveqs} and equation \eqref{eq:fngen}, and
several important features of this class of models thus determined.

An expression for the Laplace transformed distribution $\tilde{F}_n(s)$ can be
determined straightforwardly, and then inverted to find the direct formula for
$F_n(z)$ to be

\begin{equation}
    F_n(z) = \sum_{j=1}^{3 n} A_j \exp(G_{n,j}z) + \sum_{j=0}^{3 n -4} B_j
    z^j
    \label{eq:fnzexact}
\end{equation}

where $G_{n,j}$ is the $j$-th root of equation \eqref{eq:gmn} for a given
number of drivers $n$, namely

\begin{equation}
    G_{n,j} = G_n \exp(2\pi i j/3)
\end{equation}

where $i$ is the imaginary unit, and the coefficients $a_j$ and $b_j$ depend on 
$v_1,M,\mu$ and of course $n$ and $j$: in general, as they are the
coefficients of the poles of $\tilde{F}_n$, they can be determined from the
residues

\begin{equation}
    A_j = \lim_{s \rightarrow G_{n,j}} \left(
        (s-G_{n,j}) \tilde{F}_n(s)
    \right)
    \label{eq:exproot}
\end{equation}

and

\begin{equation}
    B_j = \lim_{s\rightarrow 0} \left( 
        \frac{s^{j+1}}{j!} \tilde{F}_n(s)
    \right)\;.
    \label{eq:polyroot}
\end{equation}

The precise form of the exponential growth rate of the number of micro-lesions
with $n$ drivers follows

\begin{equation}
    G_n = (8\pi M )^{1/3} nv_1-\mu = G_1 + (G_1+\mu) (n-1), \label{eq:gn1}
\end{equation}

where $G_1=(8\pi M )^{1/3} v_1-\mu$. The growth rate can thus become negative 
for sufficiently large mutation rates $\mu$. This effect is similar to what was 
found in subsection \ref{sec:singleslow}.
However, in contrast to this earlier result in which replication slowed down over 
time, resulting in fewer cells able to migrate and establish new micro-lesions, 
here the cells only change their type while their replication and migration is not 
affected. The total number of cells thus increases over time; we shall show this 
explicitly later in this section. A negative $G_n$ only means that a subpopulation 
of type $n$ cannot grow exponentially at long times: once a strain with $G_n>0$ 
emerges, it rapidly dominates the population. 

The number of micro-lesions of type $n$ can be expressed in terms of these
functions

\begin{equation}
    N_n(t) = \sum_{j=1}^{3n} a_j \frac{\exp(G_{n,j}t)-1}{G_{n,j}}  + \sum_{j=1}^{3n-3} \frac{b_j}{j} t^j
\end{equation}

as can the total volume of cells of type $n$

\begin{equation}
    V_{\mathrm{tot}, n} = \sum_{j=1}^{3n} a_j \frac{8\pi v_{n}^3}{G_{n,j}^4} \left[e^{G_{n,j} t} - 
    \left(1 + G_{n,j} t + \frac{(G_{n,j} t)^2}{2} + \frac{(G_{n,j} t)^3}{6}\right)  \right]
    + \mathcal{O}\left(1\right).
\end{equation}

In order to gain some useful insight into how rapidly drivers accumulate, we
can either calculate $\langle n \rangle$ directly from our analytical
expressions (as is done above), or we can make some approximations in order to
study the qualitative dynamics of $\langle n \rangle$. We will proceed to do
both, and to compare the results.

To accomplish this in a systematic and transparent way, consider the behaviour
of the function $\tilde{F}_n(s)$ around $s=G_n$: from equation \eqref{eq:fngen} it
follows that $\tilde{F}_n(s)$ is dominated by a pole of the form

\begin{equation}
    \tilde{F}_n(s) \cong \frac{A_n}{s-G_n}\;,
    \label{eq:surfcong}
\end{equation}

and so from equation \eqref{eq:exproot},

\begin{equation}
    A_n = \frac{(n-1)!^3(G_{n}+\mu)\left[ \left(1+\frac{\mu}{G_n}\right)^3-1\right]^{n-1}}
 {3\prod_{m=1}^{n-1} \left[n^3-m^3\right]}, \label{eq:surfacegrowthan}
\end{equation}

from which it can be seen from an inverse Laplace transform of equation \eqref{eq:surfcong} 
that $F_n(z) \cong A_n \exp(G_n z)$. 

These formulae for the age distribution are sufficient to specify the age
distribution for type-$n$ lesions: further exact results are not especially
enlightening, and so a comparison of these formulae to both simulations of the corresponding
stochastic model and approximate formulae derived from these exact results
(see chapter \ref{sec:approx1}) are discussed in detail in
\ref{sec:comparison}.

\paragraph{Surface growth with slow-down}

It is not possible for a lesion to grow constantly. Due to spatial
constraints and limited supplies of nutrients and oxygen, growth is eventually
inhibited, often stopping before tumours reach detectable
sizes \cite{folkman_what_1990,naumov_tumor_2008}. 

As an extension of the previous case which remains consistent in the limit
that the characteristic time-scale of growth rate decrease approaches zero,
suppose that the initial expansion speed of lesions with slow-down growth
obeys the same linear increase as before, so that

\begin{equation}
    v_n = n v_1
\end{equation}

with

\begin{equation}
    V_n(a) = \frac{8\pi v_n^3}{\lambda^3} \left( 1 - \left(1+\lambda a +
    \frac{(\lambda a)^2}{2}\right) e^{-\lambda a}\right)\;,
\end{equation}

\begin{equation}
    \phi_n(a) = 4 \pi M v_n^3 a^2 e^{-\lambda a}
    \;.
\end{equation}


These formulae may be substituted into the exact explicit hierarchy
equation \eqref{eq:fngen} to obtain

\begin{equation}
    \tilde{F}_n(s) = \frac{(s+\mu+\lambda)^3}{(s+\mu+\lambda)^3- 8\pi M v_0^3}
    \cdot \prod_{j=2}^n \frac{8\pi M v_{j-1}^3 \left
    [(1+\frac{\mu}{s+\lambda})^3-1\right]}{(s+\mu+\lambda)^3- 8\pi M v_j^3}
\end{equation}

the poles of which clearly occur around $s=G_n$, as is generically the case,
where


\begin{equation}
    G_n = (8\pi M )^{1/3} v_n -\mu-\lambda  = (8\pi M )^{1/3} n v_1
    -\mu-\lambda\;.
\end{equation}

The equivalent expression to equation \eqref{eq:fnzexact}, concerning the prefactors of the
series

\begin{equation}
    F_n(z) = \sum_{j=1}^n A_j \exp(G_j z) + \mathrm{other\;terms}
\end{equation}

may be found from equation \eqref{eq:exproot},

\begin{equation}
    A_n = \frac{(n-1)!^3(G_{n}+\mu+\lambda)\left[
    \left(1+\frac{\mu}{G_n+\lambda}\right)^3-1\right]^{n-1}}{3\prod_{m=1}^{n-1}
    \left[n^3-m^3\right]}.
    \label{eq:surfgrowthan2}
\end{equation}

\paragraph{Volumetric growth}

A linear increase in net growth rate must be parametrised differently for the
volumetric growth rate, and there is no well-defined surface and as such no
characteristic speed. We will use

\begin{equation}
    b_n = n b\;.
\end{equation}

We also have

\begin{equation}
    V_n(a) = \exp\left[(b_n-M)a\right]
\end{equation}

\begin{equation}
    \phi_n(a) = M V_n(a)
\end{equation}

We also assume that all $r_n(a) = \exp(-\mu a)$ as before. This formula can
only describe the fraction of cells that may become invasive correctly only if cells of
type-$n+1$ have a negligible selective advantage over cells of type-$n$, which
cannot actually be the case here: this can therefore only be approximately
correct, and the limits of this approximation are discussed in section
\ref{sec:comparison}.

The Laplace-transformed lesion age distribution from equation \eqref{eq:fngen} can be given explicitly in
this case and has the form 

\begin{equation}
    \tilde{F}_n(s) = (M\mu)^{n-1} \frac{s+M-n b}{\prod_{j=1}^n (s-jb)
    \prod_{j=1}^{n-1} (s-jb +M+\mu)}
    \label{eq:sjb}
\end{equation}

which clearly has simple poles at every $s=k b$ for $1\leq k\leq n$ and
$s=kb-M-\mu$ for every $1\leq k \leq n-1$. Consequently, the real age
distribution $F_n(z)$ can be expressed exactly by an inverse Laplace transform
of equation \eqref{eq:sjb}, so

\begin{equation}
    F_n(z) = \left[ \sum_{k=1}^n A_{n,k} e^{kbz} 
    +\sum_{k=1}^{n-1} B_{n,k} e^{(kb-M-\mu)z}
    \right] \Theta(z) + \delta_{n,1} \delta(z)
\end{equation}

which implies that the total volume of cells with $n$ drivers is

\begin{equation}
    V_{\mathrm{tot},n}(t) = \delta_{n,1} e^{(b-M)t}
    + \sum_{k=1}^n \frac{A_{n,k}}{(n-k)b-M} \left( e^{(nb-M)t}-e^{kbt} \right)
    + \sum_{k=1}^{n-1} \frac{B_{n,k}}{(n-k)b+\mu} \left( e^{(nb-M)t}-e^{(kb-M-\mu)t}\right)
    \label{eq:fnvol}
\end{equation}

where the coefficients $A_{n,k}$ and $B_{n,k}$, derived from
equation \eqref{eq:exproot} are

\begin{eqnarray}
    A_{n,k} = \left( \frac{M\mu}{b^2} \right)^{n-1} (-1)^{n-k+1} 
    \frac{(n-k)b-M}{(k-1)!(n-k)! \prod_{j=1}^{n-1}(k-j+\frac{M+\mu}{b})} \;,  \\
    B_{n,k} = \left( \frac{M\mu}{b^2} \right)^{n-1} (-1)^{n-k}
    \frac{(n-k)b+\mu}{(k-1)!(n-k-1)!\prod_{j=1}^n (k-j-\frac{M+\mu}{b})} \;.
    \label{eq:volan}
\end{eqnarray}

Although the above formulae are exact, they are rather complex, and the
asymptotic behaviour of $V_{\mathrm{tot},n}$ is not immediately apparent.
These formulae may nonetheless be used to plot and compare $V_{\mathrm{tot},n}$ 
numerically to simulations, as in section \ref{sec:comparison}: a fruitful
approximate approach is detailed in section \ref{sec:approx1}.

\subsubsection{Fitness increase with rapid plateau}

In all previous cases, all drivers carried an increasing selective advantage
over one another with an identical relative strength, meaning that cells with
$n=3$ driver mutations had a net growth rate three times that of cells with 
$n=1$. It is not possible for this pattern to continue indefinitely, as there is
an upper limit to how quickly cells can progress through the cell cycle
(detailed in chapter 1), and increases in this net growth rate presumably
occur by reductions in the rate of cell death by apoptosis in addition to
modest increases in the rate of division induced by self-signalling. A linear
increase cannot continue past the point at which there is no cell death, and
cells divide as rapidly as possible, and can at best be a rough approximation
of the truth.

Linear increases in fitness also imply that drivers accumulate exponentially
(see section \ref{sec:approx1}), which would imply in turn that the majority of
drivers emerge at a very late stage in the progression of the tumour, for
which there is very limited evidence.

In this section, we will study the case in which the selective advantage of
the first few drivers is much larger than that of any subsequent driver. The
first additional driver mutation can substantially increase the net
replication rate, to double the net growth rate of cells in the primary
lesion, and this linear increase may continue until up to the third, after
which point additional drivers have only a very weak effect. That is, for the
underlying net replication rate we will take

\begin{eqnarray}
    b_2 &=& 2 b_1 \\
    b_3 &=& 3 b_1 \\
    b_n &=& 3 b_1 + (n-3)\epsilon b_1 | n>3
\end{eqnarray}

for the volumetric growth model and for the lattice model simulations, and for
the surface growth model we will take

\begin{eqnarray}
    v_2 &=& 2 v_1 \\
    v_3 &=& 3 v_1 \\
    v_n &=& 3 v_1 + (n-3)\epsilon v_1 | n>3 \label{eq:slowvn}
\end{eqnarray}

with some $b_1 \propto v_1$, with a constant of proportionality fixed by
comparing the measured total volume in a set of Eden model simulations 
without mutation or migration to the simple
geometrical expansion $V(t) = 4\pi v_1^3 t^3 /3$.

Statistical evidence of cancer incidence rates suggest that having a
restricted number of early driver mutations which make a large contribution 
to the net replication rate of a given cell line is indeed the
case for some lung and colorectal cancers  \cite{tomasetti_2015}. 

All cancer cells in this subset of models are assumed to initially have one
driver, which enables them to outgrow surrounding tissue, and the ``additional
drivers'' we talk about here should be taken to mean the second driver
overall, and then the third, and so on.

We will study both cases in which only the first two additional drivers
substantially increase the net replication rate: there is little to be gained
from studying analytical approximations in this case, so we will restrict
ourselves to analysing a few exact formulae, and then comparing the results of
direct calculation from these to the results of simulations with the
corresponding underlying fitness landscape.

From figure \ref{fig:ini}, one can see that the average number of drivers per cell 
accumulates roughly sigmoidally, before the tumour reaches a detectable size
($V_{\mathrm{tot}}>10^9$), and consequently most drivers will emerge and
accumulate relatively early in the tumour's development. This implies that the
final tumour is much less heterogeneous upon treatment because almost all
cells have the same underlying set of driver mutations. Moreover, the total
volume grows approximately exponentially after the appearance and accumulation
of these drivers, rather than super-exponentially as in previous models: one
can clearly observe a gradual transition between the early and late
exponential growth rates in both the exact analytics and the simulations, as
evidenced in figure \ref{fig:ini}. These results are in broad agreement
with recent experimental evidence on intra-tumour heterogeneity in colorectal
cancer \cite{sottoriva_2015}.

\subsection{The generating function technique}
\label{sec:approx1}

The exact results of the previous section can already be used to calculate
relevant quantities such as the total tumour burden and the mean number of
drivers present per cell, a calculation which can be meaningfully compared to
the same quantities obtained from the stochastic model. However, this procedure is
rather opaque, and it is difficult to obtain qualitative insights into the
effect of different fitness landscapes or underlying growth models for
individual micro-lesions in this way.

To derive expressions for the total volume and mean number of drivers from equation \eqref{eq:fngen},
equation \eqref{eq:growsas} and equation \eqref{eq:totvol} more easily, and to
allow asymptotic behaviour of both of these to be studied, it would be useful to
develop an approach to systematically extract $V_{\mathrm{tot}}$ and 
$\langle n \rangle$ from the exact expressions.

First, we can notice that both the total volume and mean number of drivers
depend on expressions for the total volume of lesions of each type $n$ like so

\begin{eqnarray}
    V_{\mathrm{tot}}(t) = \sum_n V_{\mathrm{tot},n} (t) \\
    \langle n \rangle (t) = \frac{\sum_n n V_{\mathrm{tot},n}}{\sum_n
    V_{\mathrm{tot},n}} 
\end{eqnarray}

both of which are entirely determined by the sequence of functions 
$\left\{ V_{\mathrm{tot},n}\right\}$. As a result, they can be expressed in
terms of the behaviour of another function in a dummy
variable $q$ which encodes all the relevant information in the series 
$\left\{ V_{\mathrm{tot},n}\right\}$. We will call this function the generating
function $Z(t,q)$ for consistency with other literature in which closely
related techniques are used. We define $Z(t,q)$ as 

\begin{equation}
    Z(t,q) := \sum_n e^{q n} V_{\mathrm{tot},n}
    \label{eq:qtq}
\end{equation}

from which several interesting quantities can be calculated: in particular,

\begin{equation}
    V_{\mathrm{tot}} = Z(t,0)
\end{equation}

and 

\begin{equation}
    \left<n(t)\right> = \left[\frac{\partial\log Z(t,q)}{\partial
    q}\right]_{q=0}\;.
    \label{eq:avern}
\end{equation}

Other, higher-order moments and cumulants such as the variance in the number
of mutations $\sigma^2(n)$ may also be calculated by taking higher order
derivatives of $Z$, but are not interesting to us at this time.

This approach is especially useful for cases in which the precise form of $V_n$ and $F_n$
are such that $Z(t,q)$ is an analytic function in the dummy variable
$q$. Approximations of $Z$ make the 
asymptotic behaviour of $\langle n\rangle$ and $V_{\mathrm{tot}}$ much easier to investigate, 
as features of both may be determined significantly faster
than their calculation from the exact series derived in the previous
section.

\subsubsection{Linear increase in fitness}

\paragraph{Surface growth}

For human cancer cells, the mutation rate $\mu$ is considerably less than the
net growth rate $G_1$ and other $G_n$, simply because the appearance of any
new mutation at a specific locus is very unlikely in any given cell division
event. Equation \eqref{eq:surfacegrowthan}
simplifies in the relevant limit $\mu/G_1 \ll 1$ to

\begin{equation}
    A_n \approx \frac{(n-1)!^3}{3 n^{n-1} \prod_{m=1}^{n-1} \left[n^3-m^3\right]} \left(\frac{3\mu}{G_1}\right)^{n-1}
\end{equation}

so the $Z$ from \eqref{eq:qtq} is asymptotically

\begin{align}
Z(t,q) &\cong & \sum_{n=1}^\infty e^{qn} A_n \int_0^\infty e^{G_n (t-a)} \frac{4\pi}{3} v_0^3 n^3 a^3 da \cong  
    {\rm const} \cdot \sum_{n=1}^\infty e^{qn+G_1 n t} \frac{A_n}{n} \nonumber \\
	&\cong & {\rm const} \cdot e^{G_1 t} \sum_{n=1}^\infty \left(\frac{3\mu}{G_1}e^{q+G_1t}\right)^n 
    \frac{(n-1)!^3}{3 n^n \prod_{m=1}^{n-1} \left[n^3-m^3\right]}.  \label{eq:ztq3}
\end{align}

The complicated numerical prefactor in this equation can be approximated for
large $n$ as

\begin{equation}
    \frac{(n-1)!^3}{3 n^n \prod_{m=1}^{n-1} \left[n^3-m^3\right]} 
    \cong {\rm const}\cdot \frac{1}{(n-1)!} \exp\left[-n\left(\frac{\pi}{2\sqrt{3}} + \frac{3\ln 3}{2} +1\right)\right]
\end{equation}

from Stirling's approximation. Given this, it can be obtained from equation \eqref{eq:ztq3} that

\begin{equation}
    Z(t,q) \approx C \exp\left[G_1 t + q + \frac{\mu e^{G_1 t+q-\frac{\pi}{2\sqrt{3}}-1}}{\sqrt{3}G_1}\right] 
    \label{eq:ztq}
\end{equation}

and finally, using equation \eqref{eq:avern}, it follows that

\begin{equation}
    V_{\rm tot}(t) = Z(t,0) \propto \exp\left[G_1 t + \frac{\mu e^{G_1 t-\frac{\pi }{2 \sqrt{3}}-1}}{\sqrt{3} G_1}\right],
\end{equation}

and

\begin{equation}
    \left<n(t)\right> \cong 1+\frac{\mu e^{-1-\frac{\pi}{2\sqrt{3}}}}{\sqrt{3}G_1} e^{G_1 t},
    \label{eq:asymp1}
\end{equation}

so that the total volume is seen to increase faster than exponentially in this
case, driven by an exponential increase in the average number of drivers
$\langle n \rangle$.

\paragraph{Surface growth with slow-down}
\label{sec:singleslow}

In the biologically relevant limit, mutations should occur only rarely in any
given cell division event. As the probability of any type of mutation
occurring in any given cell division event should be proportional to the ratio
$\mu/G_n$, barring possibly some factors regarding turnover and ploidy, it
makes sense to ask what happens to the exact formulae in the limit that
$\mu/G_1 \ll 1$.

The prefactors from equation \eqref{eq:surfgrowthan2} reduce to

\begin{equation}
    A_n \cong \frac{(n-1)!^3}{3 n^{n-1}\prod_{m=1}^{n-1} \left[ n^3 -
    m^3\right]} \left(\frac{3\mu}{G_1+\lambda}\right)^{n-1}\;.
\end{equation}

This allows us to write

\begin{equation}
    Z(t,q) \cong \mathrm{const} \cdot \exp\left[ G_1 t + q + 
    \frac{\mu e^{(G_1+\lambda)t + q
    -\frac{\pi}{2\sqrt{3}}-1}}{\sqrt{3}(G_1+\lambda)}\right]
\end{equation}

which implies

\begin{equation}
    V_\mathrm{tot}(t) \cong \mathrm{const} \cdot \exp \left[
    G_1 t +\frac{\mu e^{(G_1+\lambda)t 
    -\frac{\pi}{2\sqrt{3}}-1}}{\sqrt{3}(G_1+\lambda)} 
    \right]
\end{equation}

\begin{equation}
    \langle n \rangle (t) \cong 1+\frac{\mu e^{(G_1+\lambda)t 
    -\frac{\pi}{2\sqrt{3}}-1}}{\sqrt{3}(G_1+\lambda)} 
\label{eq:gnslow}
\end{equation}

and the total volume can be seen to increase faster than exponentially once
again, being driven by an exponential increase in the average number of
drivers. 

In both cases, linear fitness landscapes give rise to faster-than-exponential
growth. We are not aware of any serious proposals to use such
super-exponential growth curves to successfully model the growth of actual
tumours, which motivates a look at fitness landscapes which gradually tail off
later on in this chapter.

Curiously, the average number of drivers is independent of $\lambda$.
Every appearance of $\lambda$ in the above formula is exactly cancelled out by
a corresponding $-\lambda$ term arising from the explicit expression of $G_1$.
I have not found any intuitive arguments as to why this should be the case.

\paragraph{Volumetric growth}

To obtain simpler expressions for the total volume and number of
drivers, we can first note that in real tumours migration is relatively rare,
and so the replication rate must be much larger than the migration rate, or
else cells would only move around in a diffuse mass without replicating, so
that $b\gg M$. As beneficial mutations are also very rare events, the rate
$\mu$ must also be much smaller than $b$.

As a consequence, $V_{\mathrm{tot},n}(z)$ will be dominated by the term proportional to
$A_{n,n}$ from \eqref{eq:volan} as long as $bt\gg 1$: i.e. for sufficiently long times (much longer
than the individual lesion's doubling time), $e^{bt}\gg 1$.

We can furthermore approximate $A_{n,n}$ as

\begin{equation}
    A_{n,n} \cong \left( \frac{M\mu}{b^2} \right)^{n-1} \frac{M}{n!^2} 
    \left(1+\mathcal{O}\left( \frac{M}{b} \right)\right)
\end{equation}

whereupon the generating function $Z(t,q)$ becomes

\begin{eqnarray}
    Z(t,q) \cong \frac{M}{M-\mu} e^{bt+q} (e^{-\mu t}-e^{-M t})
    \sum_{n=0}^\infty \frac{1}{n!^2} \left( \frac{M\mu}{b^2} e^{bt+q}
    \right)^n \\
    = \frac{M}{M-\mu} e^{bt+q} (e^{-\mu t}-e^{-M t}) I_0 \left( 
    \frac{2\sqrt{M\mu}}{b} e^{(bt+q)/2}
    \right)
\end{eqnarray}

where $I_0(x)$ is a modified Bessel function of the first kind. This
immediately allows us to write

\begin{equation}
    V_{\mathrm{tot}}(t) = Z(t,0) \propto e^{bt} I_0 \left(
    \frac{2\sqrt{M\mu}}{b} e^{(bt+q)/2} \right)
\end{equation}
\begin{equation}
    \langle n \rangle (t) \cong 1 + 
    \frac{e^{bt/2} \sqrt{M\mu}I_1\left(\frac{2\sqrt{M\mu}}{b} e^{bt/2}\right)}
    {b I_0\left(\frac{2\sqrt{M\mu}}{b} e^{bt/2}\right)}
    \cong 1+ \frac{\sqrt{M\mu}}{b}  e^{bt/2}  \label{eq:nvol}
\end{equation}

so that the total volume can again be seen to increase faster than
exponentially, whilst drivers accumulate exponentially, both phenomena which 
are also exhibited in the surface growth models. This therefore seems to be a
generic feature of having a linear increase in fitness with the number of
drivers: an exponential accumulation of driver mutations drives a
super-exponential increase in tumour burden.


Note that for this model to reach the same size as in the surface growth model
(around $10^{11}$ cells) in a reasonable time frame as well as having
accumulated a similar number of drivers (more than two), the actual increment
$b$ of the growth rate per driver has to be very small, and so this model is
of limited applicability.

\section{Comparison to simulations}


We compared the results obtained in section \ref{sec:analytic} with numerical
simulations of both the stochastic model defined in section \ref{sec:stochast} and
the lattice model defined in chapter \ref{ch:Lattice}. In the simplified
stochastic model, growth and the appearance of lineages carrying additional
driver mutations were both deterministic, whereas migration was left as a
stochastic process. This was intended to reflect the fact that despite the
relative complexity of all processes, migration is a relatively rare event
compared to growth. 

In the Eden-like lattice model based on that from  \cite{waclaw_spatial_2015}, 
all processes were
stochastic, but with different rates that depended on a cell's surrounding
environment. It was never the case that a cell totally surrounded by others
could migrate and found new lesions, for instance, and so the rates of both
migration and of individual lesions' growth were found to be asymptotically
proportional to their surface area.

All models attempted to make predictions for at least the total tumour burden,
denoted $V_{\mathrm{tot}}$, and for the relative frequencies of different
lineages with different numbers of accumulated driver mutations, the
populations of which are denoted $V_{\mathrm{tot},n}$. These can both be used
to calculate the mean number of additional driver mutations per cell, $\langle
n \rangle$, and other statistical measures of genetic diversity.

\subsection{Comparison of stochastic minimal model and analytics}
\label{sec:comparison}

We performed numerical simulations of the surface growth model described in
section \ref{sec:algorithm}, and treated
individual lesions of age $a$ as balls of volume $4\pi (v_n a)^3/3$. New
lesions were founded with rate $\phi_n = 4\pi M v_n^3 a^2$, a fraction $r_n(a)=\exp(-\mu
a)$ of these had the same number of driver mutations $n$ in their founding
cell, and a complementary fraction $1-r_n(a)$ of which had $n+1$.

Mutations in these simulations consisted of three initial strong drivers, and a very
small fitness advantage $\epsilon$ for any additional driver after this point.
This fitness advantage was non-zero, so that additional driver mutations did
indeed carry some selective advantage (and must by definition), but was in
practice too small to strongly affect growth on clinically relevant
timescales. The expansion speeds $v_n$ were thus given by \ref{eq:slowvn}.

\begin{figure}
\begin{center}
    \includegraphics[width=0.6\textwidth]{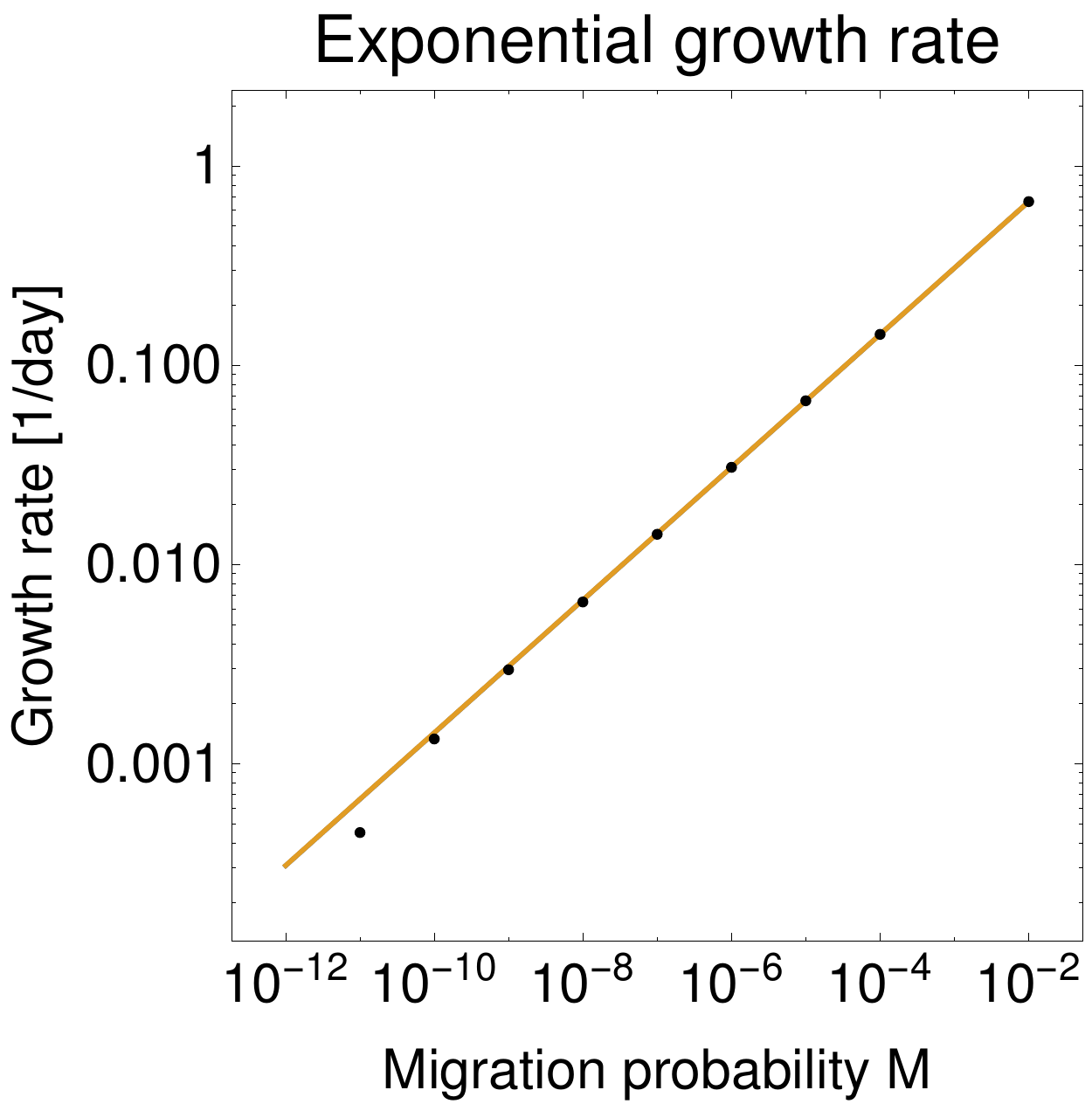}
\end{center}
\caption{\label{fig:edengr1}Plot showing the measured growth rate or the
diffuse collection of micro-lesions in the ensemble model versus migration rate
$M$. Parameter values were expansion speed $v=1.05$, with $M$ varied between
two extreme values. This plot confirms the power-law dependence with an
exponent of $1/3$.}
\end{figure}

\begin{figure}

    \begin{tabular}{lll}
    A & B & C \\
    \includegraphics[height=0.32\textwidth]{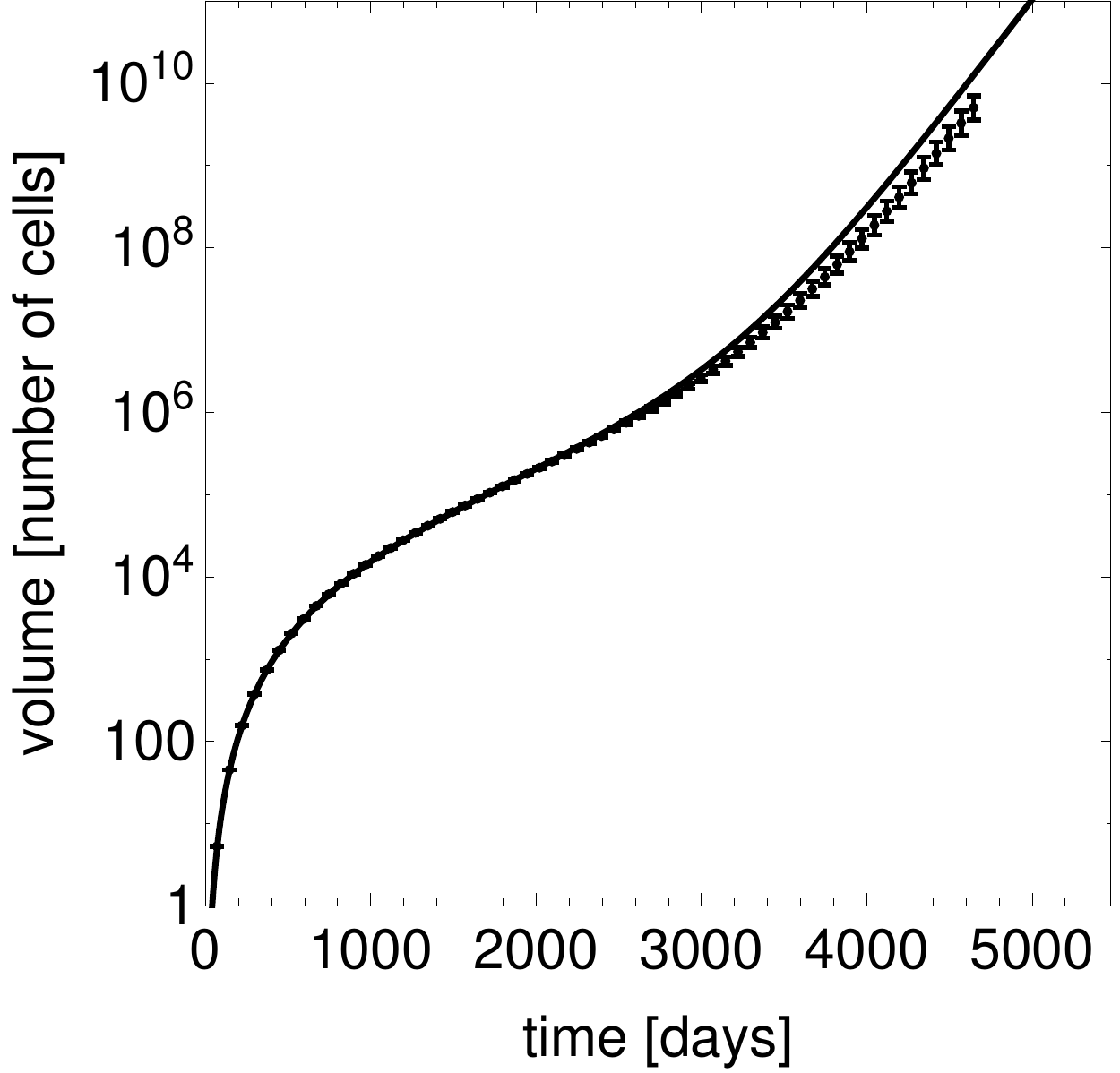} & 
    \includegraphics[height=0.32\textwidth]{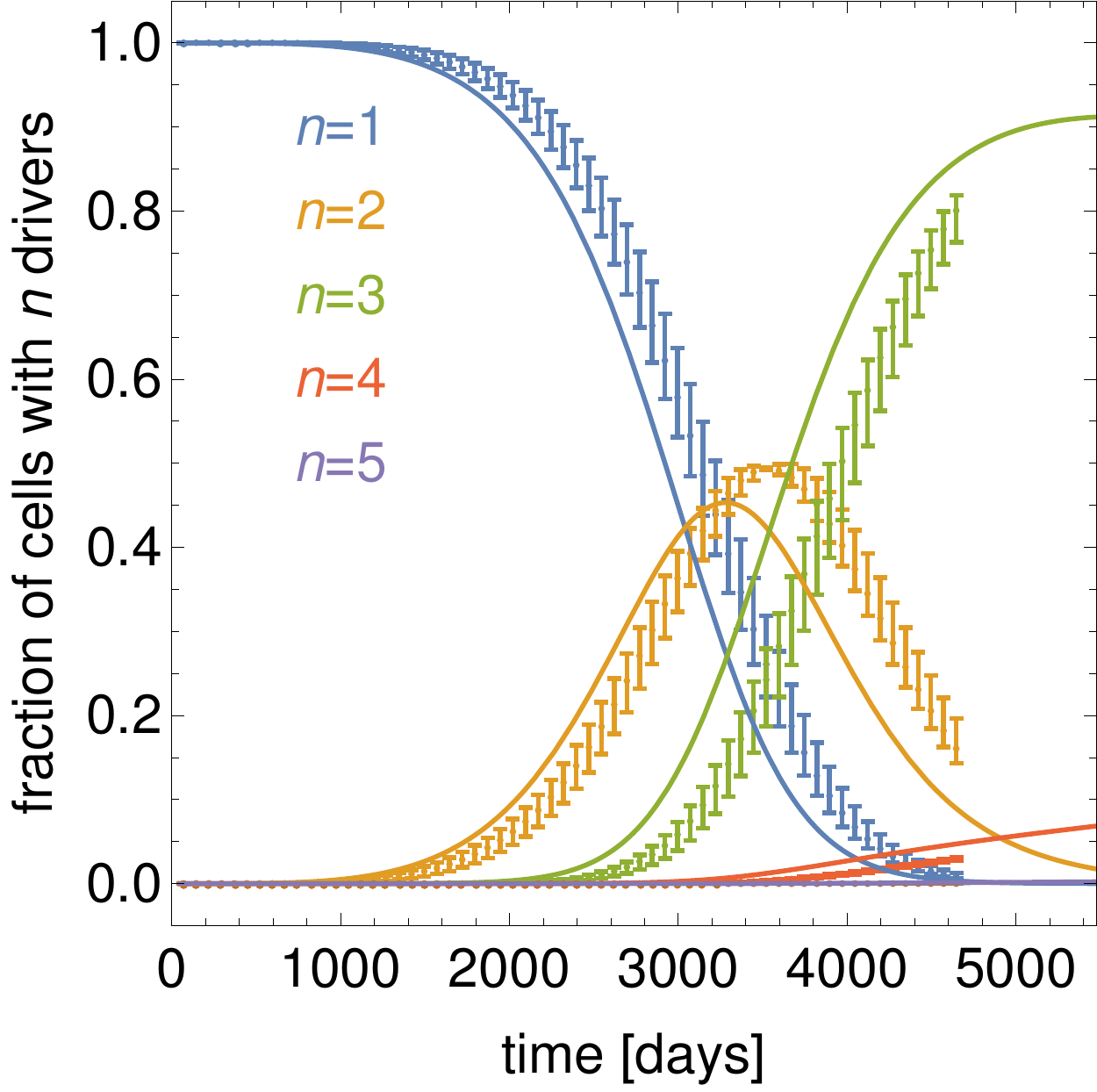} &
    \includegraphics[height=0.32\textwidth]{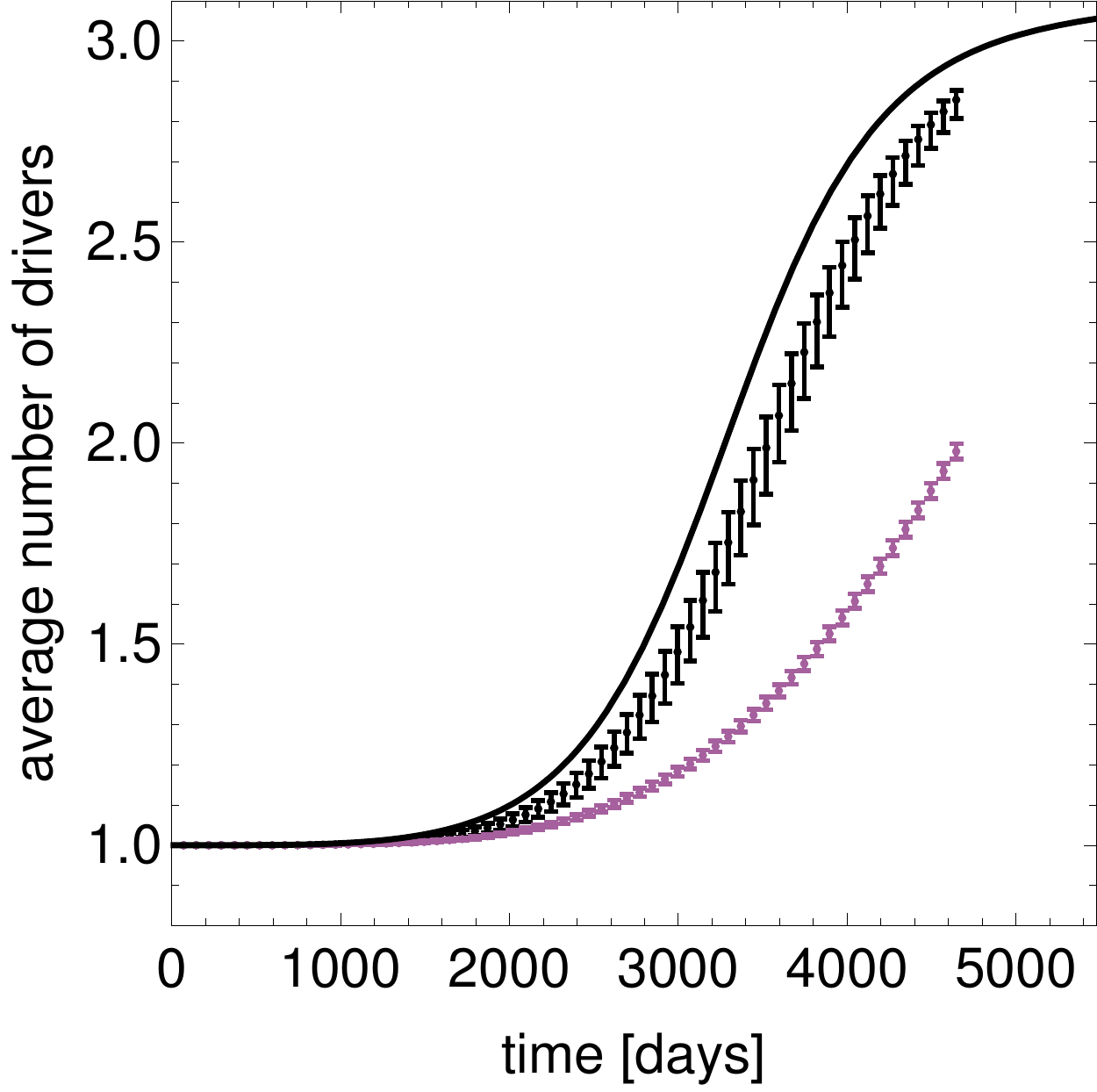} \\
    \end{tabular}

\caption{\label{fig:ini}The model with three strong drivers as in Eqs. equation \eqref{eq:slowvn}. A: The total volume of the whole tumour (black curve). B: the fraction of drivers with $n$ mutations. C: the average number of drivers (black curve). In all cases $v_1=0.015,\; M = 10^{-4},\; \mu = 2\cdot 10^{-5},\epsilon=0.01$. Data points with error bars in all panels come from stochastic simulations of the original model (Section \ref{sec:algorithm}, for which 500 replicates were performed). In panel C, two sets of data points are presented. Purple points have been obtained by calculating the average number of drivers for each simulated tumour and then taking the average over many tumours. Black points represent averaging the fractions of cells with $n$ drivers over many tumours (as in panel B), and using this to calculate the average $n$.
}

\end{figure}

In each simulation, we measured the total volume $V_{\mathrm{tot}}$, the mean number of drivers
per cell $\langle n \rangle$, the volume $V_{\mathrm{tot},n}$ of cells with a specific
number of mutations $n$, and the fraction of cells with $n$ mutations $w_n = V_{\mathrm{tot},n}/V_{\mathrm{tot}}$.
These quantities were averaged over $500$ simulations with the same underlying
parameters. Figures \ref{fig:edengr1} and \ref{fig:ini} demonstrate remarkable agreement between simulations
and the formulae derived from our analytical theory at early times: some
deviation from the analytical prediction appears at small values of $M$, 
which may be a finite-size effect. The analytical formulae also seem to 
overestimate the frequency of mutant strains, which can be seen in figure
\ref{fig:ini}.

We believe that this deviation is caused by the way in which we neglect
fluctuations in the derivation of equation \eqref{eq:goveqs}: while the neglect of
fluctuations in growth rates and mutation clearly do not have a significant
effect on the primary tumour and its cloud of secondary micro-lesions, the
effect becomes increasingly pronounced with increasing number of driver
mutations $n$. The deterministic model of equation \eqref{eq:goveqs} overestimates the
frequency of mutants of a given age and $n$ at most times, but nonetheless
predicts the growth rate of each strain with remarkable accuracy.

The effect
of fluctuations must therefore be to suppress the growth of secondary
micro-lesions when they are very small. One may well ask how this is possible,
as fluctuations are, by definition, deviations from expected behaviour, and
should not have an effect on the mean overall. This may be explained by
multiplicative noise (see section \ref{sec:multnoise}).

Other substantial disagreement, which is even present between different
calculations of what are superficially identical quantities, result from
more subtle statistical issues and properties of the distribution of mutant
proportions $w_n$.

If we first calculate the average number of drivers $\langle n \rangle$ using
the measured distribution $w_n$, calling this quantity $\langle n \rangle_{w_n}$,
and find the average this quantity with respect to many replicates of our
simulations, we find the quantity $\langle\langle n
\rangle_{w_n}\rangle_\mathrm{repl.}$, which deviates strikingly from the
analytical curve for $\langle n \rangle$ (purple points in figure \ref{fig:ini}):
call this ``Method 1''.

However, if we calculate volumes $V_n$ of cells of type $n$ for each
simulation, average this quantity over many replicates $\langle V_n \rangle_{\mathrm{repl.}}$,
and then calculate the average number of drivers $\langle n \rangle$ via 
$\langle n \rangle_{\langle w_n \rangle_{\mathrm{repl.}}} = \sum_n n \langle
V_n \rangle_{\mathrm{repl.}} / \langle
V_{\mathrm{tot}}\rangle_{\mathrm{repl.}}$ (``Method 2''), we obtain a much better agreement
(black point in figure \ref{fig:ini}). In that the second method corresponds to averaging
out randomness before calculating $\langle n \rangle$, it is arguably closer
in spirit to what we do in our analytical calculations, but there is a more
fundamental cause of the disagreement here.

Since driver mutations act to accelerate growth, in any given simulation we
can readily observe that the average number of mutations per cell is strongly
correlated with size. If we average out the proportions of cells of each type
$n$ before averaging over replicates, we treat those simulations with relatively few
drivers and corresponding \emph{much} smaller sizes with equal weight to simulations with
many drivers and enormous sizes. This acts to bias the calculation of $\langle
n\rangle$ downwards. If we average the volumes $V_n$ before
calculating proportions $w_n$ and $\langle n \rangle$, the contribution of
these smaller tumours with fewer drivers is proportional to their relatively
small size, and we see that the calculated $\langle n \rangle$ is consequently
higher.

\subsubsection{The effect of multiplicative noise}
\label{sec:multnoise}

In rare cases, fluctuations may not exactly average out when considering mean
behaviour of growth processes with multiplicative noise, and may act to
suppress growth. Depending on whether or not noise is interpreted in an Ito or
a Stratonovich sense, variations in the amplitude of stochastic noise with $n$
can have significant confounding effects, even on apparently large systems.

As a very simple illustrative model, consider the growth process

\begin{equation}
    \dot{n} = b n + \sigma \xi
\end{equation}

with $\xi$ standard white noise and $\sigma$ the scale of the fluctations. If,
as in our stochastic model, growth occurs in something like a Poisson process,
with the mean number of new cells per time step equal to $b n \;dt$, then
the variance $\sigma^2$ of this process will be equal to the mean, $\sigma^2 = b n\;dt$, 
as the mean and variance have the same scale for all Poisson-distributed random
variables. Our growth process therefore has the form

\begin{equation}
\label{eq:withnoise}
    \dot{n} = b n + \sqrt{b n} \xi\;.
\end{equation}

Our neglect of fluctuations amounts to immediately taking the noise term to
zero, as $\langle \xi \rangle=0$, so that

\begin{equation}
    \langle \dot{n} \rangle \approx b \langle n \rangle
    \label{eq:wrongapprox}
\end{equation}

However, it is not necessarily true that $\langle n^{1/2} \xi \rangle = 0$. In
fact, if we rigorously derive a Fokker-Planck equation from
equation \eqref{eq:withnoise} \emph{without} conditioning the probability distribution
on survival, we
instead recover

\begin{equation}
    \langle \dot{n} \rangle = b \langle n \rangle - \frac{b}{2} 
    \label{eq:rightapprox}
\end{equation}

with an extra $-b/2$ term representing exctinction due to stochastic fluctuations.
For identical initial conditions, say
$\langle n\rangle (0)=1$, the long-term behaviour of solutions to equation \eqref{eq:rightapprox} is
found to be lower than solutions to equation \eqref{eq:wrongapprox}, although the growth rate is identical in both cases.

Intuitively, this means that the (Stratonovich) noise gradient has a much
larger effect on small populations than large populations. Small populations
will be dominated by the random fluctuations of noise, and can only begin a
stage of steady exponential growth once they ``escape'' the noisy area of
state space: but this escape can only happen due to the stochastic dynamics of
the system itself. Eventually, the system will fluctuate out of having a small
unstable population (or else die out), so a typical history will involve some
delay before exponential growth gets underway.

This is just a heuristic argument: fluctuations which scale in the same way as
our growth rate fluctuations in a simple well-mixed can suppress growth
without changing the exponential growth \emph{rate}, but this is a simple
well-mixed model, and our above model is considerably more sophisticated with
regards to spatial structure. It may be the case that the same argument does
not hold once one carries out a fully rigorous analysis of the effect of
fluctuations on the system of partial differential equations equation \eqref{eq:goveqs}
and their corresponding Fokker-Planck equations: due to the mathematical
complexity of this task and time constraints, we cannot say for sure that the
observed disagreement is certainly due to multiplicative noise.

\section{Discussion}


In this chapter we studied models of cancer evolution which were all based on
three processes: replications of cancer cells, driver mutations which resulted
in differential increases in the net rate of replication and survival of these 
cells, and local migration which caused cells to disperse. The latter process
results in a picture of tumours in which they are formed of a large primary
and a conglomerate of micro-lesions surrounding it. The dispersal of cells in
this way has been recognized as a process by which
tumours \cite{thalhauser_selection_2010,enderling_migration_2009} and other
populations of motile cells  \cite{hallatschek_acceleration_2014} can speed up
their growth. Although our model was motivated by the process of local
migration, it is neutral to the precise details of the underlying mechanism of
cell dispersal and migration, and is potentially applicable to long-range
invasion characteristic of late-stage metastatic cancers \cite{Hallmarks}. 

The spatial structure and position of these secondary micro-lesions is not
explicitly dealt with at any stage in our formalism, and the conglomerate we
describe can be straightforwardly re-interpreted as an abstract distribution
of lesions, including distant metastases \cite{iwata2000dynamical}.

One of the strengths of the approach developed in this chapter lies in the
ease with which the average behaviour of the minimal stochastic model can be
obtained and analysed analytically, and the degree of agreement between the
minimal stochastic model, the structured population model, and the fully
stochastic lattice models. Remarkably, the inclusion of migration results in a
model which is easier to exactly solve than many spatial models of individual
tumours: as long as interactions between lesions are negligible, migration has
the effect of ``smearing out'' spatial structure, and in permitting
populations to grow exponentially, results in behaviour which is qualitatively
similar to that of well-mixed models.

Analytical solubility means that the model works for tumours of arbitrary
size, including both large masses that warrant surgical removal and
undetectable micro-lesions, and can be used to model cancer progression over a
range of timescales. The most important implications of the model deal with
two aspects of cancer evolution: growth laws, and genetic heterogeneity.

\emph{Tumour growth.} In the absence of new driver mutations and assuming 
sufficient migration, our model predicts that long-time growth is exponential.
This is also true when individual micro-lesions grow sub-exponentially, and 
even if their growth slows down over time. Given that most tumours contain 
avascular areas where the lack of oxygen and glucose inhibits proliferation 
 \cite{vaupel_blood_1989,parks_hypoxia_2016}, our model provides a plausible 
explanation how the growth of an entire tumour can still be exponential, as 
often observed experimentally for intermediate-size tumours. This phenomenon 
does not require postulating any novel or previously unknown mechanism, but 
relies only on short-range migration of cancer cells, a process that certainly 
occurs in nature, informs standard histological procedures, and is a subject
of active research \cite{stefaniejefrey1995importance,fokkelman_2016}.

If all new driver mutations carry a significant fitness advantage, and
steadily accumulate during tumour delevopment, our theory predicts
super-exponential growth. The acceleration does not have to be large, and may
arise only in tumours that are too big to arise on clinically relevant
timescales, but nonetheless we are unaware of any evidence that
super-exponential growth of this type has ever been observed in human cancer.

Exponential or faster growth is obviously unrealistic for very large tumours
for which spatial constraints and other inhibitors of growth become important.
Many models of tumour growth have been proposed that account for the sigmoidal
growth curve of many tumours \cite{rodriguez-brenes_tumor_2013}. The majority
of these models are phenomenological, and are not based on the microscopic
dynamics of cancer cells as ours is. It would be highly instructive to learn
what minimal changes would be required to our model in order to reproduce
sigmoidal growth, or if these models which do exhibit sigmoidal growth are
appropriate for tumours once local invasion or metastasis are taken into
account.

\emph{Genetic heterogeneity of tumours.} Experimental evidence gathered by a
number of different groups over the last decade strongly suggests that
cancerous masses show some degree of genetic heterogeneity, although the
precise amount and its possible importance are still a subject of debate 
 \cite{yachida2010distant,gerlinger_intratumor_2012,ling2015early}. In this
chapter, we explored this heterogeneity by calculating the size of clonal
cell lines with differing numbers of driver mutations, and showed that these
cell lines increase in population at an exponential rate. At first, only one
driver predominates, until becoming replaced by a mixture of clones with two,
three, and more drivers. The coexistence of multiple genetically distinct
cell lines in this way means that the evolution is not well represented as being
a simple counting process, in which one cell line forms the majority of the
population until a new mutant emerges and rapidly takes over after a long
time. The time to reach some average number of drivers is not a simple sum of
waiting times for consecutive mutants to emerge, and a number of distinct
cell lines are present simultaneously.

We have also shown that if each new driver increases the selective advantage of cancer cells in comparison to normal cells, the average number of driver mutations predicted by our model increases exponentially in time for all considered scenarios. This means that most drivers would accumulate late during cancer progression. There is limited evidence  \cite{ling2015early,sottoriva_2015} that this is not true. On the other hand, if only a few first drivers have significant fitness advantage, these drivers will accumulate early during growth and the tumour will become much more homogeneous. An interesting application of our model would be to predict how strong the selective advantage of new drivers can be that would still be consistent with recently postulated neutral evolution in tumours  \cite{ling2015early}.

There are several ways our model could be revised and extended. We have not
analysed the spatial distribution of drivers: it may not be feasible to do
this analytically, but as the model proposed here is easily simulated (see
section \ref{sec:comparison}), it may be practical to extend it to include
locations of micro-lesions and draw out some meaningful predictions. 

Another
interesting extension would to attempt to incorporate phenotypic plasticity
and the presence of cancer stem cells, and asking what would result if only
some small fraction of cells could replicate indefinitely, whilst the majority
of cells can only udnergo a few rounds of replication, as in Ref.
 \cite{werner_cancer_2016}. This would affect the net replication rate, and the
rate at which mutations accumulate in a single lesion $r_n$.

The analytical formulae also seem to overestimate $F_n$ and $V_n$, which we believe
is related to multiplicative noise. How this may affect equations
equation \eqref{eq:goveqs} is a question which beyond the scope of this thesis at
this time, but which has the potential to resolve the remaining discrepancies
between the analytics and simulations.

Finally, since the separability of the deterministic governing equations
equation \eqref{eq:goveqs} only depends on the validity of the infinite genome
approximation and the non-interaction of distinct lesions, it seems very
likely that this approach is extensible to evolution in more complex
landscapes than the linear chain of driver mutations that we have studied.
Particularly interesting questions surround the effect of epistatic
interactions between driver and passenger mutations \cite{bauer_cancer_2014}, and the statistical
behaviour of the model in the context of random fitness landscapes is another
open question \cite{durrett_evolutionary_2010}.

\subsection{Relevance to open questions in the field}


In the introduction, we described two broad open problems in the field of
cancer evolution and treatment: what growth curves accurately describe
complex, growing cancers and more importantly their physiological 
interpretation; and how quickly cancers evolve. Our interest in these two
problems consists of how these two aspects of cancer can be modelled in such a way
that they can be related to microscopic cell-scale determinants in a flexible,
yet minimalistic and intelligible theoretical framework.

We have found in this chapter that the total volume of many independent
lesions that constitute a cancerous ensemble can grow exponentially: when
genotypes of differing fitness are accounted for, we find that the frequencies
of each cell line grow exponentially as well, with an exponential growth rate
determined by the migration rate $\phi_n$ of solid tumours dominated by a
given cell line.

The total tumour burden in these cases is therefore a sum of multiple
exponential growth curves, despite the fact that none of the distinct 
lesions in the ensemble grow exponentially.

The widespread usage of exponential growth as an heuristic by clinicians and a
simple basis for other models by applied mathematicians may not, as a result,
be totally baseless: at least in the case of metastatic cancer for a limited
period of time, the paradoxical popularity of these models may not be
paradoxical at all, but the simplest workable model in the context of a
metastatic cascade.

On the second point, the ``speed'' of evolution has been shown to be related
to the effective fitness landscape generated by the differential growth and
invasion of lesions comprised of different cell lines. Quantified by the average
number of drivers $\langle n \rangle$ present in a cell picked at random from 
the ensemble of tumours, steady linear increases in additive fitness were
found to correspond to accelerating, super-exponential growth and an
exponential increase in mean number of drivers $\langle n \rangle$. Linear
increases in fitness of underlying cells, whether this is quantified in terms
of expansion speed for solid spheroids or simply growth rate for well mixed
lesions, all correspond to accelerating growth and accelerating selection:
this corresponds with the ``late sweep'' class of models described in the
introduction.

Given that super-exponential growth has not (to the best of our knowledge)
been observed in actual clinical studies of metastatic cancers, this rather
generic finding warrants a look at other underlying landscapes.

If the fitness is bounded from above, say by physiological constraints on how
rapidly cells can divide, then a plausible alternative would be for only the
first few drivers to carry a substantial advantage, followed by a mostly-flat
plateau. In this case, we found that after an initial burst of growth and
selection, the metastatic ensemble underwent an extended phase of exponential
growth. The growth rate of this extended phase may be quite slow, as it is
determined by the frequency of migration rather than the underlying cell
division time. The average number of drivers increased sigmoidally, before
settling down to a much slower rate of increase for an extended length of
time.

This scenario displays more realistic growth curves than the simple linear
increase popular with modellers, but also predicts that most selection occurs
in an early burst: this early set of selective sweeps still occurs after
initiation, and is driven by migration, in contrast to ``Big Bang'' like
models. This more realistic scenario of fitness landscapes with a plateau
therefore correspond to an intermittent acceleration and deceleration of
evolution by selection, and an ``early sweep''.

Finally, in all cases multiple drivers were found to be present at any one
time, which meant that the simplifying assumption of a succession of hard
sweeps such as in the work of I. Bozi\v{c} \emph{et al.} was not valid in general, resulting
in slightly different growth rates of $\langle n \rangle(t)$ for the same underlying fitness function
(among other less important details). The most plausible set of parameters we
have looked at in this chapter therefore seem to point to an early selective
sweep.

\subsection{Future work}

A number of simplifying assumptions have made our research tractable, but
limited its application. Evolution does not occur on linear landscapes,
comprised of a simple succession of drivers: there are multiple, branching
pathways available, and epistasis (roughness in the fitness function) can
strongly affect dynamics.

In addition, we have noted the effect of noise on our growing structured
populations, but a fully stochastic development of structured population
models is rather beyond the scope of this thesis. Some initial attempts at
formulating such stochastic models \cite{greenman2016kinetic} and providing a
means of building solutions have been attempted, but are not yet at the stage that
useful conclusions can be drawn from analytics alone. The effect of stochastic
fluctuations on the expected behaviour of structured population models is 
therefore one area of research that may prove relevant and fruitful in the future.

There are also a number of interesting experimental issues. The actual
detection and timing of selective sweeps is strongly frustrated by the lack of
dynamical data in most cases, which raises the question of how the occurrence
of selection can be reliably detected in real tumours. One serious proposal
for gaining useful dynamical data is the use of neutral mutations in a lesion
as a molecular clock: if these occur at a constant rate, the number of
expected drivers can be related to the number of passengers. This could be a
potentially useful proxy for the missing dynamical information. One attempt to
do this, and use neutral passenger mutations as an indirect measure of time,
assumes that cell divisions occur at a constant rate, with some given expected
number of passenger mutations occurring at each division event: the total expected
number of passenger mutations $n_p$ present after some time $t$ is then

\begin{equation}
    n_p = \nu t /T
    \label{eq:passengers}
\end{equation}

where $\nu$ is in essence the expected number of new neutral mutations per
division, and $T$ the average cell cycle time \cite{bozic2010accumulation}.
This formula could be used to express our results in terms of the number of
passenger mutations, a more easily measurable quantity than time, in a similar
way to earlier work. However, this assumes the rate of cell division is
constant over time: if it varies continually in some way,
equation \eqref{eq:passengers} will not be valid. More generally, one can say that
$n_p$ should be proportional to the number of cell divisions $n_D$,

\begin{equation}
    n_p = \nu n_D(t)\;.
\end{equation}

Although the precise form of $n_D(t)$ is not known (and is an independent
assumption from any made so far in our framework), the absolute lower bound is
the minimum number of divisions needed to produce a total tumour size of
$V_{\mathrm{tot}}$: this minimum number of divisions is clearly
$\mathrm{log}_{2}(V_{\mathrm{tot}})$, so

\begin{equation}
    n_p > \nu \mathrm{log}(V_{\mathrm{tot}})/\mathrm{log}(2)\;.
    \label{eq:ndmin}
\end{equation}

\begin{figure}
\begin{center}
    \begin{tabular}{ll}
    A & B \\
    \includegraphics[width=0.4\textwidth]{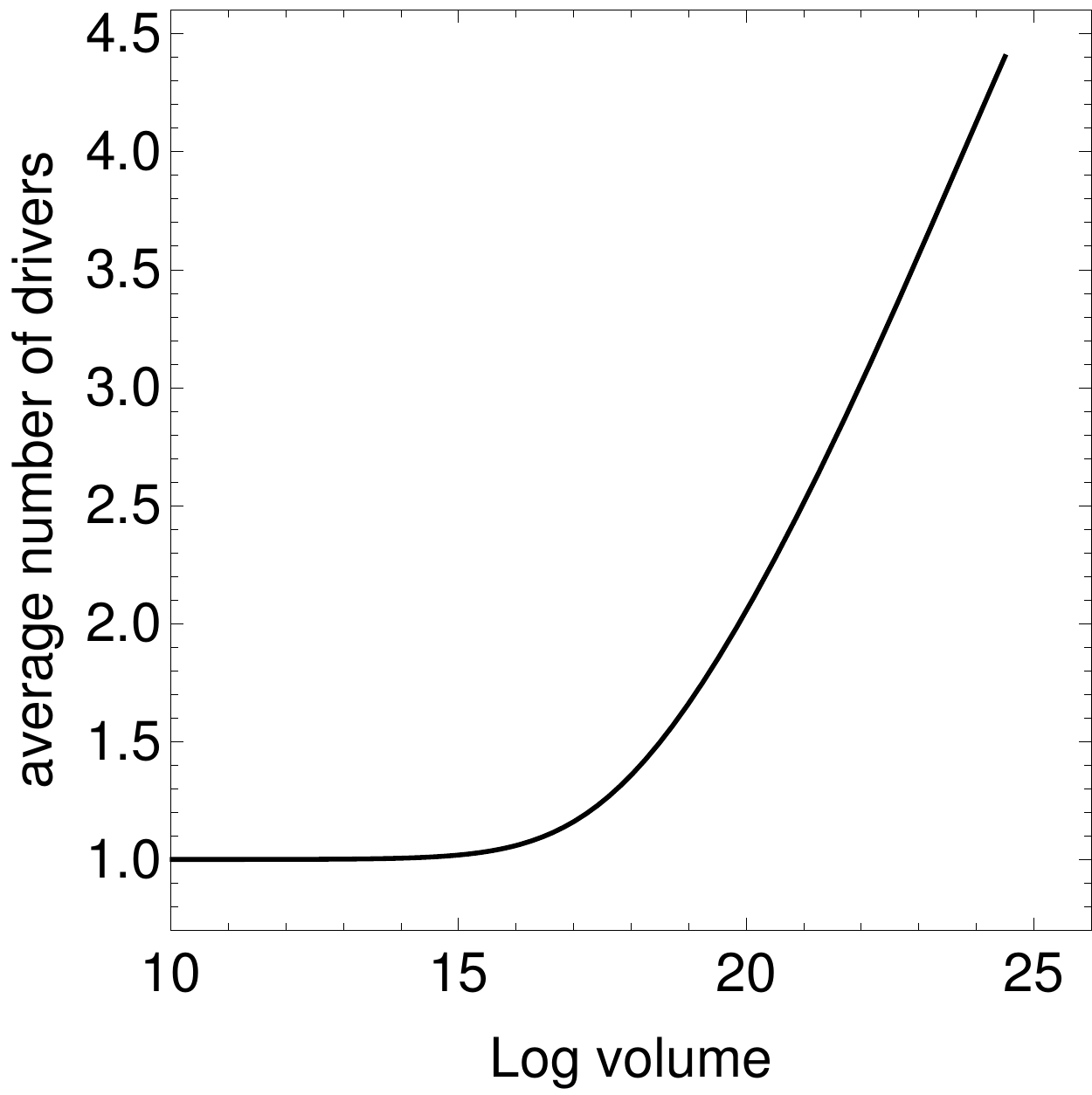} &
    \includegraphics[width=0.4\textwidth]{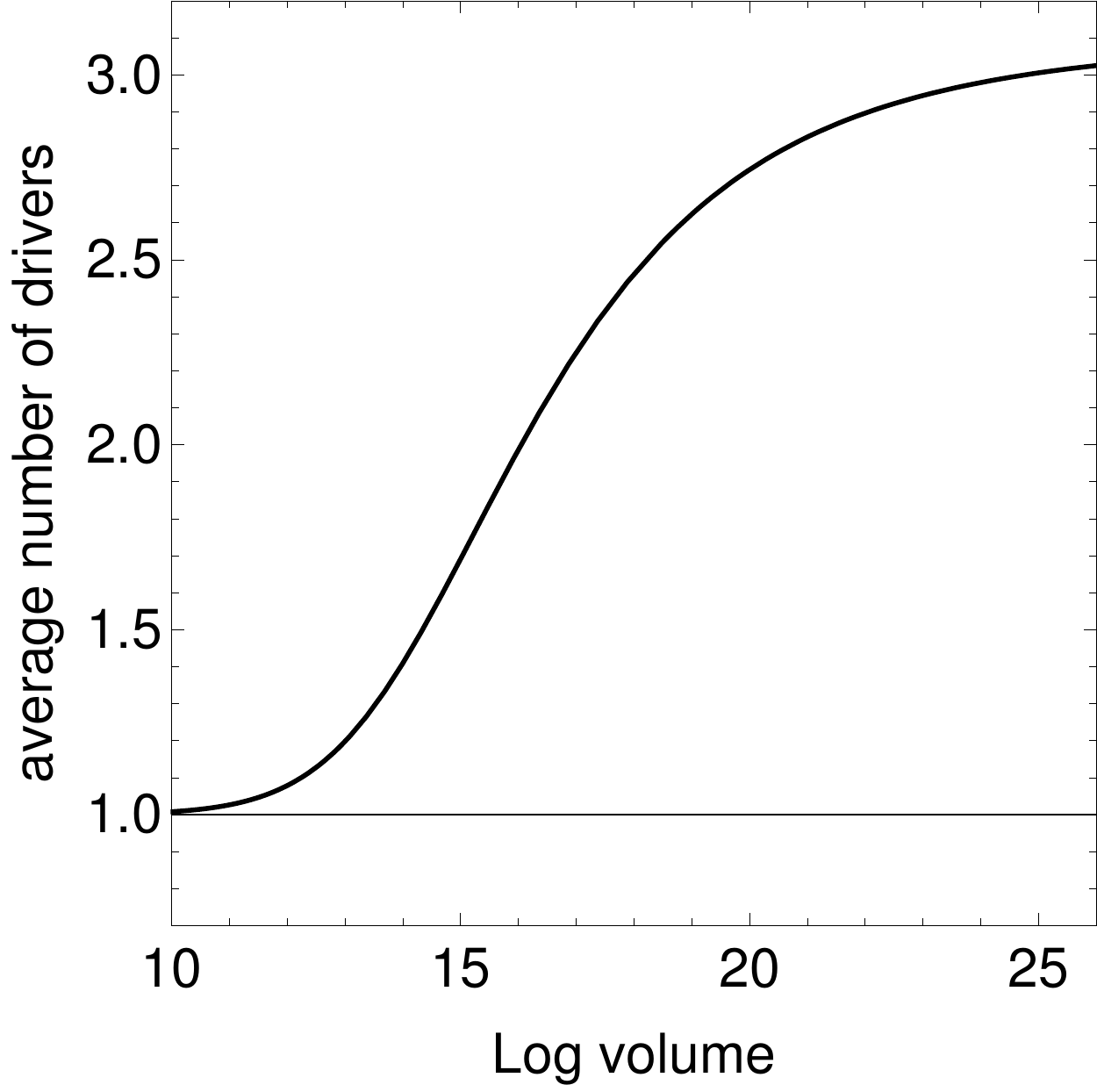} \\
    \end{tabular}
    \caption{\label{fig:convexconcave}Figure showing qualitative differences in
    plots of the average number of drivers per cell $\langle n \rangle$ versus
    the logarithm of the total tumour burden, demonstrating A. a clear convex
    curve for the late sweep model (gradualistic fitness landscape) and B. a
    warped sigmoidal curve for the soft sweep model (fitness landscape with plateau).}
\end{center}
\end{figure}

The result of plotting the average number of drivers $\langle n \rangle$
against $\mathrm{log}(V_\mathrm{tot})$, which is related to the lower bound on
$n_p$ from equation \eqref{eq:ndmin} can be seen in \ref{fig:convexconcave}.

In light of this, we can also use our above theory to produce parametric plots of
average number of drivers against the logarithm of the population size: the
latter quantity is a lower bound on the number of cell division events, and
hence also on the number of potential passenger mutations. There is still a
qualitative difference between the predicted curves for the linear fitness
increase and the fitness increase with a sharp plateau: if selective sweeps occur
late, we see a convex curve. If they occur early, we see a warped sigmoidal curve.


However, this approach has the problems that it leaves aside the actual
detection of driver mutations themselves, and furthermore depends on assuming
well-mixed models (which may be justified in some sense: see above) and a
constant rate of appearance of passenger mutations as do most versions of
molecular clocks and coalescent theory. Extensions of structured population
models and the stochastic model underlying ours here are clearly necessary to
make the connection to this experimental observable.
\chapter{Resistance to low-dose combination chemotherapy}
\label{ch:clinical}


\section{Background}


The main areas of interest in this thesis are simple models of population-level
behaviour of evolving cancers. We have attempted to derive predictions for 
growth curves and the pace of
evolution, especially from models which can be related to the biophysical
properties of individual cells. 
In relation to this, the questions of how quickly cancers can evolve, and 
how this is affected by migration and invasion are
needed to understand how quickly a metastatic cancer can gain a 
mutation which enables it to resist a specific type of therapy. One application
and opportunity for testing our work can be found in clinical studies of the
incidence of resistance to chemotherapy. We contributed analytical techniques 
and offered a number of possible interpretations of the results of a clinical 
study on the resistance to combined chemotherapy in pancreatic adenocarcinoma. 

The usually accepted therapeutic regimen is the maximally tolerated dose
concept: the highest possible non-lethal dose of a cytotoxic chemical is
administered, the purpose being that since the cytotoxin is usually chosen to
kill cells at the point of mitosis (division), it will have a
disproportionately large effect on the cancer cells, which divide more often
than normal cells. Administering the maximally tolerated dose attempts to kill
off more of the cancer cells than normal cells, which are also inevitably
affected. Of particular note are cells of the intestinal lining and some
immune cells, which have an especially short cell cycle.

Although for many cancers, this treatment regimen has represented a
significant advance in survival and recovery, pancreatic cancer has stubbornly
resisted improvements in outlook after many attempts at devising an effective
therapy. In large part, this is due to the location of the pancreas. Cancers
of internal organs tend to have longer periods during which they are
effectively asymptomatic before detection, and reach a stage at which they 
have several metastases by the time they are detected. Pancreatic cancers are usually inoperable
by the time they are detected and diagnosed, and treatment is generally palliative
  \cite{bond2012pancreatic}. The search for more effective treatments, and the twin search
for the reasons for previous treatment regimens' shortcomings, is an area of
intense research. Progress in the field has been slow, with extended periods
in which no new treatment regimen showed any clinically significant
improvement \cite{conroy2011folfirinox}.

In the early 2010s, combinations of multiple drugs administered at maximally
tolerated doses in the same
course (megadose combination chemotherapy) were found to have better outcomes than single drug maximally tolerated
dose (MTD) regimens, which constituted the first significant improvement in
more than a decade \cite{conroy2011folfirinox}. However, this and subsequent improvements 
are still ultimately palliative, and in most cases the disease ends in 
relapse and eventual failure. Median overall survival is rarely longer than a
year \cite{von2011gemcitabine,conroy2011folfirinox}.
This resistance was hypothesised to be due to the presence of some small
population of cells with some mutation which conferred resistance to the
drug combinations that were explored. When the treatment was applied, there
was a strong selective pressure for these subclones to expand, and outpace the
susceptible population \cite{sausen2015clinical,bozic2013evolutionary}. In a straightforward evolutionary process, the cancer
acquires resistance and regrows. If this hypothesis were true, it should be
very unlikely that resistance to many different chemotherapeutic agents with
different mechanisms of action could evolve by this mechanism. The chances
that a single mutation could confer resistance to many different drugs at once
are, at least in theory, very small \cite{bozic2013evolutionary}.

To test this theory, a novel multi-drug therapy at lower doses than MTD or
megadose combination chemotherapy was designed and tested by Dr. L.A. Diaz. 
We developed a simple
mathematical model for observations of antigen concentrations,
and measures of survival compared to the parameters of this mathematical
model. In addition, and of particular relevance to this thesis, the inferred
frequency of resistant subclones could also be analysed with this model, which
was used to make some comments about the potential mechanism and suggestions
for future research.

\subsection{Protocol and mechanisms of action}

Twenty-eight patients were administered the low dose combination therapy over
a 21-day cycle, and a further ten patients were administered the same doses
over a 28-day cycle. Four cytotoxic agents were administered in cyclical 
succession: gemcitabine, docetaxel, 
capecitabine, and cisplatin. The combination was selected due to the clinicians' previous
experience with
similar combinations in megadose therapy (gemcitabine, docetaxel and
capecitabine) and for the heterogeneity of the combination of four. Both the
mechanisms of action and chemical structures are, on the whole, very
different, which should in theory minimise the chances of cross-resistance
emerging. To see how, it is worthwhile to quickly review some of what is known about the
pharmacology of the drugs in question.

Of the four drugs that were used, capecitabine and gemcitabine deserve some
mention as having the most similar mechanisms of action as well as some structural
similarities. Both chemicals interrupt the synthesis of DNA, in capecitabine's
case by metabolism to the molecule 5-fluorouracil (5-FU), which also happens
to be synthesised and administered as a distinct chemotherapeutic agent. After
metabolism to 5-FU, the 5-FU binds to thymidilate synthase, which prevents it
from carrying out its normal function of synthesising thymidine \cite{longley20035}. Thymidine is
necessary to produce thymine, one of the base pairs and ``building blocks'' of
DNA, and without it cells are unable to replicate, resulting in what is known 
as thymineless death \cite{longley20035}. Gemcitabine, on the other hand,
is incorporated into the DNA strand during synthesis after being
phosphorylated, which results in an irreversibly ``broken'' DNA strand being
produced, and DNA synthesis is effectively ``aborted'' \cite{mini2006cellular}.
Both capecitabine and gemcitabine interfere with DNA replication, and are both
fluoridated nucleoside analogues. They therefore both have similar uptake routes, being
transported into the cell along the same pathways as
nucleosides, such as hENT1 \cite{galmarini2001nucleoside,zhang2007role}.
Cross-resistance for these two would not be especially surprising.

Docetaxel, by contrast, binds to microtubules rather than interfering with DNA
synthesis. Cell division cannot continue without replicated DNA and other
components being separated into the new pair of daughter cells by
microtubules, and so docetaxel blocks cell division \cite{yvon1999taxol}.
Docetaxel is transported into cells across cell membranes by a number of
distinct transmembrane transporter in the solute carrier
family \cite{baker2009pharmacogenetic}, and is structurally a taxane. It is a
considerably larger molecule than capecitabine or
gemcitabine \cite{mastropaolo1995crystal}.


Finally, cisplatin directly damages DNA, which triggers cell death by
apoptosis as cells progress through the cell cycle \cite{galluzzi2012molecular}. 
As cancer cells progress through the cell cycle with fewer checks and balances on their activity, they
are theoretically more strongly affected and killed more than normal cells,
in a similar statistical way to other chemotherapeutic agents. Cisplatin is a
small molecule and an inorganic compound, in contrast to the other drugs in
the combination which are as a rule organic. The molecule only becomes toxic
as such when the chloride groups are replaced with water molecules (aquation),
which happens essentially as soon as the molecule enters the cytoplasm, having
been transported into the cell \cite{galluzzi2012molecular,el1999reactions}.
Cisplatin is transported into cells via similar routes to other metal-based 
small molecules and ions, by the transmembrane pump copper transporter
1 \cite{arnesano2013updated}. Resistance to cisplatin in pancreatic cancer is 
quite common, with some fraction of the cancer cell population appearing to
acquire resistance during the course of therapy, and many patients
intrinsically resistant to cisplatin from the start \cite{galluzzi2012molecular}. 
Both of these observations will be returned to shortly, but it should be
emphasised that the mechanism of action, uptake pathway, and chemistry of
cisplatin are all quite distinct from any other drug in the combination.


With the possible exception of gemcitabine and capecitabine, these agents have
widely different mechanisms of action and pharmacology. By design, just as the
components of cellular machinery that interact with and are damaged by each drug have relatively
little in common, mutations that alter one piece of this machinery are
unlikely to have an effect on different components. In this way, the
likelihood of cross-resistance emerging is supposed to be minimised. All of
the patients were newly diagnosed and had had no previous treatment for the
disease, and so there should be no selective pressure for resistance before 
their treatment. The drug combination was in
evolutionary terms essentially novel, and pre-existing resistance should have
been exceptionally unlikely.

\section{Analysis}

Over the course of the therapy, concentrations of the tumour marker CA19-9 in the bloodstream, an antigen 
associated with pancreatic tumour masses, were measured at 3 week intervals. CA19-9 
is an antigen, a molecule that the immune system ``recognises'' and responds
to \cite{MolBioCell}, which is often used as an indicator of the sizes of
tumours and the effect of therapy \cite{ni2005clinical}. Measuring
concentrations of a known molecule in the bloodstream only involves a blood
test, rather than a medical imaging scan, which would be a more direct method
of measuring tumour size, and can be performed more frequently and cheaply.

\begin{figure}
\begin{center}
    \includegraphics[width=0.6\textwidth]{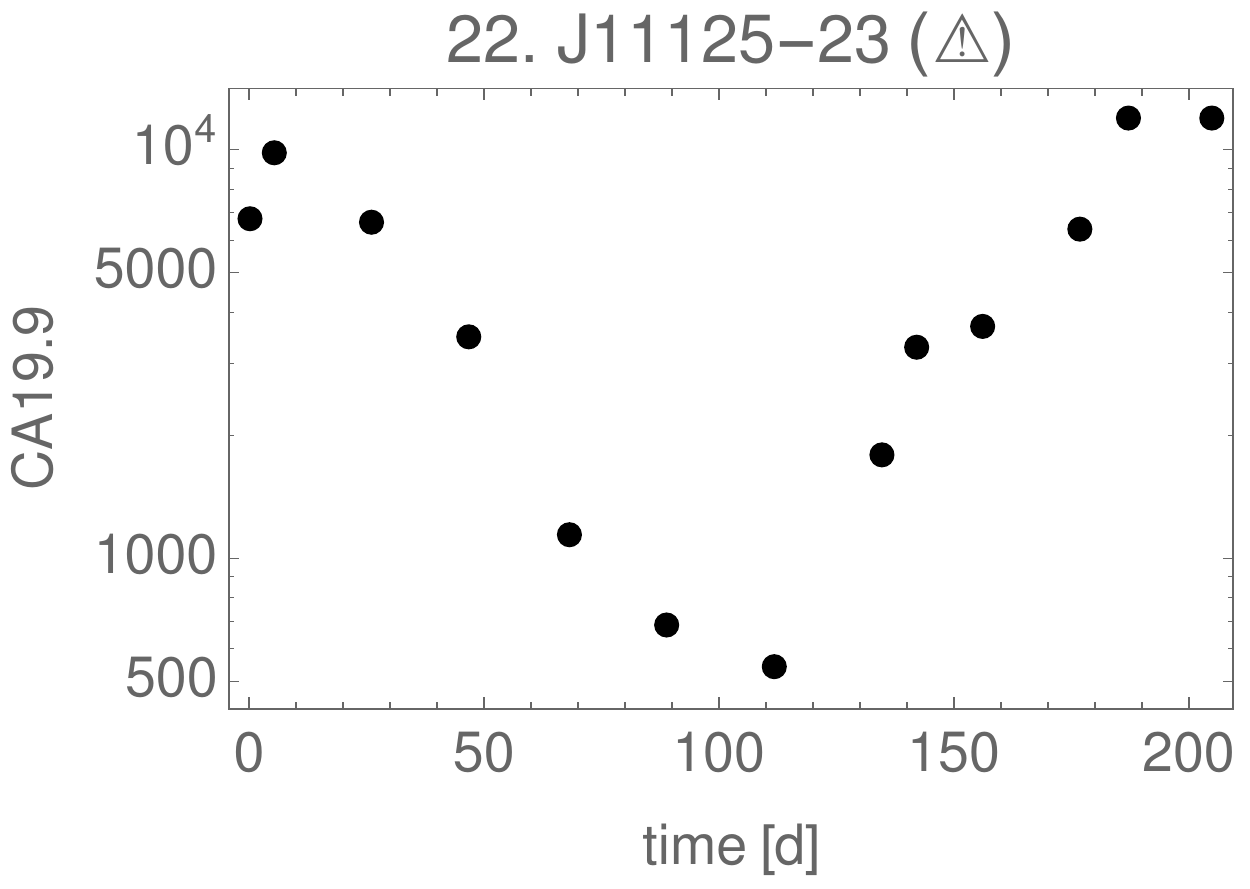}
    \caption{\label{fig:ca199} A plot of a typical CA19-9 concentration during the course of chemotherapy
    showing two characteristic phases: an initial decline followed by a rebound after
    a few months. The simplest interpretation of this is that the underlying
    burden of cancer cells declines, and then regrows. It is tempting to interpret
    this plot as representing a resistant and sensitive sub-population of cancer
    cells, but some care must be taken to account for the noisy and fluctuating
    nature of CA19-9 and its relationship to inflammation (see main text).}
\end{center}
\end{figure}

Our goal was to use
the measurements of CA19-9 levels to infer the underlying tumour burden, and 
the proportion of resistant cells present at the onset of chemotherapy. But, as
CA19-9 is not a direct measure of tumour size, but rather reflects something
more akin to inflammation of the pancreas \emph{due} to the tumour, there are
some issues in interpreting CA19-9 levels that have to be addressed.
One well-known problem with the use of CA19-9 as a marker is that while it is
a highly sensitive indicator of tumour sizes in some cases, it is not very
specific overall. Levels can be high when no cancerous cells are present, in cases such
as pancreatitis, jaundice, and other inflammatory disorders of the pancreas;
and a substantial minority of patients do not
express CA19-9 at all \cite{ni2005clinical,ogawa1998case,goonetilleke2007systematic}. 
The ``signal'' of tumour size carried by CA19-9 levels is therefore very
noisy, and affected by natural variations and the complex inflammatory response
of the immune system in ways that are difficult to quantify. Given that
useful quantitative measurements of the strength of this ``inflammatory noise'' do not 
(to the best of our knowledge) exist, we had to devise some way of classifying
the experimental data into sets where the levels could be meaningfully
interpreted in terms of tumour burden and sets where the signal was apparently
being swamped by noise or was otherwise on the threshold of detection.

\subsection{Immunological noise and over-interpretation}
%

We developed two four-parameter mathematical models, and fitted to the CA19-9 data. The
first model was a simple sum of two exponentials,

\begin{equation}
    c_{\mathrm{exp.}}(t) = a \mathrm{e}^{-d t} + c \mathrm{e}^{b t}
\end{equation}

which can be simply interpreted as representing the CA19-9 secretion of two 
sub-populations of tumour cells, one of which is sensitive to the therapy and
dies off at some exponential rate $d$, and the other of which is resistant to
the therapy for one reason or another and regrows at some rate $b$. We will
henceforth refer to this as the ``exponential model''.

This model has a number of plausible motivations. As discussed in Chapter
\ref{ch:intro}, simple exponential growth is a very popular model with
clinicians, essentially because it is the simplest possible model, and which in
many cases does not disagree with actual observations. Despite the debate
regarding which of the many postulated models is the most appropriate for real 
tumours, we have also seen in Chapters \ref{ch:Lattice} and \ref{ch:minimal}
that exponential growth can emerge even in scenarios when there are complex
spatial constraints on the growth of tumours, and when individual lesions grow much
slower than exponentially.  This can be facilitated by cell migration and
local invasion. The escape of cancer cells to distinct, nearby locations can
lift constraints on growth that would otherwise force power-law or
decelerating growth with some plateau, and permit exponential growth for an
extended period of time.

These were all cases of late stage cancers, with several liver metastases visible
on CT scans in most (about 80\%) cases according to the interpretation of other 
clinicians working on this study. However, as
these patients were all clinically naive, no prior treatments or surgeries had been
attempted at this stage, and there was no record of whether or not
a histologist had found the surgical margin to be clear, an important
prognostic test that would have been performed following surgery and a smoking gun
indicator for whether or not the type of local migration we set out to model
in Chapters \ref{ch:Lattice} and \ref{ch:minimal} was occurring \cite{stefaniejefrey1995importance}.

The parameters $a$ and $c$ of the multiple exponential model have natural interpretations
as proportional to the populations of the sensitive and resistant cell
populations, and $d$ and $b$ their respective death and regrowth rates. One can calculate 
the inferred fraction of resistant cells $f_R$,

\begin{equation}
    f_R = \frac{c}{a+c}
    \label{eq:FR}
\end{equation}

and the approximate time-scale $T$ needed for the tumour to regrow to some lethal
size $N$,

\begin{equation}
    T = \frac{\mathrm{ln}(N/c)}{b}
    \label{eq:regrowth}
\end{equation}

which can be seen to be rather insensitive (in theory) to $N$ and $c$ and more
sensitive to $b$, which is likely to be an important predictor of survival
after recurrence.

This exponential model was fitted to each patient's dataset, and the parameter
values used to calculate values for the resistant fraction $f_R$.
However, whether or not these parameters can actually be interpreted
\emph{meaningfully} for a given dataset depends on how ``trustworthy'' the CA19-9
data is in this case. In some cases, the levels were so low as to be on the
threshold of detection, and in others fluctuated so widely that there was no
obvious trend. As previously discussed, CA19-9 is associated with
inflammation, and the effect of the complex dynamics of the immune
system is very difficult to quantify (and quite beyond the scope of this
thesis). Some test for how meaningful the signal extracted by fitting the
multiple exponential model is necessary, but the exact dynamics of noise are
not known, so this test should ideally be independent of specified ``noise
levels'' or detailed postulates about the dynamics of CA19-9 in the absence of
cancer (as we are ignorant about both of these). What we want is a
relatively systematic way of avoiding over-interpreting data which is known to
include some highly complicated confounding processes.

The simple test that we arrived at was to compare the performance
of the relatively well-motivated multiple exponential model to a null model
with the same number of parameters. If the exponential
model had a worse fit to the model than the null model, this was interpreted
as it failing to extract meaningful information in this case. Only cases in
which the multiple exponential model did better than the null model would be
considered as having meaningful interpretations.

The second model, or ``null model'', was defined by

\begin{equation}
    c_{\mathrm{null}}(t) = \exp(a + b t + c t^2 + d t^3)
    \label{eq:nullmodel}
\end{equation}

and by design has no meaningful interpretation nor correspondence with other
proposed models of tumour growth outlined in chapter \ref{ch:intro}. It has the same number of
parameters as the multiple exponential model, but there is no reason to
believe that such a form would be a good model of anything on either an
empirical or \emph{a priori} basis. 

Cubic polynomials have the property that they minimize the amount of
``bending'' in a curve or surface, and arise naturally in Euler-Bernoulli beam
theory and engineering \cite{kreyszig2010advanced,landau1986theory}. Their use
as splines, for interpolation between points in a dataset, derives directly
from the use of flexible beams (the original meaning of ``spline'') to model 
smooth, mechanically stable curves in shipbuilding \cite{schoenberg1964spline}. 
There is no simple mechanism for motivating \eqref{eq:nullmodel} in this way.
Perhaps some complex chemical kinetics can turn up something of the correct
form after some abuse, but this is more of a post-hoc rationalization
than a genuine motivation in the absence of a compelling underlying mechanism.

But it isn't too difficult to motivate \eqref{eq:nullmodel} without reference
to splines. The idea is that higher-order polynomials will provide
progressively better models for $\log c$. A polynomial will be on equal
footing with a more mechanistic model when it has the same number of
parameters, so a cubic polynomial is the natural null choice for comparing to
a four-parameter mechanistic model.

As there is no reason to expect this model to accurately reflect the data
except coincidence, any dataset to which the null model had a better fit than
the multiple exponential model (as quantified by reduced chi-squared statistic) was
considered to be swamped by immunological ``noise'' or else was on the
threshold of detection. We consider datasets to be swamped by this noise if

\begin{equation}
    \chi^2_r (\mathrm{exp.}) \leq \chi^2_r (\mathrm{null})
\end{equation}

and are therefore over-interpreted by the exponential model, thus ``failing'',
and otherwise are considered to ``pass''. We should emphasise that this is a
test of how plausible and trustworthy the exponential model is for a given
dataset, and not a test for noise in the underlying data itself.

This difference-in-$\chi^2_r$ test
came the closest of several candidate tests to being independent of
quantitative thresholds, which due to empirical
ignorance would necessarily be arbitrary. This allowed us to comment on
which datasets were ``reliable'' and which were over-interpreted by the
multiple exponential model.

\subsection{Theoretical fraction of resistant cells}
\label{sec:branching}

The theory that motivated the therapy and allowed relatively precise
estimates of the probability that multi-drug resistance could emerge by cells'
acquiring several independent neutral mutations, was developed by I. Bozi\v{c}, T.
Antal \emph{et al.}. It is formulated in terms of
branching processes, and is essentially a generalization of the Luria-Delbr\"{u}ck
model of evolution to include cell death, which is an important complication
in the case of cancer cells \cite{antal2011exact,bozic2013evolutionary,luria1943mutations}.

From the theory of branching processes, the fraction of resistant cells that 
appear before therapy (while these mutations are neutral)
$f_R$ depends on the point mutation rate $p_\mu \approx 10^{-9}$ 
(for most cancers \cite{bozic2013evolutionary}), on the number of
different mutations that can result in resistance to a given drug $n_i$ or to
several drugs simultaneously $n_{ij\cdots}$, and (weakly) on the size of the
tumour at the start of therapy $N$. In the case of two drugs, labelled $1$ and $2$,
for which there are $n_1$ and $n_2$ possible mutations that cause resistance
to each alone and $n_{12}=0$ (no possible mutations that cause cross-resistance),
the expected fraction of resistant cells at the start of therapy $f_R$ is on
the order of

\begin{equation}
    f_R \approx (\log(N))^2 n_1 n_2 p_\mu^2
    \label{eq:nocrossres}
\end{equation}

neglecting other pre-factors of order $1$ that arise in the detailed 
theory \cite{bozic2013evolutionary}. When there are $n_{12}$ mutations that can 
cause cross-resistance, $f_R$ is instead

\begin{equation}
    f_R \approx \log(N) n_{12} p_\mu
    \label{eq:withcrossres}
\end{equation}

which is much larger. Suppose upon the start of treatment that a tumour
consists of $10^{10}$ cells, and only $1$ mutation can confer
resistance to each drug, but there is no cross-resistance mutations: $f_R$ is then
$\approx 5\times 10^{-16}$, a fraction so small that we should never see
cross-resistance emerge in this way. If instead there is one mutation that can
confer cross-resistance, $n_{12}=1$, then $f_R \approx 2 \times 10^{-8}$, and
we can expect around $100$ cross-resistant cells to be present in the tumour.

Even if some cross-resistance, in particular between capecitabine and
gemcitabine, could plausibly be granted by one mutation, the therapy still
makes use of three distinct classes of drugs which are heterogeneous enough that no
single mutation should grant cross-resistance. In theory, $n_{1234} = 0$, 
calling gemcitabine and capecitabine drugs $3$ and $4$, let $n_1 = n_2 = 1$
and $n_{34}=1$, and we have

\begin{equation}
    f_R \approx (\log(N))^3 n_1 n_2 n_{34} p_\mu^3 \approx 10^{-23}
\end{equation}

so that, again, cross-resistance should essentially never happen. 
Really, the different numbers of possible mutations $n_i$ are not precisely
known, and may well be much larger than $1$. Pessimistically, if
$n_1=n_2=n_{34}=1000$, then 

\begin{equation}
    f_R \approx 10^{-14}
\end{equation}

which is still so low compared to the number of cells present that it should
- in principle - never be observed.

Let us summarise the working hypothesis underlying low-dose combination chemotherapy 
in simple terms: if there are no mutations which confer cross-resistance to all $4$
drugs simultaneously, and all relevant mutations are selectively neutral prior to
therapy, and resistance can only be acquired by the appearance of
genetic mutations, then $f_R$ should be unmeasurably small. If it is not unmeasurably 
small, at least one of these postulates has to be wrong.

\section{Results and discussion}


\subsection{Inferred fraction of resistant cells}

\begin{figure}[h]
\begin{center}
    \includegraphics[width=1.1\textwidth]{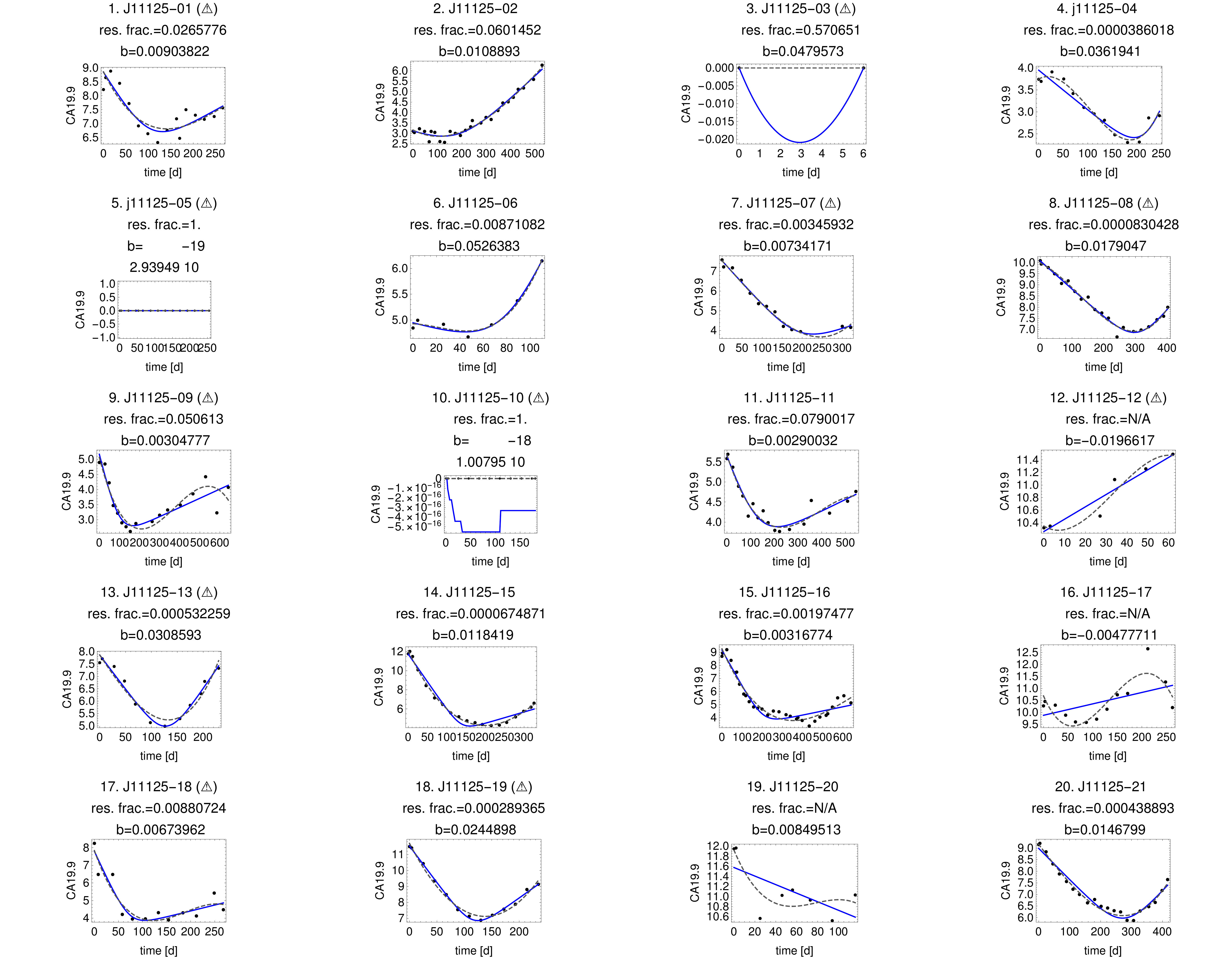}
    \caption{\label{fig:ca199-1}Plots of CA19-9 levels in patients 1 --- 20, showing the best
    fit for the multiple exponential model (blue curve) along with the
    inferred resistant fraction $f_R$ and regrowth rate $b$, and the best fit for the null model
    (grey dashed curve). Cases in which the multiple exponential model had a worse fit
    than the null model (failing our test for over-interpretation) are labelled
    with warning signs.}
\end{center}
\end{figure}

\begin{figure}[h]
\begin{center}
    \includegraphics[width=1.1\textwidth]{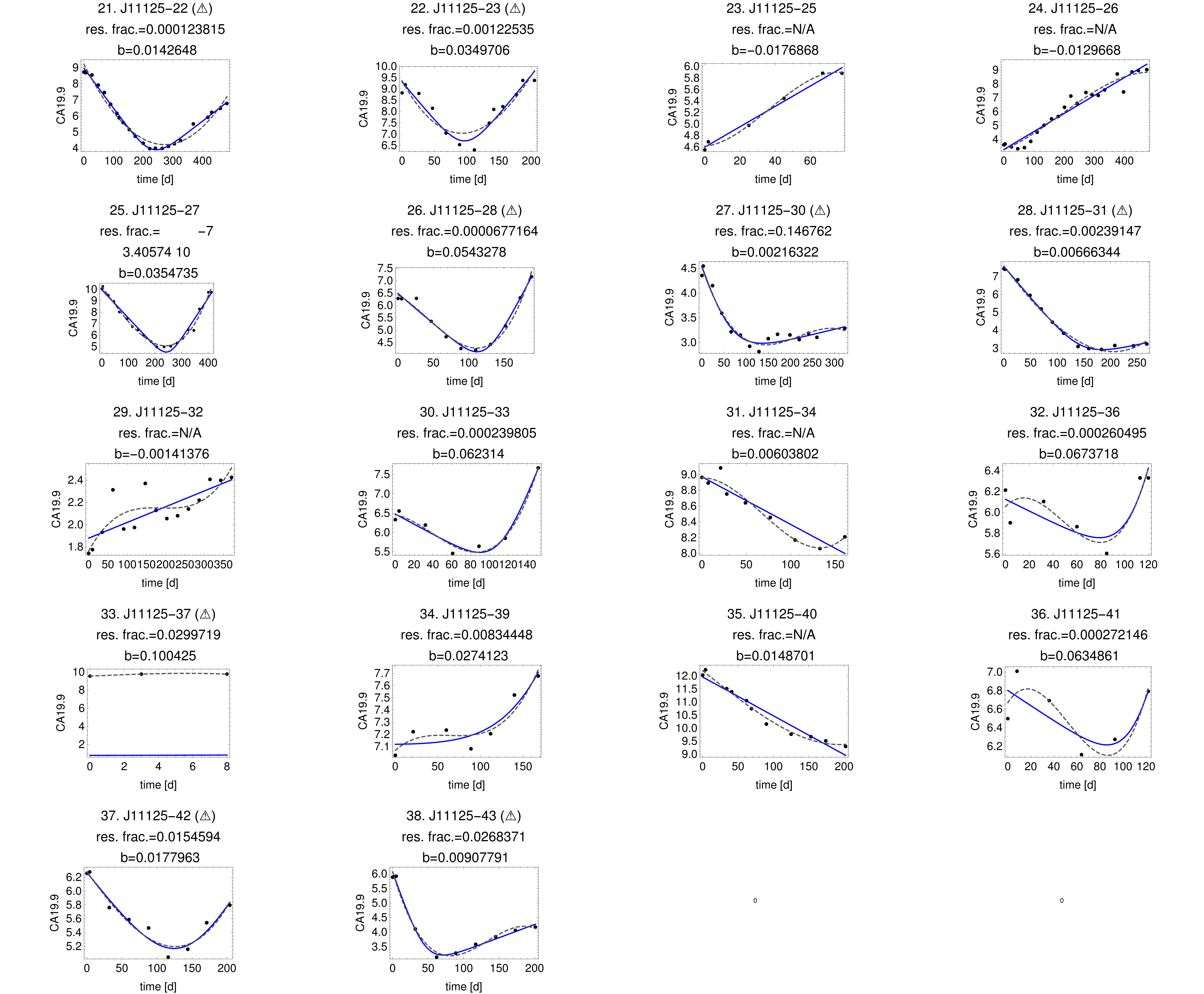}
    \caption{\label{fig:ca199-2}Plots of CA19-9 levels in patients 21 --- 38, showing the best
    fit for the multiple exponential model (blue curve) along with the
    inferred resistant fraction $f_R$ and regrowth rate $b$, and the best fit for the null model
    (grey dashed curve). Cases in which the multiple exponential model had a worse fit
    than the null model (failing our test for over-interpretation) are labelled
    with warning signs.}
\end{center}
\end{figure}

Out of all 38 datasets analysed, 19 were rejected: about half. This is what 
we might expect if both models had equally bad fits to datasets with a very
low ``signal-to-noise ratio''. However, there are qualitative differences in
the shape of datasets which show a clear decline and regrowth phase, and those
with CA19-9 concentrations so low as to be on the verge of detectability (such
as patients 3, 5, 10 and 33, figures \ref{fig:ca199-1} and \ref{fig:ca199-2}).
Many of the datasets which were rejected as noisy by
the difference-in-$\chi^2_r$ test were indeed qualitatively noisy, or had flat CA19-9 concentrations
on the lower limit of detectability. On the other hand, some datasets that we
expected to pass the test did not, such as patients 7 and 8 (see figure \ref{fig:ca199-1}). 
This difference between the qualitative shape and results of the quantitative
test may be due to the fact that the null model can more easily fit wildly
varying large concentrations than the exponential model, and more datasets are
rejected than would be under a more specific test. However, some advances in
dynamical models of inflammatory responses are required before a well-founded
test to distinguish immunological noise can be developed.

The median length of time that patients survived before succumbing to their recurrent
cancer, or median overall survival, was 13 months. Individual overall survival
was found to be closely correlated to the relapse time calculated from
equation \eqref{eq:regrowth} having set the unknown ``tumour size at relapse'' $N$ to
be equal to the initial CA19-9 concentration $a+c$. The time to relapse is closely
related to the regrowth rate $b$, but not very sensitive to the tumour size at
relapse $N$, depending on it only logarithmically. That equation \eqref{eq:regrowth} 
should be a decent predictor of the actually observed relapse time is not especially surprising
from a mathematical perspective, but bears mentioning as the clinicians on the
study found this to be practically useful.

Although the clinicians who participated in the study found the multiple
exponential model's measurement of the regrowth rate $b$ very useful, the
applied mathematicians and biophysicists were more interested in the inferred
fraction of resistant cells, $f_R$. Since one of the motivations for this
combination was to minimise the likelihood of cross-resistance, the inferred 
fraction of resistant cells is an indicator of the proportion of cells
that are cross-resistant at the start of therapy, and measurement of $f_R$ can
provide a stringent test of the working hypothesis that cross-resistance
arises from purely evolutionary means. That is, it is acquired by novel
genetic mutations and requires several distinct genetic mutations (to the
different pathways relevant to the distinct mechanisms of action of each
drug).

\begin{figure}
\begin{center}
    \includegraphics[width=0.6\textwidth]{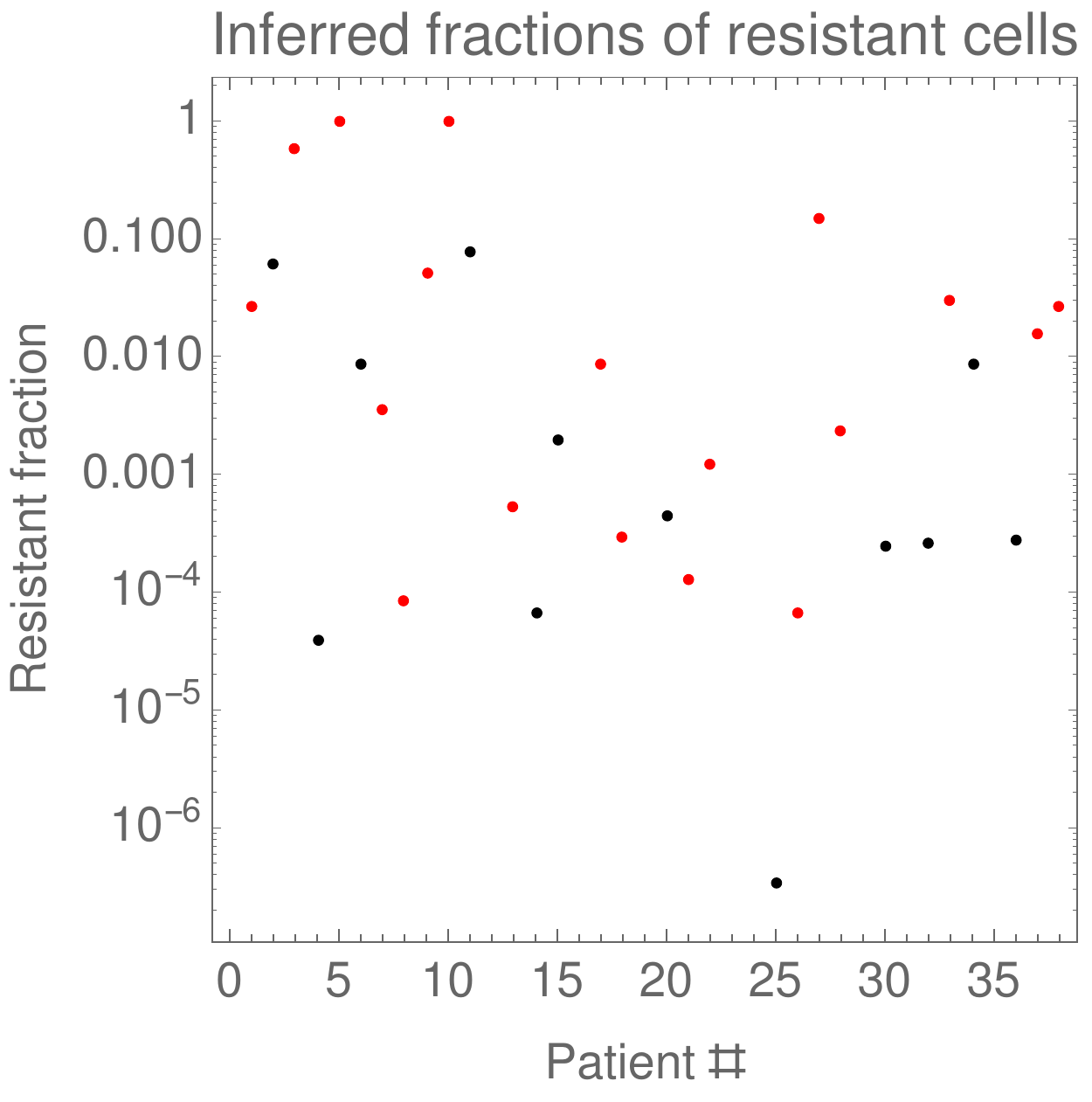}
    \caption{\label{fig:resfracs} A plot of the inferred fractions of
    resistant cells for each patient. Only those datasets for which $f_R$ 
    was well-defined were plotted. Fits which failed our test for
    over-interpretation are represented by red points, fits that passed are
    represented with black points. Note the logarithmic
    scale: the range of observed values spans several orders of magnitude.
    }
\end{center}
\end{figure}

The inferred fraction of resistant cells $f_R$ varies substantially between
patients (see figure \ref{fig:resfracs}), but in all cases was many orders of magnitude higher than
theoretically expected. The lowest value observed was $3.4\times 10^{-7}$
(patient 25, see figure \ref{fig:resfracs}, and \ref{fig:ca199-2}), while the highest values were $1$
(i.e. all cells seemed to be resistant, patients 12, 25 and 26), although we 
should point out that for many of these particularly
high values, the multiple exponential model fails our test for
over-interpretation. The highest value of $f_R$ for which the multiple
exponential model was considered meaningful was $7.9\times 10^{-2}$ (patient 11, see figure 
\ref{fig:ca199-1}), which is still many orders of magnitude higher than was theoretically expected.
Median values of $f_R$ and $b$ were $2.98\times 10^{-3}$ and $1.44\times 10^{-2}\;\mathrm{day}^{-1}$
respectively. If we restrict this analysis to the datasets for which the multiple
exponential model did not fail our test for over-interpretation of noise, we
instead find $3.99\times 10^{-4}$ and $2.46\times 10^{-3}\;\mathrm{day}^{-1}$.

Other research on multi-drug resistance also estimates $f_R$ to be much higher
than what one would expect on the basis of the above working hypothesis,
in the range $5\times 10^{-5}$ to $3\times 10^{-4}$, consistent with
our inferences from the best-fit model parameters above \cite{hyo2015studying}.
However, it is important to note that this study analysed resistance to two
drugs only, erlotinib and crizotinib, two protein kinase inhibitors which 
have very different mechanisms of action to any in the combination studied here. 
In fact, these drugs target specific signalling proteins, whereas the combination 
chemotherapy analysed here is much less selective.

As $f_R$ was clearly not unmeasurably small, and the disease recurred in a
resistant form in the majority of cases, it must be the case that at least one
of the postulates underlying the working hypothesis behind the therapy is wrong. 
Either there \emph{is} in fact a mutation which can confer resistance to all four 
drugs simultaneously, the relevant single-drug resistance mutations are not
selectively neutral prior to therapy, or cells can become resistant without having acquired 
genetic mutations, or some combination of the above. We shall explore each of
this possibilities in turn.

\subsection{Molecular mechanisms of single-drug resistance}

Before considering possible mechanisms of multi-drug resistance that can arise
from a single mutation, let's review some of what is known about mechanisms of
single-drug resistance, as it is a simpler case which can inform and constrain
our discussion of the possible mechanisms of cross-resistance. As cisplatin
resistance is very common when used in isolation \cite{galluzzi2012molecular}, and cisplatin is
the best studied of the drugs in the combination, we will focus primarily on
the mechanisms of action and transport into and out of the cell.

Cisplatin's primary mechanism of action as an anticancer agent is by damaging
DNA. This triggers a DNA damage response in the cell's signalling pathways,
inducing apoptosis (programmed cell death) \cite{galluzzi2012molecular}. Before
any damage can be done by cisplatin's reaction with the DNA itself, it must
be present in the nucleus of the cell \cite{el1999reactions}. It is also only
activated by a reaction with the cytoplasm \cite{el1999reactions}. 
It therefore has to have been transported into the cell sufficiently quickly (and out 
sufficiently slowly)
that it has an appreciable concentration inside the cell \cite{arnesano2013updated}. 
Examples of uptake and efflux pathways are via copper transporter 1 (CTR1) and MRP2, one
of the ATP-binding cassette (``ABC'') transporters \cite{galluzzi2012molecular,arnesano2013updated}.
Resistance to cisplatin can therefore consist of a reduced response to
DNA damage, decreased activation of the cisplatin complex by aquation, and lower concentrations of
cisplatin inside the cell by modified uptake and efflux \cite{galluzzi2012molecular,arnesano2013updated}.

Changes in efflux mediated by the ABC transporters like MRP2 have been studied
for some time, due to the potential importance of these transporters as
markers for responsiveness to therapy and as potential targets of new
drugs \cite{yamasaki2011role}. At one time, they were even thought to be almost 
exclusively responsible for the emergence of multi-drug resistance \cite{gillet2010mechanisms}. 
However, although inhibitors of ABC transporters do reduce chemoresistance to a degree, they do not
have a very strong affect with regards to restoring cells' sensitivity to
cisplatin. As a result, it is suspected that other factors must be involved in
the development of cisplatin resistance. Either other efflux pumps insensitive
to the inhibitors tested, or changes in uptake and action in the cell \cite{jin2005reversal}.

Evidence for CTR1's role in cisplatin uptake seems more straightforward, and includes:

\begin{itemize}
    \item Mice in which the
    transmembrane pump CTR1 has been knocked out are much more resistant to the
    cytotoxic effects of cisplatin \cite{ishida2002uptake}.
    \item Cells with higher than normal copper
    concentrations (copper shares an uptake pathway with cisplatin, CTR1) are
    more resistant to cisplatin \cite{more2010role}.
    \item Cells with lowered copper concentrations are more
    sensitive to cisplatin \cite{ishida2010enhancing}.
\end{itemize}

All of which seems to indicate that CTR1 transports cisplatin into the cell,
and less cisplatin is transported when CTR1 is absent or busy.

Finally, changes in signal transduction pathways could provide a way for
cancer cells to continue to survive and divide even in the presence of high
concentrations of cisplatin. Out of a preponderance of partially understood
mechanisms, the ability of cells to divide in the presence of DNA damage by
inhibiting apoptosis \cite{galluzzi2012molecular} and the ability of all cells
to break down and recycle individual damaged components, autophagy, bear
particular mention \cite{bao2015induction}. A mutation in one of the protein
kinases that triggers apoptosis (programmed cell death) could plausibly result
in a line of cancer cells which doggedly divide even with cisplatin-damaged
DNA, although we should note that this mechanism of resistance is specific to
cisplatin's mode of action. Autophagy, on the other hand, is a non-specific
response to stress that is latent even in normal cells, and though its
relevance to cisplatin resistance is not well understood, there are
results which indicate that there is an inverse relationship between cisplatin
resistance and the inhibition of autophagy \cite{galluzzi2012molecular,bao2015induction}.

To summarise, proposed mechanisms that may contribute to cisplatin resistance
include increased efflux and cisplatin removal from the cell, decreased
uptake of cisplatin into the cell, and weakened cellular response to DNA
damage and toxicity.

\subsection{Possible mechanisms of cross-resistance}

As this is the first time that this specific combination has been tested
clinically, and consequently the first time that cross-resistance to the four
agents involved has been observed, the mechanism behind this resistance is
basically unknown and as a result we can at present only speculate as to its
true nature. On the basis of the stochastic theory of Antal, Bozi\v{c} and
others, the cross-resistance is extremely unlikely to be due to cells' acquiring multiple
independent mutations by virtue of the
heterogeneity of the drugs' mechanisms of action 
\cite{bozic2013evolutionary,antal2011exact}.
however,
the anomalously high fractions of resistant cells $f_R$ can be explained in
principle by there being many different mutations which can provide resistance
to individual drugs, with $n_i$ on the order of $10^4$. Given present
uncertainties about the mechanism of action of several of these drugs, this
possibility cannot be conclusively ruled out at this time.

Besides invoking such high values of $n_i$, the
alternatives to the working hypotheses underlying this prediction can be
tentatively sorted into two groups: mechanisms which involve cross-resistance being
\emph{acquired} by one (or very few) novel mutation, and mechanisms in which a
sub-population of cells in which a latent \emph{intrinsic} resistance is
activated without requiring new genetic alterations as such.
The heterogeneity in structure, transport pathways, and mechanisms of action
between the drugs in the combination makes it unlikely that a mutation
affecting one transport protein, or indeed any one of the specific mechanisms
or cellular responses to each drug will result in resistance to others in the
combination. A mutation that allows cells to avoid apoptosis triggered by
cisplatin's damage of cellular DNA seems unlikely to have any relevance to the
formation of microtubules blocked by docetaxel, for example. This presents a
difficult puzzle to attempts to explain the observed frequencies of
cross-resistance in terms of either acquired single mutations or intrinsic
cellular responses. We will first discuss possible mechanisms of \emph{acquired}
cross-resistance (caused by mutations), and then discuss different mechanisms of 
\emph{intrinsic} cross-resistance (not caused by mutations as such).

\paragraph{Acquired resistance}
\label{sec:possible}

One possibility for cross-resistance to be acquired by a single mutation
is if there were a protein kinase that interacted
with every drug's uptake and efflux pumps, perhaps downregulating the uptake
pumps and upregulating the efflux pumps, acting as part of a
signal transduction pathway that is activated during a relatively
non-specific cellular response to toxicity and stress. When this protein
kinase was activated by another above it in the pathway, the effect of its
simultaneously closing many intake transporters and opening many efflux
transporters would be a kind of emergency evacuation of cytotoxins from the
cell, as though an alarm had been sounded. A mutation which caused this kinase
to take on its active conformation all the time would lead to broad
changes in membrane transport, which would be consistent with
cross-resistance. This hypothetical ``alarm kinase'', if it exists at all,
may be some undiscovered molecule yet to be characterised, or may
simply be a hitherto obscure function of a protein that has already been
partially characterised in the literature. In either case, there are some
simple criteria that this protein kinase would have to meet:

\begin{itemize}
    \item It should be sensitive to cytotoxic stress on the cell (possibly
    associated with inflammation, autophagy or the heat shock response)
    \item It should interact with many different pump and transporter proteins
    or should be ``upstream'' of them in many distinct signalling pathways
    \item The mutated protein should lose the specificity to stress but interact
    with transport proteins similarly
\end{itemize}

The existence of such an alarm kinase could provide one mechanism for cells 
to acquire broad cross-resistance with a single mutation. This single mutation would plausibly be
neutral in the absence of combination chemotherapy, as there would be no
selective advantage associated with excluding high concentrations of
cytotoxins in this case. We should emphasise, nonetheless, that this is
speculative. A single neutral mutation could arise through chance as in the
branching process theory developed by Antal et al. detailed above in section
\ref{sec:branching}, or else could arise as a genetic hitch-hiker \cite{antal2011exact}.

While we are not aware of any one protein kinase that definitely satisfies
all of the above criteria, after surveying the literature the best educated
guess as to the identity of this alarm kinase that we were not able to 
conclusively rule out was JNK1 (also known as MAPK8). JNK1 is a signalling
protein which is associated with stress and inflammation, interacts with
several distinct signalling pathways, and there are results suggesting that 
activation of JNK1 is associated with drug resistance and conversely that 
inhibition of JNK1 is associated with reversal of 
resistance \cite{hayakawa2003activation,fallahi2015systematic}. Furthermore,
JNK inhibition also results in P-glycoprotein (or MDR1), an important efflux pump, being
downregulated, which sensitizes some lines of cancer cells to cisplatin, one of
the drugs in the combination \cite{liu2016inhibition}. However, these
results should be interpreted cautiously, as there is at least one contrary
experimental result in the literature, and it may be the case that the
specific cancers and cell lines studied are not representative of the specific
type of cancer in our own study \cite{lagadinou2008c}. We are also not aware of
any results regarding common mutated forms of JNK1, in particular whether or
not mutants are permanently in the ``active'' conformation:
continual failure to identify any single mutant protein kinase associated with
cross-resistance would strongly suggest that the alarm kinase hypothesis was
false.

While there are other conceivable ways that multi-drug responses can arise
without such an alarm kinase, it does provide a simple explanation for the
non-specific (in that several very different drugs are involved) but unified 
(in that they are all cytotoxic and resisted by the cell \emph{somehow})
response observed in this study.

Alternatively, it may be the case that several distinct mutations are indeed
required, but for some unknown reason are not neutral before the therapy is
applied. This would throw out the framework in section \ref{sec:branching} and
fractions of resistant cells may be much higher, perhaps approaching fixation
(that is, $f_R \approx 1$). While it seems unlikely to us at first glance that
alterations in the specific mechanisms of uptake, efflux and the targets of four 
specific drugs should also have some adaptive selective advantage in the
absence of these drugs, enough is unknown about them that this cannot be
conclusively ruled out --- for example, modified efflux pump activity could
be a side-effect of an advantageous mutation controlling proliferation via
some complex gene-gene or protein-protein interactions. Furthermore, this 
explanation does not need hypothetical uncharacterised functions of protein 
kinases to be invoked.

Either hypothesis - that resistance is acquired through the mutation of a
single alarm kinase, or several independent drivers - would be conclusively
falsified by demonstrating that cross-resistance could be induced in a
line of cancer (or indeed normal) cells without any additional mutations. This
would indicate that cross-resistance was an intrinsic response, rather than
acquired.

\paragraph{Intrinsic resistance}

It may also be the case that resistance to several drugs at once can be
induced in otherwise genetically normal cells, or simply that the drugs fail
to have an effect for reasons unrelated to genetic alterations. This class of
possible mechanisms attempt to explain the anomalously high resistant
fractions in terms of intrinsic resistance. Cells resist the effect of the
drug combination without inheriting the ability to do so as such, due to
differences in regulation of normal genes or in their microenvironment. 

Perhaps the simplest way in which this kind of apparent ``resistance'' can
come about is by poor vascularization. If there is very poor penetration of
the tumour by capillaries, drugs present in the bloodstream will find it
difficult to reach high concentrations in the tumour cells by diffusion from
the bloodstream alone. Poor vascularization, and the associated depletion of
oxygen around these areas of the tumour, is known to be a factor in multi-drug
resistance, but may be mediated by more complex cellular responses
(e.g. involving ABC transporters) rather than
the simple fact that less agent will be present \cite{gillet2010mechanisms}.

Another way that cells could develop resistance to the combination of drugs is
if there were a latent response to exclude cytotoxins from the cell by 
altering membrane transport, and this latent response was
activated by signals from other cells. This generic response to cytotoxicity could possibly be
triggered by the activation of an ``alarm kinase'', but the possibility 
of this being controlled by a less centralised pathway cannot be ruled out. 
This class of mechanisms could work in concert with acquired resistance. The 
effective number of resistant cells can be amplified substantially, and one resistant cell
could ``shield'' a number of other sensitive cells from the effects of the therapy
by activating a latent intrinsic response.

It would be desirable to make some quantitative predictions as to how many
sensitive cells could be shielded by one resistant cell by such a mechanism. The observed
$f_R$ of around $10^{-4}$ would then be determined by the product of the
underlying tiny core population of truly resistant cells $f_R^{\star}$ and this
shielding ratio, $S$. The tiny resistant core $f_R^{\star}$ could take on some
small value consistent with T. Antal and I. Bozi\v{c}'s theory based on branching
processes, say $f_R^{\star}$ in the range $10^{-14}-10^{-8}$, and since

\begin{equation}
    f_R = f_R^{\star} S\;,
\end{equation}

then values of $S$, the number of sensitive cells shielded by one cell in the
resistant core population, would then have to be at least $10^4$ or larger to
account for clinical observations. To understand whether or not this is an
achievable value of $S$, we will have to look at more explicit models of
signalling mechanisms.

\begin{figure}
\begin{center}
    \includegraphics[width=0.8\textwidth]{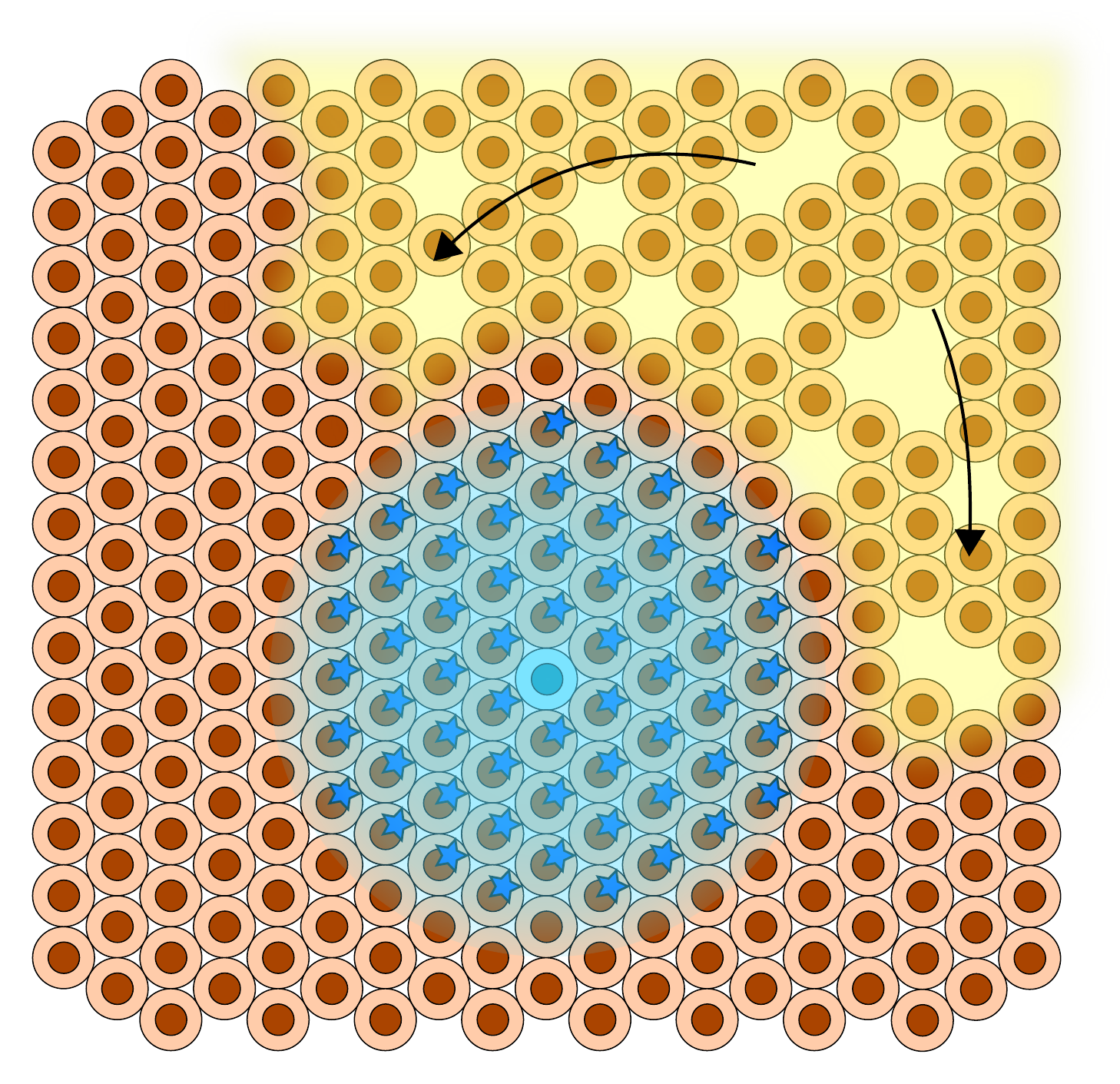}
    \caption{\label{fig:cytokine}A diagram showing one possible mechanism by
    which the apparent number of resistant cells can be much larger than the
    true number of underlying resistant mutants. Resistant mutants ``shield''
    sensitive cells by producing a small signalling molecule (represented by blue haze), which triggers
    sensitive cells to alter their transmembrane pump activity in response and
    slow their uptake of the cytotoxin diffusing across the tumour from top right
    (represented by yellow haze). The small molecule in question may be
    a cytokine, and the response related to inflammatory response in normal
    tissue: CA19-9 is also known to be elevated by inflammation.
    Resistant mutant cell in blue, sensitive cells with modified uptake/efflux
    tagged with blue stars.}
\end{center}
\end{figure}

First, consider the case that this latent resistance is activated by the
diffusion of a small molecule such as a cytokine. The molecule is produced by
the core resistant population, diffuses through the surrounding sensitive
cells, triggering their latent intrinsic resistance program, and gradually
breaks down over time (see figure \ref{fig:cytokine} for an illustration). 
If this molecule is only produced by core resistant
cells, is not transported actively, having a diffusion coefficient $D_c$, and has a degradation 
rate $\lambda$, then its steady-state concentration in the tumour $\rho_c$ obeys the
simple differential equation

\begin{equation}
    D_c \Delta \rho_c - \lambda_c \rho_c = 0
    \label{eq:cytokine}
\end{equation}

outside the core population of cells from which it is released. While one can
explicitly solve this equation analytically in a few cases (such as one
dimension, spherical symmetry, \emph{et cetera}), we won't reproduce exact
solutions here. Instead, we will be satisfied to observe that solutions of
equation \eqref{eq:cytokine} have a characteristic length scale $L_c$

\begin{equation}
    L_c = \sqrt{\frac{D_c}{\lambda_c}}
\end{equation}

which can be interpreted as the effective distance that the signalling molecule
can travel before breaking down. A single resistant cell can then shield
sensitive cells in a volume on the order of $L_c^3$, and so if a typical cell
has size $d_{\mathrm{cell}}=10\;\mu \mathrm{m}$, the shielding ratio $S$ will be

\begin{equation}
    S = \left(\frac{L_c}{d_{\mathrm{cell}}}\right)^3\;.
    \label{eq:cytokineshield}
\end{equation}

Many cytokines have diffusion coefficients on the order of
$10^{-6}\;\mathrm{cm}^2\mathrm{s}^{-1}$,
and degradation rates on the order of $10^{-3}\;\mathrm{s}^{-1}$,
which correspond to $L_c \approx 300\;\mu\mathrm{m}$ \cite{ao2006microdialysis,cheong2006transient}.
Via equation \eqref{eq:cytokineshield}, given a cell size of around
$10\;\mu\mathrm{m}$, this corresponds to a shielding ratio of $S\approx
3\times 10^{4}$, the correct order of magnitude for this to be a viable
explanation. While far more research would have to be done to exclude other
explanations, this simple argument shows that shielding of sensitive cells 
by resistant cells is not ruled out by what is currently known about the
dynamics of cell signalling.


The hypothesis that cytokine signalling could result in cross-resistance, which
is at first glance compatible with current experimental knowledge, could be
ruled out by demonstrating the presence of cross-resistance in the absence of
signalling by cytokines or other small molecules, perhaps with a knock-out or
inhibition experiment (on mice) \cite{jones2016cytokines}.

\begin{figure}
\begin{center}
    \includegraphics[width=0.8\textwidth]{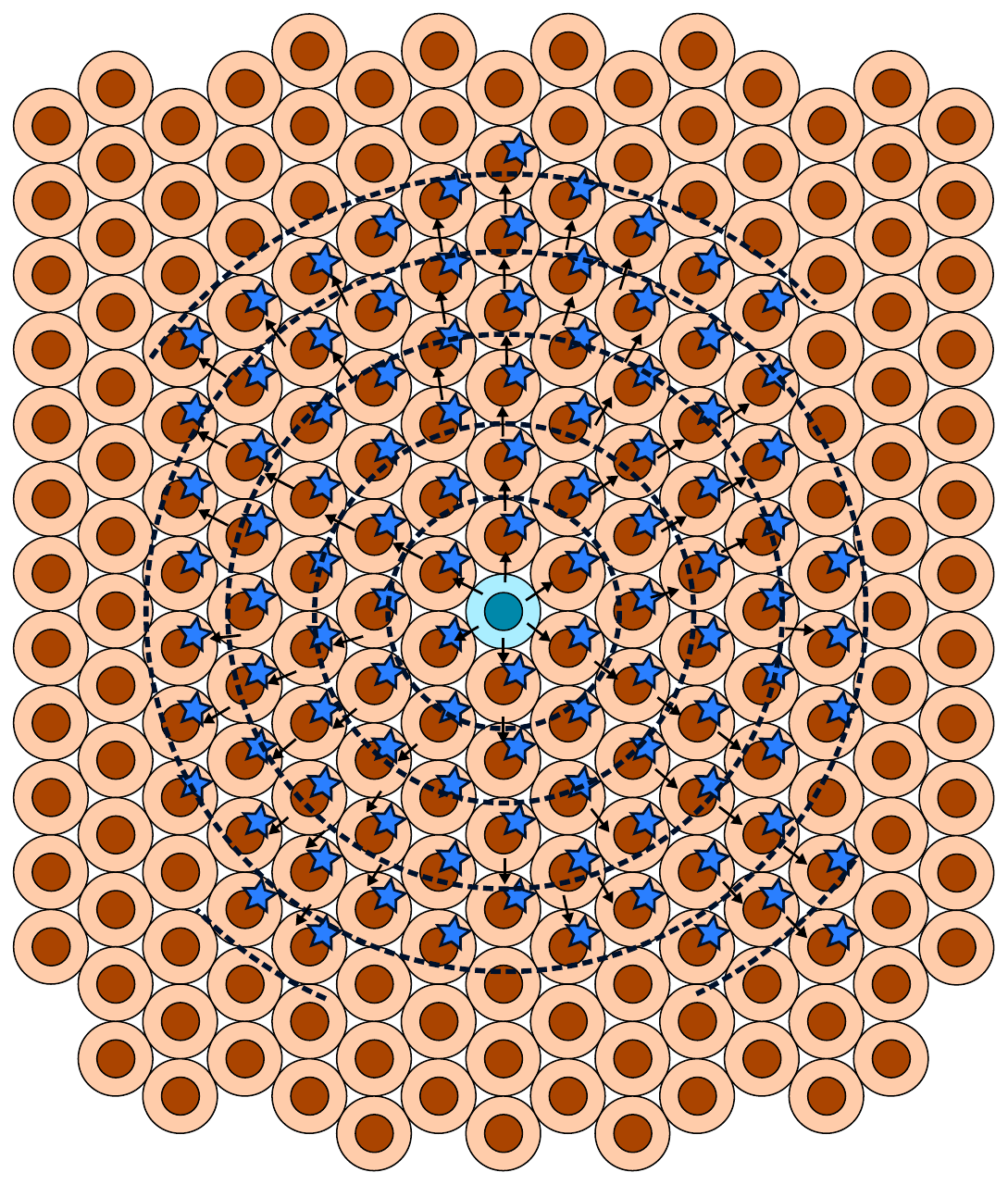}
    \caption{\label{fig:signalwave}A diagram showing one possible mechanism by
    which a small number of resistant mutants can trigger a cascade of
    resistance in otherwise sensitive cells. A travelling wave of cell-to-cell signalling through
    transmembrane proteins and tight junctions triggers a change in
    transmembrane pump activity and slows the uptake of cytotoxin in cells.
    Candidate molecules that may be associated with this travelling wave are Connexin-42
    and JNK1.
    Mutant cells in blue, sensitive cells with modified uptake/efflux
    tagged with blue stars, and the travelling wave of resistance to cytotoxin
    represented by successive wavefronts, dashed black line.}
\end{center}
\end{figure}

Larger shielding ratios can be achieved with other signalling mechanisms.
Rather than being mediated by a small signalling molecule, cells are able to
transmit signals by coming into biochemical contact with cells they are
adjacent to. Conformational changes inreceptors on the membranes of two adjacent cells can trigger
signalling cascades when they come into adhesive contact \cite{MolBioCell} (see
figure \ref{fig:signalwave}). A given signal can then propagate through a population of connected cells in a
wavelike manner, with characteristic speeds, in which newly activated cells 
on the wavefront triggering a new front of cells in front of them. Wavelike 
phenomena in tissues are well-attested, and the activation patterns associated 
with them are readily monitored by tracking second messengers such as
calcium \cite{gilkey1978free,kupferman1997analytical}.

Shielding models based on wavelike propagation differ from models based on
diffusion of a small molecule in the crucial regard that the distance the
signal can travel is not limited by the degradation of a small molecule. 
It follows that the number of cells that can be shielded could be much larger.
Naively, one
might suspect that arbitarily large volumes of tumour cells could be reached
by the signal, but this is probably not the case. The propagation of calcium
waves depends on a reliable response of the cells to activation, and
occasional random failure of some of these cells due to
variability in response (among other imaginable factors) will attenuate the signal. 

Although the mechanism
underlying the attenuation was not studied in detail, an experiment to
measure the distance that calcium-dependent waves could travel before
becoming attenuated to the point of undetectability put the attenuation distance at about 
$1\;\mathrm{mm}$. We should note, however, that this was in a very different
type of tissue than ductal pancreatic cancer, and may also be affected
by geometry \cite{uhrenholt2007propagation}. This places an experimental
estimate on the upper limit of shielding range by this mechanism.

An attenuation distance of approximately $1\;\mathrm{mm}$ and a typical cell
size of $10\;\mathrm{\hbox{\textmu} m}$ imply a shielding ratio 

\begin{equation}
    S \approx \left(\frac{1\;\mathrm{mm}}{10\;\mathrm{\hbox{\textmu} m}}\right)^3
    = 10^6
\end{equation}

via equation \eqref{eq:cytokineshield}. This is indeed much higher
than in the small-molecule proposal.

In order to actually test this possibility, at least one candidate molecule
and an interesting novel prediction are in order. The current experimental
literature on calcium-dependent signalling in cancer seems to point to IP3R as
another important second messenger in these pathways, but plays such diverse 
roles even in normal tissues that its presence wouldn't be specific to this
proposed mechanism \cite{yoshida1997structure}. An alternative candidate, which
transmitted signals from cell to cell via gap junctions, is Connexin-42.
If it is the case that wavelike signalling is behind cross-resistance, then
the fact that regions which can be reached by these waves must be spatially
contiguous could be used to exclude this hypothesis. If only one spatially
connected region of a tumour was found to have established multi-drug
resistance, this would be consistent with the wavelike hypothesis; whereas if
many distinct regions were found to have independently established
cross-resistance or it was otherwise diffused throughout he tumour in a way
not obviously related to space, this would effectively rule out wavelike
signalling as responsible.

The fact that the resistant subpopulation regrows exponentially seems to
suggest that well-mixed, and therefore spatially diffuse growth 
of the resistant subpopulation is occurring. However, it is not possible
on the basis of our clinical data alone to rule out the possibility that this
is due to the resistant population being well-perfused and growing
exponentially albeit in a spatially limited region. An experiment which
could detect cross-resistant regions and comment on their topology, perhaps
via advances in medical imaging or analysis of tumour biopsies, could provide
a stringent test of this hypothesis.

\section{Concluding remarks}


Despite positive remarks about our mathematical model's usefulness from
practising clinicians who organised and carried out the experimental work of 
the study, the inferred fraction of resistant subclones $f_R$ was several
orders of magnitude higher than what was expected on the basis of a
mathematically rigorous theory that assumed as little as possible about
unknown biology. This inconsistency between the
experimental result and the theoretical prediction indicates that at least one
of the postulates of the underlying theory is wrong, a very surprising result
which presents a major interpretative challenge.

The relevant postulates of the mathematical theory are
that: 

\begin{enumerate}
    \item Resistance to chemotherapy is acquired by the accumulation of novel
    genetic mutations,
    \item Several distinct mutations are required, as the 
    mechanisms of action of the drugs do not overlap,
    \item The relevant
    mutations are selectively neutral prior to the application of the therapy. 
\end{enumerate}

Different
alternative mechanisms can be speculated about as explanations in each case,
but much additional research would need to be done to exclude even a few of
the abundance of possible explanations. Of particular interest is the
possibility of poorly-characterised interactions between known protein kinases
bringing about resistance to several agents simultaneously, whether this is by
one of these mutating or a yet-to-be-discovered response to toxicity latent
even in normal cells.

Each of our proposed mechanisms to explain the anomalously high $f_R$ can be
clarified and made more concrete with educated guesses as to the identity of
candidate molecules which may be involved. However, as many of our candidate
molecules are known to have many different interactions, many of which are
still poorly characterised, experimental evidence that one of them is correlated with
cross-resistance does not in itself constitute evidence for our purported
mechanism. More stringent tests that could specifically exclude each proposal,
and ideally novel predictions of more rigorous quantitative formulations
thereof, are necessary to clarify which if any are responsible for the
unexpectedly high rates of cross-resistance observed. We have tried to put
forward at least one way to experimentally falsify each proposal.

In conclusion, this analysis has highlighted very serious holes in the current
understanding of multi-drug resistance in pancreatic cancer, in particular
regarding the specificity of cellular responses to cytotoxins, and the
possibility of overlapping mechanisms. While it may be the case that one
yet-to-be-identified mutated protein can bring about resistance to drugs with
widely different pharmacological properties and cellular transport mechanisms,
there are a large number of other possible mechanisms by which intrinsic resistance could
emerge which cannot be ruled out on the basis of current experimental
knowledge, and would need to be ruled out by future research in order to
shed light on the true mechanism of cross-resistance observed here.

\chapter{Conclusion}




The main object of this thesis was to explore the relationships between cell
motility, tumour growth, and evolution in advanced cancers. Although motility
and invasion is known to be of critical importance for growth and clinical
outcomes, there is a lack of research which is both precise, clinically
applicable, and simple enough to be intelligible and
transparent to ordinary humans. We intended to develop and apply a
series of minimalistic models with a view to establishing how changes in
migration could affect the speed at which cancers evolve, and furthermore we
attempted to constrain and test these models' results empirically.

There are a number of more specific questions that we set out to address. What
mechanical and kinetic properties determine the speed of cell migration, and
the frequency with which motile cells are released by a tumour? What effect do
the resulting changes in cell migration have on the growth curve of cancers?
How do differences in growth rates and fitness determine the pace of
evolution? And finally, how rapidly do cancers grow and evolve in reality?


To address how the mechanical properties of cancer cells affected how easily
they could migrate through tissues, we performed simulations of how motile
cells responded to applied forces while embedded in cellular monolayers. These
simulations treated cells as isotropic elastic spheres, modelled the transition
between different cell types as a stochastic process, and measured how quickly
the cells moved when subjected to a constant traction force. The two cell
types were treated as static, non-invasive epithelial cells and motile invasive 
mesenchymal type cells, with the invasive cells assigned a different elastic
stiffness. The effect of adjusting the elastic stiffness was that more elastic
and easily deformed cells could move faster through tissue, and stiffer cells
moved slower through tissue: elastic stiffness therefore affects cells'
drag coefficients, at least in cellular monolayers.

We also found that making invasive cells arbitrarily stiff did not result in
an arbitrarily high resistance to motion. We interpreted this as representing
the fact that even an ideally stiff sphere surrounded in a deformable medium
would have its resistance to motion limited by the surrounding medium,
similarly to the case of a very hard metal sphere falling through a viscous 
liquid.


These mechanistic simulations were very computationally intensive,
 so were not suitable for simulations of larger tumours.
To examine the effect of changes in cell migration speed, changes in rates of
type switching, and changes in cell division rate on the growth curves of
large tumours, we developed a lattice model in which cells could transition between
a non-motile type that stayed at one lattice site and divided with an
adjustable stochastic rate, and a motile type which hopped between sites with
a constant speed and randomly changed direction with some rate $\alpha$.

These
simulations naturally showed how spheroidal tumours could emerge from a simple
set of rules, and that migration enabled limited periods of exponential
growth. They also demonstrated that the rate of this exponential growth was
not only affected by the rate at which cells switched to motile types, which
was expected, but also by the rate at which they switched back to non-motile
types, and furthermore by the migration speed. These latter two dependencies
were unexpected and rather complex, but could be interpreted in terms of their
effect on the proportion of motile cells present at any one time and the delay
between a cell's escape and its founding of a new tumour.


Another lattice model was developed which abstracted cell migration entirely,
treating migration events as instantaneous and extremely distant from the
originating tumour. The new secondary lesions which were founded as a result
of these events could therefore be thought of as developing independently of
the originating primary. These simulations also included the appearance of new
mutations, and treated cell division as a stochastic process. These
simulations established that ensembles of tumours do indeed grow exponentially
if the individual lesions' growth do not interfere with each other, and that
the rate of this exponential growth can be related to the proportion of cells
which migrate and found new lesions.

The roughness of the surfaces of the spheroids in this lattice model was also
important, resulting in a more complex dependence between migration and growth
rates and a slower speed at which new advantageous mutants can grow across
the surface than previously estimated by other authors.


To understand how this last lattice model could support exponential growth and
how the pace of evolution could be estimated, we developed a coarse-grained
analytical theory which treated individual lesions as continuous ``blobs''
with pre-determined growth curves, and migration and mutation as stochastic events 
which occurred with some specified rates. The rationale for this was that
invasion and new mutations occur much more rarely than cell division. Both numerical realisations of the
stochastic process, and a mean-field approximation could be studied: they
corroborated the idea that exponential growth could occur as long as
individual lesions were isolated from each other, and showed how the
accumulation of advantageous mutations could be related to the underlying
growth curves of lesions of different genetic backgrounds. Some questions
remain about the accuracy of the approximation scheme we used, and we were not
able to answer them definitively.


Finally, we attempted to model how tumour size developed over time in
pancreatic cancer patients during a new form chemotherapy designed to have
very few possible mutations that could cause resistance. We found that in many
cases a sum of two simple exponential functions provided a very good fit, but
the inferred number of resistant cells was many orders of magnitude higher
than predicted by other authors on the study. This revealed serious
deficiencies in the current understanding of pancreatic cancer, as the
prediction was made on the basis of a mathematical model which did not make
any controversial assumptions with respect to the underlying biology, and we
were forced to consider alternative hypotheses for the mechanisms by which
resistance to chemotherapy emerges.


Although they weren't related in a strongly systematic way, we went
to some effort to ensure that the conclusions of one 
model could inform the inputs of the following model. For
example, the mechanistic simulation predicts cell motility speed, which is an
important input to the run-and-tumble lattice model. This lattice model
predicts growth curves of tumours, which were an important input to the
analytical ensemble model, and the ensemble model made predictions for the
proportion of advantageous mutants, which was one possible interpretation of
the fraction of resistant cells observed in the clinical study. In this way,
future work in a related area may be able to chain together improvements on
our minimal models to estimate the effect of changing individual cell
properties like elasticity on how rapidly new mutant strains emerge. We should
emphasise that this is at this point rather speculative, and the number of
possible factors affecting the macroscopic outcomes is probably much larger
than the ones we have chosen to study in this thesis. 

The usefulness of our approach, of making models that are ``as simple as
possible but no simpler'', and of trying to connect our calculations to
experimental observables, is not only that the resulting mathematical models
are open to experimental refutations, as is a prerequisite for all science, but
also that the whole of our theoretical work can be understood and in principle 
reproduced by a single person. This is
in contrast to many trends in current scientific work, which rely on models
and analyses so complex that their reproduction requires cooperation between
many different specialists on a wide scale.

Of course, the experimental study
to which we contributed necessarily required a large collaboration, as the
actual practical business of conducting it was intrinsically difficult. This does not mean,
however, that theory should follow suit. I believe that we have
demonstrated how theory which balances minimalism and empirical
relevance still has the potential to give a unified picture of scientific
knowledge where this is possible, and more importantly to reveal unexpected
holes where this is not. If the point of science is to
advance human understanding, then science cannot depend on black boxes.

\bibliographystyle{unsrt}
\bibliography{../bibliography/refs}

\end{document}